# Design Of Drug-Like Protein-Protein Interaction Stabilizers Guided By Chelation-Controlled Bioactive Conformation Stabilization

Francesco Bosica[a,b], Sebastian Andrei[b], João Filipe Neves[c,d], Peter Brandt[a], Anders Gunnarsson[e], Isabelle Landrieu[c,d], Christian Ottmann[b,f] and Gavin O'Mahony*,[a]

[a] F. Bosica, Dr. P. Brandt, Dr. G. O'Mahony
Research and Early Development, Cardiovascular, Renal and Metabolism, BioPharmaceuticals R&D
AstraZeneca
Gothenburg, Sweden
E-mail: gavin.omahony@astrazeneca.com

[b] F. Bosica, Dr. S. Andrei, Prof. C. Ottmann
Laboratory of Chemical Biology, Department of Biomedical Engineering and Institute for Complex Molecular Systems (ICMS)
Eindhoven University of Technology
Den Dolech 2, 5612 AZ Eindhoven, The Netherlands

[c] Dr. J. F. Neves, Dr. I. Landrieu
ERL9002 Integrative Structural Biology
CNRS
50 Avenue de Halley, 59658 Villeneuve d'Ascq, Lille, France

[d] Dr. J. F. Neves, Dr. I. Landrieu
Department U1167 RID-AGE Risk Factors and Molecular Determinants of Aging-Related Diseases
Univ. Lille, Inserm, CHU Lille, Institut Pasteur de Lille
1 Rue du Professeur Calmette 59800 Lille, France

[e] Dr. A. Gunnarsson
Discovery Sciences, BioPharmaceuticals R&D
AstraZeneca
Gothenburg, Sweden

[f] Prof. C. Ottmann
Department of Chemistry
University of Duisburg-Essen
Universitätsstrasse 7, 45117 Essen, Germany

Supporting information for this article is given via a link at the end of the document.

**Abstract:** The protein-protein interactions (PPIs) of 14-3-3 proteins are a model system for studying PPI stabilization. The complex natural product Fusicoccin A stabilizes many 14-3-3 PPIs but is not amenable for use in SAR studies, motivating the search for more drug-like chemical matter. However, drug-like 14-3-3 PPI stabilizers enabling such study have remained elusive. An X-ray crystal structure of a PPI in complex with an extremely low potency stabilizer uncovered an unexpected non-protein interacting, ligand-chelated $Mg^{2+}$ leading to the discovery of metal ion-dependent 14-3-3 PPI stabilization potency. This originates from a novel chelation-controlled bioactive conformation stabilization effect. Metal chelation has been associated with pan-assay interference compounds (PAINS) and frequent hitter behavior, but chelation can evidently also lead to true potency gains and find use as a medicinal chemistry strategy to guide compound optimization. To demonstrate this, we exploited the effect to design the first potent, selective and drug-like 14-3-3 PPI stabilizers.

## Introduction

Protein-protein interactions (PPIs) are a major class of macromolecular interactions which control the function of many proteins.[1] The protein interactome is crucial to many biological processes, is implicated in many diseases[2-4] and contains many promising intervention points for development of therapeutics.[5-6] Inhibition is the most advanced mode of action for PPI modulation, with several rationally-designed PPI inhibitors in clinical testing or approved as drugs.[7] In contrast, PPI stabilization is comparatively unexplored.[8] Several natural products have been shown *post hoc* to operate by stabilization of PPIs,[9] but there are few examples of rationally designed small-molecule, drug-like PPI stabilizers. This is partly due to incomplete understanding of the structural and kinetic principles driving stabilization, making robust screen design difficult and resulting in a paucity of tractable starting points for PPI stabilizer development.[10-12]

14-3-3 proteins are non-enzymatic adapter proteins that bind as dimers to phosphorylated client proteins, controlling client activity and cellular fate.[13-14] They are involved in many cellular processes such as cell cycle regulation and apoptosis,[15] subcellular localization[16] and enzymatic activity regulation.[17] There are 7 human 14-3-3 isoforms, structurally composed of 9 helices forming an amphipathic binding groove which binds 14-3-3 consensus motifs on phosphorylated client proteins.[13] The natural product Fusicoccin A (**1**, Figure 1a) and related fusicoccanes[18] stabilize 14-3-3 PPIs with many phosphorylated partners.[19-20] Ternary complex crystal structures show that they occupy a well-defined binding pocket ("FC pocket") at the interface between 14-3-3 and the binding partner (Figure 1b).
The fusicoccanes' synthetic complexity makes them unsuitable for a molecular matched pair approach[21] to generate structure-activity relationship (SAR) principles or to study PPI stabilization selectivity. We therefore aimed to identify synthetically tractable, drug-like small-molecule starting points for systematic investigation of 14-3-3 PPI stabilization. In the course of this, we also discovered a novel chelation-controlled ligand conformational stabilization effect which had profound effects on compound potency, resulting in metal ion-assisted small molecule PPI stabilization. Chelation of metals by ligands of interest has been associated with assay interference and frequent hitter behavior, especially in PPI inhibition assays based on AlphaScreen technology.[22] Based on this observation, compounds containing potential chelating moieties tend to be filtered out from screening collections either prior to screening or in order to triage large screening data sets prior to data analysis, as an extension to the originally-reported PAINS filters.[23]

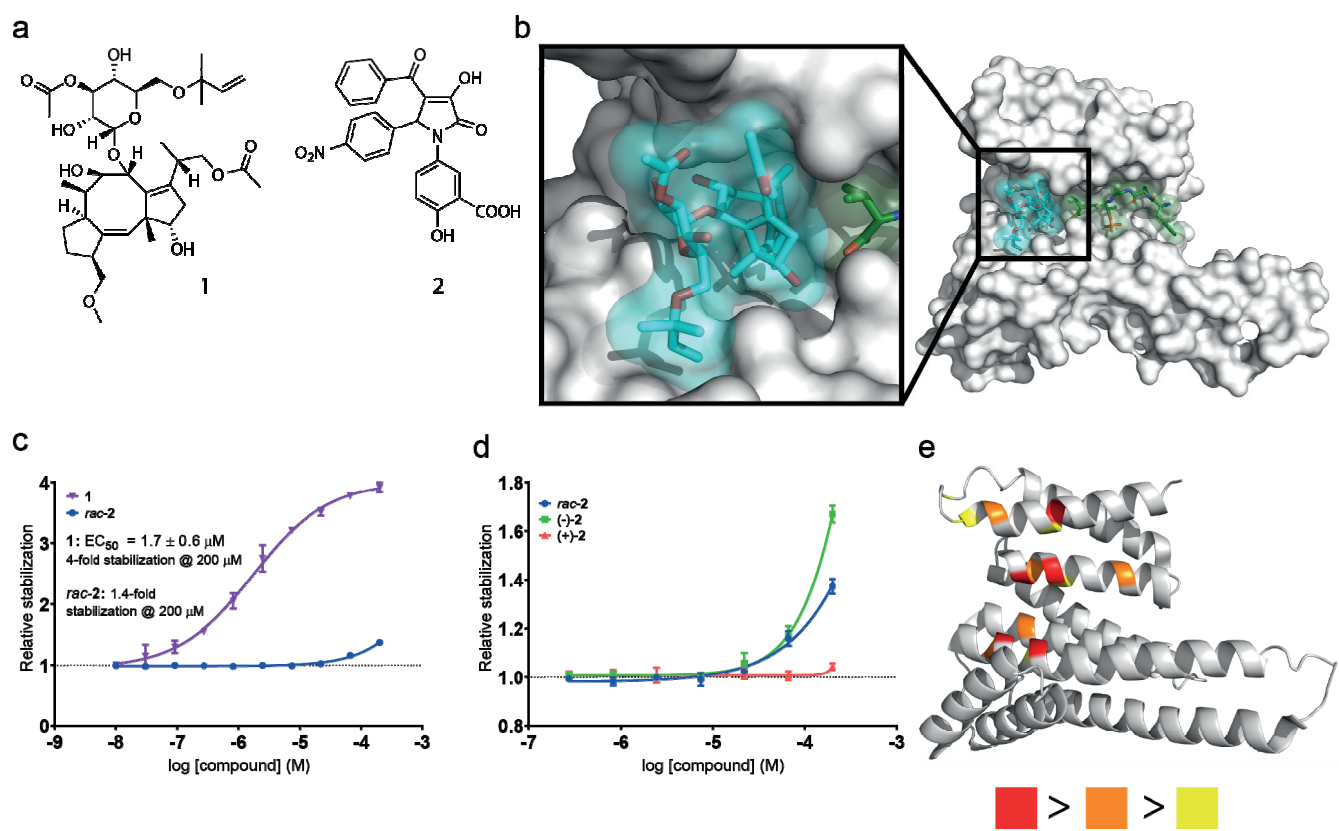

**Figure 1. Identification of 2 as a low potency 14-3-3/ERα PPI stabilizer.** (a) Chemical structures of Fusicoccin A **1** and Pyrrolidone1 **2**. (b) Crystal structure of the ternary complex between 14-3-3σ (white surface), ERα pT$^{594}$ phosphopeptide **3** (green sticks) and **1** (blue sticks) rendered from PDB 4JDD and highlighting the "FC pocket". (c) Comparison of 14-3-3ζ/ERα(pT$^{594}$) PPI stabilization activity of **1** (purple inverse triangles) and *rac*-**2** (blue circles) in an FP assay, using 50 nM 14-3-3ζ and 10 nM ERα(pT$^{594}$) phosphopeptide FITC-**3**. "Relative stabilization" (y-axis) is the mean fold-increase of FP signal over baseline (i.e. interaction between 14-3-3ζ and FITC-**3** alone). The error bars in all plots indicate +/- SD (n =3). (d) Comparison of PPI stabilization activity of *rac*-**2**, (-)-**2** (green squares) and (+)-**2** (red triangles) in 14-3-3/ERα(pT$^{594}$) FP assay. (e) The $^{1}$H-$^{15}$N TROSY-HSQC peak intensities most affected by binding of (-)-**2** correspond to 14-3-3 residues lining the FC pocket (mapped on the crystal structure of 14-3-3σ rendered from PDB 1YZ5). The 5 most affected residues are colored red, the next 5 orange and the next 5 yellow. I/I$_o$ values and colour coding is provided in Fig S6.

In this report, we show that metal ion chelation may in some cases lead to true potency gains and allow identification of hits that otherwise would be discarded as of insufficient potency. In our case, this manifested itself as a metal-ion assisted PPI stabilization, where the addition of bivalent metal ions led to up to two orders of magnitude increase in compound potency.

We furthermore exploited this effect to design metal-independent PPI stabilizers by mimicking the chelation with intramolecular hydrogen bonds, leading to compounds with high PPI stabilization potency and which are insensitive to metal ion concentration. The resulting compounds enable the systematic investigation of 14-3-3 PPI stabilization SAR and selectivity and further development of small-molecule 14-3-3 PPI stabilizers with potency rivaling that of the natural product **1**. We also propose that this type of ligand-specific conformational effect is a potential source of false negatives and should be considered when analyzing screening data and interpreting SAR for chelation-competent ligands as well as during the analysis and triage of high-throughput screen output.

## Results

### Racemic Pyrrolidone1 (*rac*-**2**) is a weak stabilizer of the canonical 14-3-3/ERα(pT$^{594}$) PPI

Binding of 14-3-3 to the ERα pT$^{594}$ phosphosite has been shown to prevent estradiol-induced activation of ERα by preventing ERα dimerisation.[24] By stabilizing the 14-3-3/ERα interaction, **1** was shown to enhance the 14-3-3-mediated inhibition of ERα activity and therefore the 14-3-3/ERα interaction is of interest as a potential target for development of ERα–dependent breast cancer therapies. To test compounds for 14-3-3 PPI stabilization, we selected the ERα pT$^{594}$ phosphosite (ERα(pT$^{594}$)) as the phosphorylated 14-3-3 binding partner, due to the availability of a ternary X-ray crystal structure with **1**. Pyrrolidone1 (**2**, Figure 1a) has previously been reported as a low potency stabilizer of the PPI between 14-3-3 and the plant plasma membrane

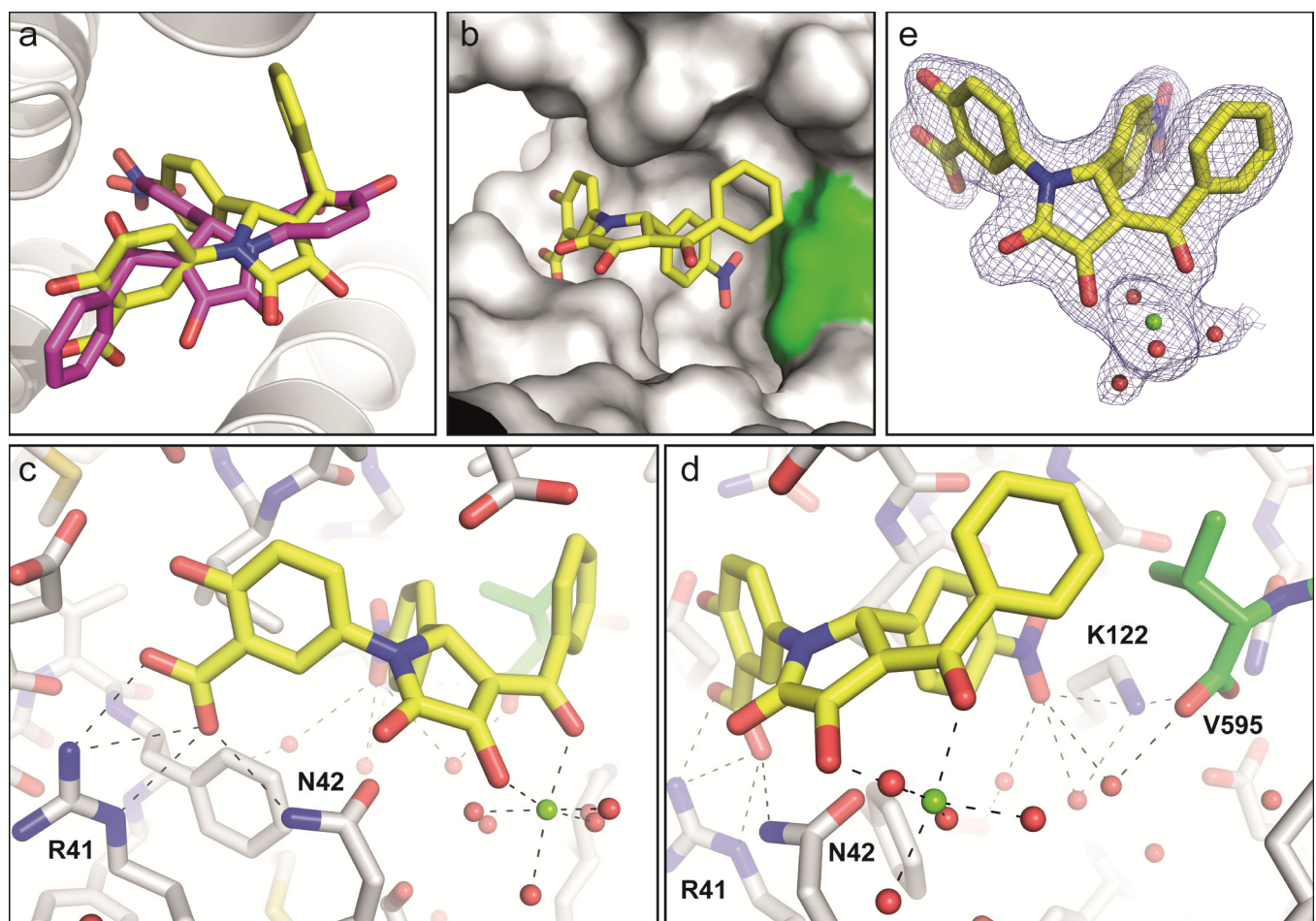

**Figure 2. X-ray crystal structure of 14-3-3/ERα(pT$^{594}$)/(R)-2 ternary complex and identification of (R)-2-chelated Mg$^{2+}$ ion.** (a) Comparison of binding mode of **2** from 3.25 Å 14-3-3/PMA2 complex structure (magenta sticks, assigned as (S), rendered from PDB 3M51) with revised binding mode derived from the 1.85 Å 14-3-3/ERα(pT$^{594}$) complex structure (yellow sticks, assigned (R) absolute configuration, PDB 6TJM). (b) Binding mode of (R)-**2** in the FC pocket, showing surface contributions from 14-3-3 (grey surface) and ERα(pT$^{594}$) phosphopeptide **3** (green surface). (c) Details of interactions of salicylate moiety of (R)-**2**. (d) Details of interactions of nitrophenyl moiety of (R)-**2**, showing the nitro group acting as H-bond acceptor from 14-3-3 Lys122 and water-mediated interaction with the C-terminus of ERα(pT$^{594}$) phosphopeptide **3** (green). (e) Additional ligand-associated electron density (2F$_{obs}$-F$_{cal}$ map contoured at 1 σ) corresponding to a fully-hydrated Mg$^{2+}$ ion (shown in green) chelated by the vinylogous carboxylate moiety of (R)-**2**.

H$^{+}$-ATPase 2 (PMA2).[25] Since **1** stabilizes both the 14-3-3/PMA2 and 14-3-3/ERα(pT$^{594}$) PPIs[24, 26] and **2** binds to the FC pocket of the non-canonical 14-3-3/PMA2 complex (the PMA2 peptide used was non-phosphorylated), we decided to investigate **2** as a potential stabilizer of the 14-3-3/ERα(pT$^{594}$) complex. We developed fluorescence polarization (FP) and surface plasmon resonance (SPR) assays based on 14-3-3ζ and ERα pT$^{594}$ phosphopeptides (**3** or FITC-**3**, sequence AEGFPApT$^{594}$V-COOH).

The $K_d$ of the 14-3-3/**3** interaction was 227 ± 14 nM (Figure S1) as measured by SPR. In a 14-3-3/ERα(pT$^{594}$) FP assay (based on a labelled derivative of **3**, FITC-**3**), **2** showed very low activity (1.4-fold stabilization at 200 μM, Fig. 1c) compared to **1** (EC$_{50}$ 1.7 ± 0.6 μM, 4-fold stabilization at 200 μM). In the 14-3-3/ERα(pT$^{594}$) SPR assay, a comparable EC$_{50}$ was determined for **1** (EC$_{50}$ 1.8 ± 0.2 μM), while the PPI stabilization effect of **2** was undetectable, with only binding of **2** to 14-3-3 protein being observed (Figure S2, S3). The enantiomers of **2** were then separated by chiral HPLC. In the 14-3-3/ERα(pT$^{594}$) FP and SPR assays, (-)-**2** was found to be more active than rac-**2** and (+)-**2** was shown to be inactive (Figure 1d and S9). (-)-**2** was shown to be stable with respect to epimerization under basic conditions (see Supporting Information).

NMR experiments were conducted using 14-3-3σΔC[27] and **3** to determine if (-)-**2** binds to the FC pocket of the 14-3-3/ERα(pT$^{594}$) complex. WaterLOGSY experiments[28] were performed with (-)-**2** in the presence and absence of either 14-3-3σΔC or the 14-3-3σΔC/**3** complex. Positively phased $^1$H signals for (-)-**2** confirmed binding to 14-3-3σΔC alone (Figure S4, red spectrum) and the 14-3-3σΔC/**3** complex (Figure S4, green spectrum). Protein-based NMR experiments were then carried out to determine the binding site of (-)-**2** on 14-3-3σΔC. By observing changes in the intensity ratio (I/I$_0$) of resonances in the fully-assigned $^1$H-$^{15}$N TROSY-HSQC spectrum of 14-3-3σΔC[29] in the presence (I) or absence (I$_0$) of (-)-**2** (Figure S5, S6), a mapping of the residues most affected by (-)-**2** binding (Figure 1e) was generated, indicating that the binding site of (-)-**2** corresponds to the FC pocket.[25]

Phosphopeptide **3** induced significant chemical shift perturbations in the 14-3-3σΔC $^1$H-$^{15}$N TROSY-HSQC spectrum, with some resonances showing characteristics of a slow exchange regime on the NMR time scale (Figure S7), which is compatible with the high affinity observed by SPR. These strong chemical shift perturbations prevented a comprehensive determination of the binding site of (-)-2 to the 14-3-3σΔC/3 complex by NMR. Nonetheless, two resonances with chemical shifts minimally but specifically affected by the presence of (-)-2 were identified. These signals, corresponding to G171 and I219 (Figure S8), lie in the FC pocket and were already affected by the presence of (-)-2 alone, indicating that (-)-2 binds to the FC pocket of both 14-3-3σΔC alone and in the 14-3-3σΔC/3 complex.

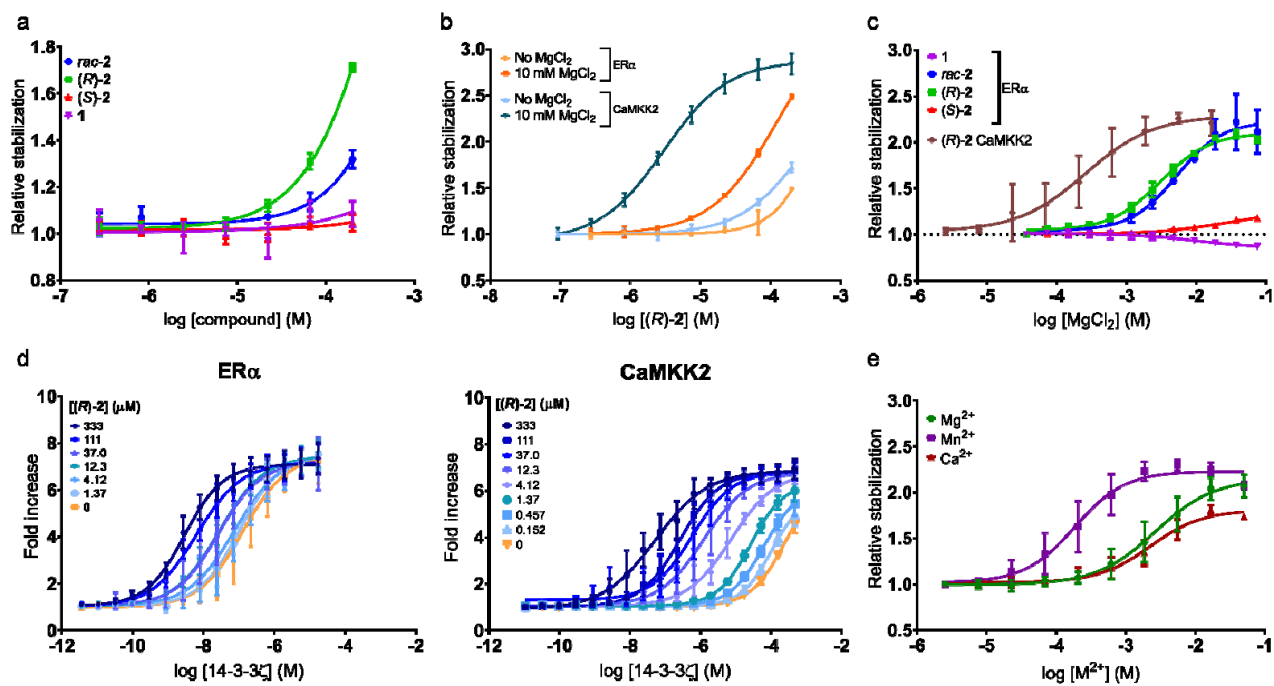

**Figure 3.** Metal ion-dependent stabilization of 14-3-3/ERα(pT$^{594}$) and 14-3-3/CaMKK2(pS$^{100}$) PPIs by (R)-2. (a) Stabilization of the 14-3-3/CaMKK2(pS$^{100}$) PPI measured in FP assay. *Rac*-2 (blue circles), (R)-2 (green squares) are stabilizers while (S)-2 (red triangles) and 1 (purple inverse triangles) are not. (b) The addition of 10 mM MgCl$_2$ increases the apparent potency of (R)-2 for 14-3-3/ERα(pT$^{594}$) (light orange circles vs dark orange squares) and 14-3-3/CaMKK2(pS$^{100}$) (light blue triangles vs dark blue inverse triangles) PPI stabilization. "Relative stabilization" (y-axis, panels a, b and c) is defined in Figure 1. (c) Mg$^{2+}$ increases PPI stabilization efficacy of 10 μM *rac*-2 (blue dots) and (R)-2 (green squares) whereas 1 and (S)-2 are unaffected (50 nM 14-3-3ζ/10 nM FITC-3 or 30 μM 14-3-3ζ/10 nM FAM-4). "Relative stabilization" (y-axis) is mean fold-increase of signal at a given [Mg$^{2+}$] over baseline (no added Mg$^{2+}$). (d) Increase in apparent $K_d$ of 14-3-3/ERα(pT$^{594}$) (10 nM FITC-3, left panel) or 14-3-3/CaMKK2(pS$^{100}$) (10 nM FAM-4, right panel) PPIs by FP with increasing (R)-2 concentration, in the presence of 10 mM Mg$^{2+}$. "Fold increase" (y-axis) is the mean fold-increase of signal over baseline (no (R)-2). (e) Ca$^{2+}$ (red triangles) and Mn$^{2+}$ (purple squares) also increase apparent efficacy of (R)-2 in 14-3-3/ERα(pT$^{594}$) FP assay.

**(R)-2 is the active enantiomer of 2: reassignment of binding mode and absolute configuration by X-ray crystallography**

Having confirmed by NMR that (-)-2 binds to the FC pocket of the 14-3-3/ERα(pT$^{594}$) complex, we set out to obtain an X-ray crystal structure of the 14-3-3σ/ERα(pT$^{594}$) phosphopeptide 3/(-)-2 ternary complex. By co-crystallization, we obtained crystals which diffracted reproducibly at a resolution of 1.85 Å. The 3.25 Å X-ray crystal structure of the ternary complex of (S)-2 bound to the *Nicotiana tabacum* 14-3-3 like protein C/PMA2 complex[25] (Figure 2a, magenta sticks) was initially used to model the binding orientation of 2. However, it was not possible to model (S)-2 into the observed ligand-derived electron density of the 14-3-3σΔC/3/(-)-2 ternary complex. The 1.85 Å resolution was sufficient to model the ligand *ab initio*, confirming that (-)-2 binds to the FC pocket as well as unambiguously assigning the (R) absolute configuration to (-)-2 (Figure 2a, yellow sticks). This assignment agreed with that obtained by vibrational circular dichroism on (+)-2 (assigned (S), see Figure S23).

Testing of (-)-2 and (+)-2 by SPR on the 14-3-3η/PMA2 PPI (Figure S10) showed that only (-)-2 stabilized this complex, as was the case for 14-3-3/ERα(pT$^{594}$). The fact that only (-)-2 stabilizes both PPIs, together with the higher resolution of the 14-3-3σΔC/3/(-)-2 ternary complex crystal structure compared to the published 14-3-3η/PMA2/2 structure supports our revised binding mode of 2 (Figure 2a) and reassignment of the absolute configuration for the active 14-3-3 PPI stabilizing enantiomer of 2 as (R) (from this point in the text, (-)-2 and (+)-2 will be referred to as (R)-2 and (S)-2 respectively).

Overall, (R)-2 binds in a T-shaped conformation, with each of its phenyl rings pointing into separate sub-pockets (Figure 2b) and with all three phenyl rings oriented orthogonally to the pyrrolidone ring. The salicylate moiety of (R)-2 occupies a narrow, largely negatively charged cleft (Figure 2c) not exploited by 1 (Figure S11). The carboxylate group participates in a bidentate interaction with Arg41 and accepts a H-bond from Asn42. The carbonyl of the pyrrolidone ring participates in an attractive antiparallel interaction with the sidechain of Asn42.[30] The nitro group accepts a H-bond from Lys122 (Figure 2d) which is also in direct contact with the C-terminus of 3, thus (R)-2 bridges 14-3-3 and 3 via polar interactions.

Unexpectedly, additional electron density was observed in conjunction with the ligand (Figure 2e), corresponding to a fully hydrated Mg$^{2+}$ ion (presumably derived from the Mg$^{2+}$-containing crystallization buffer) chelated by the vinylogous carboxylate moiety of (R)-2. A p$K_a$ of 3.2 was measured (see Supporting Information) for the vinylogous carboxylate moiety of *rac*-2, supporting the observation of the magnesium vinylogous carboxylate salt in the ternary complex.

**(R)-2 but not 1 stabilizes the 14-3-3/CaMKK2(pS$^{100}$) PPI**

Given the relatively high affinity of the 14-3-3/**3** interaction, we sought a 14-3-3 binding partner with a binary complex of similar overall structure to the 14-3-3/ERα(pT$^{594}$) structure (i.e. with a vacant FC pocket) but with lower 14-3-3 affinity, in order to investigate the dynamic range of PPI stabilization. We reasoned that such an interaction would also be useful to evaluate the 14-3-3 PPI stabilization specificity of (R)-**2**. pS$^{100}$ of calcium/calmodulin-dependent kinase kinase 2 (CaMKK2)[31] was identified as a suitable candidate. Unlike the pT$^{594}$ ERα phosphosite from which phosphopeptide **3** is derived, the N-terminal domain of CaMKK2 pS$^{100}$ phosphosite is not a canonical mode III 14-3-3 binder.[32] However, the crystal structure of 14-3-3ζ in complex with a phosphopeptide derived from the CaMKK2 pS$^{100}$ site (sequence RKLpS$^{100}$LQER, PDB ID: 6EWW) indicates that it mimics a canonical mode III binder and that the FC pocket in this complex is largely unoccupied and available for potential binding of stabilizers such as (R)-**2**.[31]

We determined the binding affinity of CaMKK2 pS$^{100}$ 13-mer phosphopeptide **4**[31] to 14-3-3ζ by SPR (estimated $K_d$ 112 ± 14 μM, Figure S12) and found it significantly lower than for 14-3-3ζ/**3** ($K_d$ 227 ± 14 nM). An FP assay was then developed based on a FAM-labelled derivative of **4** (FAM-**4**). The stabilization of the 14-3-3/CaMKK2(pS$^{100}$) PPI by rac-, (R)- and (S)-**2** and natural product **1** was then determined using by FP and SPR assays (Figure 3a and S13). Interestingly, despite an apparently vacant FC pocket, **1** exhibited only borderline statistically significant stabilization of the 14-3-3/CaMKK2(pS$^{100}$) PPI by FP (p 0.060) and SPR (p 0.038) at 200 μM (purple inverted triangles). Following the same stereochemical dependence as for the 14-3-3/ERα(pT$^{594}$) PPI, (R)-**2** was a more effective stabilizer of the 14-3-3/CaMKK2(pS$^{100}$) PPI than **1** (green squares), while (S)-**2** showed no significant stabilization activity (red triangles).

**(R)-2 exhibits metal ion-dependent 14-3-3 PPI stabilization potency**

With stabilization by (R)-**2** confirmed in two 14-3-3 PPIs, we investigated a potential role for the crystallographically-observed Mg$^{2+}$ chelation by (R)-**2**. Performing the FP and SPR assays in the presence of Mg$^{2+}$ led to an apparent increase in stabilization potency and maximum efficacy of (R)-**2** in both the 14-3-3/ERα(pT$^{594}$) and 14-3-3/CaMKK2(pS$^{100}$) PPIs (Figure 3b and S14, S15). We then carried out Mg$^{2+}$ concentration-response experiments in the 14-3-3/ERα(pT$^{594}$) and 14-3-3/CaMKK2(pS$^{100}$) FP assays at a fixed concentration (10 μM) of rac-**2**, (R)-**2**, (S)-**2** and **1** (Figure 3c). The 14-3-3/ERα(pT$^{594}$) PPI stabilization efficacy of (R)-**2** and rac-**2** (green squares and blue circles respectively) and 14-3-3/CaMKK2(pS$^{100}$) PPI stabilization efficacy of (R)-**2** (brown diamonds) was shown to be magnesium concentration-dependent. (S)-**2** is not a PPI stabilizer and hence was unaffected by Mg$^{2+}$ concentration (red triangles). The 14-3-3/ERα(pT$^{594}$) PPI stabilization effect of **1** (purple inverse triangles) was also unaffected by varying Mg$^{2+}$ concentration, indicating that the Mg$^{2+}$ effect is ligand-specific and not due to effects on 14-3-3 protein or the phosphopeptide (see also Figure S16, S17).

The addition of 10 mM MgCl$_2$ afforded full concentration-response curves for (R)-**2**, allowing the determination of apparent EC$_{50}$ values. In the 14-3-3/CaMKK2(pS$^{100}$) FP assay, adding 10 mM MgCl$_2$ led to an apparent EC$_{50}$ of 3.2 ± 0.3 μM and 2.9-fold stabilization (Figure 3b). The PPI stabilizing potency and efficacy of the Mg$^{2+}$ salt of (R)-**2** on the 14-3-3/CaMKK2(pS$^{100}$) PPI is comparable to that of **1** on the 14-3-3/ERα(pT$^{594}$) PPI (EC$_{50}$ 1.7 ± 0.6 μM, 3.9-fold stabilization, Figure 1c). Taking the relative stabilization effect of (R)-**2** at 200 μM in the absence of added Mg$^{2+}$ as a baseline (1.5- and 1.7-fold for 14-3-3/ERα(pT$^{594}$) and 14-3-3/CaMKK2(pS$^{100}$) PPIs respectively), the addition of 10 mM Mg$^{2+}$ leads to 7-fold and 99-fold increases in apparent potency (Figure 3b).

Exploiting the potentiating effect of Mg$^{2+}$ on the PPI stabilization potency of (R)-**2**, we investigated the apparent affinity increase between 14-3-3 and phosphorylated binding partners induced by increasing concentrations of (R)-2 in the 14-3-3/ERα(pT$^{594}$) and 14-3-3/CaMKK2(pS$^{100}$) FP assays (Figure 3d). At the highest concentration of (R)-2 tested (333 μM), the observed affinity of the 14-3-3/ERα(pT$^{594}$) PPI (left panel) was increased 57-fold and 3,700-fold for the 14-3-3/CaMKK2(pS$^{100}$) PPI (right panel).

To assess whether this metal ion-assisted 14-3-3 PPI stabilization effect was specific to Mg$^{2+}$, we tested by FP the effect of adding increasing concentrations of MnCl$_2$, CaCl$_2$ and ZnCl$_2$ to a fixed concentration of 14-3-3, FITC-**3** and (R)-**2** on the overall 14-3-3/ERα(pT$^{594}$) PPI stabilization efficacy (Figure 3e and S17). The potency of the assistance effect of Mg$^{2+}$ and Ca$^{2+}$ were comparable, while Mn$^{2+}$ was approximately 15-fold more effective, with its maximum assistance effect achieved at less than 1 mM MnCl$_2$.[33] All three metals show similar overall stabilization efficacy in this setting, affording approximately 2-fold stabilization of the interaction. Addition of Zn$^{2+}$ caused precipitation of phosphopeptide FITC-**3** precluding measurement of its effects.

**Mimicry of the chelate by cyclization or intramolecular H-bonds affords metal independent 14-3-3 PPI stabilizers**

Based on the 14-3-3/ERα(pT$^{594}$)/(R)-**2** ternary complex crystal structure, the origin of the metal ion potentiation effect on the PPI stabilization potency of **2** was assumed to be due to solution phase stabilization of the crystallographically observed binding conformation of (R)-**2**. To test this hypothesis, we decided to mimic the conformational restriction of the Mg$^{2+}$ chelate of (R)-**2** by moieties that would not depend on the chelation of metal ions. Cyclisation of rac-**2** and its analogues to afford pyrazole derivatives such as **5** has previously been reported as a strategy to improve their 14-3-3 PPI stabilization activity.[34] However, this was based on a binding mode (Figure 2b, magenta sticks) determined from a low-resolution crystal structure that we have now shown to be erroneous. Rac-**5** was reported to be a better stabilizer of the 14-3-3/PMA2 complex than rac-**2**. A crystal structure of a similar bicyclic analogue of **5** in the 14-3-3/PMA2 complex was reported, however (S)-**2** (the inactive absolute configuration) was used to model **5** into the observed electron density, leading to difficulties in interpreting SAR due to incorrect stereochemistry and binding mode assumptions.

In addition, based on the 6-membered Mg$^{2+}$-containing chelate ring observed (Figure 2e), we reasoned that an intramolecular hydrogen bond contained in a 6-membered pseudo-ring, such as in the regioisomeric vinylogous amides **6** and **9**, would be a better chelate mimic than the 5-membered pyrazole ring. Reaction of rac-**2** with aqueous ammonia in acetic acid under microwave heating afforded only vinylogous amide rac-**6** in 20% yield, with no rac-**9** observed (Scheme 1). This contrasts with literature reports that 4-aroyl substituents favor regiospecific endocyclic nucleophilic attack by sterically unencumbered amines to afford the regioisomers corresponding to **9**.[35-40] Reaction of *rac*-**2** with p-methoxybenzylamine under microwave heating afforded *rac*-**7** (6%) and *rac*-**8** (26%). Removal of the PMB group of *rac*-**8** using TFA afforded *rac*-**9** in 53% yield. The single enantiomers (*R*)-**6** and (*R*)-**9** were prepared analogously to *rac*-**9**. Treatment of (*R*)-**2** with p-methoxybenzylamine yielded the two regioisomeric PMB-protected vinylogous amides (*R*)-**7** (10%) and (*R*)-**8** (16%), which on deprotection with TFA afforded (*R*)-**6** and (*R*)-**9** in 10% and 51% yield, respectively. (*R*)-**5** was prepared in 47% yield from (*R*)-**2** by reaction with hydrazine.

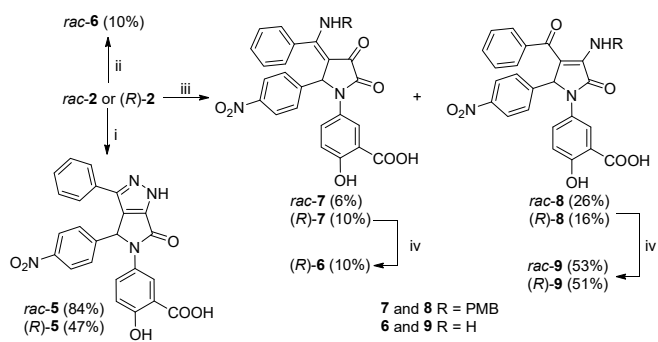

**Scheme 1. Synthesis of conformationally-restricted analogues of 2.** i) H$_2$NNH$_2$, AcOH, 2 h, 120 °C. ii) aq. NH$_3$, AcOH, 2h, 120 °C. iii) PMBNH$_2$, 2 h, 120 °C. iv) TFA, 30 min, 120 °C.

(*R*)-**5**, (*R*)-**6** and (*R*)-**9** were then tested by FP and SPR for their 14-3-3/ERα(pT$^{594}$) and 14-3-3/CaMKK2(pS$^{100}$) PPI stabilization effect (Figure 4a,b and S19). Gratifyingly, (*R*)-**6** (inverted blue triangles) and (*R*)-**9** (pink circles) were more potent and efficacious stabilizers of both PPIs than (*R*)-**2** (without added Mg$^{2+}$, green squares). Pyrazole (*R*)-**5** (orange triangles) was somewhat more potent and efficacious than (*R*)-**2** alone but was less active than either (*R*)-**6** or (*R*)-**9**, suggesting that a 6-membered ring is preferable to a 5-membered ring. In the 14-3-3/CaMKK2(pS$^{100}$) PPI, (*R*)-**2**+10 mM Mg$^{2+}$ is still the most potent, suggesting that the ordered water molecules of the Mg$^{2+}$-(*R*)-**2** chelate may contribute to binding affinity/PPI stabilization in this case. As expected, increasing concentrations of Mg$^{2+}$ did not affect the 14-3-3/ERα(pT$^{594}$) PPI stabilization efficacy of *rac*-**5**, **6** or **9** (Figure 4c, see Figure S18 for PPI stabilization data in absence of Mg$^{2+}$) as they lack the chelation-competent vinylogous carboxylate moiety. The salicylate moiety of **2**, **5**, **6** and **9** can potentially chelate metals, but the fact that **5**, **6** or **9** did not show metal-dependent potency (Figure 4c) suggests that the salicylate moiety does not affect the potency of these compounds in our system. Therefore the effect of chelation by the salicylate moiety can be discounted for our purposes.

**Potency increase upon metal chelation is due to stabilization of bioactive conformation**

In the 14-3-3/ERα(pT$^{594}$)-bound conformation of **2** (Figure 2a), the O$^-$ and carbonyl oxygen of the vinylogous carboxylate moiety are in a *syn* relationship (*syn*-(*R*)-**2**, Figure 5a). In solution, and in the absence of chelatable metal ions, repulsive electrostatic allylic strain was expected to disfavor this conformation relative to the *anti* conformation (*anti*-(*R*)-**2**, Figure 5a). Comparing the DFT calculated gas phase free energies of *syn*- and *anti*-(*R*)-**2** (see Supporting Information) shows the protein-bound and chelation-stabilized *syn*-(*R*)-**2** conformation to be significantly higher energy. Accounting for solvation effects by using two different implicit solvation models for water reduced the energy difference between *syn*- and *anti*-(*R*)-**2** (Table 1).

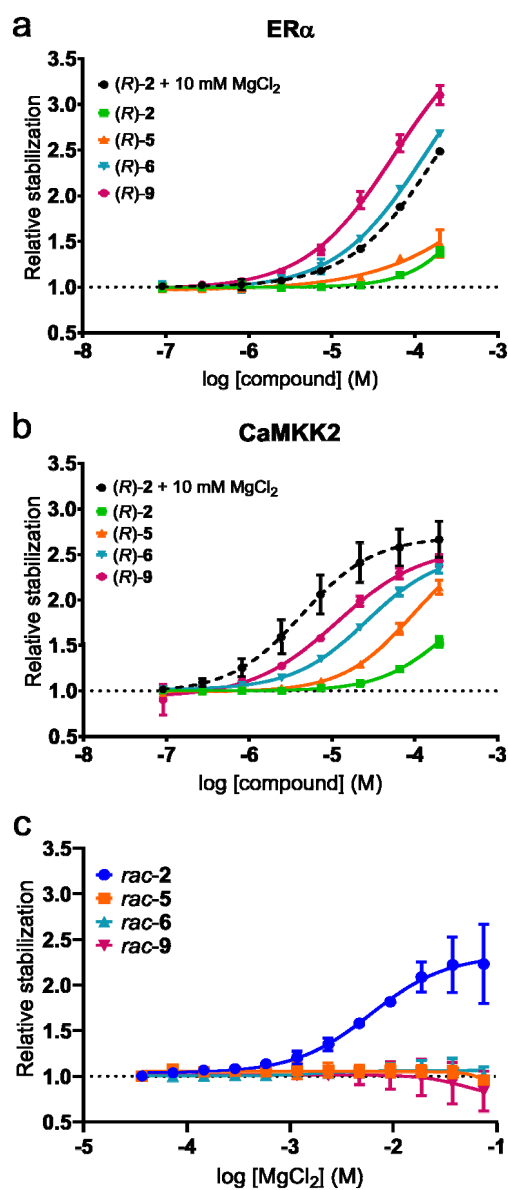

**Figure 4. Metal ion-independent 14-3-3 PPI stabilizers.** (a,b) Stabilization of the 14-3-3/ERα(pT$^{594}$) and 14-3-3/CaMKK2(pS$^{100}$) PPIs by (*R*)-**6** and -**9** as measured by FP. "Relative stabilization" (y-axes) is defined in Figure 1. (c) Mg$^{2+}$ concentration-response in 14-3-3/ERα(pT$^{594}$) FP assay (50 nM 14-3-3ζ/10 nM FITC-**3** and 100 μM *rac*-**2**, -**5**, -**6** or -**9**), showing no effect of Mg$^{2+}$ on activity of *rac*-**5**, -**6** or -**9** (orange squares, pale blue triangles and inverted pink triangles respectively). "Relative stabilization" (y-axis) refers to mean fold increase of FP signal of a given [Mg$^{2+}$] over signal observed in absence of added [Mg$^{2+}$].

Both confirm that the *syn*-(*R*)-**2** conformation is not favored in water. Therefore, the *anti*-(*R*)-**2** conformation is predicted to be present in significant amounts in the solution phase in the absence of Mg$^{2+}$ or other metals. For the vinylogous amides (*R*)-**6** and (*R*)-**9**, both solvation models favor the corresponding *syn*- conformations (i.e. corresponding to the protein bound conformation of **2**) as the lowest energy solution conformation, as expected and as borne out by their increased, metal ion-independent 14-3-3 PPI stabilization compared to (*R*)-**2**.

**Table 1.** Calculated relative energies for *syn* and *anti*-conformations of (*R*)-**2**, -**6** and -**9**.

| Cpd | Conformation of exocyclic carbonyl[a] | Free energies in water | |
|---|---|---|---|
| | | B3LYP-D3 PBF[b] | M06-2X-D3 SM6[c] |
| (*R*)-**2** | *syn* | 0 | 2.2 |
| (*R*)-**2** | *anti* | 0.4 | 0 |
| (*R*)-**6** | *syn* | 0.0 | 0.0 |
| (*R*)-**6** | *anti* | 5.9 | 7.5 |
| (*R*)-**9** | *syn* | 0.0 | 0.0 |
| (*R*)-**9** | *anti* | 2.9 | 2.6 |

[a] The conformation of the salicylate was selected according to the crystal structure of (*R*)-**2** (Figure 5). [b] PBF = Poisson Boltzmann Finite element method; a solvation model. [c] SM6 = Solvation Model 6.

To confirm the calculations, 1D ROE difference NMR experiments were performed on *rac*-**2**, *rac*-**6** and *rac*-**9** (Figures S20-S22) to determine their preferred solution conformations. Irradiation of protons of either the benzoyl or nitrophenyl rings of *rac*-**2** showed only a small resonance transfer between these rings, suggesting that they prefer not to be in close spatial proximity and corresponding to the *anti* conformation (Figure 5a). On the other hand, strong resonance transfer was observed between the protons of the benzoyl and nitrophenyl rings for both *rac*-**6** and *rac*-**9**, indicating close spatial proximity suggestive of the syn conformation induced by intramolecular hydrogen bonding. Precipitation precluded determination by NMR of the conformational effects of $Mg^{2+}$ on *rac*-**2**, but these may be inferred from the X-ray crystal structure (Figure 5b).

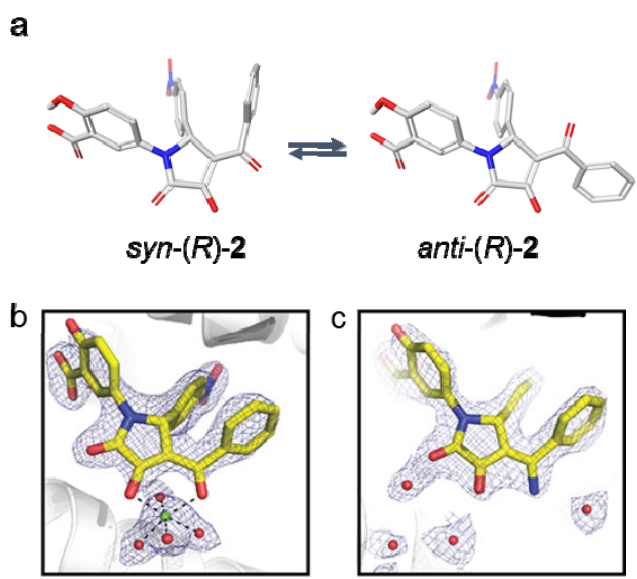

**Figure 5. Conformational analysis of 2 and binding mode of (*R*)-6 in 14-3-3/ERα(pT594) complex.** (a) Comparison of *syn* and *anti* conformations of the vinylogous carboxylate moiety of (*R*)-**2**. Calculated free energies of (*R*)-**2** conformations show that protein-bound conformation *syn*-(*R*)-**2** is not favored in solution. (b) and (c), Binding modes and $2F_{obs}-F_{cal}$ electron density (contoured at 1 σ) of the magnesium complex of (*R*)-**2** (panel b, PDB 6TJM) and (*R*)-**6** (panel c, PDB 6TL3), shown in the same orientation of the 14-3-3/ERα(pT594) complex, indicating absence of chelated metal ion in (*R*)-**6** structure and intramolecular hydrogen bond-stabilized conformation of (*R*)-**6** that mimics *syn*-(*R*)-**2** conformation.

*Rac*-**6** was then selected for crystallization studies in the 14-3-3/ERα(pT594) complex. A 2.45 Å resolution X-ray crystal structure of the ternary 14-3-3σΔC/**3**/(*R*)-**6** complex was obtained (Figure 5c), confirming the absolute configuration of the active enantiomer as (*R*) and a binding mode analogous to (*R*)-**2**. No additional ligand-associated electron density attributable to a chelated metal ion was observed, indicating that the intramolecular hydrogen bond successfully stabilizes the *syn* conformation as demonstrated computationally.

## Discussion

Ligand conformational restriction is a widely used strategy to increase potency, and minimizes the entropic loss observed on ligand binding if the preferred solution and bound conformations differ.[41-43] This is typically achieved by cyclisation, introducing intramolecular hydrogen-bonding or by exploiting steric/stereoelectronic effects. Ligand chelation by solution phase metal ions can now be added to this list of strategies. However, given the lack of control over metal ion concentration in medicinally-relevant settings (*e.g.* intracellularly) and the requirement that the binding site accommodate a metal ion and associated ordered water molecules, it may prove to be difficult to easily and reliably exploit this effect in drug substances.

On the other hand, metal contamination has been reported to be a cause of false positives due to assay interference[44] or by ligands complexing to bioactive metal centers.[45] To our knowledge, no incidences of such ligand-specific conformational effects due to metal ions have been reported. Given that many assays require the presence of metal ions (such as $Ca^{2+}$, $Mg^{2+}$ or $Zn^{2+}$) and high intracellular concentrations of some of these metals, such conformational effects should be considered as a potential cause of difficult to interpret assay data for chelation competent ligands. In principle, such behavior could lead to false negatives in screening campaigns and should be considered when analyzing screening data, especially in the case of conflicting results from samples that may have been subjected to more or less rigorous purification procedures which could result in different metal ion content. This effect could also be at play if unexpected discrepancies are observed between potencies in biochemical and cellular assays, where there may be differences in metal ion concentration. It has been proposed that chelating moieties should be used in substructure alerts to filter out compounds with potential to cause assay interference and frequent hitters.[22] However, the application of such filters needs to be considered on a case by case basis, with the assay technology and particular target in question taken into account. As we have shown, in some cases metal chelation can act as a ligand conformational probe and provide valuable hints as to the optimizability of low potency hits.

## Conclusion

We have shown that by leveraging the serendipitously discovered $Mg^{2+}$-chelate and resulting chelation-controlled stabilization of bioactive ligand conformation, (R)-**2** can be optimized to afford stabilizers of canonical 14-3-3 PPIs with potency rivaling natural product **1**. We also show that this metal-assisted PPI stabilization effect is due to solution phase stabilization of the bioactive conformation of **2** and can be mimicked by an intramolecular H-bond to afford metal-independent PPI stabilizers such as (R)-**6** and (R)-**9**. Compared to **1**, these compounds have the advantage of higher synthetic tractability and stabilize a 14-3-3 PPI that is refractive to stabilization by **1** and they should be useful tool compounds for the study of 14-3-3 PPI stabilization. This work gives new insights into the SAR of small molecule, non-natural product-derived 14-3-3 PPI stabilization and provides opportunities for structure-based drug design to identify new, small molecule 14-3-3 PPI stabilizers. More broadly, this work indicates that ligand-specific conformational effects due to metal ion chelation should be considered during the interpretation of assay and screening data, especially for chelation-competent ligands.

## Acknowledgements

The crystallographic data collection was performed at the Deutsches Elektronen-Synchrotron (DESY, PETRA III) of Hamburg (Germany). The authors acknowledge Anna Jonson and Kristina Öhlén of the Separation Science Group at AstraZeneca Gothenburg for help with chiral HPLC, Richard Lewis of the NMR Group at AstraZeneca Gothenburg for the VCD analysis and the interpretation of the 1D selective ROE NMR spectra, Frederik Wågberg and Johan Wernevik for assistance with HRMS determination for compounds (R)-**6** and (R)-**9**. Anaïs Noisier is thanked for her support in the peptide synthesis. François-Xavier Cantrelle is also thanked for NMR data acquisition and technical advice. This work is supported by the Initial Training Network TASPPI, funded by the H2020 Marie Curie Actions of the European Commission under Grant Agreement 675179. The NMR facilities at Univ. Lille were funded by the Nord Region Council, CNRS, Institut Pasteur de Lille, the European Community (ERDF), the French Ministry of Research and the University of Lille and by the CTRL CPER co-funded by the European Union with the European Regional Development Fund (ERDF), by the Hauts de France Regional Council (contract n° 17003781), Métropole Européenne de Lille (contract n° 2016_ESR_05), and French State (contract n° 2017-R3-CTRL-Phase 1). We acknowledge support for the NMR facilities from TGE RMN THC (CNRS, FR-3050) and FRABio (Univ. Lille, CNRS, FR-3688).

**Keywords:** Chelates • Drug design • Medicinal chemistry • Protein-protein interaction stabilization • 14-3-3

[1] J. A. Wells, C. L. McClendon, *Nature* **2007**, *450*, 1001-1009.
[2] J. Menche, A. Sharma, M. Kitsak, S. D. Ghiassian, M. Vidal, J. Loscalzo, A. L. Barabasi, *Science* **2015**, *347*, 1257601.
[3] M. Vidal, M. E. Cusick, A. L. Barabasi, *Cell* **2011**, *144*, 986-998.
[4] T. Ideker, R. Sharan, *Genome Res.* **2008**, *18*, 644-652.
[5] G. Zinzalla, D. E. Thurston, *Future Med. Chem.* **2009**, *1*, 65-93.
[6] S. Jaeger, P. Aloy, *IUBMB Life* **2012**, *64*, 529-537.
[7] D. E. Scott, A. R. Bayly, C. Abell, J. Skidmore, *Nat. Rev. Drug Discovery* **2016**, *15*, 533-550.
[8] P. Thiel, M. Kaiser, C. Ottmann, *Angew. Chem. Int. Ed.* **2012**, *51*, 2012-2018.
[9] F. Giordanetto, A. Schafer, C. Ottmann, *Drug Discov Today* **2014**, *19*, 1812-1821.
[10] V. Azzarito, K. Long, N. S. Murphy, A. J. Wilson, *Nat. Chem.* **2013**, *5*, 161-173.
[11] E. Sijbesma, K. K. Hallenbeck, S. Leysen, P. J. de Vink, L. Skora, W. Jahnke, L. Brunsveld, M. R. Arkin, C. Ottmann, *J. Am. Chem. Soc.* **2019**, *141*, 3524-3531.
[12] S. Surade, T. L. Blundell, *Chem. Biol.* **2012**, *19*, 42-50.
[13] A. Aitken, *Semin. Cancer Biol.* **2006**, *16*, 162-172.
[14] H. Fu, R. R. Subramanian, S. C. Masters, *Annu. Rev. Pharmacool. Toxicol.* **2000**, *40*, 617-647.
[15] H. Hermeking, A. Benzinger, *Semin. Cancer Biol.* **2006**, *16*, 183-192.
[16] A. J. Muslin, H. Xing, *Cell. Signalling* **2000**, *12*, 703-709.
[17] T. Obsil, R. Ghirlando, D. C. Klein, S. Ganguly, F. Dyda, *Cell* **2001**, *105*, 257-267.
[18] L. M. Stevers, E. Sijbesma, M. Botta, C. MacKintosh, T. Obsil, I. Landrieu, Y. Cau, A. J. Wilson, A. Karawajczyk, J. Eickhoff, J. Davis, M. Hann, G. O'Mahony, R. G. Doveston, L. Brunsveld, C. Ottmann, *J. Med. Chem.* **2018**, *61*, 3755-3778.
[19] M. B. Yaffe, K. Rittinger, S. Volinia, P. R. Caron, A. Aitken, H. Leffers, S. J. Gamblin, S. J. Smerdon, L. C. Cantley, *Cell* **1997**, *91*, 961-971.
[20] B. Coblitz, M. Wu, S. Shikano, M. Li, *FEBS Lett.* **2006**, *580*, 1531-1535.
[21] P. W. Kenny, J. Sadowski, in *Chemoinformatics in Drug Discovery* (Eds.: R. Mannhold, H. Kubinyi, G. Folkers, T. I. Oprea), Wiley VCH, **2005**, pp. 271-285.
[22] K. Schorpp, I. Rothenaigner, E. Salmina, J. Reinshagen, T. Low, J. K. Brenke, J. Gopalakrishnan, I. V. Tetko, S. Gul, K. Hadian, *J. Biomol. Screening* **2014**, *19*, 715-726.
[23] J. B. Baell, G. A. Holloway, *J. Med. Chem.* **2010**, *53*, 2719-2740.
[24] I. J. De Vries-van Leeuwen, D. da Costa Pereira, K. D. Flach, S. R. Piersma, C. Haase, D. Bier, Z. Yalcin, R. Michalides, K. A. Feenstra, C. R. Jimenez, T. F. de Greef, L. Brunsveld, C. Ottmann, W. Zwart, A. H. de Boer, *Proc. Natl. Acad. Sci. U. S. A.* **2013**, *110*, 8894-8899.
[25] R. Rose, S. Erdmann, S. Bovens, A. Wolf, M. Rose, S. Hennig, H. Waldmann, C. Ottmann, *Angew. Chem. Int. Ed.* **2010**, *49*, 4129-4132.
[26] M. Wurtele, C. Jelich-Ottmann, A. Wittinghofer, C. Oecking, *EMBO J.* **2003**, *22*, 987-994.
[27] V. Obsilova, P. Herman, J. Vecer, M. Sulc, J. Teisinger, T. Obsil, *J. Biol. Chem.* **2004**, *279*, 4531-4540.
[28] C. Dalvit, G. Fogliatto, A. Stewart, M. Veronesi, B. Stockman, *J. Biomol. NMR* **2001**, *21*, 349-359.
[29] J. F. Neves, I. Landrieu, H. Merzougui, E. Boll, X. Hanoulle, F. X. Cantrelle, *Biomol. NMR Assignments* **2019**, *13*, 103-107.
[30] F. H. Allen, C. A. Baalham, J. P. M. Lommerse, P. R. Raithby, *Acta Crystallogr. Sect. B: Struct. Sci.* **1998**, *54*, 320-329.
[31] K. Psenakova, O. Petrvalska, S. Kylarova, D. Lentini Santo, D. Kalabova, P. Herman, V. Obsilova, T. Obsil, *Biochim. Biophys. Acta Gen. Subj.* **2018**, *1862*, 1612-1625.
[32] S. Ganguly, J. L. Weller, A. Ho, P. Chemineau, B. Malpaux, D. C. Klein, *Proc. Natl. Acad. Sci. U. S. A.* **2005**, *102*, 1222-1227.
[33] C. W. Bock, A. K. Katz, G. D. Markham, J. P. Glusker, *J. Am. Chem. Soc.* **1999**, *121*, 7360-7372.
[34] A. Richter, R. Rose, C. Hedberg, H. Waldmann, C. Ottmann, *Chem. – Eur. J.* **2012**, *18*, 6520-6527.
[35] M. N. Armisheva, N. A. Kornienko, V. L. Gein, M. I. Vakhrin, *Russ. J. Gen. Chem.* **2011**, *81*, 1893-1895.
[36] L. F. Gein, V. L. Gein, I. A. Kylosova, Z. G. Aliev, *Russ. J. Org. Chem.* **2010**, *46*, 252-254.
[37] V. L. Gein, M. N. Armisheva, N. A. Kornienko, L. F. Gein, *Russ. J. Gen. Chem.* **2014**, *84*, 2270-2272.
[38] V. L. Gein, N. L. Fedorova, E. B. Levandovskaya, M. I. Vakhrin, *Russ. J. Org. Chem.* **2011**, *47*, 95-99.
[39] V. L. Gein, N. N. Kasimova, *Russ. J. Gen. Chem.* **2005**, *75*, 254-260.

[40] V. L. Gein, N. N. Kasimova, Z. G. Aliev, M. I. Vakhrin, *Russ. J. Org. Chem.* **2010**, *46*, 875-883.

[41] Z. Fang, Y. Song, P. Zhan, Q. Zhang, X. Liu, *Future Med. Chem.* **2014**, *6*, 885-901.

[42] A. D. G. Lawson, M. MacCoss, J. P. Heer, *J. Med. Chem.* **2018**, *61*, 4283-4289.

[43] Y. Zheng, C. M. Tice, S. B. Singh, *Bioorg. Med. Chem. Lett.* **2017**, *27*, 2825-2837.

[44] J. C. Hermann, Y. Chen, C. Wartchow, J. Menke, L. Gao, S. K. Gleason, N. E. Haynes, N. Scott, A. Petersen, S. Gabriel, B. Vu, K. M. George, A. Narayanan, S. H. Li, H. Qian, N. Beatini, L. Niu, Q. F. Gan, *ACS Med. Chem. Lett.* **2013**, *4*, 197-200.

[45] F. E. Morreale, A. Testa, V. K. Chaugule, A. Bortoluzzi, A. Ciulli, H. Walden, *J. Med. Chem.* **2017**, *60*, 8183-8191.

**Entry for the Table of Contents**

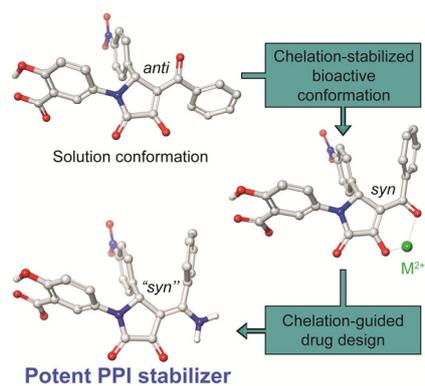

**No PAINS, No Gain.** Chelation is associated with pan-assay interference compounds (PAINS) but can also lead to true potency gains and be exploited to guide medicinal chemistry optimization. Bivalent metal ions increased the potency of a marginally active 14-3-3 PPI stabilizer up to 100-fold via stabilization of the bioactive ligand conformation. Mimicry of this by intramolecular H-bonds lead to the first potent, drug-like 14-3-3 PPI stabilizers.

SUPPLEMENTARY INFORMATION

METAL-ASSISTED SMALL MOLECULE PROTEIN-PROTEIN INTERACTION STABILIZATION BY CHELATION-CONTROLLED BIOACTIVE LIGAND CONFORMATION STABILIZATION AND RATIONAL DESIGN OF 14-3-3 PROTEIN-PROTEIN INTERACTION STABILIZERS

Francesco Bosica,[†,#] Sebastian Andrei,[#] João Filipe Neves,[‡] Peter Brandt,[†] Anders Gunnarsson,[⊥] Isabelle Landrieu,[‡] Christian Ottmann,[#,%] and Gavin O'Mahony[†,*]

TABLE OF CONTENTS

# I. Supplementary Figures

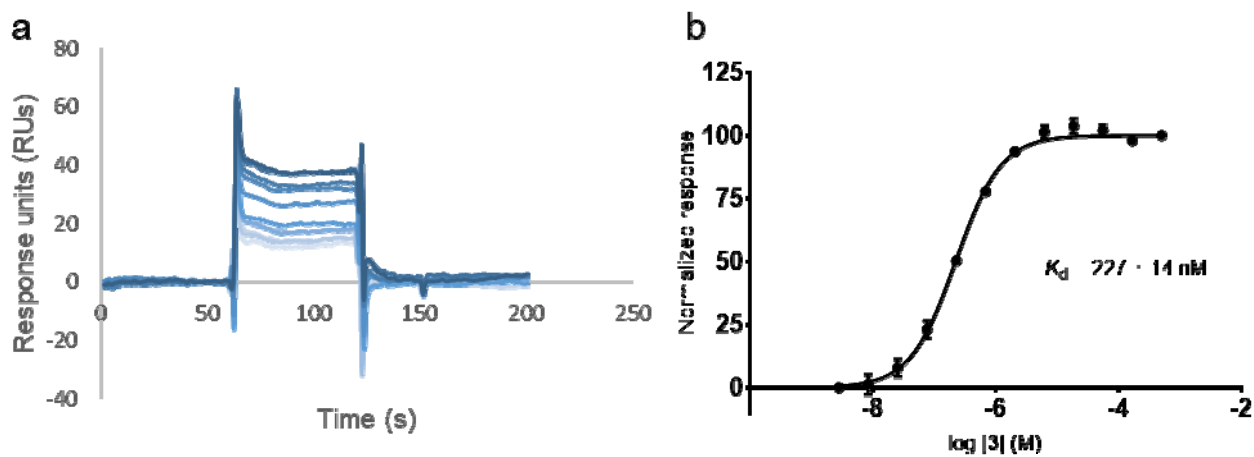

**Supplementary Figure 1.** Determination of affinity ERα-derived phosphopeptide **3** for 14-3-3ζ by SPR. a) Representative example of an SPR sensorgram in which increasing concentrations of **3** were flowed over immobilized 14-3-3ζ. The response units (RUs) achieved (y axis) are presented as a function of time in seconds (x axis). For each curve, the RU values at equilibrium response were extracted and fitted in a dose-response curve using a four-parameter logistic model (4PL), shown in (b), against the log of the molar concentration of **3** (mean ± SD, n = 3). To account for the variations in protein immobilization between runs, the equilibrium RU values were normalized to a 0-100 % interval, where 0% is baseline response and 100% is the mean curve plateau value. Experiment was performed in 1:3 dilution series from an initial concentration of **3** of 500 μM.

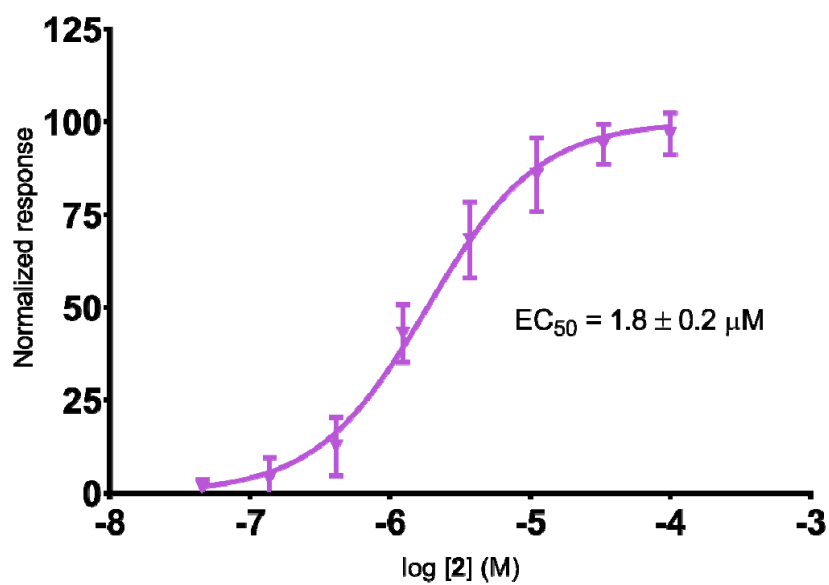

**Supplementary Figure 2** Stabilization effect of Fusicoccin A **1** for the 14-3-3ζ/**3** complex, measured by SPR. **1** was titrated (1:3 dilution series, initial concentration 100 μM) in the presence of 50 nM Erα-derived phosphopeptide **3** and surface-immobilized 14-3-3ζ. Binding affinity was estimated to be $EC_{50}$ = 1.8 ± 0.3 μM by curve fitting using a four-parameter logistic model (4PL). Error bars show standard deviation from the mean for each data point (n = 3). To account for the variations in protein immobilization between runs, the equilibrium RU values were normalized to a 0-100 % interval, where 0% is baseline response and 100% is the mean curve plateau value.

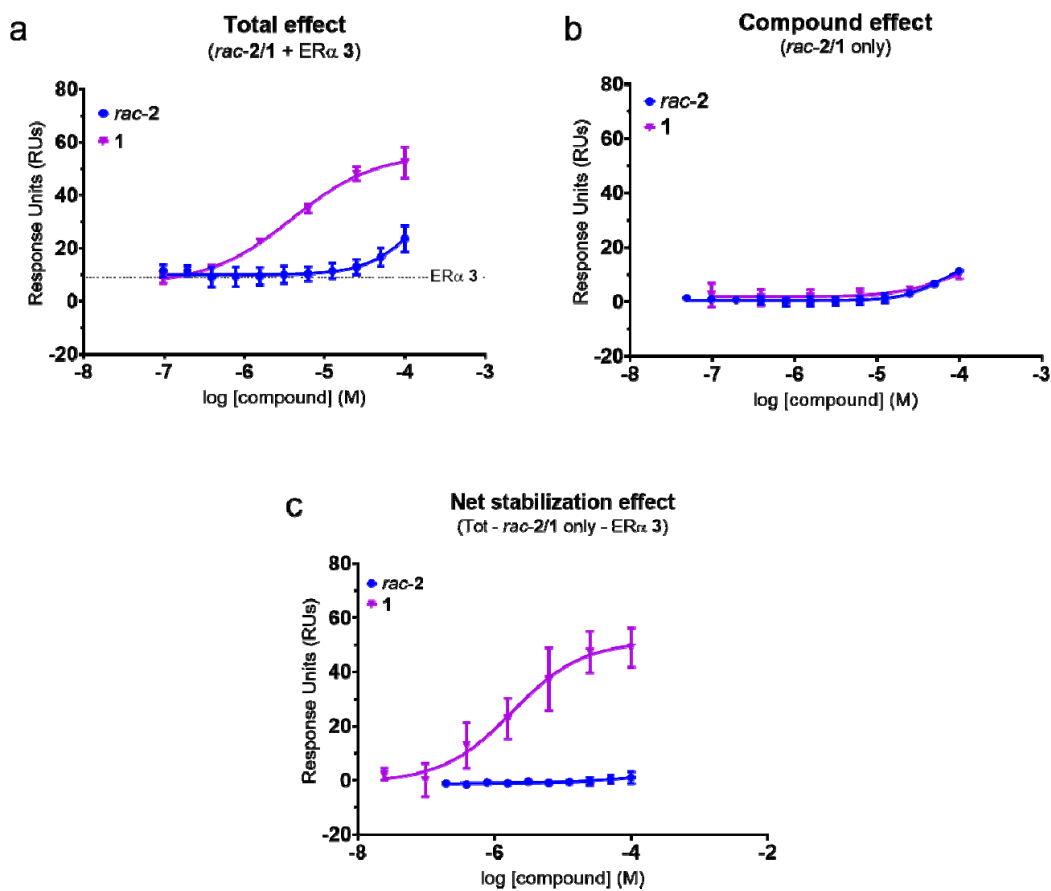

**Supplementary Figure 3**. Comparison of the stabilization effect of *rac*-**2** and **1** for the 14-3-3ζ/ERα complex measured by SPR (in the absence of added $Mg^{2+}$). Compound (*rac*-**2** or **1**) concentration (x axis) is plotted as the log of the compound concentration in molar. (a) "Total effect" is the total RUs afforded by adding compound (*rac*-**2** or **1**) to immobilised 14-3-3ζ in the presence of 50 nM **3** ("**3** only", indicated by dashed line, 9.4 ± 1.6 RUs), i.e. (affinity for 14-3-3ζ + affinity for 14-3-3ζ/**3** complex + stabilization of 14-4-4ζ/**3** interaction), (b) Determination of contribution of compound affinity for immobilized 14-3-3ζ protein in the absence of peptide **3**, c) "Net stabilization effect" (ΔRUs) is defined as (total RUs from 14-3-3ζ/**3**/compound)-((RUs from 14-3-3ζ/**3** interaction)+(RUs from 14-3-3ζ/compound interaction)).

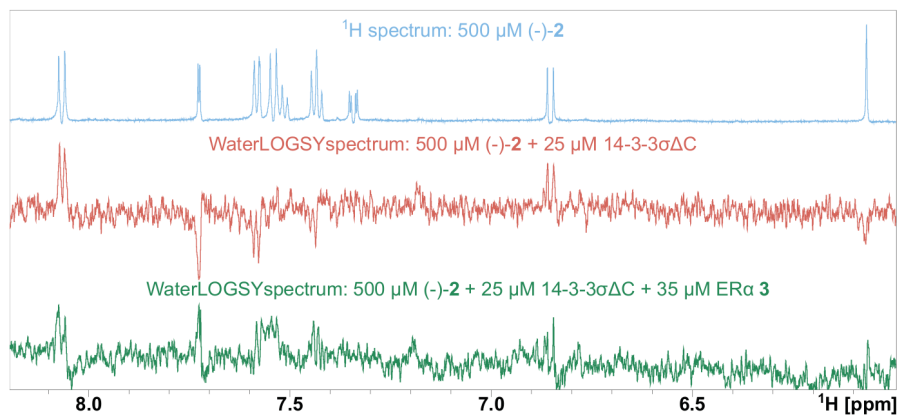

**Supplementary Figure 4**. WaterLOGSY NMR experiments indicate that (-)-**2** binds to 14-3-3σΔC both in the absence and presence of **3**. $^1$H spectrum (blue) and WaterLOGSY spectra of 500 μM (-)-**2** in the presence of either 25 μM 14-3-3σΔC (red) or 25 μM 14-3-3σΔC+35 μM phosphopeptide **3** (green).

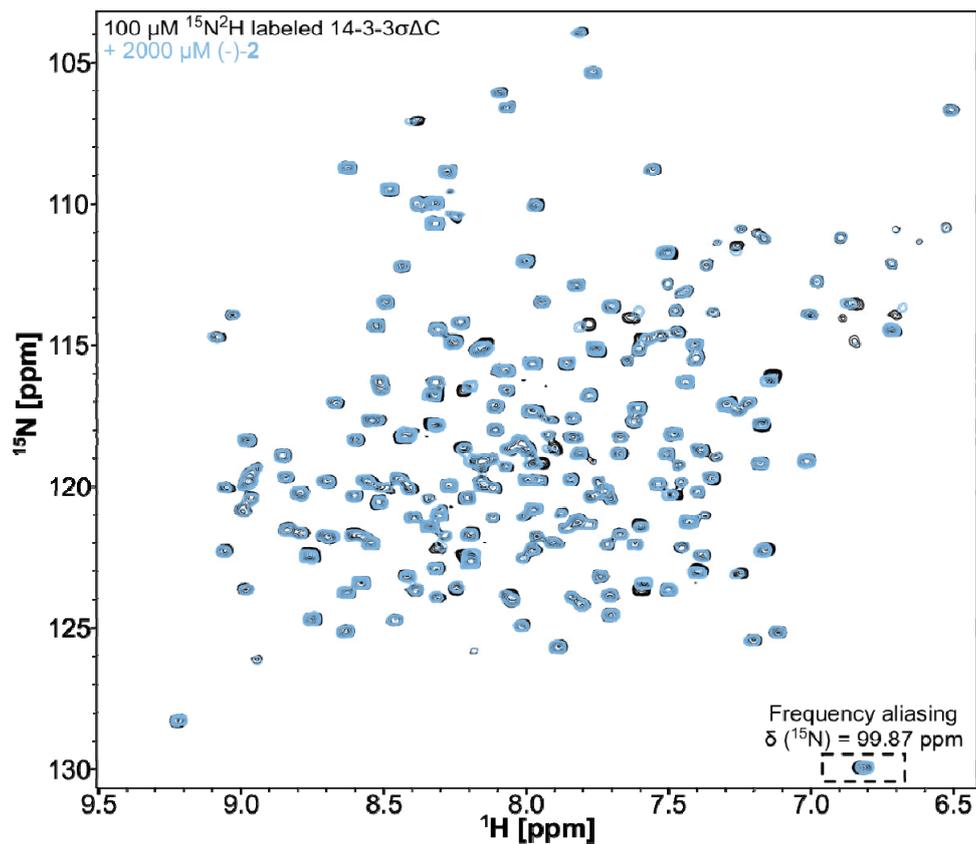

**Supplementary Figure 5.** $^{1}$H-$^{15}$N TROSY-HSQC spectra of 100 µM $^{15}$N$^{2}$H labeled 14-3-3σΔC alone (black), or in the presence of 2000 µM (-)-**2** (superimposed in blue).

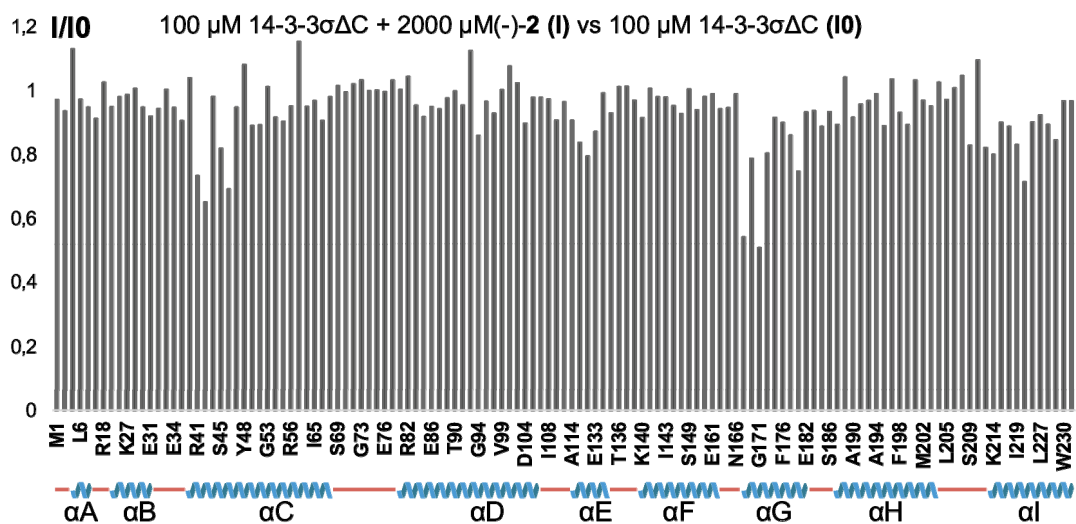

**Supplementary Figure 6.** Binding site of (-)-**2** on 14-3-3σΔC can be identified by reporting the intensity ratio (I/I$_0$) for each pair of corresponding resonances in the analyzed spectrum (I) compared to the control spectrum (I$_0$). Plot of the I/I$_0$ values of $^1$H-$^{15}$N correlation peak intensities in the spectrum of 100 μM $^{15}$N$^2$H labeled 14-3-3σΔC in the presence of 2000 μM (-)-**2** (I), compared to corresponding resonances in the reference spectrum of 100 μM 14-3-3σΔC (I$_0$) (y axis) *versus* 14-3-3σΔC amino acid sequence (x axis, not proportional to sequence length). A total of 131 correlation peak intensity ratios are shown. The helices of 14-3-3σΔC are identified below the x axis as blue cartoons, while disordered regions are represented by red lines.

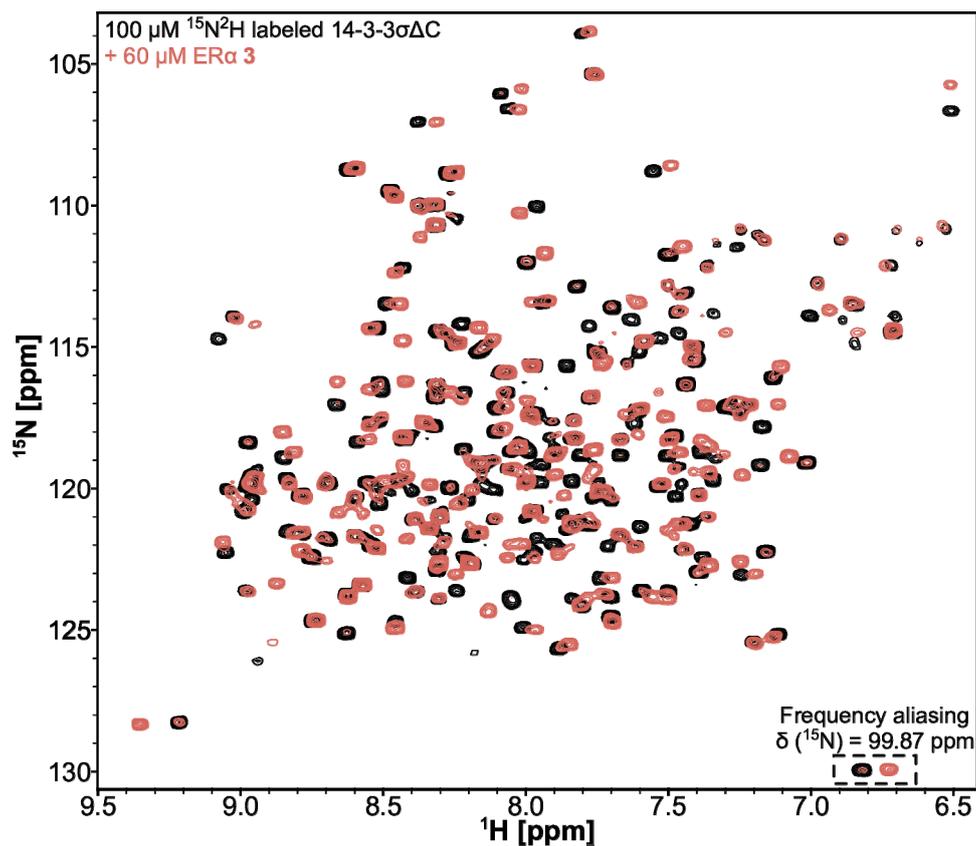

**Supplementary Figure 7.** $^1$H-$^{15}$N TROSY-HSQC spectra of 100 µM $^{15}$N$^2$H labeled 14-3-3σΔC alone (black), or in the presence of 60 µM ERα phosphopeptide **3** (superimposed in red). Note that for some resonances, both the free and the bound form are observed, suggesting a slow exchange regime on the NMR time scale.

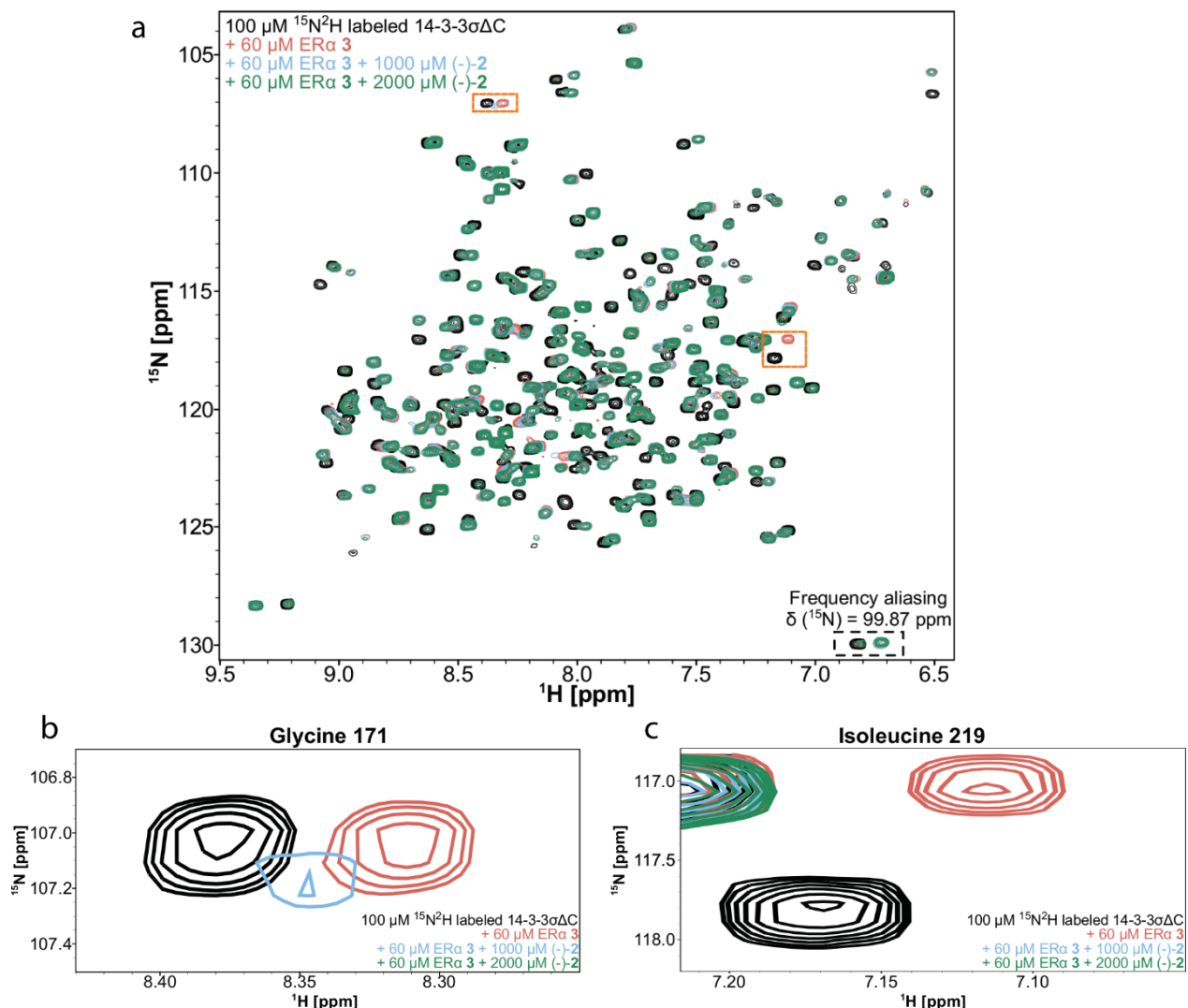

**Supplementary Figure 8.** $^1$H-$^{15}$N TROSY-HSQC experiments show that (-)-2 binds in the FC pocket. a) $^1$H-$^{15}$N TROSY-HSQC spectra of 100 µM $^{15}$N$^2$H labeled 14-3-3σΔC alone (black), or in the presence of: 60 µM ERα phosphopeptide **3** (superimposed in red); 60 µM ERα phosphopeptide **3** + 1000 µM (-)-**2** (superimposed in blue) or 60 µM ERα phosphopeptide **3** + 2000 µM (-)-**2** (superimposed in green). The spectral regions delimited by orange dashes are enlarged in b and c. (b,c) Overlaid enlarged spectral regions showing the resonances corresponding to G171 (b) and I219 (c). 1000 µM (-)-**2** induced broadening of the resonance of G171 (blue spectrum), while 2000 µM (-)-**2** induced broadening beyond detection (green spectrum). For I219, broadening beyond detection of the resonance (blue spectrum) was observed already at 1000 µM (-)-**2**.

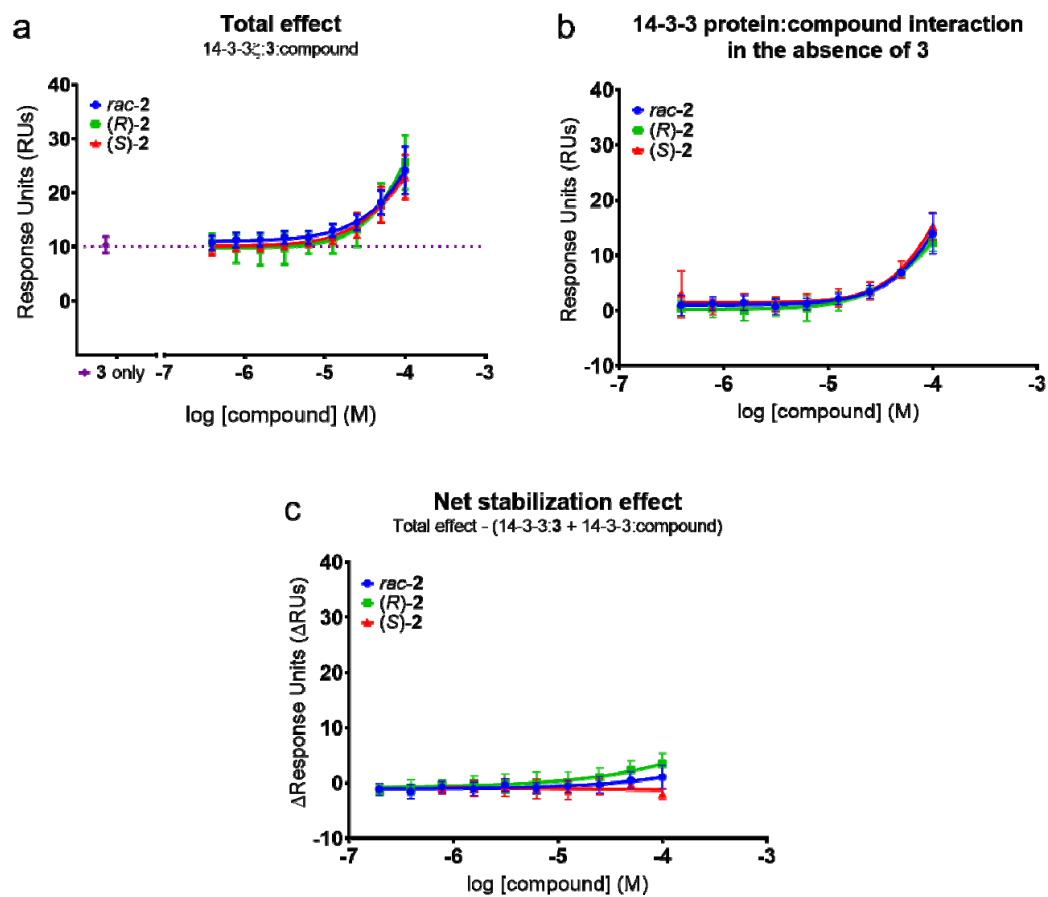

**Supplementary Figure 9**. Comparison of the stabilization effect of *rac*-**2**, (*R*)-**2** and (*S*)-**2** for the 14-3-3ζ:ERα phosphopeptide **3** complex, measured by SPR (in the absence of added $Mg^{2+}$). Compound (*rac*-, (*R*)- or (*S*)-**2**) concentration (x axis) is plotted as the log of the compound concentration in molar. (a) "Total effect" is the total RUs afforded by adding compound (*rac*-, (*R*)- or (*S*)-**2**) to immobilized 14-3-3ζ in the presence of 50 nM ERα phosphopeptide **3** ("**3** only", indicated by dashed line, 10.4 ± 1.2 RUs), i.e. (affinity for 14-3-3ζ + affinity for 14-3-3ζ/**3** complex + stabilization of 14-4-4ζ/**3** interaction), (b) Determination of compound (*rac*-, (*R*)- or (*S*)-**2**) affinity for immobilized 14-3-3ζ protein in the absence of peptide **3**, (c) "Net stabilization effect" (ΔRUs) is defined as (total RUs from 14-3-3ζ/**3**/compound)-((RUs from 14-3-3ζ/**3** interaction)+(RUs from 14-3-3ζ/compound interaction)).

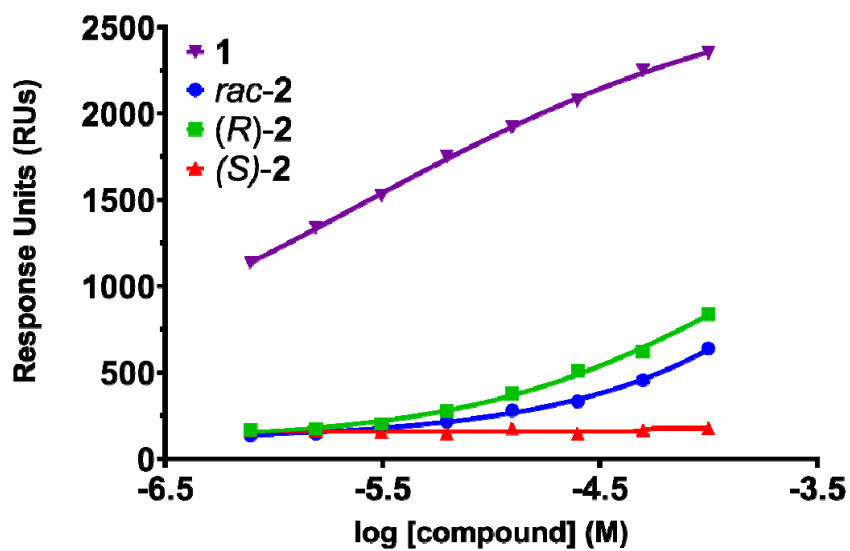

**Supplementary Figure 10**. Measurement by SPR of the stabilization of the 14-3-3η:PMA2 complex by **2** (blue dots), (*R*)-**2** (green squares), (*S*)-**2** (red triangles) and **1** (purple inverse triangles) in the presence of 10 mM $MgCl_2$. Response units (RUs) are plotted *vs* the log of compound concentration (M). In the experiment, PMA2 peptide was immobilised on a CMD200M chip *via* EDC/NHS coupling chemistry, at 8000 RUs. 14-3-3η (10 μM) and increasing concentrations of compound (1:2 dilution series, 100 μM initial concentration, n = 1) were premixed and injected at a flowrate of 20 μL/min and 20 °C for or 120 s in running buffer, followed by a single injection of 0.5% SDS for 60 s, as regeneration step.

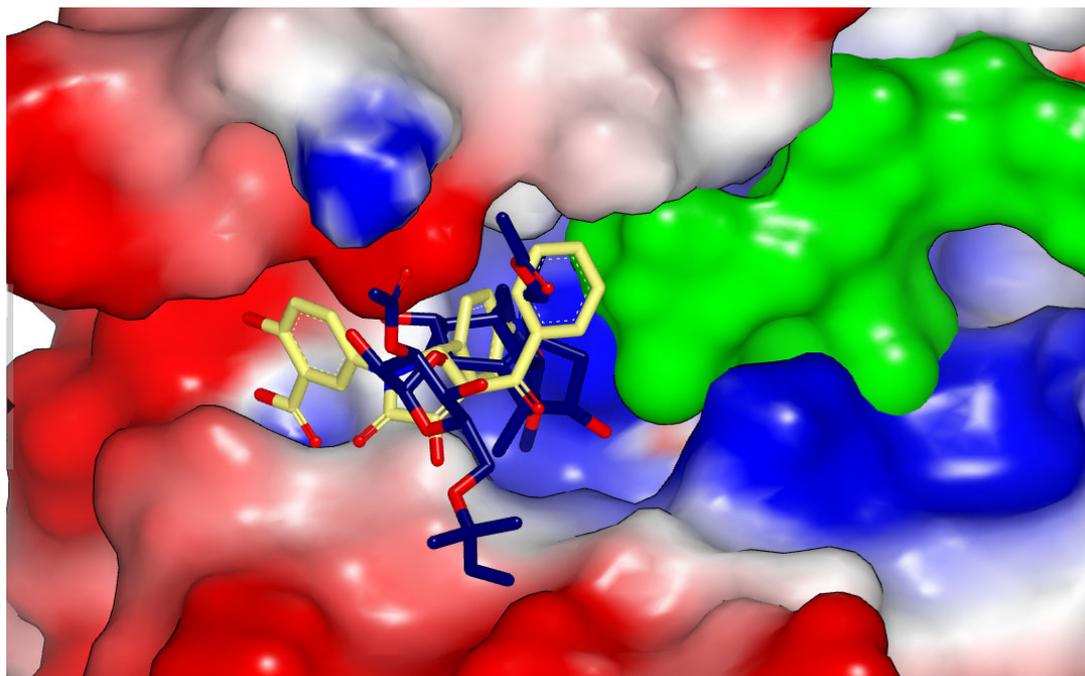

**Supplementary Figure 11**. Comparison of the binding modes in the FC pocket of Fusicoccin A **1** (from PDB 4JDD, dark blue sticks) and (*R*)-**2** (yellow sticks) in complex with 14-3-3σΔC/ERα phosphopeptide **3** (14-3-3 σΔC protein surface coloured according to electrostatic potential, ERα phosphopeptide **3** surface coloured green) showing that the salicylate moiety of (*R*)-**2** occupies a subpocket not exploited by Fusicoccin A **1**.

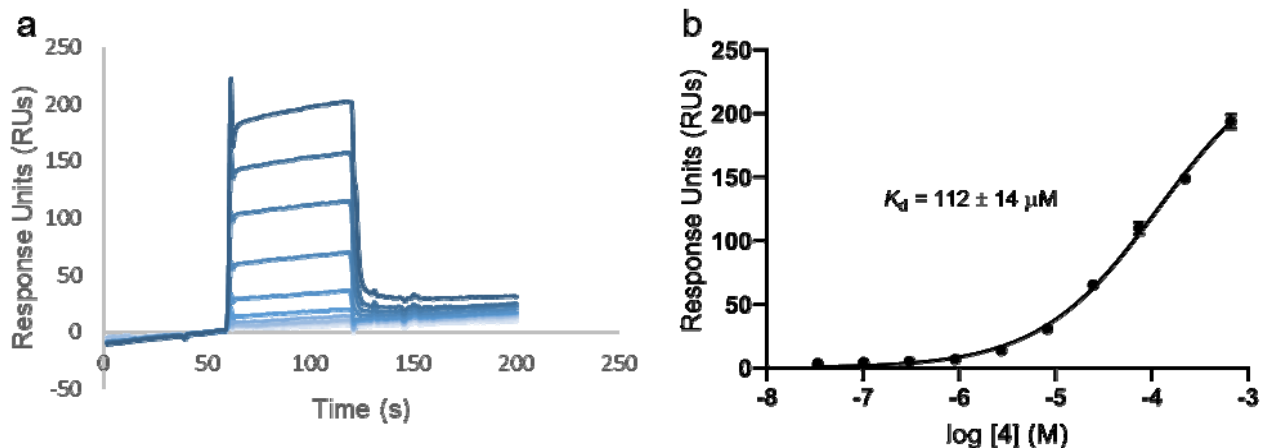

**Supplementary Figure 12**. Determination of affinity of CaMKK2 phosphopeptide **4** for 14-3-3ζ by SPR (a) Representative example of an SPR sensorgram in which increasing concentrations of **4** were flowed over immobilized 14-3-3ζ. The Response Units (RUs) achieved (y axis) are presented as a function of time in seconds (x axis). For each curve, the values at equilibrium response (i.e. binding coverage) where extrapolated and fitted in a dose-response curve using a four-parameter logistic model (4PL), shown in (b), against the log of the molar concentration of **4** (mean ± SD, n = 3). Titration was performed in 1:3 dilution series from an initial concentration of **4** of 667 μM.

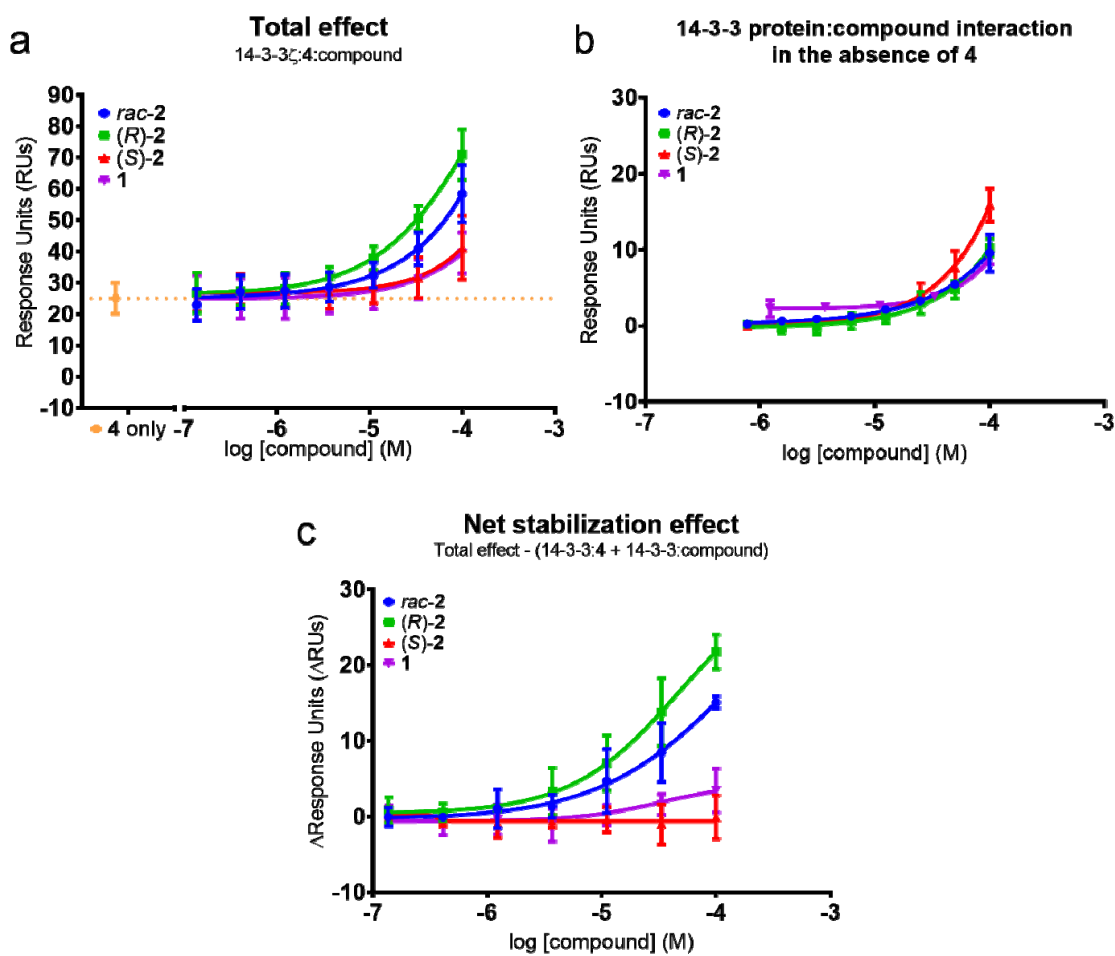

**Supplementary Figure 13** Comparison of the stabilization effect of **1** and *rac*-, (*R*)- or (*S*)-**2** for the 14-3-3ζ/CaMKK2 complex measured by SPR (in the absence of added $Mg^2$). Compound concentration (x axis) is plotted as the log of the compound concentration in molar. (a) "Total effect" is the total RUs afforded by adding compound (**1**, *rac*-, (*R*)- or (*S*)-**2**) to immobilised 14-3-3ζ in the presence of 30 μM **4** ("**4** only", indicated by dashed line, 25.1 ± 4.9 RUs), i.e. (affinity for 14-3-3ζ + affinity for 14-3-3ζ/**4** complex + stabilization of 14-3-3ζ/**4** interaction), (b) Determination of contribution of compound affinity for immobilized 14-3-3ζ protein in the absence of peptide 3, (c) "Net stabilization effect" (ΔRUs) is defined as (total RUs from 14-3-3ζ/**4**/compound)-((RUs from 14-3-3ζ /**4** interaction)+(RUs from 14-3-3ζ/compound interaction)).

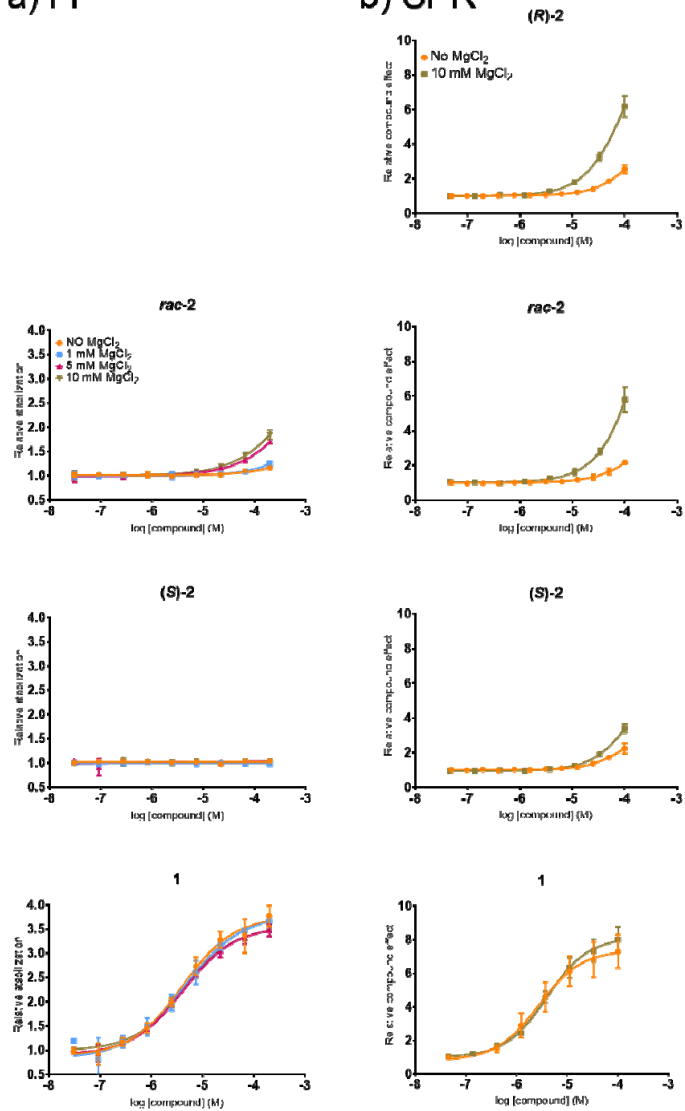

**Supplementary Figure 14**. Determination of the effect of MgCl$_2$ on the stabilization of *rac*-**2**, (*R*)-**2,** (*S*)-**2** and **1** towards the 14-3-3ζ/ERα complex measured by FP (a) and SPR (b). (a) Concentration-response of compound in FP assay (10 nM FITC-**3** and 50 nM 14-3-3ζ) in the absence of MgCl$_2$ (orange circles), 1 mM MgCl$_2$ (light blue squares), 5 mM MgCl$_2$ (red triangles) or 10 mM MgCl$_2$ (gold inverse triangles). "Relative stabilization" (y-axis) is the mean fold-increase of FP signal over baseline (i.e. FP signal from interaction between 14-3-3ζ and ERα phosphopeptide FITC-**3** alone), (b) Concentration-response of compound in SPR assay in the absence of MgCl$_2$ (orange circles) or with 10 mM MgCl$_2$ (gold squares). Compound titration was performed in the presence of 50 nM ERα phosphopeptide **3.** "Relative compound effect" is the mean-fold-increase of SPR signal over baseline (without correcting for compound binding to 14-3-3ζ alone). The error bars indicate +/- SD (n =3).

S25

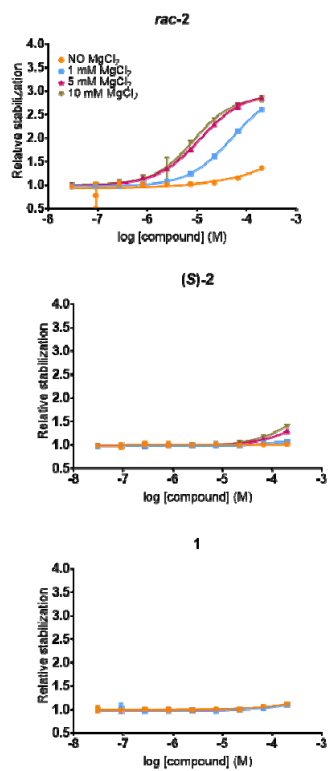
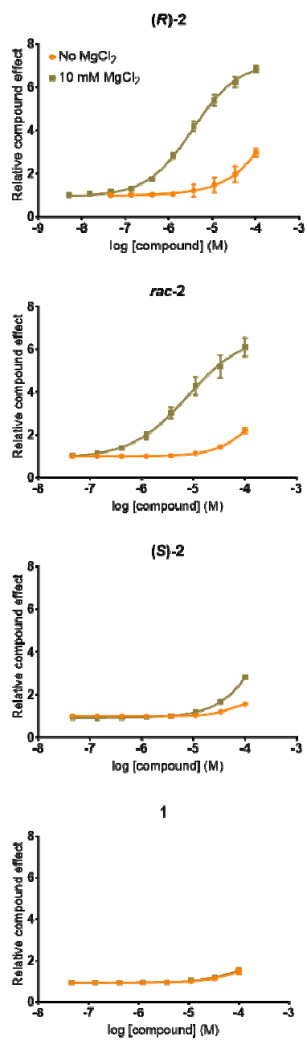

**Supplementary Figure 15**. Determination of the effect of MgCl$_2$ on the stabilization of **2**, (*S*)-**2** and **1** towards the 14-3-3ζ/CaMKK2 complex measured by FP (a) and SPR (b). a) Concentration-response of compound in FP assay (10 nM FAM-**4** and 30 μM 14-3-3ζ) in the absence of MgCl$_2$ (orange circles), 1 mM MgCl$_2$ (light blue squares), 5 mM MgCl$_2$ (red triangles) or 10 mM MgCl$_2$ (gold inverse triangles). "Relative stabilization" (y-axis) is the mean fold-increase of FP signal over baseline (i.e. FP signal from interaction between 14-3-3ζ and CaMKK2 phosphopeptide FAM-**4** alone), b) Concentration-response of compound in SPR assay in the absence of MgCl$_2$ (orange circles) or with 10 mM MgCl$_2$ (gold squares). Compound titration was performed in the presence of 30 μM CaMKK2 phosphopeptide **4**. "Relative compound effect" is the mean-fold-increase of SPR signal over baseline (without correcting for compound binding to 14-3-3ζ alone). The error bars indicate +/- SD (n =3).

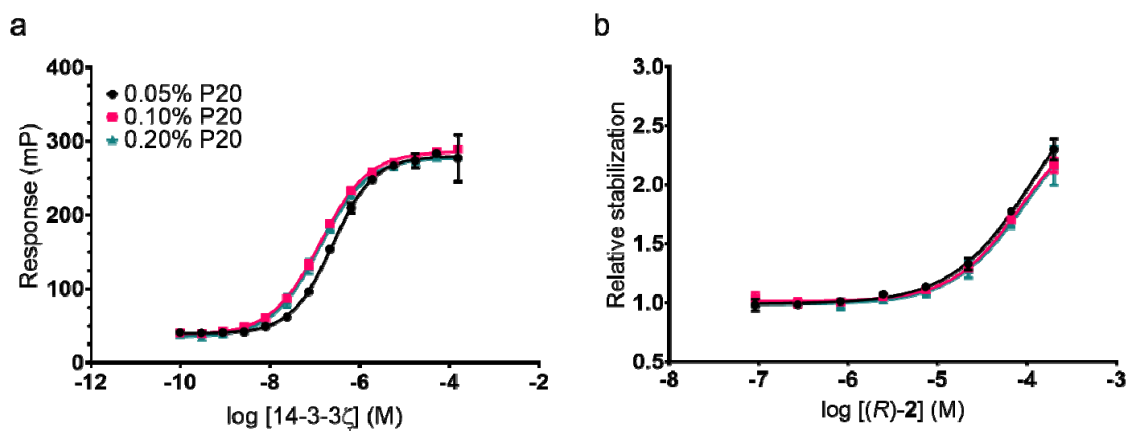

**Supplementary Figure 16**. Test for assay interference induced by aggregation. Determination of effect of different detergent (Tween20 – P20) concentrations on the affinity of the ERα phosphopeptide FITC-**3** for 14-3-3ζ (panel a) and on the stabilization effect promoted by (*R*)-**2** for the 14-3-3ζ:ERα complex (panel b) measured by FP and in the presence of 10 mM $Mg^{2+}$. (A) The FP response (mP, y axis) achieved is plotted against the log of the 14-3-3ζ concentration (in molar) at different detergent concentrations: 0.05% (black dots), 0.10% (pink squares) and 0.20% (teal triangles). (B) "Relative stabilization" (expressed as mean fold-increase of FP signal over baseline, i.e. interaction between 14-3-3ζ and ERα phosphopeptide FITC-**3** alone) is plotted *versus* increasing concentrations of (*R*)-**2**. The error bars indicate +/- SD (n =3).

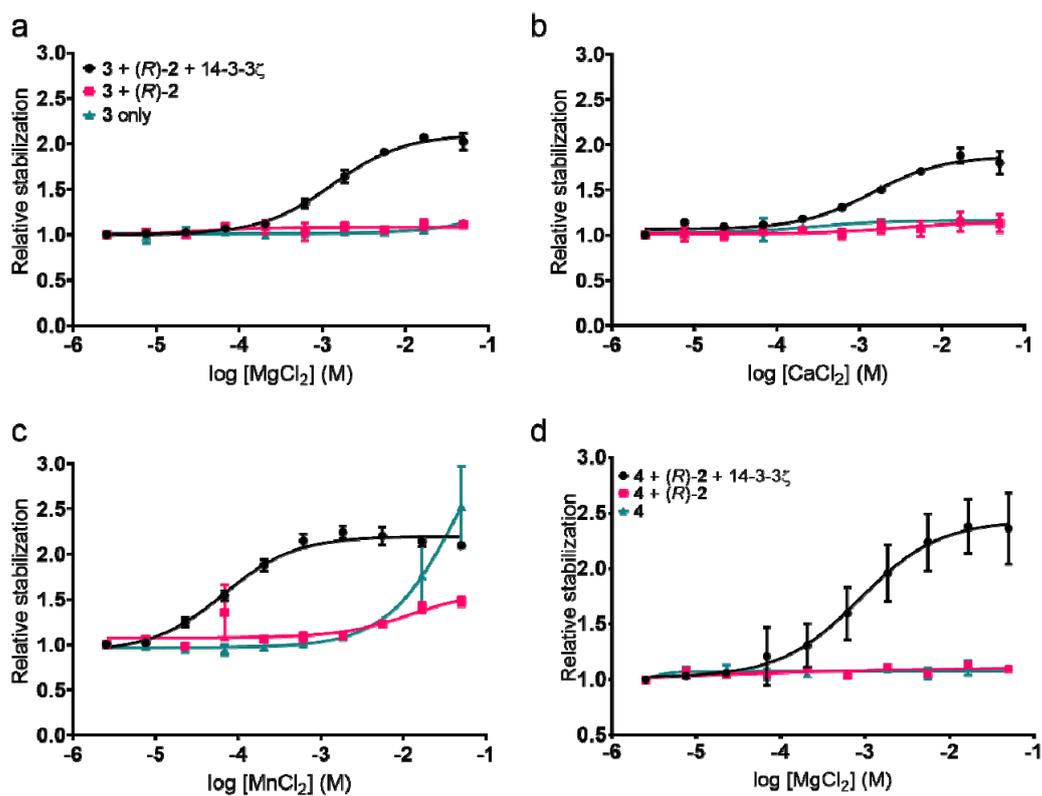

**Supplementary Figure 17**. Effect of bivalent metal ions on stabilisation of 14-3-3/ERα and 14-3-3/CaMKK2 PPIs by (*R*)-**2** in FP assay and various counter-screens. "Relative stabilization" (y-axis) is the mean fold-increase of FP signal over baseline (i.e. interaction between 14-3-3ζ and phosphopeptides FITC-**3** or FAM-**4** alone). Relative stabilization (black circles) is plotted *versus* metal ion concentration (1:3 dilution series, 50 mM initial concentration) at fixed concentrations of (*R*)-**2** (10 μM), phosphopeptide (10 nM FITC-**3** in panels a-c and FAM-**4** in panel d) and 14-3-3ζ (50 nM in panels a-c, 30 μM in panel d). For the 14-3-3ζ/ERα complex magnesium (panel a), calcium (panel b) and manganese (panel c) were tested, while for the 14-3-3ζ:CaMKK2 complex only magnesium was tested (panel d). To rule out protein independent effects, and/or unspecific binding, counter-screens with compound + peptide only (pink squares) and compound alone (teal triangles) were performed.

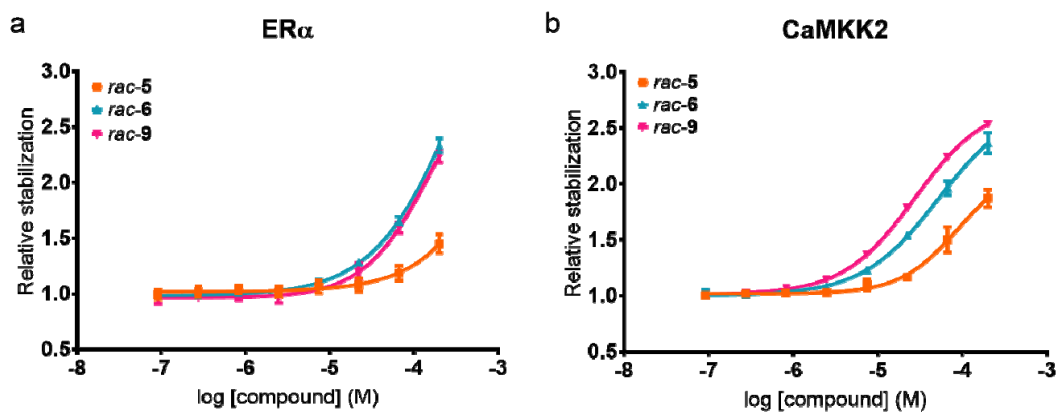

**Supplementary Figure 18**. Stabilization effects of *rac*-**5**, *rac*-**6** and *rac*-**9** (conformationally restricted analogues of *rac*-**2**) measured in FP assays in the absence of added $MgCl_2$. (a) Stabilization of the 14-3-3/ERα PPI (10 nM FITC-**3**, 50 nM 14-3-3ζ), (b) Stabilization of the 14-3-3ζ/CaMKK2 PPI (10 nM FAM-**4**, 30 μM 14-3-3ζ). "Relative stabilization" (y-axis) refers to mean fold increase of FP signal at a given [compound] over signal observed in absence of compound. The error bars indicate +/- SD (n =3).

5-[3-benzoyl-4-hydroxy-2-(4-nitrophenyl)-5-oxo-2H-pyrrol-1-yl]-2-hydroxy-benzoic acid (**2**)

¹H NMR (600 MHz, D₂O) δ 8.00 (d, *J* = 8.8 Hz, 2H, H19 and H21), 7.74 (d, *J* = 2.7 Hz, 1H, H23), 7.58 – 7.65 (m, 2H, H13 and H17), 7.54 (t, *J* = 7.4 Hz, 1H, H15), 7.50 (d, *J* = 8.8 Hz, 2H, H18 and H22), 7.46 (t, *J* = 7.6 Hz, 2H, H14 and H16), 7.33 (dd, *J* = 8.7, 2.7 Hz, 1H, H27), 6.86 (d, *J* = 8.7 Hz, 1H, H26), 6.06 (s, 1H, H3).

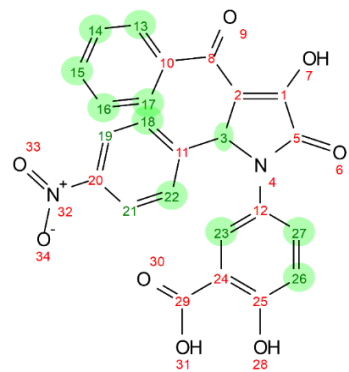

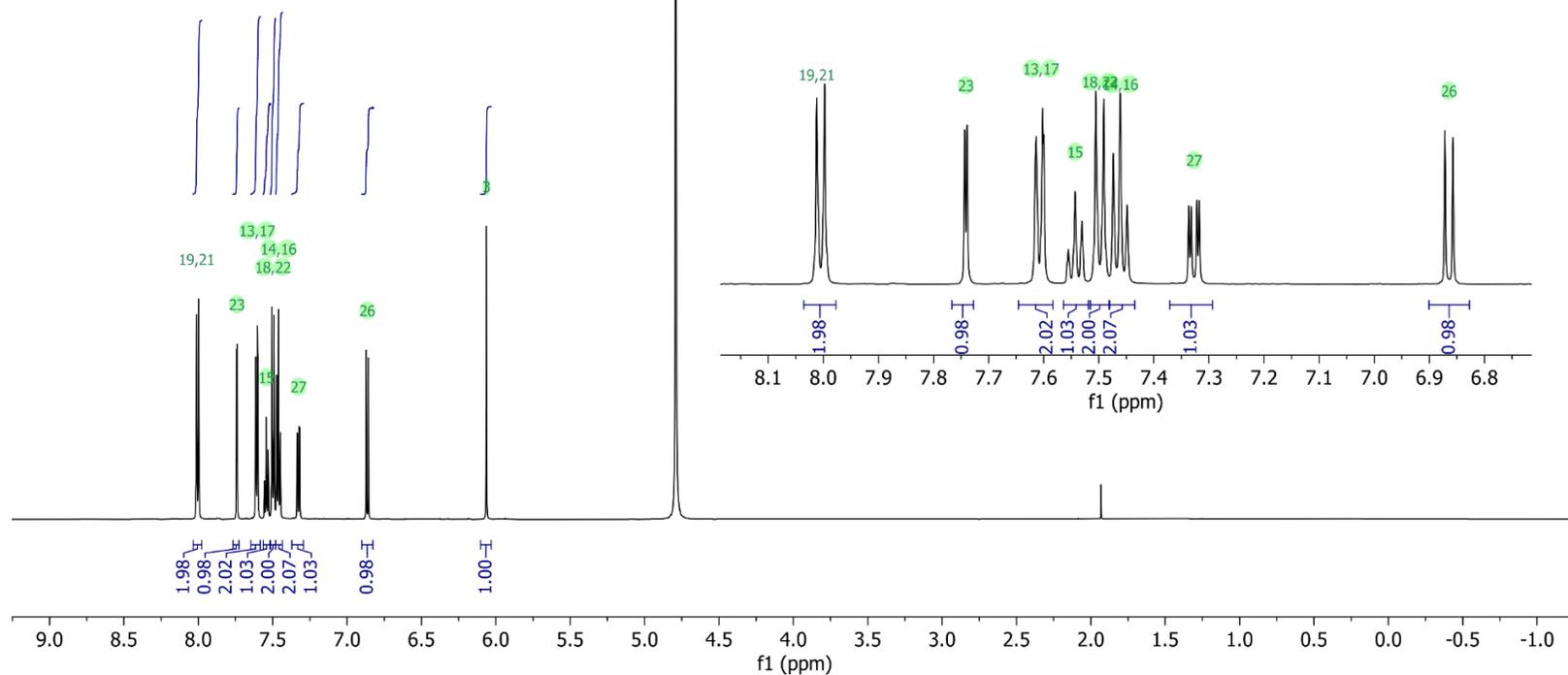

**Supplementary Figure 19a**. ¹H NMR spectrum of *rac*-**2** in D₂O, with assignments. *Rac*-**2** (2.5 mg, 5.43 µmol) was dissolved in 1 mM NaOH in D₂O (109 µL, 10.86 µmol) transferred into a 3 mm NMR tube and D₂O added to bring final volume to 160 µL.

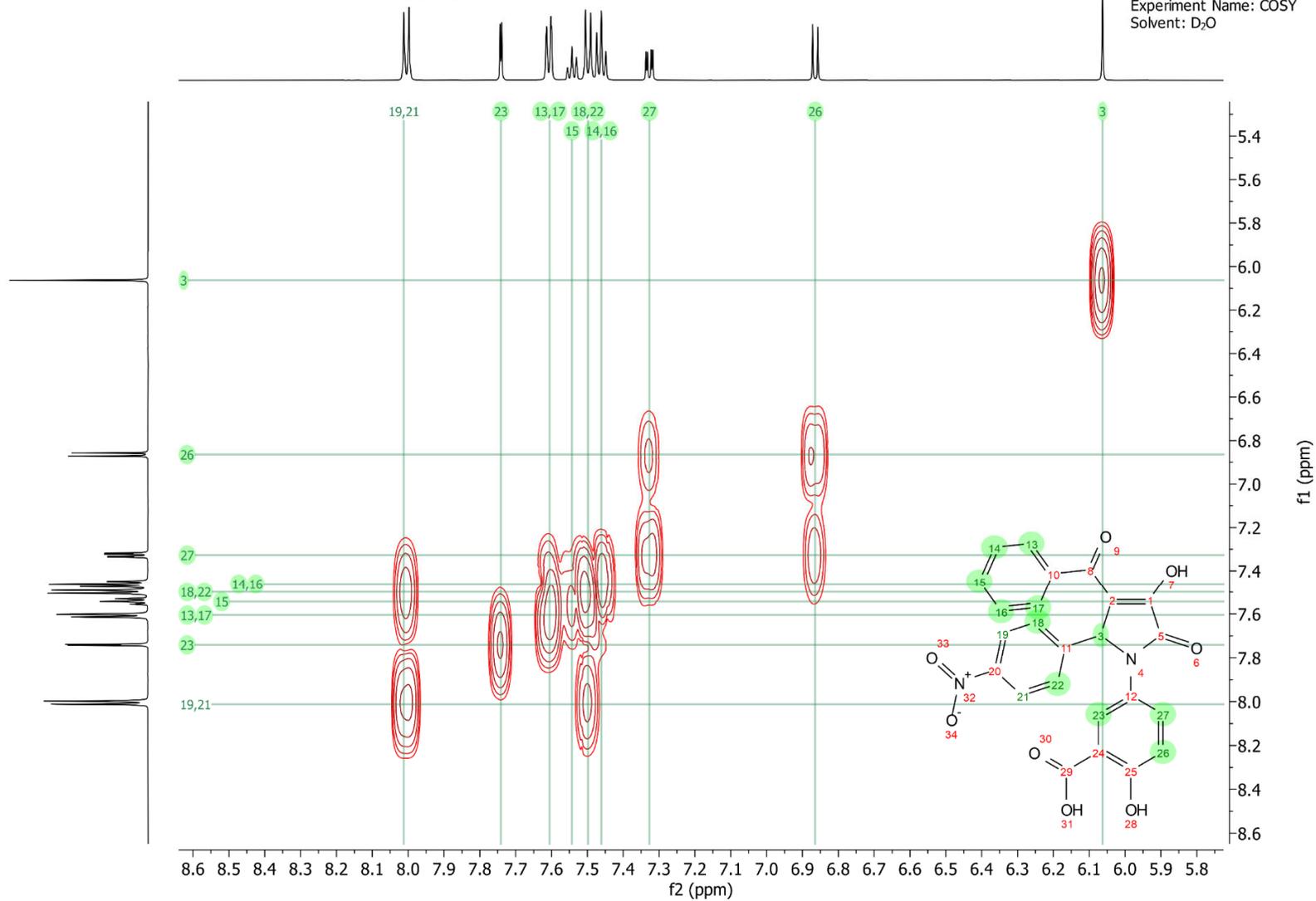

**Supplementary Figure 19b**. COSY NMR of *rac*-**2.**

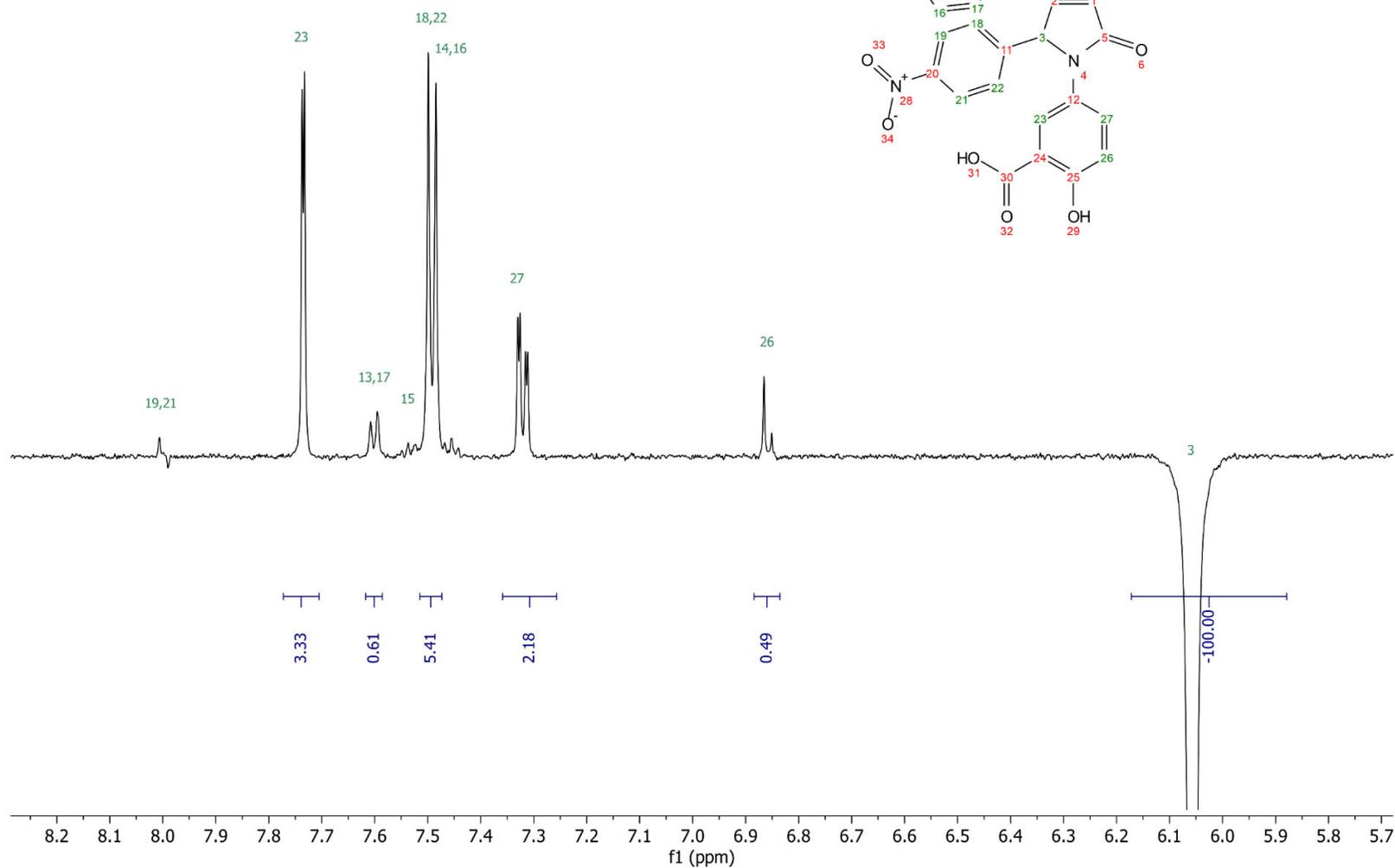

Supplementary Figure 19c. 1D selective ROESY of *rac*-2 (excitation of H3, 6.06 ppm).

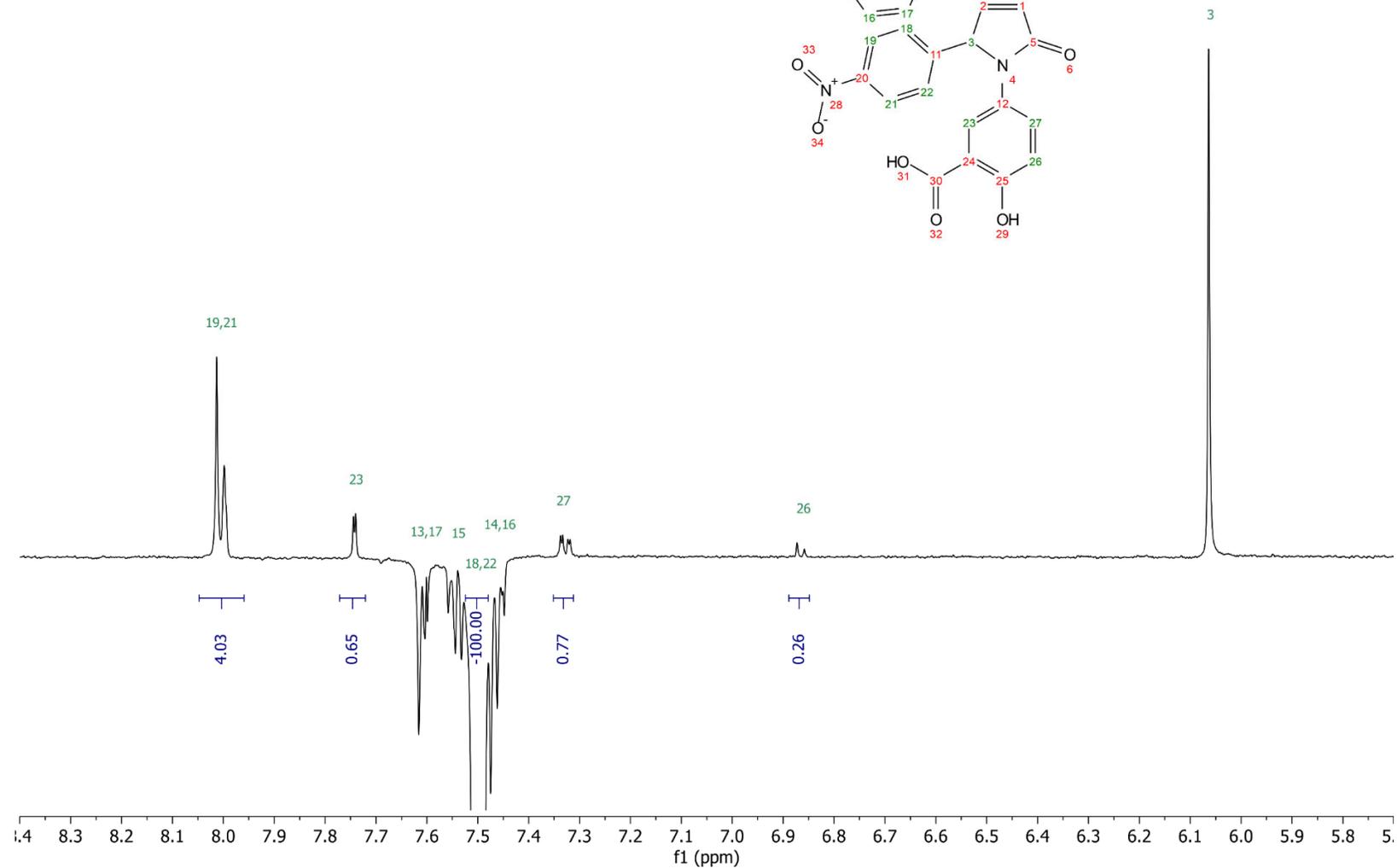

**Supplementary Figure 19d**. 1D selective ROESY of *rac*-**2** (excitation of H18-H22, 7.50 ppm).

S33

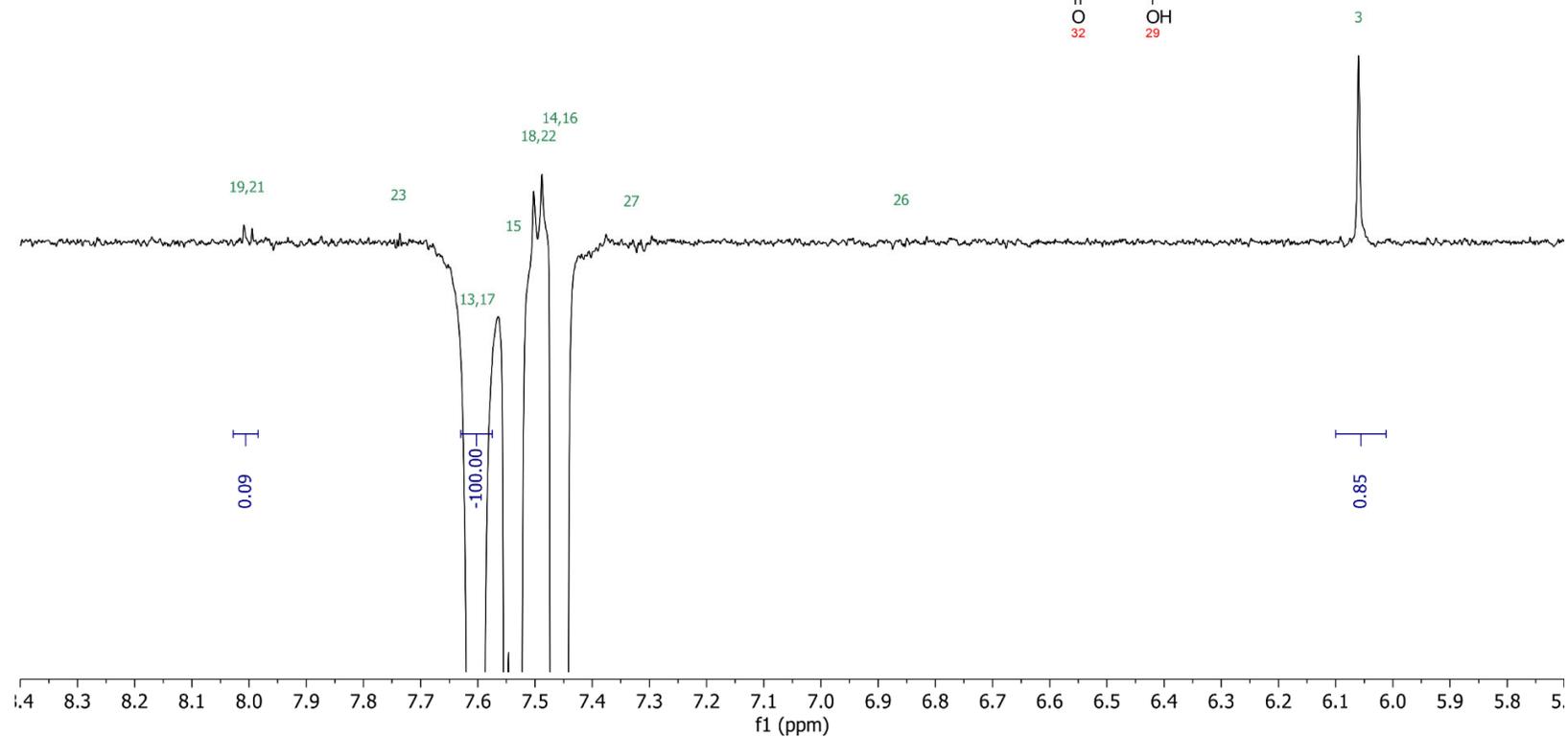

**Supplementary Figure 19e**. 1D selective ROESY of *rac*-**2** (excitation of H13-H17, 7.61 ppm).

5-[4-amino-3-benzoyl-2-(4-nitrophenyl)-5-oxo-2H-pyrrol-1-yl]-2-hydroxy-benzoic acid (**9**)

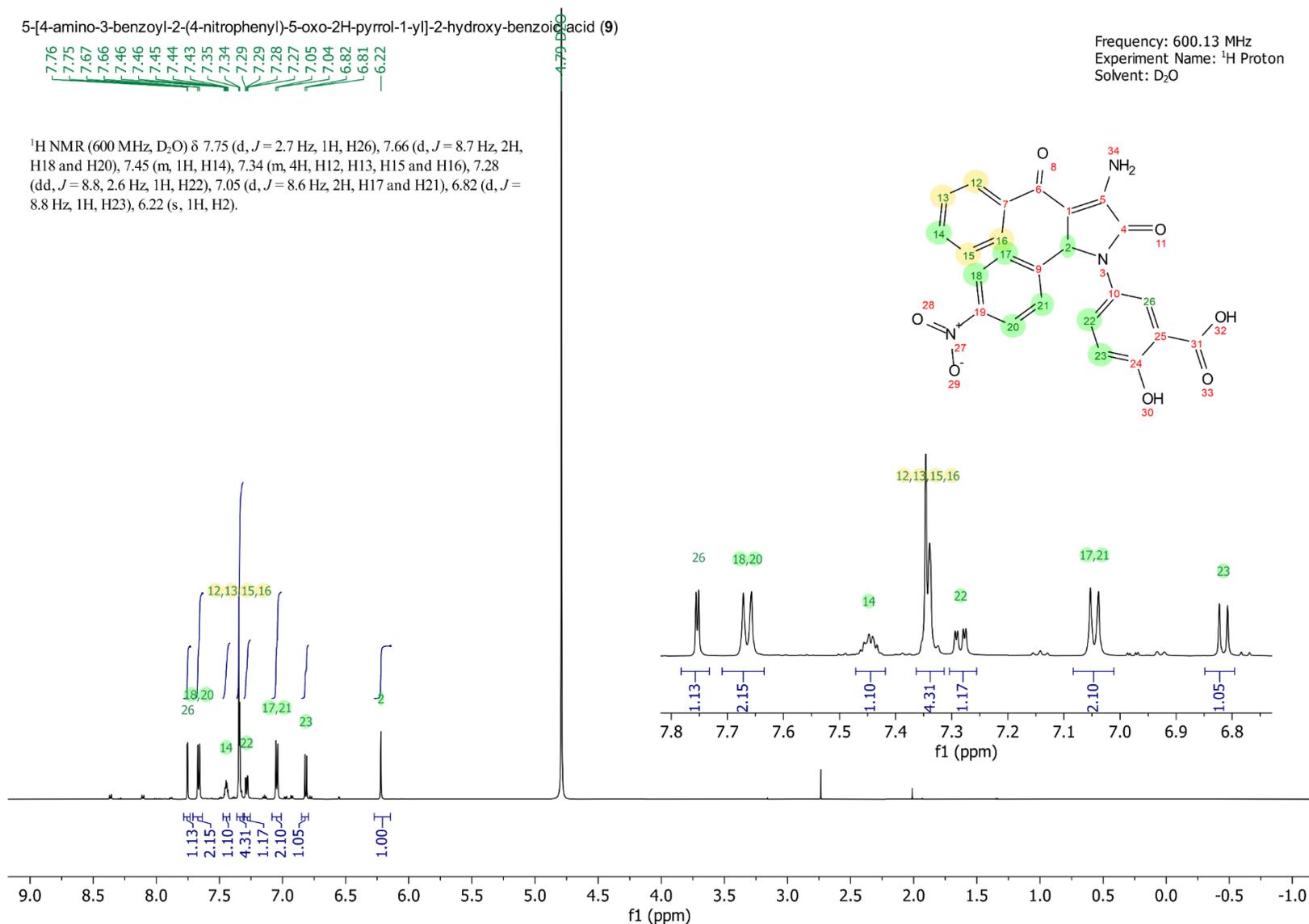

¹H NMR (600 MHz, D₂O) δ 7.75 (d, *J* = 2.7 Hz, 1H, H26), 7.66 (d, *J* = 8.7 Hz, 2H, H18 and H20), 7.45 (m, 1H, H14), 7.34 (m, 4H, H12, H13, H15 and H16), 7.28 (dd, *J* = 8.8, 2.6 Hz, 1H, H22), 7.05 (d, *J* = 8.6 Hz, 2H, H17 and H21), 6.82 (d, *J* = 8.8 Hz, 1H, H23), 6.22 (s, 1H, H2).

**Supplementary Figure 20a**. ¹H NMR of *rac*-**9** in D₂O solvent, with assignments. *Rac*-**9** (3.0 mg, 6.53 μmol) was dissolved in 1 mM NaOH in D₂O (131 μL, 13.06 μmol) transferred into a 3 mm NMR tube and D₂O was added to bring final volume to 160 μL.

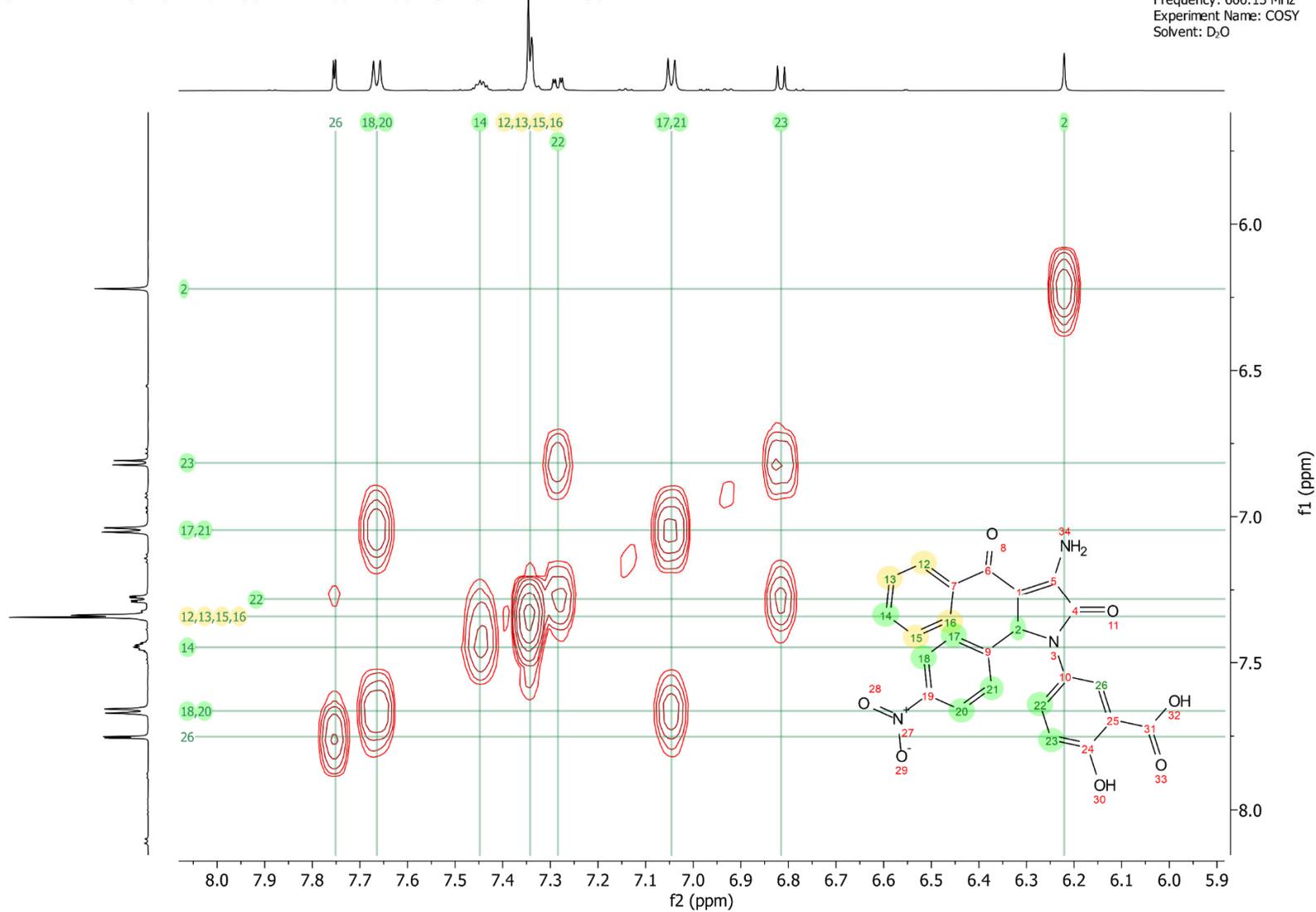

**Supplementary Figure 20b**. COSY NMR of *rac*-**9.**

5-[4-amino-3-benzoyl-2-(4-nitrophenyl)-5-oxo-2H-pyrrol-1-yl]-2-hydroxy-benzoic acid (**9**)

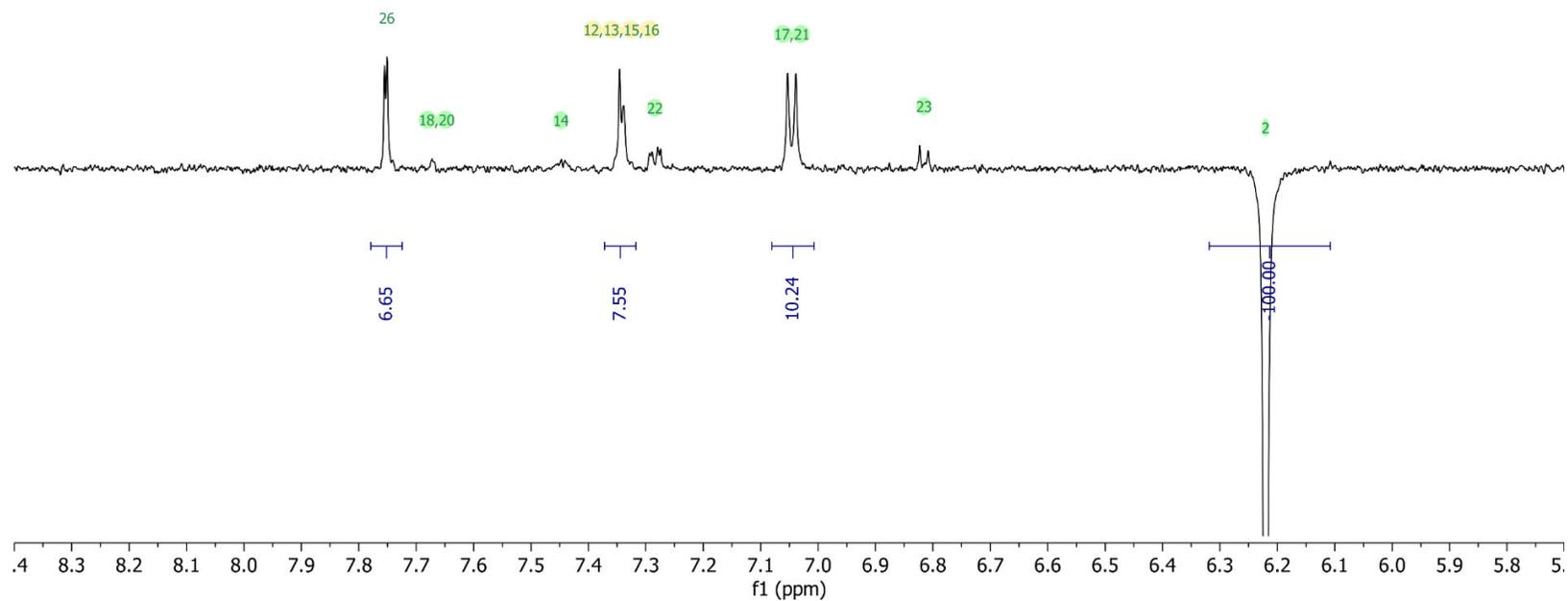

**Supplementary Figure 20c**. 1D selective ROESY of *rac*-**9** (excitation of H2, 6.22 ppm).

5-[4-amino-3-benzoyl-2-(4-nitrophenyl)-5-oxo-2H-pyrrol-1-yl]-2-hydroxy-benzoic acid (**9**)

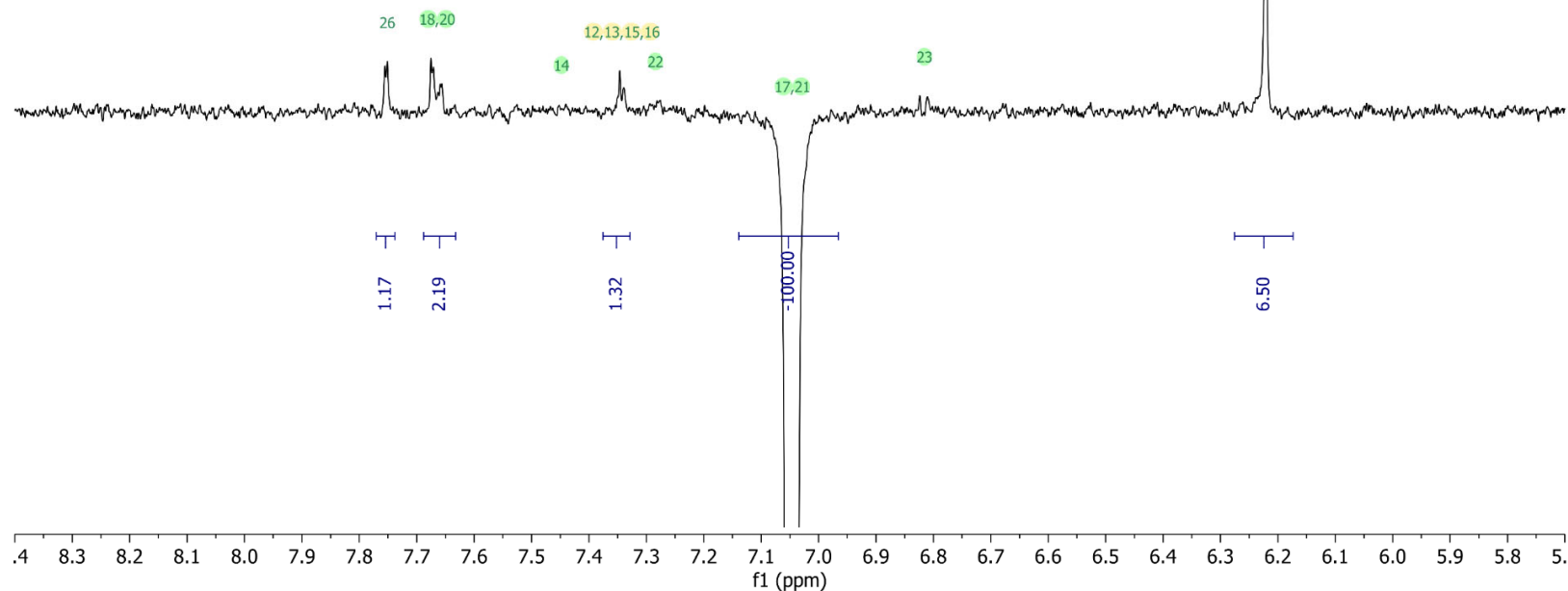

**Supplementary Figure 20d**. 1D selective ROESY of *rac*-**9** (excitation of H17-H21, 7.05 ppm).

S38

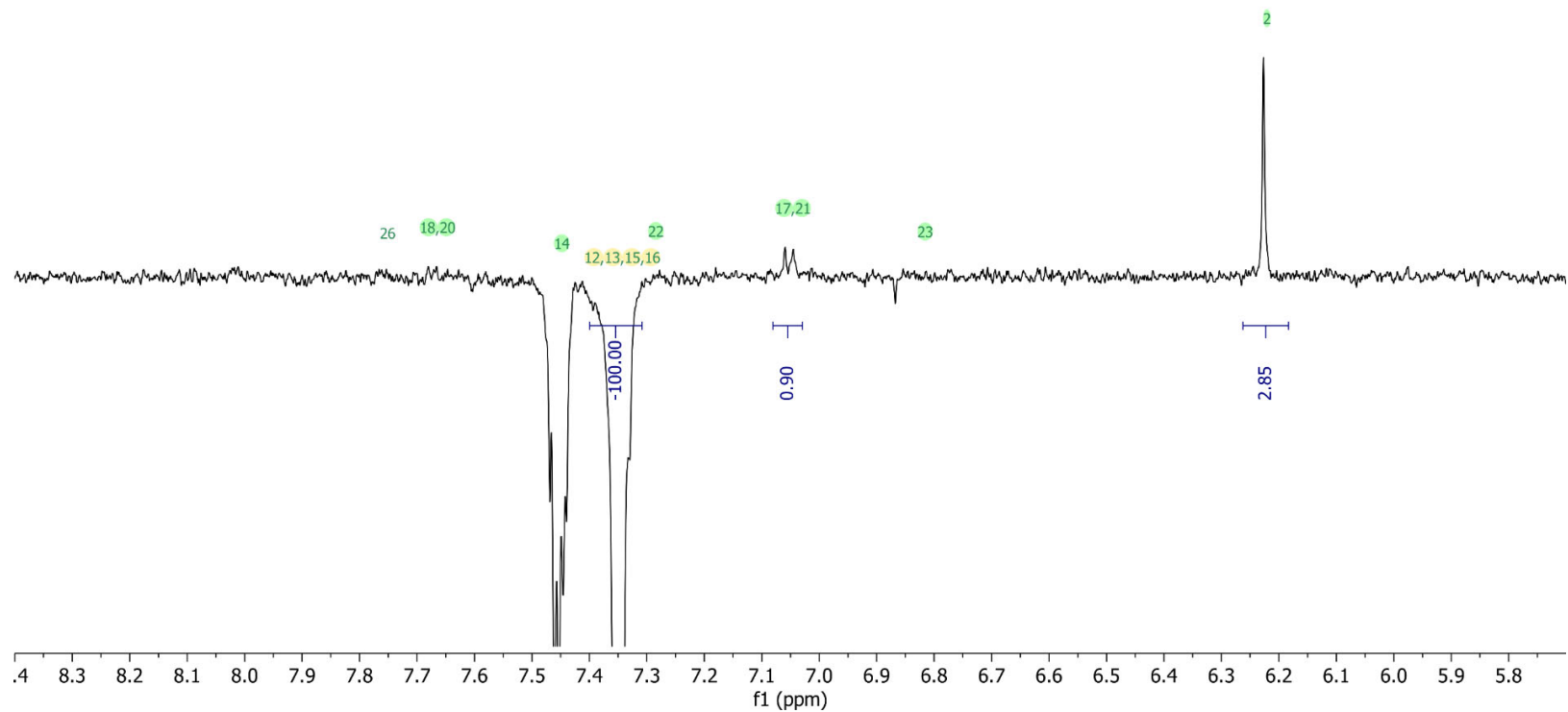

**Supplementary Figure 20e**. 1D selective ROESY of *rac*-**9** (excitation of H12-H16, H13-H15 overlapping, 7.34 ppm).

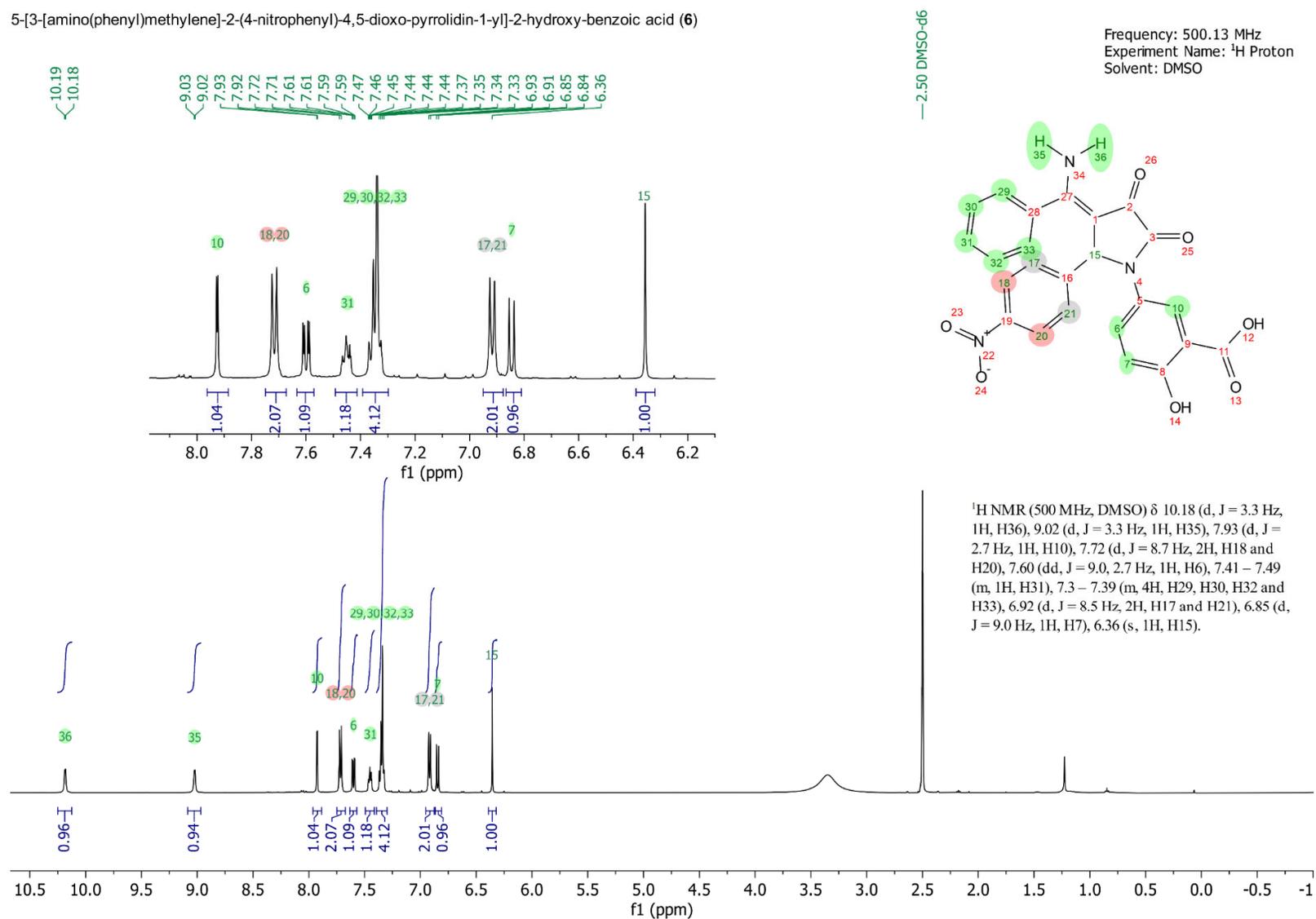

**Supplementary Figure 21a**. ¹H NMR of *rac*-**6** with in DMSO-d6 solvent, with assignments. *Rac*-**6** (5.2 mg, 11 μmol) was dissolved in DMSO-d6 (160 μL) and transferred into a 3 mm NMR tube. For full spectrum assignments, refer to NMR spectra of **6** in the spectroscopic data section (pages S70-S73).

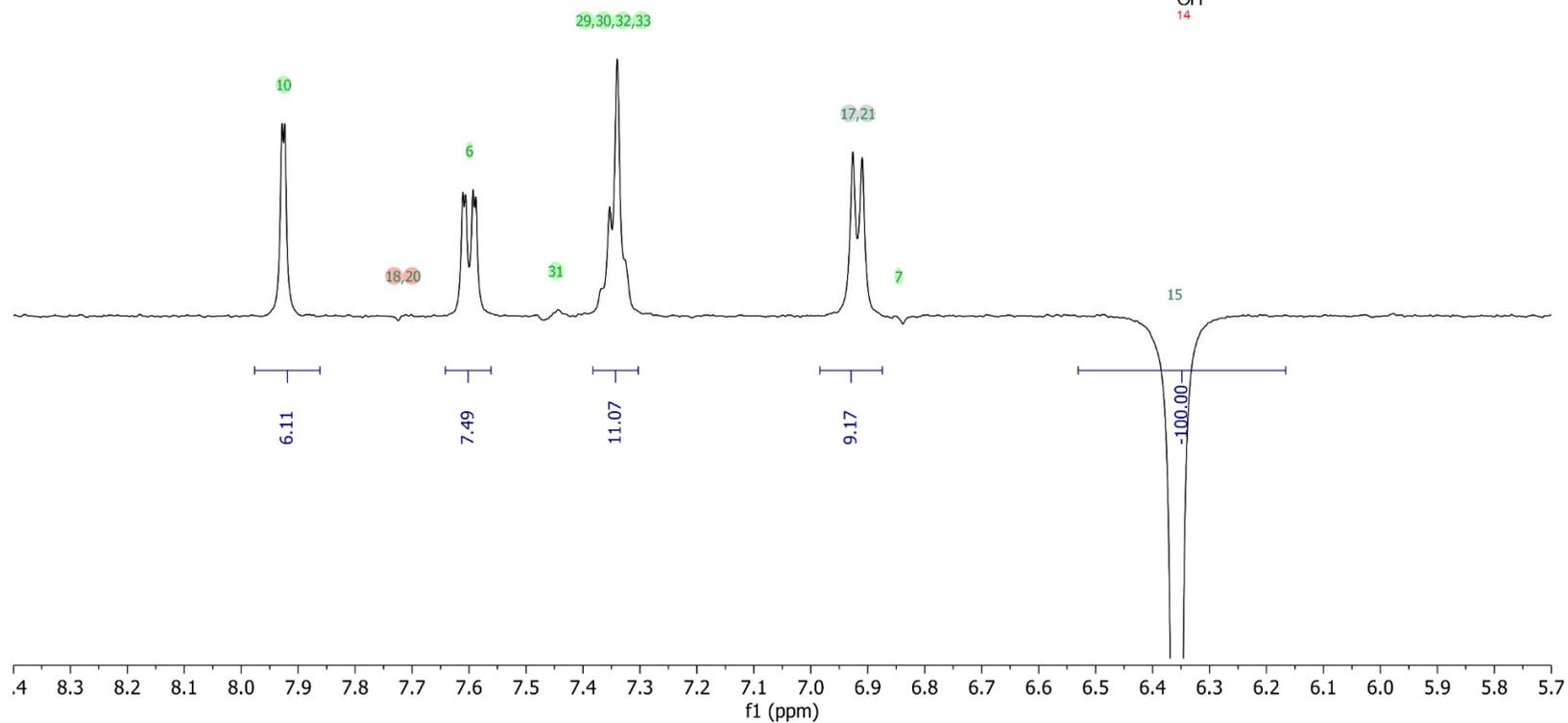

**Supplementary Figure 21b**. 1D selective ROESY of *rac*-**6** (excitation of H15, 6.36 ppm).

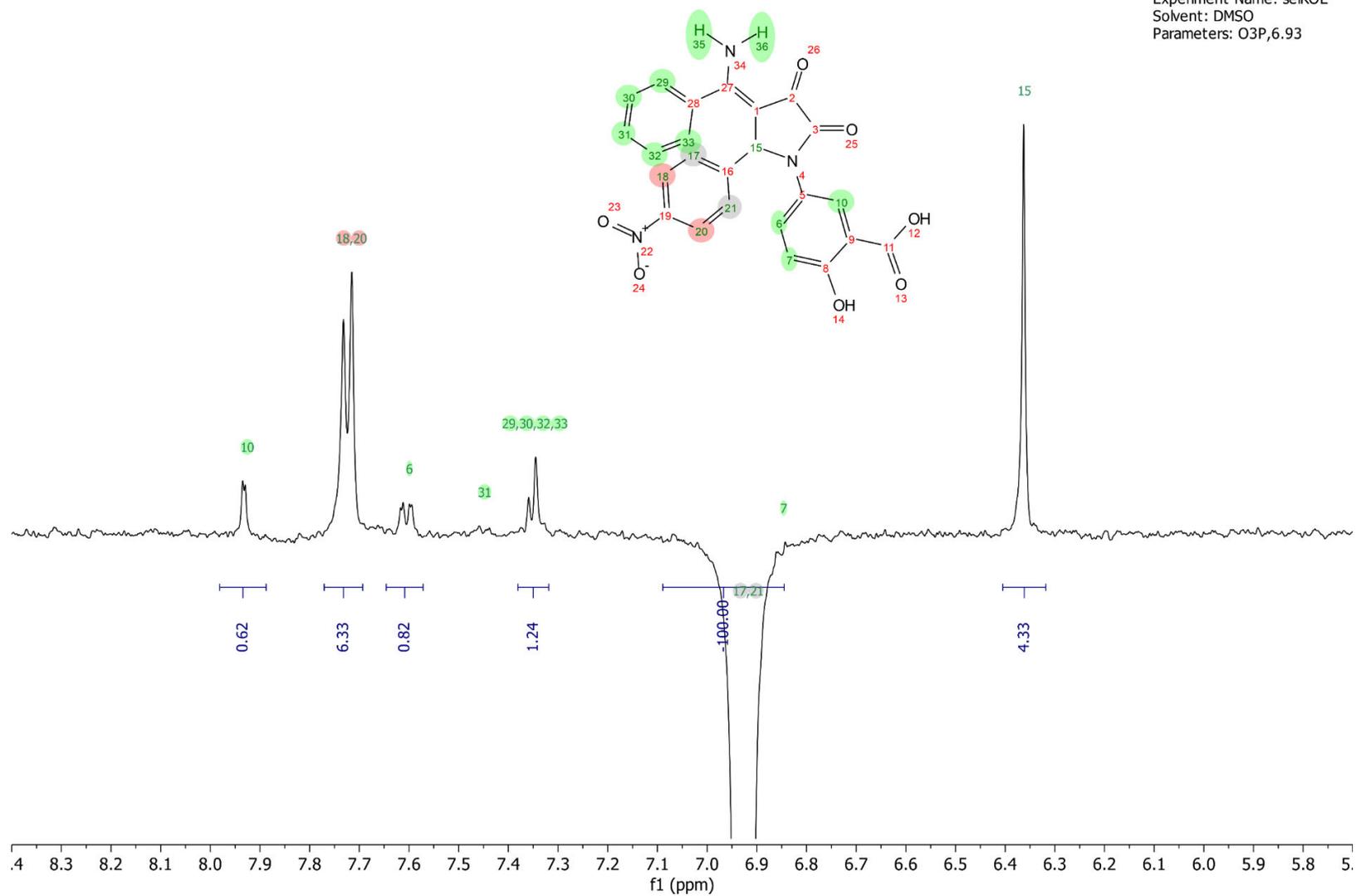

**Supplementary Figure 21c**. 1D selective ROESY of *rac*-**6** (excitation of H17-H21, 6.93 ppm).

## II. NMR data

### IIa) Production of $^{15}$N$^{2}$H labeled 14-3-3σΔC for $^{1}$H-$^{15}$N TROSY-HSQC NMR spectroscopy.

The $^{15}$N$^{2}$H labeled 14-3-3σΔC (ΔC17, cleaved after T231) for NMR studies was expressed in E. coli BL21 (DE3) cells transformed with a pProExHtb vector carrying the cDNA to express an N-terminally His$_6$-tagged human 14-3-3σΔC. Bacterial cells were grown in 1 L of deuterated M9 minimal medium supplemented with 2 g/L $^{12}$C$_6$$^{2}$H$_7$ Glucose, 1 g/L $^{15}$N Ammonium Chloride, 0.4 g/L Isogro $^{15}$N$^{12}$C$^{2}$H Powder – Growth Medium (Sigma Aldrich) and 100 µg/mL ampicillin. The recombinant protein was then purified from the bacterial extract by affinity chromatography using a Ni-NTA column (GE Healthcare). The His$_6$-tag was further cleaved by the TEV protease. The protein was finally dialyzed overnight at 4°C against NMR buffer (100 mM Sodium Phosphate, pH 6.8, 50 mM NaCl), concentrated, aliquoted, flash frozen and stored at -80 °C. A detailed protocol was previously published can be found at ref [1].

### IIb) $^{1}$H-$^{15}$N TROSY-HSQC NMR experiments

$^{1}$H-$^{15}$N TROSY-HSQC (Transverse Relaxation Optimized Spectroscopy - Heteronuclear Single Quantum Coherence Spectroscopy) spectra were acquired at the temperature of 32 °C in 3 mm tubes (sample volume 200 µL) using a 900 MHz Bruker Avance Neo spectrometer, equipped with a cryoprobe. All samples were prepared in a buffer containing 100 mM sodium phosphate, 50 mM NaCl, pH 6.8, 4% (v/v) DMSO-d6, 1 mM DTT, EDTA-free protease inhibitor cocktail (Roche, Basel, Switzerland) and 10% (v/v) D$_2$O. The experiments were recorded with 3072 complex data points in the direct dimension and 128 complex data points in the indirect dimension, with 184 scans per increment. For the evaluation of the binding of (-)-**2** to 14-3-3σΔC, spectra of $^{15}$N$^{2}$H labeled 14-3-3σ 100 µM were recorded in the presence and absence of 2000 µM (-)-**2**. For the evaluation of the binding of (-)-**2** to the 14-3-3σΔC/ERα complex, spectra of $^{15}$N$^{2}$H labeled 14-3-3σΔC 100 µM with 60µM ERα peptide **3** were recorded in the presence and absence of 1000 µM and 2000 µM (-)-**2**. For the evaluation of the binding of (+)-**2** to 14-3-3σΔC, spectra of $^{15}$N$^{2}$H labeled 14-3-3σ 100 µM were recorded in the presence and absence of 2000 µM (+)-**2**. The reference for the $^{1}$H chemical shift was relative to TMSP (trimethylsilylpropanoic acid) while $^{15}$N chemical shift values were referenced indirectly. Assignments of the backbone resonances of $^{15}$N$^{2}$H labeled 14-3-3σΔC were previously reported.[1] Spectra were collected and processed with Topspin 4.0 (Bruker Biospin, Karlsruhe, Germany) and analyzed with Sparky 3.12 (T. D. Goddard and D. G. Kneller, SPARKY 3, University of California, San Francisco).

### IIc) WaterLOGSY NMR experiments

WaterLOGSY spectra were acquired at the temperature of 16 °C in 5 mm tubes (sample volume 530 µL) using a 600 MHz Bruker Avance III HD spectrometer equipped with a CPQCI cryogenic probe. The spectra were recorded with 32768 complex data points, with 384 scans per increment and with a mixing time of 1.7s (acquisition time of 35 minutes). All samples were prepared in a buffer containing 100 mM sodium phosphate, 50 mM NaCl, pH 6.8 and 10% (v/v) D$_2$O. The final concentration of DMSO-d6 was 2% (v/v) and was kept constant for all experiments. To examine the binding of both (-)-**2** and (+)-**2** to either 14-3-3σ alone or in complex to the ERα peptide, WaterLOGSY spectra were recorded on solutions containing each of the enantiomers of **2** at 500 µM in the presence of 25 µM 14-3-3σΔC alone or together with 35 µM ERα peptide **3**, respectively. A $^{1}$H spectrum with water-suppression was additionally recorded for each

|  | 14-3-3σΔC/ERα/(*R*)-2 | 14-3-3σΔC/ERα/(*R*)-6 |
|---|---|---|
| PDB code | 6TJM | 6TL3 |

The sample. reference $^1$H chemical shift was relative for the to TMSP (Trimethylsilylpropanoic acid). Spectra were collected, processed and analyzed with Topspin 3.6 (Bruker Biospin, Karlsruhe, Germany).

## III.   Protein purification and X-ray crystallography

*Protein purification*

Prior to purification, the cell pellets were thawed and resuspended in 10 mL/g pellet lysis buffer (25 mM Tris, pH = 8.0, 150 mM NaCl, 5% v/v glycerol, 10 mM imidazole, 4 mM BME and 1 mM PMSF). The cells were then lysed twice by homogenization using an EmulsiFlex-C3 homogenizer. The lysate was incubated with benzonase (Merck Millipore) for 15 minutes and then centrifuged at 20000 g for 15 minutes. The supernatant was applied in overnight circulation at 4 °C to a 5 mL HisTrap column pre-equilibrated with 20 column volumes (CV) lysis buffer. The column was then washed with 20 CV wash buffer (25 mM Tris, pH = 8.0, 300 mM NaCl, 5% v/v glycerol, 25 mM imidazole and 4 mM BME) and the protein eluted with 40 mL elution buffer (20 mM HEPES, pH 8.0, 100 mM NaCl, 5% v/v glycerol, 250 mM imidazole and 4 mM BME). The protein was then pipetted into a SpectrumLabs Spectra/Por 10000 Da MWCO dialysis bag and dialysed overnight at 4 °C against dialysis buffer (25 mM HEPES, pH = 8.0, 100 mM NaCl, 4 mM BME, 2 mM MgCl2). For the σΔC protein, 1:500 mg/mg TEV protease was added to the dialysis bag. The full length proteins were then concentrated to ~50 mg/mL using 10000 Da MWCO Amicon spinfilters, aliquoted, flash frozen in liquid nitrogen and stored at -80 °C until further usage. The σΔC protein was instead applied to a 5 mL HisTrap column pre-equilibrated with 20 CV dialysis buffer. The flowtrough was captured and concentrated to ~50 mg/mL using 10000 Da MWCO Amicon spinfilters. The concentrated σΔC protein was then applied to a HiLoad superdex 75 16/60 SEC column using an Äkta FPLC apparatus. The fractions containing protein were then pooled, concentrated to ~50 mg/mL protein, flash frozen in liquid nitrogen and stored at -80 °C until further use.

*Crystallography – data collection and analysis*

X-ray diffraction data for the 14-3-3σΔC/ERα/(*R*)-**2** complex was collected at 100 K at the p11 beamline of the PETRA-III synchrotron of the DESY facility in Hamburg, Germany using a Pilatus 6M-F detector.[2,3] X-ray diffraction data for the 14-3-3σΔC/ERα/(*R*)-**6** complex was collected at 100 K on a Rigaku Micromax-003 sealed tube X-ray source and a Dectris Pilatus 200K detector. The data was indexed, integrated, scaled and merged using xia2 DIALS.[4] Phasing was done by molecular replacement using Phaser[5] and 4JC3 as a starting model and was followed by iterative rounds of refinement and manual model building using Phenix.Refine[6] and Coot[7] respectively. Model validation was performed using MolProbity.[8] Figures were created using PyMol.

| | Data collection | | |
|---|---|---|---|
| | Resolution (Å)[a] | 1.85 (1.85 – 1.88) | 2.46 (2.50 – 2.45) |
| | Space group | C222 | C222 |
| | Cell parameters (Å)[b] | a = 62.84, b = 152.31, c = 76.79<br><br>α = β = γ = 90 ° | a = 63.69, b = 152.23, c = 76.26<br><br>α = β = γ = 90 ° |
| | $R_{merge}$[a,b] | 0.042 (1.17) | 0.17 (0.64) |
| | Average $I/\sigma_{(I)}$[a,b] | 24.0 (1.0) | 7.72 (2.03) |
| | $CC_{1/2}$ (%)[a,b,c] | 100 (80.3) | 99.1 (86.3) |
| | Completeness (%)[a,b] | 99.9 (99.1) | 99.8 (100) |
| | Redundancy[a,b] | 12.5 (11.6) | 6.2 (6.6) |
| | **Refinement** | | |
| | Number of protein/solvent/ligand atoms | 3770/127/50 | 3749/194/49 |
| | $R_{work}/R_{free}$ (%) | 19.7/21.8 | 21.0/25.9 |
| | Unique reflections used in refinement | 31888 | 13882 |
| | R.m.s. deviations from ideal values bond lengths (Å) / bond angles (°) | 0.006/0.617 | 0.006/0.600 |
| | Average protein/solvent/ligand B-factor (Å$^2$) | 52.5/55.7/47.7 | 36.4/40.7/34.21 |
| | Ramachandran favored (%) | 98.22 | 97.25 |
| | Ramachandran allowed (%) | 1.78 | 2.75 |
| | Ramachandran outliers (%) | 0 | 0 |

[a] Number in parentheses is for the highest resolution shell
[b] As reported by xia2 DIALS.
[c] $CC_{1/2}$ = Pearson's intradataset correlation coefficient, as described by Karplus and Diederichs.[9]

## IV. Chemistry section

### IVa) General information

All solvents and reagents were obtained from commercially available sources and used without further purification. The microwave syntheses were performed in a Biotage Initiator with an external surface IR probe. Flash column chromatography was carried out on prepacked silica gel columns supplied by Biotage and using Biotage automated flash systems with UV detection.

UHPLC-MS experiments were performed using a Waters Acquity UHPLC system combined with a SQD mass spectrometer. The UHPLC system was equipped with both a BEH C18 column 1.7 μm 2.1×50 mm in combination with a 46 mM $(NH_4)_2CO_3/NH_3$ buffer at pH 10 and a HSS C18 column 1.8 μm 2.1×50 mm in combination with 10 mM formic acid or 1 mM ammonium formate buffer at pH 3. The mass spectrometer used ESI+/- as ion source. UPLC was also carried out using a Waters UPLC fitted with Waters QDa mass spectrometer (Column temp 40°C, UV = 190–400 nm, MS = ESI with pos/neg switching) equipped with a Waters Acquity BEH 1.7 μm 2.1×100 mm in combination with either 0.1% formic acid in water, 0.05% TFA in water or 0.04% $NH_3$ in water. The flow rate was 1 mL/min.

Preparative HPLC was performed by Waters Fraction Lynx with ZQ MS detector on either a Waters Xbridge C18 OBD 5 μm column (19×150 mm, flow rate 30 mL/min or 30×150 mm, flow rate 60 mL/min) using a gradient of 5–95% MeCN with 0.2% $NH_3$ at pH 10 or a Waters SunFire C18 OBD 5 μm column (19×150 mm, flow rate 30 mL/min or 30×150 mm, flow rate 60 mL/min) using a gradient of 5–95% MeCN with 0.1 M formic acid or on a Gilson Preparative HPLC with a UV/VIS detector 155 on a Kromasil C8 10 μm column (20 × 250 mm, flow rate 19 mL/min, or 50 × 250 mm, flow rate 100 mL/min) using a varying gradient of ACN with 0.1% formic acid (FA) in water or 0.2% trifluoroacetic acid (TFA) in water or 0.2% acetic acid (AcOH) in water or 0.2% ammonia ($NH_3$) in water. Molecular mass (HR-ESI-MS) was recorded using a Shimadzu LCMS-2020 instrument (ESI+). Purity of all test compounds was determined by LCMS. All screening compounds had a purity >95%.

General $^1H$ NMR spectra were recorded on a Bruker Avance II, III, AV300, AV400 or AVIII500 spectrometer at a proton frequency of 400, 500 or 600 MHz at 25 °C or at a temperature and frequency stated in each experiment. $^{13}C$ NMR spectra were recorded at 101 MHz or 126 MHz.

The chemical shifts (δ) are reported in parts per million (ppm) with residual solvent signal used as a reference ($CD_2Cl_2$ at 5.32 ppm for $^1H$ NMR and 53.84 ppm for $^{13}C$ NMR, $(CD_3)_2SO$ at 2.50 ppm for $^1H$ NMR and 39.52 ppm for $^{13}C$ NMR, $CDCl_3$ at 7.26 ppm for $^1H$ NMR and 77.16 ppm for $^{13}C$ NMR). Coupling constants (J) are reported as Hz. NMR abbreviations are used as follows: br = broad, s = singlet, d = doublet, t = triplet, q = quartet, m = multiplet. Protons on heteroatoms such as COO<u>H</u> protons are only reported when detected in NMR and can therefore be missing.

The PMA2 peptide (C-terminal, 52 mer, amino acids 905-956) expression and purification protocol is reported at ref[10].

### IVb) Synthetic procedures and compound characterization

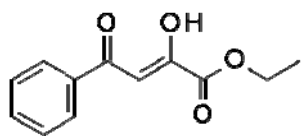

**Ethyl 2,4-dioxo-4-phenyl-butanoate.**
In a 250 mL RBF, acetophenone (2.92 mL, 24.97 mmol) was dissolved in THF (100 mL) and the resulting solution cooled down to 0 °C. Sodium ethanolate (13.98 mL, 37.45 mmol) was then added dropwise and the reaction allowed to stir for 15min at 0 °C. Diethyl oxalate (3.73 mL, 27.47 mmol) was finally added dropwise, the cooling bath removed, and the reaction allowed to stir overnight at rt.
The reaction was quenched with 1M HCl (50 mL). The resulting suspension was poured into a separatory funnel and the crude product was extracted with DCM (3x). The combined organic layers were dried using a phase separator and solvent was removed under reduced pressure.
The crude product was purified by preparative HPLC (40-80% ACN in H$_2$O/ACN/AcOH 95/5/0.2 buffer over 20 minutes). Collected fractions were freeze-dried, to give ethyl 2,4-dioxo-4-phenyl-butanoate (4.25 g, 77%) as a yellow solid.
$^1$H NMR (400 MHz, CDCl$_3$) δ 7.99 (d, J = 7.4 Hz, 2H), 7.60 (t, J = 7.4 Hz, 1H), 7.50 (t, J = 7.6 Hz, 2H), 7.07 (s, 1H), 4.39 (q, J = 7.1 Hz, 2H), 1.41 (t, J = 7.1 Hz, 3H). $^{13}$C NMR (101 MHz, CDCl$_3$) δ 190.82, 169.89, 162.29, 135.00, 133.89, 129.00, 127.99, 98.05, 62.72, 14.21. Matches with a previously reported characterization.[11]

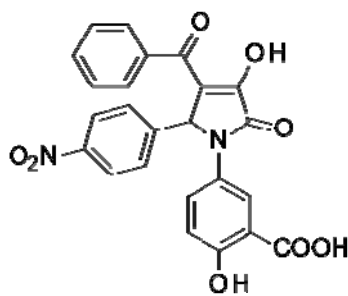

**5-[3-benzoyl-4-hydroxy-2-(4-nitrophenyl)-5-oxo-2H-pyrrol-1-yl]-2-hydroxy-benzoic acid (2).**
In a 20 mL vial, to a solution of ethyl 2,4-dioxo-4-phenylbutanoate (603 mg, 2.74 mmol) in AcOH (7 mL), 4-nitrobenzaldehyde (422 mg, 2.74 mmol) and 5-amino-2-hydroxybenzoic acid (441 mg, 2.74 mmol) were added. The vial was capped and heated at 120 °C for 180 min in a single node microwave reactor. The pressure monitored was 1 bar.
The mixture was diluted with diethyl ether and filtered. The residue was washed with diethyl ether, dried under reduced pressure, to give compound **2** (543 mg, 43%) as a pale yellow solid.
$^1$H NMR (400 MHz, DMSO) δ 7.97 – 8.18 (m, 3H), 7.66 – 7.77 (m, 5H), 7.52 – 7.61 (m, 1H), 7.45 (t, J = 7.6 Hz, 2H), 6.91 (d, J = 8.9 Hz, 1H), 6.46 (s, 1H). $^{13}$C NMR (101 MHz, DMSO) δ 189.01, 171.17, 164.47, 158.76, 151.04, 147.20, 144.40, 137.87, 132.70, 130.55, 129.26, 128.71, 128.17, 127.49, 124.91, 123.51, 119.08, 117.57, 113.01, 60.83. HRMS (ESI) *m/z* [M + H]$^+$ calcd for C$_{24}$H$_{16}$N$_2$O$_8$: 461.0985, found: 461.0974. Matches with a previously reported characterization.[11]

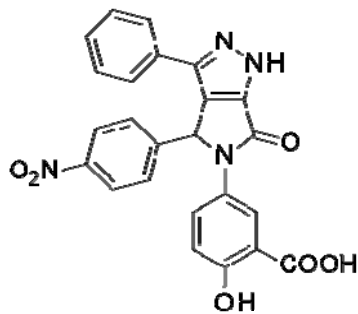

**2-hydroxy-5-[4-(4-nitrophenyl)-6-oxo-3-phenyl-1,4-dihydropyrrolo[3,4-c]pyrazol-5-yl]benzoic acid (5).**

In 20 mL vial, compound **2** (98 mg, 0.21 mmol) was suspended in AcOH (6 mL). Hydrazine (35% in water) (0.025 mL, 0.28 mmol) was added, while stirring at room temperature. The vial was capped and heated at 120 °C for 120 min in a single node microwave reactor. The pressure monitored was 1 bar. After solvent removal, the crude mixture was dissolved in EtOAc and washed with HCl aq 1M (3x). The organic layer was dried using a phase separator and concentrated under reduced pressure. The residue was purified by preparative HPLC (15-35% acetonitrile in $H_2O$/ACN/$NH_3$ 95/5/0.2 buffer over 20 minutes), to give compound **5** (82 mg, 84.0%) as a off-white solid.

$^1$H NMR (500 MHz, DMSO) δ 8.07 (d, *J* = 8.7 Hz, 2H), 7.97 (d, *J* = 2.6 Hz, 1H), 7.61 – 7.71 (m, 3H), 7.59 (d, *J* = 7.5 Hz, 2H), 7.37 (t, *J* = 7.5 Hz, 2H), 7.29 (t, *J* = 7.3 Hz, 1H), 7.02 (s, 1H), 6.93 (d, *J* = 8.9 Hz, 1H). $^{13}$C NMR (126 MHz, DMSO) δ 171.30, 161.19, 159.21, 150.07, 147.35, 144.55, 136.83, 130.95, 129.28, 129.04, 128.78, 128.04, 127.54, 125.93, 124.87, 123.92, 117.19, 114.57, 59.33. HRMS (ESI) *m/z* [M + H]$^+$ calcd for $C_{24}H_{16}N_4O_6$ : 457.1148, found: 457.1147. Matches with a previously reported characterization.[11]

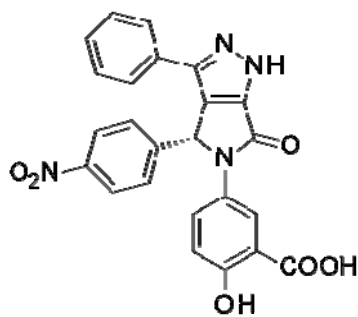

**(*R*)-2-hydroxy-5-[4-(4-nitrophenyl)-6-oxo-3-phenyl-1,4-dihydropyrrolo[3,4-c]pyrazol-5-yl]benzoic acid ((*R*)-5).**

In 5 mL vial, compound **(*R*)-2** (110 mg, 0.24 mmol) was suspended in AcOH (3 mL). Hydrazine (35% in water) (0.028 mL, 0.31 mmol) was added, while stirring at room temperature. The vial was capped and heated at 120 °C for 120 min in a single node microwave reactor. The pressure monitored was 1 bar. After solvent removal, the residue was purified by preparative HPLC (25-65% acetonitrile in $H_2O$/ACN/FA 95/5/0.2 buffer over 20 minutes), to give compound **(*R*)-5** (51 mg, 46.8 %) as a off-white solid.

$^1$H NMR (500 MHz, DMSO) δ 8.02 – 8.11 (m, 2H), 7.94 (d, J = 2.7 Hz, 1H), 7.6 – 7.69 (m, 3H), 7.58 (d, J = 7.4 Hz, 2H), 7.39 (br, 2H), 7.25 – 7.35 (br, 1H), 7.02 (s, 1H), 6.92 (d, J = 8.9 Hz, 1H). $^{13}$C NMR (126 MHz,

DMSO) δ 171.31, 161.13, 158.70, 149.91, 147.34, 144.46, 136.75, 131.55, 129.24, 129.07, 128.83, 128.58, 127.46, 125.89, 125.71, 124.80, 123.91, 117.47, 113.08, 59.16. HRMS (ESI) *m/z* [M + H]$^+$ calcd for C$_{24}$H$_{16}$N$_4$O$_6$ : 457.1148, found: 457.1158. [α]$_D^{20}$: +20.0 (*c* 0.125, MeOH).

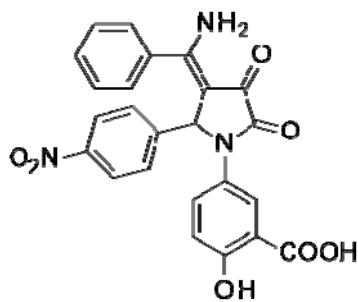

**5-[3-[amino(phenyl)methylene]-2-(4-nitrophenyl)-4,5-dioxo-pyrrolidin-1-yl]-2-hydroxy-benzoic acid (6).**
In a 5 mL vial, to a suspension of **2** (100 mg, 0.22 mmol) in AcOH (3 mL), ammonium hydroxide (0.328 mL, 2.17 mmol) was added. The vial was capped and heated at 120 °C for 120 min in a single node microwave reactor. The pressure monitored was 1 bar. After solvent removal, the residue was purified by preparative HPLC (10-50% acetonitrile in H$_2$O/ACN/FA 95/5/0.2 buffer over 20 minutes), to give compound **6** (20 mg, 20.0 %) as a yellow solid.

$^1$H NMR (500 MHz, DMSO) δ 10.18 (d, *J* = 3.3 Hz, 1H, H36), 9.02 (d, *J* = 3.3 Hz, 1H, H35), 7.93 (d, *J* = 2.7 Hz, 1H, H10), 7.72 (d, *J* = 8.7 Hz, 2H, H18 and H20), 7.60 (dd, *J* = 9.0, 2.7 Hz, 1H, H6), 7.41 – 7.49 (m, 1H, H31), 7.3 – 7.39 (m, 4H, H29, H30, H32 and H33), 6.92 (d, *J* = 8.5 Hz, 2H, H17 and H21), 6.85 (d, *J* = 9.0 Hz, 1H, H7), 6.36 (s, 1H, H15). $^{13}$C NMR (151 MHz, DMSO) δ 178.00, 171.13, 164.05, 163.17, 158.93, 146.64, 146.27, 133.49, 130.86, 130.52, 128.56, 128.44, 127.85, 127.63, 125.25, 122.77, 117.31, 113.11, 105.53, 58.89. HRMS (ESI) *m/z* [M + H]$^+$ calcd for C$_{24}$H$_{17}$N$_3$O$_7$ : 460.1145, found: 460.1147.

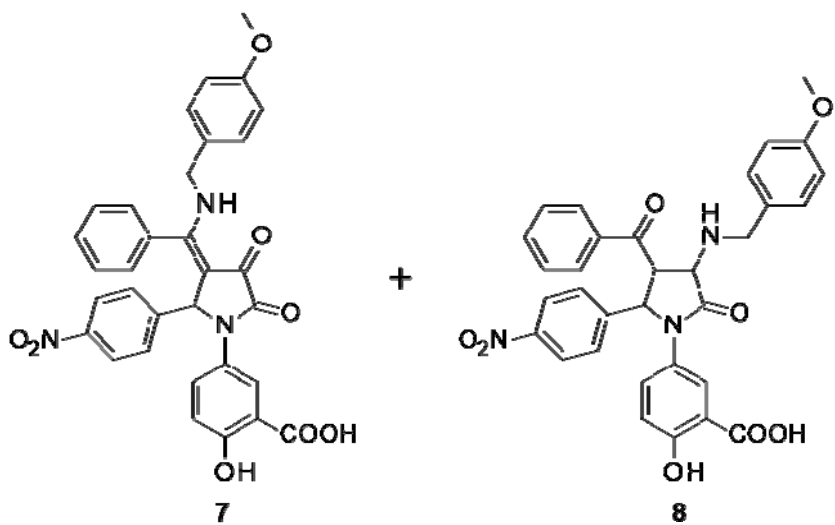

**2-hydroxy-5-[3-[[(4-methoxyphenyl)methylamino]-phenyl-methylene]-2-(4-nitrophenyl)-4,5-dioxo-pyrrolidin-1-yl]benzoic acid (7) and 5-[3-benzoyl-4-[(4-methoxyphenyl)methylamino]-2-(4-nitrophenyl)-5-oxo-2H-pyrrol-1-yl]-2-hydroxy-benzoic acid (8)**
In a 20 mL vial, to compound **2** (536 mg, 1.16 mmol) in AcOH (5 mL), (4-methoxyphenyl)methanamine (1.086 mL, 8.15 mmol) was added. The vial was capped and heated at 120 °C for 120 min in a single node microwave reactor. The pressure monitored was 1 bar. After solvent removal, the residue was purified by preparative HPLC (45-85% acetonitrile in H$_2$O/ACN/TFA 95/5/0.2 buffer over 20 minutes), to give compound **7** (40 mg, 5.93%) and compound **8** (174 mg, 25.8%), both as yellow solids.

**7**: $^1$H NMR (500 MHz, DMSO) δ 11.33 (t, J = 6.2 Hz, 1H), 7.88 (d, J = 2.7 Hz, 1H), 7.73 (d, J = 8.7 Hz, 2H), 7.62 (br, 2H), 7.55 (dd, J = 9.0, 2.7 Hz, 1H), 7.45 (t, J = 7.5 Hz, 1H), 7.07 (d, J = 8.6 Hz, 3H), 6.89 (d, J = 8.6 Hz, 2H), 6.75 – 6.85 (m, 3H), 6.51 (br, 1H), 6.05 (s, 1H), 4.31 (dd, J = 15.0, 5.3 Hz, 1H), 4.19 (dd, J = 14.9, 7.2 Hz, 1H), 3.73 (s, 3H).$^{13}$C NMR (126 MHz, DMSO) δ 177.19, 171.19, 164.30, 163.32, 158.85, 158.80, 146.47, 146.33, 130.99, 130.14, 129.07, 128.79, 128.38, 128.10, 127.92, 126.97, 125.23, 122.93, 117.41, 114.12, 112.83, 106.90, 104.55, 59.09, 55.12, 47.10. HRMS (ESI) *m/z* [M + H]$^+$ calcd for C$_{32}$H$_{25}$N$_3$O$_8$ : 580.1720, found: 580.1722.

**8**: $^1$H NMR (600 MHz, DMSO) δ 8.96 (br, 1H), 7.93 (d, J = 2.7 Hz, 1H), 7.80 (d, J = 8.7 Hz, 2H), 7.58 (dd, J = 9.0, 2.7 Hz, 1H), 7.37 – 7.45 (m, 3H), 7.32 (t, J = 7.6 Hz, 2H), 7.25 (br, 2H), 7.07 (d, J = 8.7 Hz, 2H), 6.90 (d, J = 8.4 Hz, 2H), 6.85 (d, J = 8.9 Hz, 1H), 6.47 (s, 1H), 4.55-5.19 (br d, 2H), 3.73 (s, 3H). $^{13}$C NMR (126 MHz, DMSO) δ 190.23, 171.08, 163.78, 159.12, 158.48, 146.59, 145.24, 140.16, 131.23, 130.82, 129.02, 128.94, 128.16, 127.13, 126.98, 125.85, 123.00, 117.35, 113.90, 113.39, 111.37, 62.05, 55.07, 45.10. HRMS (ESI) *m/z* [M - H]$^-$ calcd for C$_{32}$H$_{25}$N$_3$O$_8$ : 578.1564, found: 578.1559.

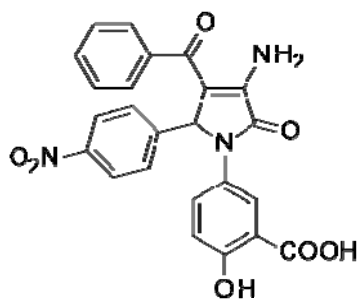

**5-[4-amino-3-benzoyl-2-(4-nitrophenyl)-5-oxo-2H-pyrrol-1-yl]-2-hydroxy-benzoic acid (9).**
In a 5 mL vial, compound **8** (50 mg, 0.09 mmol) was dissolved in trifluoroacetic acid (1 mL, 0.09 mmol). The vial was capped and heated at 120 °C for 30 min in a single node microwave reactor. The pressure monitored was 3 bar. After solvent removal, the residue was purified by automated flash chromatography on a Biotage® KP-SIL 10 g column (40-100% of heptane in EtOAc + 2% FA over 18CV), to give compound **9** (21 mg, 53.0%) as a pale yellow solid.

$^1$H NMR (500 MHz, DMSO) δ 11.15 (br, 1H, H32), 7.98 (d, J = 2.5 Hz, 1H, H26), 7.80 (d, J = 8.4 Hz, 2H, H18 and H20), 7.76 (br, 2H, H34), 7.65 (dd, J = 8.9, 2.5 Hz, 1H, H22), 7.48 (d, J = 7.7 Hz, 2H, H12 and H16), 7.43 (t, J = 7.1 Hz, 1H, H14), 7.34 (t, J = 7.5 Hz, 2H, H13 and H15), 7.13 (d, J = 8.5 Hz, 2H, H17 and H21), 6.89 (d, J = 8.9 Hz, 1H, H23), 6.58 (s, 1H, H2). $^{13}$C NMR (126 MHz, DMSO) δ 190.40, 171.19, 163.75, 158.92, 148.27, 146.61, 145.29, 140.12, 131.26, 130.73, 128.99, 128.23, 127.31, 126.88, 125.48, 123.05, 117.54, 112.96, 110.22, 62.13. HRMS (ESI) *m/z* [M + H]$^+$ calcd for C$_{24}$H$_{17}$N$_3$O$_7$ : 460.1145, found: 460.1137.

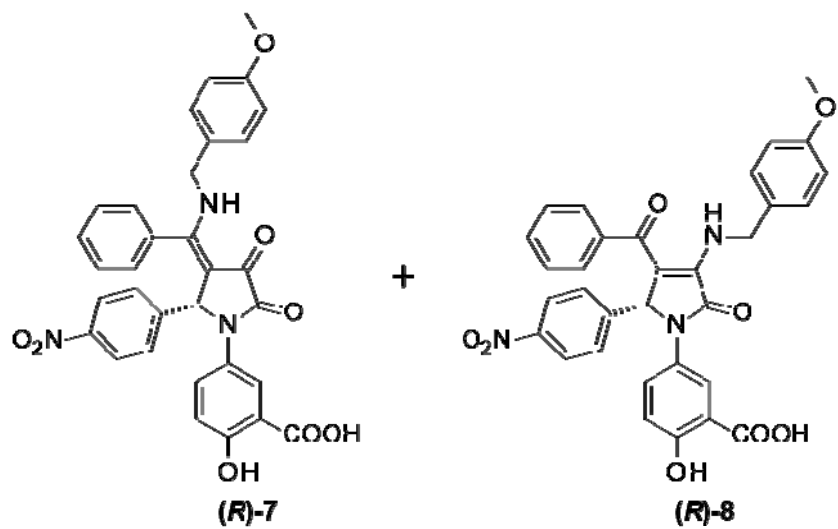

**(R)-2-hydroxy-5-[3-[[(4-methoxyphenyl)methylamino]-phenyl-methylene]-2-(4-nitrophenyl)-4,5-dioxo-pyrrolidin-1-yl]benzoic acid ((R)-7) and (R)-5-[3-benzoyl-4-[(4-methoxyphenyl)methylamino]-2-(4-nitrophenyl)-5-oxo-2H-pyrrol-1-yl]-2-hydroxy-benzoic acid ((R)-8).**

In a 20 mL vial, to compound (R)-**2** (78 mg, 0.17 mmol) in AcOH (2 mL), (4-methoxyphenyl)methanamine (0.158 mL, 1.19 mmol) was added. The vial was capped and heated at 120 °C for 120 min in a single node microwave reactor. The pressure monitored was 1 bar. After solvent removal, the residue was purified by preparative HPLC (45-85% acetonitrile in $H_2O$/ACN/TFA 95/5/0.2 buffer over 30 minutes). Collected fractions were freeze-dried, to give compounds (R)-**7** (10 mg, 10.2%) and (R)-**8** (16 mg, 16.3%), both as yellow solids. (R)-**7**: m/z (ESI-MS): $[M+H^+]^+$ calculated mass = 580.2, observed = 580.4. (R)-**8**: m/z (ESI-MS): $[M-H^+]^-$ calculated mass = 578.2, observed = 578.3 Compounds were not further characterized but used directly for the next step.

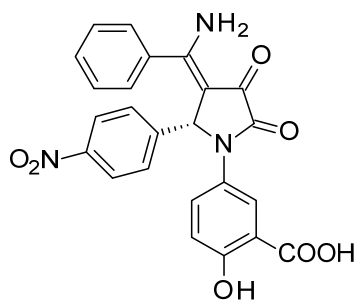

**(R)-5-[3-[amino(phenyl)methylene]-2-(4-nitrophenyl)-4,5-dioxo-pyrrolidin-1-yl]-2-hydroxy-benzoic acid ((R)-6).**

In a 5 mL vial, compound (R)-**7** (10 mg, 0.02 mmol) was dissolved in trifluoroacetic acid (1 mL, 0.09 mmol). The vial was capped and heated at 120 °C for 30 min in a single node microwave reactor. The pressure monitored was 2 bar. After solvent removal, the residue was purified by preparative HPLC (25-50% acetonitrile in $H_2O$/ACN/TFA 95/5/0.2 buffer over 20 minutes). Collected fractions were freeze-dried, to give compound (R)-**6** (3.2 mg, 40.5%, 91.8% ee) as a pale yellow solid.

$^1$H NMR (600 MHz, DMSO) δ 11.11 (br, 1H), 10.18 (br, 1H), 8.99 (br, 1H), 7.93 (d, J = 2.7 Hz, 1H), 7.72 (d, J = 8.4 Hz, 2H), 7.61 (dd, J = 9.0, 2.7 Hz, 1H), 7.42 – 7.48 (m, 1H), 7.3 – 7.39 (m, 4H), 6.93 (br, J = 5.9 Hz, 2H),

6.86 (d, J = 9.0 Hz, 1H), 6.35 (s, 1H). $^{13}$C NMR (151 MHz, DMSO) δ 177.92, 171.14, 164.01, 163.18, 158.82, 146.61, 146.28, 133.48, 130.98, 130.52, 128.56, 128.42, 127.95, 127.65, 125.22, 122.78, 117.36, 112.82, 105.49, 58.85. HRMS (ESI) *m/z* [M + H]$^+$ calcd for C$_{24}$H$_{17}$N$_3$O$_7$ : 460.1145, found: 460.1156.

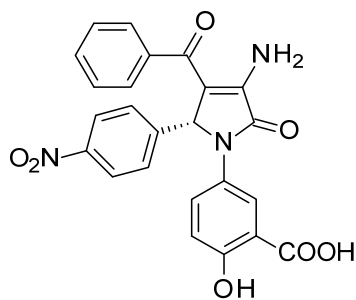

**(*R*)-5-[4-amino-3-benzoyl-2-(4-nitrophenyl)-5-oxo-2H-pyrrol-1-yl]-2-hydroxy-benzoic acid ((*R*)-9).**
In a 5 mL vial, compound **(*R*)-8** (16 mg, 0.03 mmol) was dissolved in trifluoroacetic acid (1 mL, 0.09 mmol). The vial was capped and heated at 120 °C for 30 min in a single node microwave reactor. The pressure monitored was 2 bar. After solvent removal, the residue was purified by preparative HPLC (35-65% acetonitrile in H$_2$O/ACN/TFA 95/5/0.2 buffer over 20 minutes). Collected fractions were freeze-dried, to give compound (*R*)-**9** (6.4 mg, 50.5%, 89.8% ee) as a pale yellow solid.
$^1$H NMR (600 MHz, DMSO) δ 11.13 (br, 1H), 7.97 (d, J = 2.7 Hz, 1H), 7.81 (d, J = 8.8 Hz, 2H), 7.73 (br, 2H), 7.65 (dd, J = 9.0, 2.7 Hz, 1H), 7.46 – 7.5 (m, 2H), 7.43 (t, J = 7.4 Hz, 1H), 7.34 (t, J = 7.6 Hz, 2H), 7.13 (d, J = 8.8 Hz, 2H), 6.89 (d, J = 9.0 Hz, 1H), 6.57 (s, 1H). $^{13}$C NMR (151 MHz, DMSO) δ 190.36, 171.11, 163.71, 158.87, 148.19, 146.60, 145.25, 140.08, 131.21, 130.67, 128.95, 128.18, 127.28, 126.83, 125.45, 123.00, 117.49, 112.94, 110.21, 62.13. HRMS (ESI) *m/z* [M + H]$^+$ calcd for C$_{24}$H$_{17}$N$_3$O$_7$ : 460.1145, found: 460.1153.

## IVc) Chiral separation, VCD analysis and racemisation studies of 2

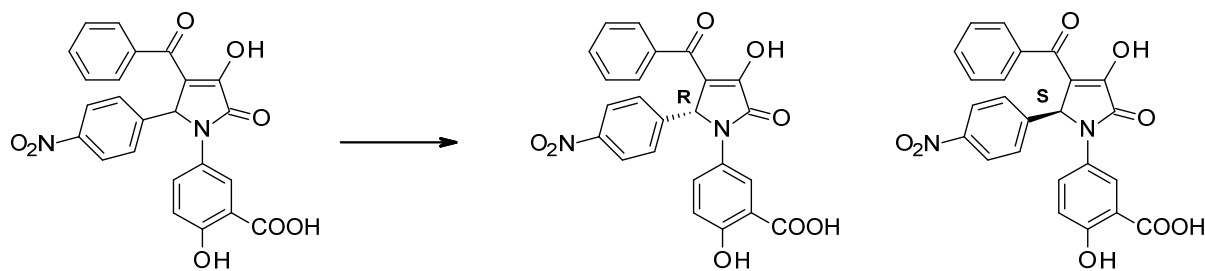

**Chiral separation**

(***R*** and ***S***)-5-[3-benzoyl-4-hydroxy-2-(4-nitrophenyl)-5-oxo-2H-pyrrol-1-yl]-2-hydroxy-benzoic acid (-)-2 and (+)-2).

The enantiomers of compound **2** (1.2 g, 2.61 mmol) were separated by chiral column chromatography on a Chiralpak IC (250x20 mm, 5 μm) column. 50 mg (50 mg/mL in EtOH/TEA 10:0.1) were injected and eluted with 100% EtOH/TEA (100:0.1), 120 bar at 25 °C, a flow rate of 12 mL/min and detected at 270 nm. The first eluted compound was collected and evaporated to give compound (-)-**2** (759 mg, 97.3% ee) as a yellow solid. $[\alpha]_D^{20}$: -96.8 (*c* 0.5, MeOH). $^1$H NMR (500 MHz, DMSO) δ 8.01 – 8.11 (m, 3H), 7.68 – 7.77 (m, 5H), 7.52 – 7.59 (m, 1H), 7.44 (t, *J* = 7.7 Hz, 2H), 6.92 (d, *J* = 8.9 Hz, 1H), 6.47 (s, 1H). $^{13}$C NMR (126 MHz, DMSO) δ 189.13, 171.32, 164.58, 158.88, 151.15, 147.28, 144.48, 137.95, 132.80, 130.64, 129.35, 128.82, 128.26, 127.58, 124.99, 123.61, 119.18, 117.68, 113.08, 60.92. HRMS (ESI) *m/z* [M + H]$^+$ calcd for $C_{24}H_{16}N_2O_8$: 461.0985, found: 461.0966.

The second eluted compound was collected and evaporated to give compound (+)-**2** (644 mg, 99.6% ee) as a yellow solid. $[\alpha]_D^{20}$: +103.6 (*c* 0.5, MeOH). 1H NMR (500 MHz, DMSO) δ 8.02 – 8.08 (m, 3H), 7.67 – 7.77 (m, 5H), 7.56 (tt, J = 7.0, 1.3 Hz, 1H), 7.41 – 7.49 (m, 2H), 6.91 (d, J = 8.9 Hz, 1H), 6.47 (s, 1H).13C NMR (126 MHz, DMSO) δ 189.10, 171.32, 164.60, 158.91, 151.27, 147.27, 144.53, 137.97, 132.77, 130.56, 129.34, 128.82, 128.24, 127.54, 125.00, 123.61, 119.12, 117.64, 113.23, 60.92. HRMS (ESI) *m/z* [M + H]$^+$ calcd for $C_{24}H_{16}N_2O_8$: 461.0985, found: 461.0987.

| **SSL Report** | **Request type** | **Separation of isomers** |
|---|---|---|

## Sample Information

| Sample ID | EN Number | Incoming amount | Compound name | Project name |
|---|---|---|---|---|
| *Rac*-2 | | 1.2 g | | |

## Preparative Conditions

| Incoming *rac*-2 | | |
|---|---|---|
| **Column** | **Dimensions (mm)** | **Particle Size (µm)** |
| Chiralpak IC | 250 x 20 | 5 |
| **Mobile Phase** | | **Gradient** |
| EtOH/TEA 100/0.1 | | |
| **Flow (ml/min)** | **Detection (nm)** | **Temperature (C)** |
| 12 | 270 | RT |
| **Injected Amount** | **Injection Volume** | **Cycle time (min)** |
| 50 mg | 1 ml | |
| **Sample Conc. (mg/ml)** | **Instrument** | **Operator** |
| 50 mg/ml in EtOH/TEA | Semi 3 | |
| **Date** | | |
| 2019-02-07 10:25:28 | | |
| **Comment** | | |
| | | |
| **Outgoing** Peak 1 (P1) and peak 2 (P2) | | |

## Analytical Conditions

| Column | Dimension (mm) | Particle Size (µm) |
|---|---|---|
| Chiralpak IB-N3 | 150 x 4.6 | 3 |
| **Mobile Phase** | | |
| 25% EtOH/DEA 100/20mM in $CO_2$, 120 bar | | |
| **Flow (ml/min)** | **Detection (nm)** | **Temperature (C)** |
| 3.5 | 247 | 40 |
| **Gradient** | **Instrument** | **Sample concentration (mg/ml)** |
| | | |

| Sample ID | Purity | Chiral purity | Weight |
|---|---|---|---|
| Peak:1 | | 97.3 ee | 700 mg |

| Sample ID | Purity | Chiral purity | Weight |
|---|---|---|---|
| Peak:2 | | 99.6 ee | 600 mg |

**Sample Name: P1**

Date Acquired: 2019-02-07 11:04:32

| | |
|---|---|
| Column: | Chiralpak IB N-3 |
| Mobile Phase A: | CO$_2$ |
| Mobile Phase B: | EtOH/DEA 100/20mM |
| Gradient: | 25% B, 120 bar |
| Temperature: | 40°C |
| Sample Concentration: | Sample in EtOH |
| Column ID: | |
| Column Dimension: | 150 * 4.6 |
| Particle Size: | 3 |
| Injection volume: | 2.00 ul |
| Flow: | 3.5 ml/min |
| Wavelength: | PDA Spectrum PDA |
| Vial: | 1:F,6 |

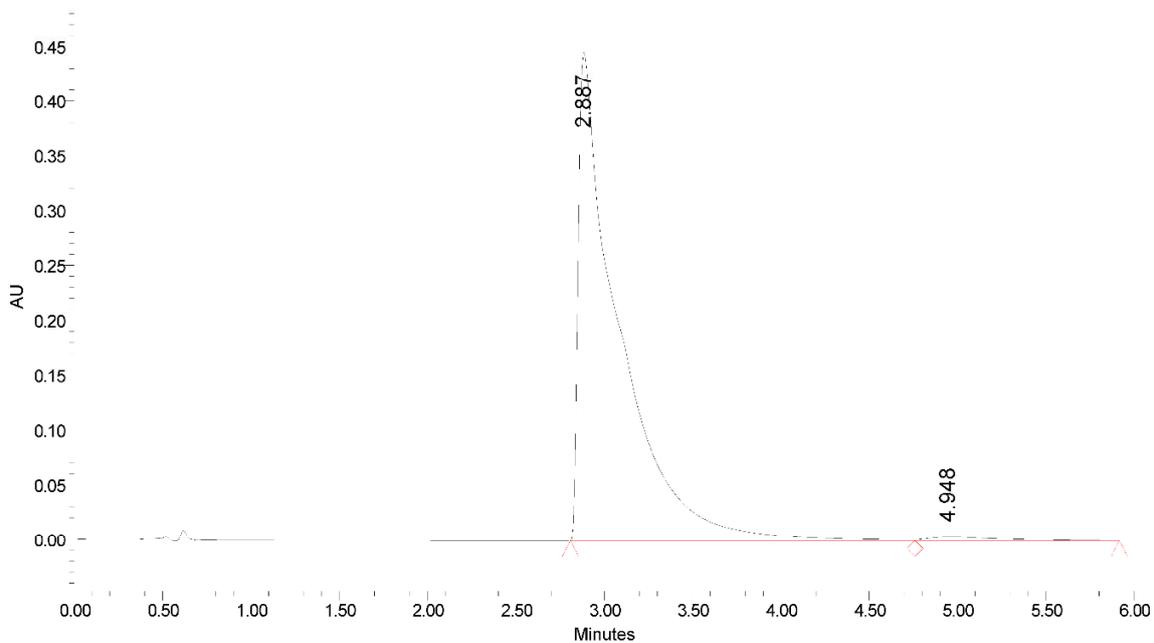

| | Retention Time | Area | % Area | k' | N | USP Resolution |
|---|---|---|---|---|---|---|
| 1 | 2.887 | 7119325 | 98.63 | 0.000 | 778.5 | |
| 2 | 4.948 | 98593 | 1.37 | 0.714 | 763.8 | 3.042 |

ee = 97.3

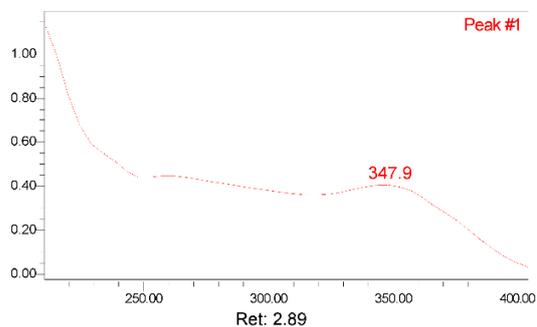

Peak #1 — Ret: 2.89 — 347.9

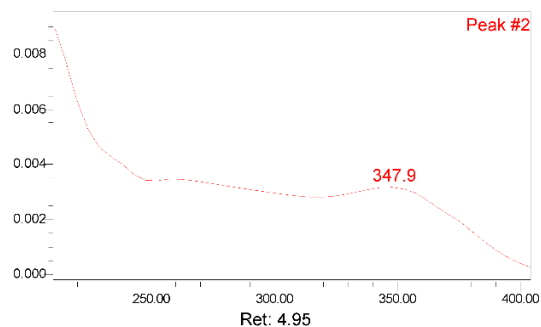

Peak #2 — Ret: 4.95 — 347.9

Report Method: ee_spektrum
Acq Method Set: 120bar_B2_25_40deg

Injection Id 6700

Sample Name: P2  
Date Acquired: 2019-02-07 10:58:08

| | |
|---|---|
| Column: | Chiralpak IB N-3 |
| Mobile Phase A: | $CO_2$ |
| Mobile Phase B: | EtOH/DEA 100/20mM |
| Gradient: | 25% B, 120 bar |
| Temperature: | 40°C |
| Sample Concentration: | Sample in EtOH |
| Column ID: | |
| Column Dimension: | 150 * 4.6 |
| Particle Size: | 3 |
| Injection volume: | 2.00 ul |
| Flow: | 3.5 ml/min |
| Wavelength: | PDA Spectrum PDA |
| Vial: | 1:F,7 |

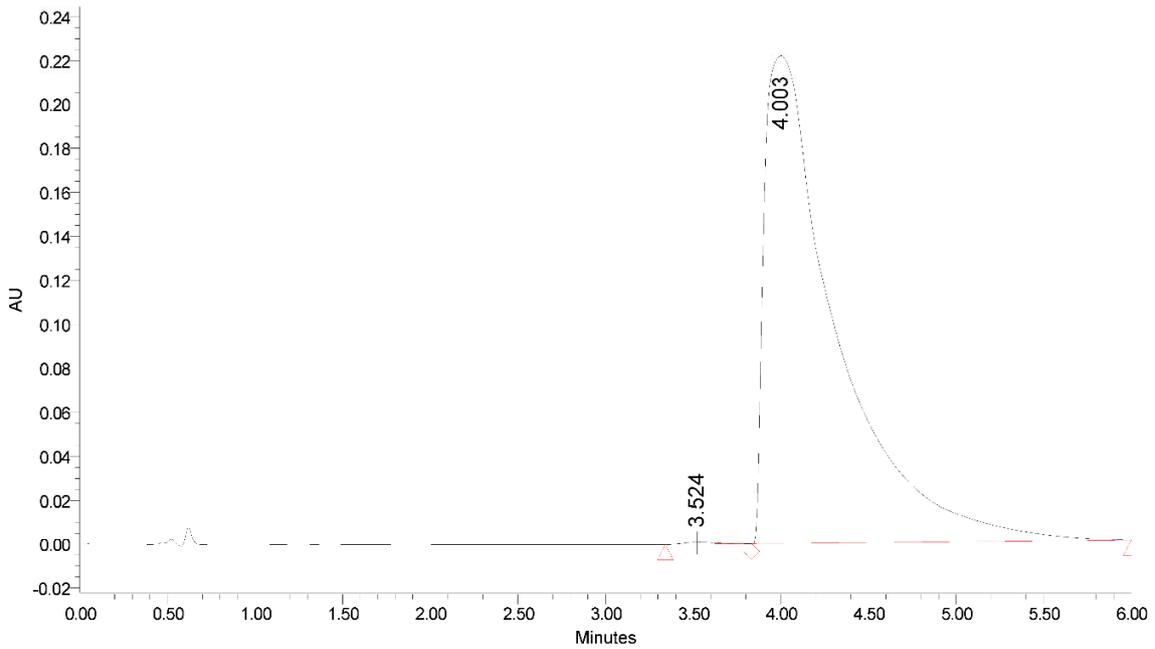

| | Retention Time | Area | % Area | k' | N | USP Resolution |
|---|---|---|---|---|---|---|
| 1 | 3.524 | 11990 | 0.19 | 0.000 | 1627.9 | |
| 2 | 4.003 | 6374643 | 99.81 | 0.136 | 598.5 | 0.878 |

ee = 99.6

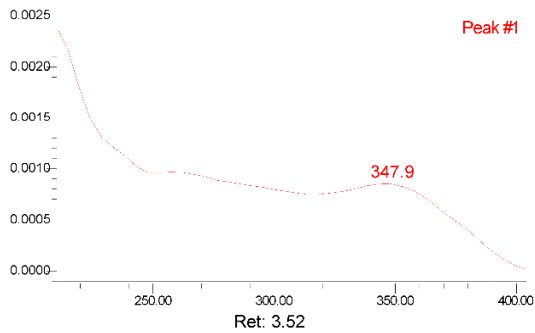

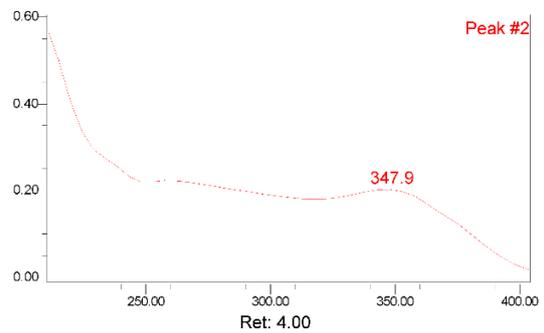

Report Method: ee_spektrum  
Acq Method Set: 120bar_B2_25_40deg

Injection Id 6693

**Sample Name:** *rac*-2  
**Date Acquired:** 2019-01-31  09:49:02

| | |
|---|---|
| **Column:** | Chiralpak IC |
| **Mobile Phase A:** | |
| **Mobile Phase B:** | EtOH/TEA 100/0.1 |
| **Gradient:** | 100%B |
| **Temperature:** | 25°C |
| **Sample Concentration:** | Sample in EtOH |
| **Column ID:** | |
| **Column Dimension:** | 150*4.6 mm |
| **Particle Size:** | 3 |
| **Injection volume:** | 15.00 ul |
| **Flow:** | 0.8 ml/min |
| **Wavelength:** | 2998 PDA 269.0 nm |
| **Vial:** | 2:a,1 |

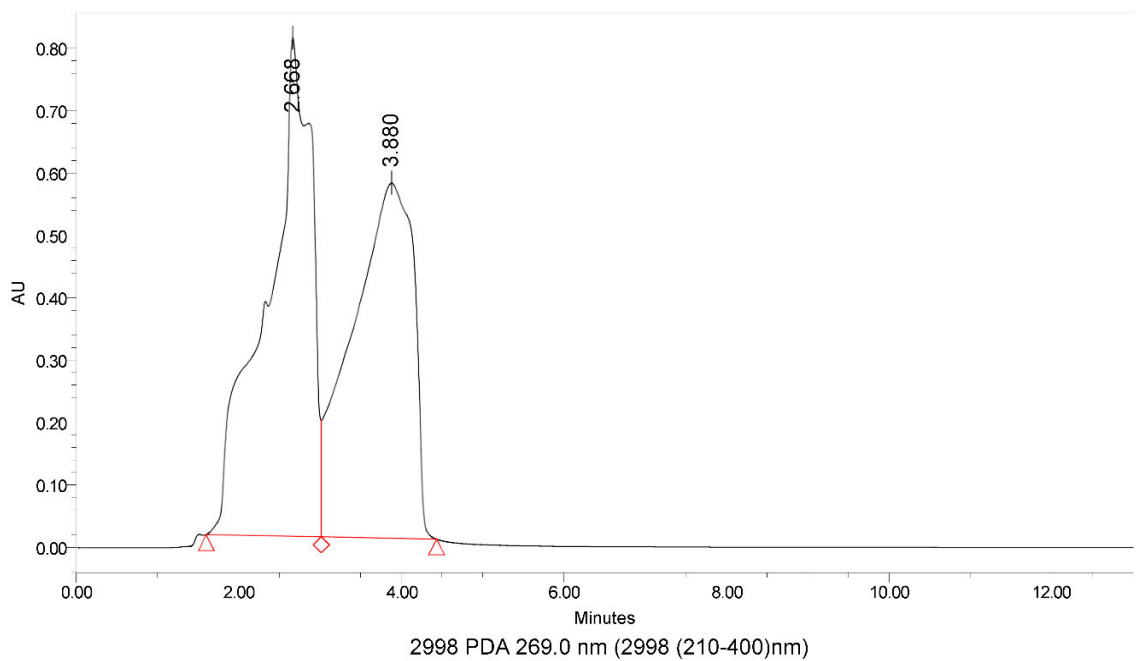

2998 PDA 269.0 nm (2998 (210-400)nm)

| | Retention Time | Area | % Area | k' | N | USP Resolution | Width @ 50% |
|---|---|---|---|---|---|---|---|
| 1 | 2.668 | 30714976 | 50.92 | 0.000 | 103.7 | | 5.238618e-001 |
| 2 | 3.880 | 29604704 | 49.08 | 0.454 | 93.3 | 0.760 | 9.176087e-001 |

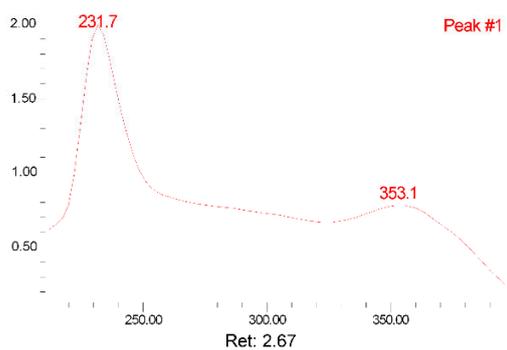

Ret: 2.67

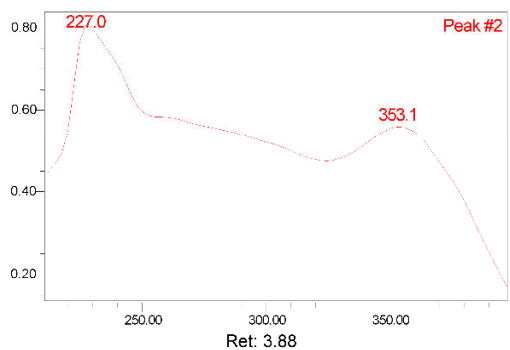

Ret: 3.88

**Sample Name:** *rac*-2

Date Acquired: 2019-02-07 10:39:18

| | | |
|---|---|---|
| **Column:** | Chiralpak IB N-3 | **Column ID:** |
| **Mobile Phase A:** | $CO_2$ | **Column Dimension:** 150 * 4.6 |
| **Mobile Phase B:** | EtOH/DEA 100/20mM | **Particle Size:** 3 |
| **Gradient:** | 25% B, 120 bar | **Injection volume:** 5.00 ul |
| | | **Flow:** 3.5 ml/min |
| **Temperature:** | 40°C | **Wav elength:** PDA Spectrum PDA |
| **Sample Concentration:** | Sample in EtOH | **Vial:** 1:f,5 |

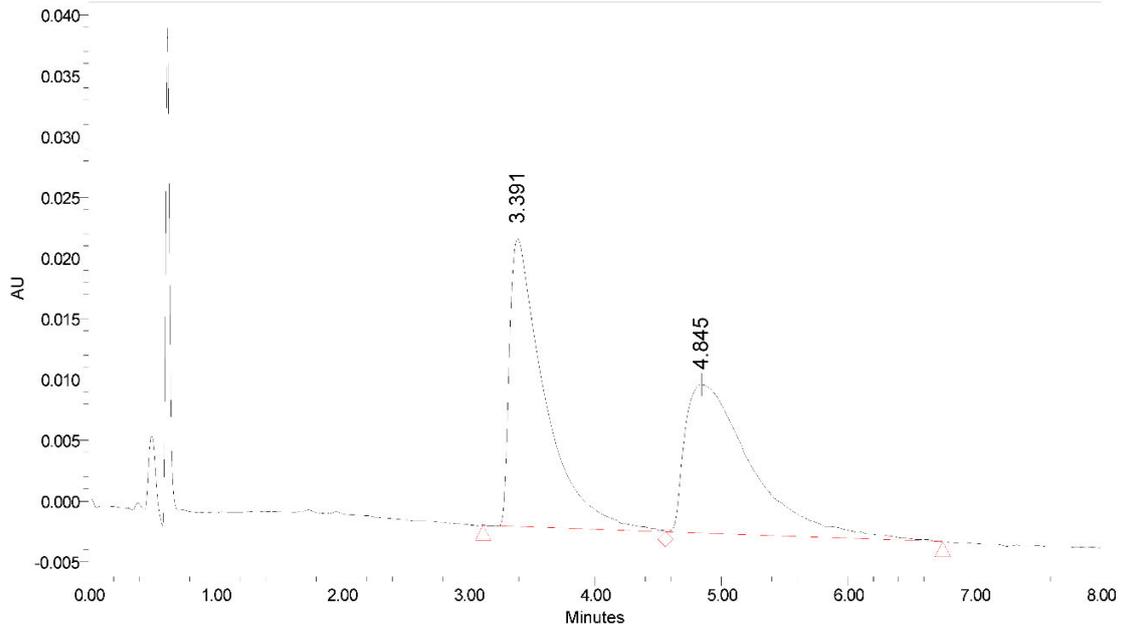

| | Retention Time | Area | % Area | k' | N | USP Resolution | Width @ 50% |
|---|---|---|---|---|---|---|---|
| 1 | 3.391 | 457983 | 50.68 | 0.000 | 743.8 | | 2.710256e-001 |
| 2 | 4.845 | 445739 | 49.32 | 0.429 | 545.1 | 2.015 | 5.344113e-001 |

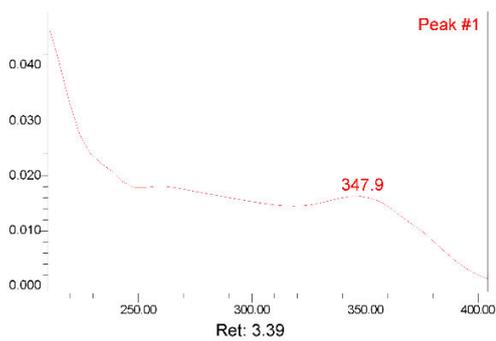

Peak #1, Ret: 3.39, 347.9

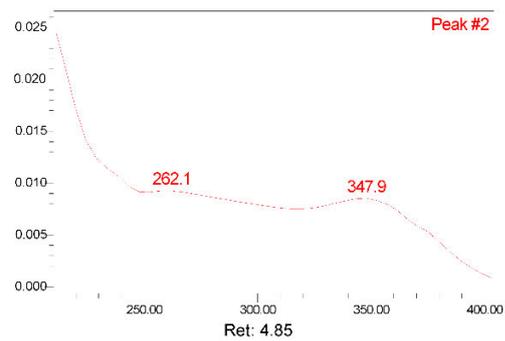

Peak #2, Ret: 4.85, 262.1, 347.9

Report Method: normal_spektrum
Acq Method Set: 120bar_B2_25_40deg

Injection Id 6677

## Vibrational Circular Dichroism (VCD)

*Summary*

VCD analysis was performed on (+)-**2** only. Although not conclusive, the data collected suggest that (+)-xx1 is likely to be the (S) enantiomer. All spectra, experimental and simulated, are shown in Figure 1.

*Experimental*

(+)-**2** (5.0 mg) was dissolved in 110 μl $CDCl_3$. Approximately 90 μl of the solution was transferred to a 0.100 mm $BaF_2$ cell and VCD spectra acquired for 12 hours in a Biotools ChiralIR2X instrument. The resolution was 4 $cm^{-1}$. A VCD spectrum was collected on a $CDCl_3$ blank in the same cell to act as a baseline reference and subtracted from the experimental spectrum.

*Computational Spectral Simulation*

A Monte Carlo molecular mechanics search for low energy geometries was conducted for the S enantiomer. MacroModel within the Maestro graphical interface (Schrödinger Inc.) was used to generate 123 starting coordinates for conformers. All conformers within 5 kcal/mole of the lowest energy conformer were used as starting point for density functional theory (DFT) minimizations within Gaussian09. Optimized structures, harmonic vibrational frequencies/intensities, VCD rotational strengths, and free energies at STP (including zero-point energies) were determined at B3LYP/6-31G* level of theory. Three conformations were found that contributed over 10% to the Boltzmann distribution. An in-house built program was used to fit Lorentzian line shapes (12 $cm^{-1}$ line width) to the computed spectrum of a Boltzmann distributed average thereby allowing direct comparisons between simulated and experimental spectra.

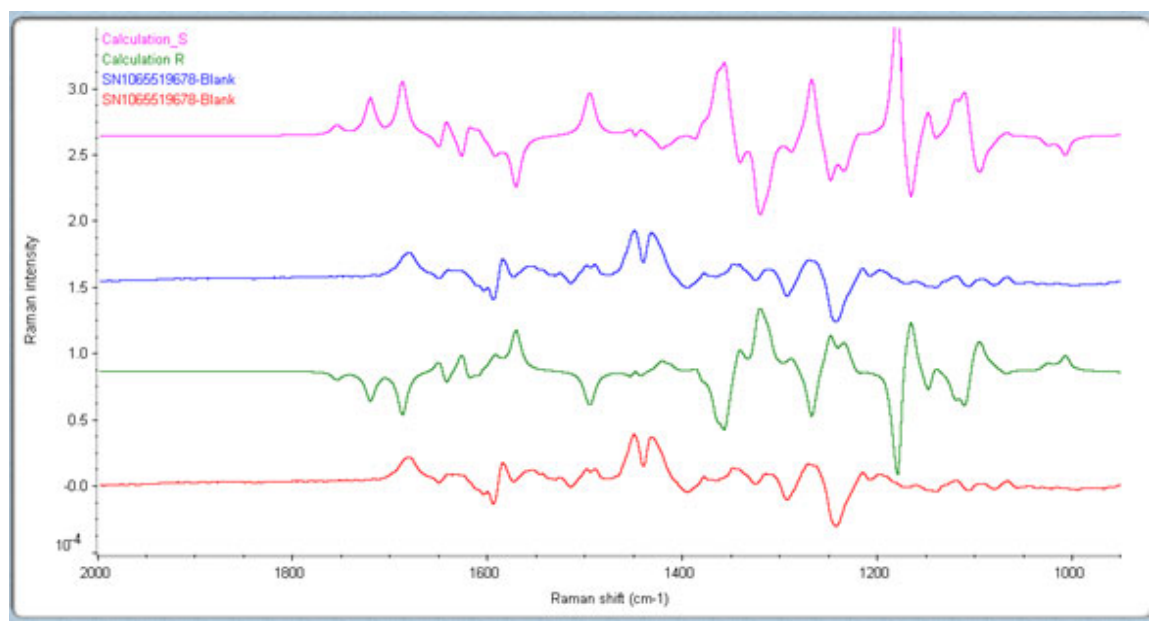

**Supplementary Figure 22.** Comparison of calculated and experimental spectra: the calculated spectrum for the *(S)* enantiomer is shown in pink and the *(R)* enantiomer (by inversion) in green. There is a reasonable, but not high certainty match between the experimental spectrum of (+)-**2** and the calculated spectrum for the *(S)* enantiomer, particularly in the region 1200-1300 cm$^{-1}$.

## Racemisation studies

Sample was dissolved in EtOH, then triethylamine (TEA, 10 eq) was added and the resulting solution was stirred overnight at room temperature. The reaction mixture was then transferred into a vial and analyzed directly.

| SSL Report | Request type | Racemisation test |
|---|---|---|

**Sample Information**

| Sample ID | EN Number | Incoming amount | Compound name | Project name |
|---|---|---|---|---|
| rac-**2**, (R)-**2** | | 2 mg | | |

**Analytical Conditions**

| Column | Dimension (mm) | Particle Size (µm) |
|---|---|---|
| Chiralpak IB-N3 | 150 x 4.6 | 3 |

| Mobile Phase |
|---|
| 30% EtOH/DEA 100/20mM in $CO_2$, 120 bar |

| Flow (ml/min) | Detection | Temperature (C) |
|---|---|---|
| 3,5 | | 40 |

| Gradient | Instrument | Sample concentration (mg/ml) |
|---|---|---|
| | | |

| Sample ID | Purity | Chiral Purity | Weight |
|---|---|---|---|
| Peak 1 (R)-**2** | | | |
| Comments | | | |
| Peak 2 (S)-**2** | | | |
| Comments | | | |

## Sample Name: *rac*-**2**

| | | |
|---|---|---|
| Column: | Chiralpak IB N-3 CO2 | |
| Mobile Phase A: | EtOH/DEA 100/20mM | |
| Mobile Phase B: | 30% B, 120 bar | |
| Gradient: | | |
| Temperature: | 40°C | |
| Sample Concentration: | Sample in EtOH | |

| | |
|---|---|
| Column ID: | |
| Column Dimension: | 150 * 4.6 |
| Particle Size: | 3 |
| Injection volume: | 10.00 ul |
| Flow: | 3.5 ml/min |
| Wavelength: | PDA Spectrum PDA |
| Vial: | 1:b,3 |

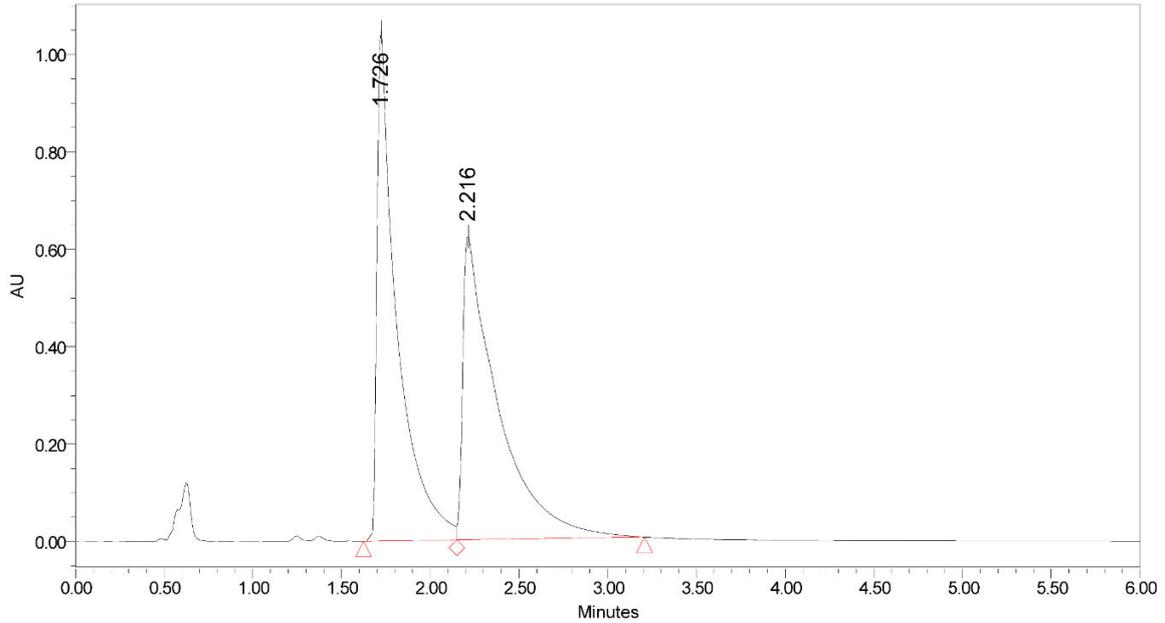

Channel PDA Spectrum; Channel Description PDA Spectrum (210-400)nm

| | Retention Time | Area | % Area | k' | N | USP Resolution | Width @ 50% |
|---|---|---|---|---|---|---|---|
| 1 | 1.726 | 8200105 | 49.19 | 0.000 | 1321.5 | | 9.595719e-002 |
| 2 | 2.216 | 8469435 | 50.81 | 0.284 | 518.2 | 1.692 | 1.805067e-001 |

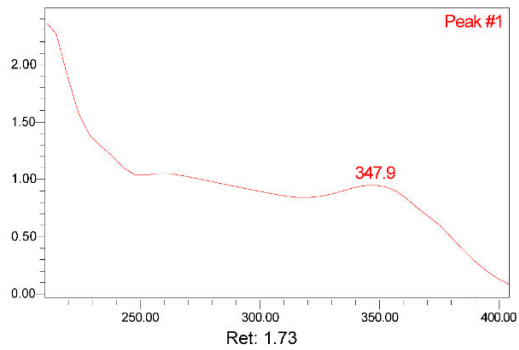

Peak #1 — 347.9 — Ret: 1.73

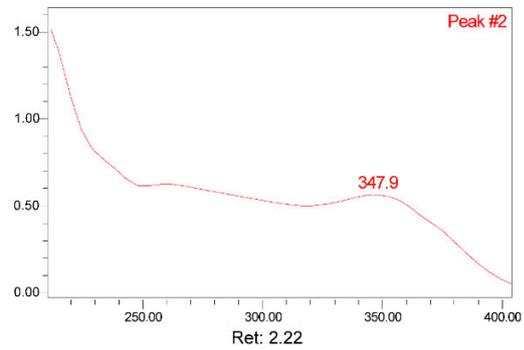

Peak #2 — 347.9 — Ret: 2.22

Report Method: normal_spektrum_x
Acq Method Set: 120bar_B2_30_40deg

Injection Id 29152

S64

Sample Name: (*R*)-**2**

| | | |
|---|---|---|
| **Column:** | Chiralpak IB N-3 | **Column ID:** |
| **Mobile Phase A:** | CO2 | **Column Dimension:** 150 * 4.6 |
| **Mobile Phase B:** | EtOH/DEA 100/20mM | **Particle Size:** 3 |
| **Gradient:** | 30% B, 120 bar | **Injection volume:** 10.00 ul |
| | | **Flow :** 3.5 ml/min |
| **Temperature:** | 40°C | **Wavelength:** PDA Spectrum PDA |
| **Sample Concentration:** | Sample in EtOH | **Vial:** 1:b,1 |

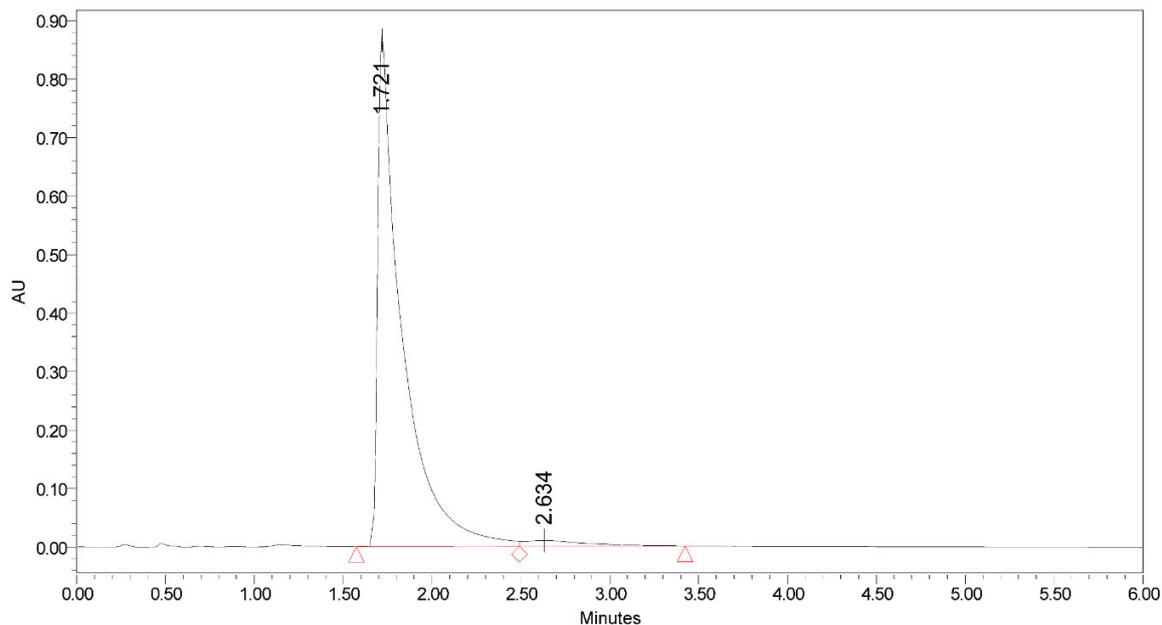

Channel PDA Spectrum; Channel Description PDA Spectrum (210-400)nm

| | Retention Time | Area | % Area | k' | N | USP Resolution | Width @ 50% |
|---|---|---|---|---|---|---|---|
| 1 | 1.721 | 8292807 | 97.21 | 0.000 | 957.5 | | 1.155378e-001 |
| 2 | 2.634 | 237826 | 2.79 | 0.530 | | | |

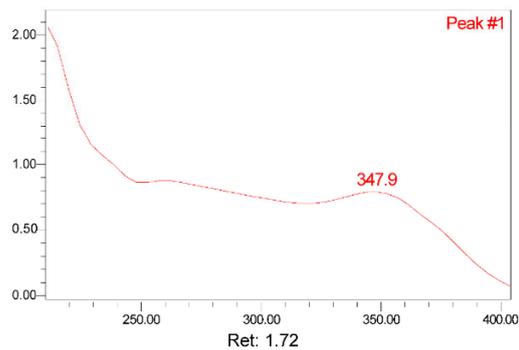
Ret: 1.72

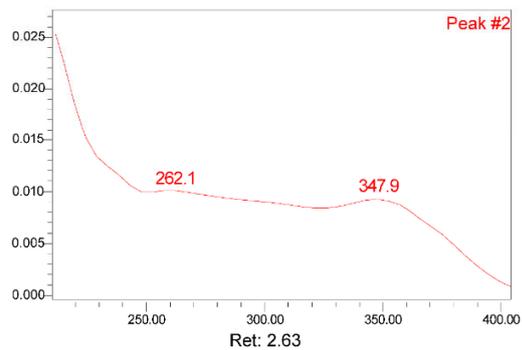
Ret: 2.63

## Sample Name: (*R*)-**2** + TEA

| | | | | |
|---|---|---|---|---|
| **Column:** | Chiralpak IB N-3 | | **Column ID:** | |
| **Mobile Phase A:** | CO2 | | **Column Dimension:** | 150 * 4.6 |
| **Mobile Phase B:** | EtOH/DEA 100/20mM | | **Particle Size:** | 3 |
| **Gradient:** | 30% B, 120 bar | | **Injection volume:** | 10.00 ul |
| | | | **Flow :** | 3.5 ml/min |
| **Temperature:** | 40°C | | **Wavelength:** | PDA Spectrum PDA |
| **Sample Concentration:** | Sample in EtOH | | **Vial:** | 1:B,2 |

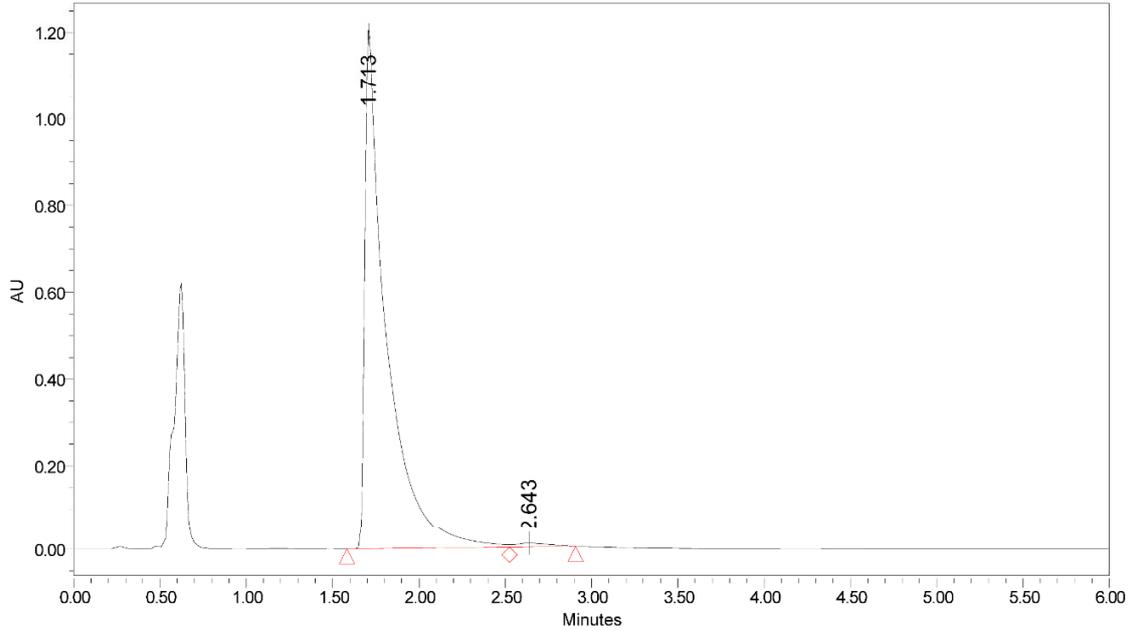

| | Retention Time | Area | % Area | k' | N | USP Resolution | Width @ 50% |
|---|---|---|---|---|---|---|---|
| 1 | 1.713 | 10566923 | 98.90 | 0.000 | 1178.8 | | 1.036627e-001 |
| 2 | 2.643 | 117647 | 1.10 | 0.543 | | | |

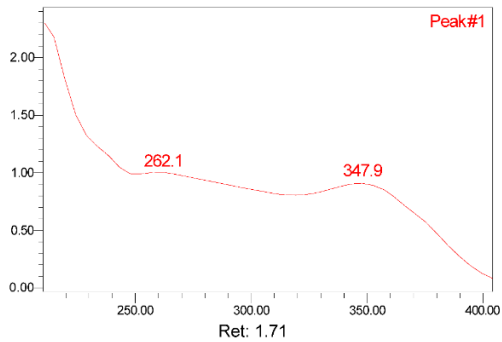

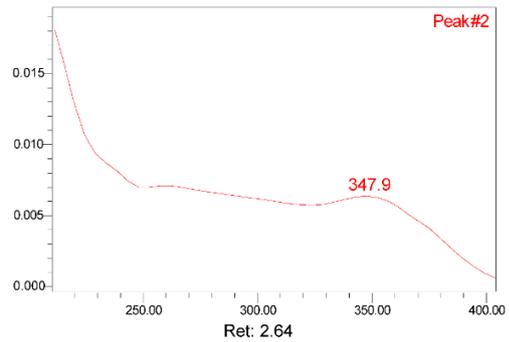

## IVd) FITC-3 peptide synthesis

*General information*

Fmoc-amino acids were purchased from Chem-Impex International, Inc., with the following side-chain
protection: Fmoc-Ala-OH, Fmoc-Glu(OtBu)-OH, Fmoc-Gly-OH, Fmoc-Phe-OH, Fmoc-Pro-OH, Fmoc-Thr(PO(Obzl)OH)-OH, Fmoc-Val-OH. L-Amino acids were used in every case. Fluorescein 5-isothiocyanate (5-FITC) was purchased from Sigma-Aldrich. Diisopropylethylamine (DIPEA) was purchased from Sigma-Aldrich, 1-[bis(dimethylamino)methylene]-1H-1,2,3-triazolo[4,5-b]pyridinium 3-oxid hexafluorophosphate (HATU) was purchased from Chem-Impex International, Inc. 2-chlorotrityl chloride resin (200-400 mesh) from Chem-Impex International, Inc.
Peptide synthesis was monitored by reversed-phase (RP) ultra-performance liquid chromatography tandem mass spectrometer (UPLC-MS). Analytical RP-UPLC-MS was performed on a Waters Acquity UPLC system (PDA, sample manager, sample organiser, column oven modules) and Waters SQD2 mass spectrometer using the following column: Waters Acquity CSH C18 column, 130 Å, 1.7 μm, 50 x 2.1 mm at a flow rate of 0.5 ml/min at 45 °C. A linear gradient of mobile phase: A = $H_2O$ + 10 mM formic acid (FA), 1 mM ammonia and 0.03% trifluoroacetic acid (TFA) and B = ACN/$H_2O$ 95/5 (vol/vol-%) + 10 mM FA, 1 mM ammonia and 0.03% TFA was used with detection at 220 nm.

*Resin loading*

In a peptide reactor 2-chlorotrityl chloride resin 0.8 mmol/g (0.2 mmol, 250 mg) was swollen in $CH_2Cl_2$ for 10 min, then the solvent was drained. Fmoc-L-Val-OH (1 eq) was dissolved in $CH_2Cl_2$ and DIPEA (3 eq) was added. The clear solution was added to the resin which was agitated for 10 min. Additional DIPEA (7 eq) was added and the resin was agitated for further 45 min. The remaining trityl groups were capped adding MeOH (0.8 μL/mg of resin), the resin was agitated for 10 min. The mixture was drained and the resin beads washed with DMF (3 x) and $CH_2Cl_2$ (3 x).

*Fmoc cleavage*

The N-terminal Fmoc protecting group was removed with a 20% solution of piperidine in DMF (2 x 5 min). The mixture was drained and the resin beads washed with DMF (3 x) and $CH_2Cl_2$ (3 x).

*Peptide elongation*

Fmoc-L-AA-OH (4 eq) and HATU (4 eq) were dissolved in DMF, then DIPEA (6 eq) was added. After a pre-activation period of 2 min, the mixture was added to the resin, which was agitated for 45-60 min. The mixture was drained and the resin beads washed with DMF (3 x) and $CH_2Cl_2$ (3 x). Successful coupling was indicated by ninhydrin test. Coupling of the spacer Fmoc-6-aminoexanoic acid (Fmoc-6-Ahx-OH, 4 eq) was performed using the same procedure as above. The peptide was finally reacted with Fluorescein-5-isothiocyanate (5-FITC, 4 eq) and DIPEA (6 eq) for 2h in the dark to yield fluorescent-immobilized peptidyl-resin.

*Cleavage from resin*

The resin-bound peptide was cleaved from resin using a solution of TFA/TIS/H$_2$O/EDT (94:1:2.5:2.5). After shaking for 60 min, the TFA solution was filtered and the reaction mixture was poured into cold diethyl ether. Upon precipitation, the suspension was centrifuged, the precipitate dissolved in ACN/H$_2$O and lyophilized.

*Purification*

The compound was purified by preparative HPLC on a Kromasil C8 column (10 μm 250x20 ID mm) using a gradient of 43-63% ACN in H$_2$O/ACN/FA 95/5/0.2 buffer, over 20 minutes with a flow of 19 mL/min. The compounds were detected by UV at 220nm. Collected fractions were lyophilized, to yield the purified peptide FITC-**3** (26 mg, 9.47%) as a yellow solid, in ca. 95% purity as estimated by analytical UPLC.

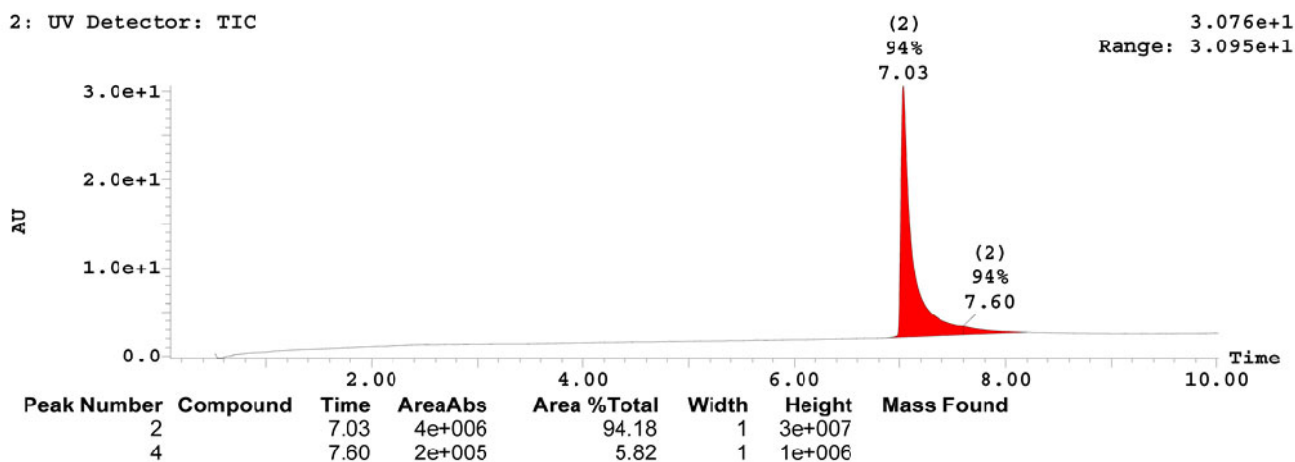

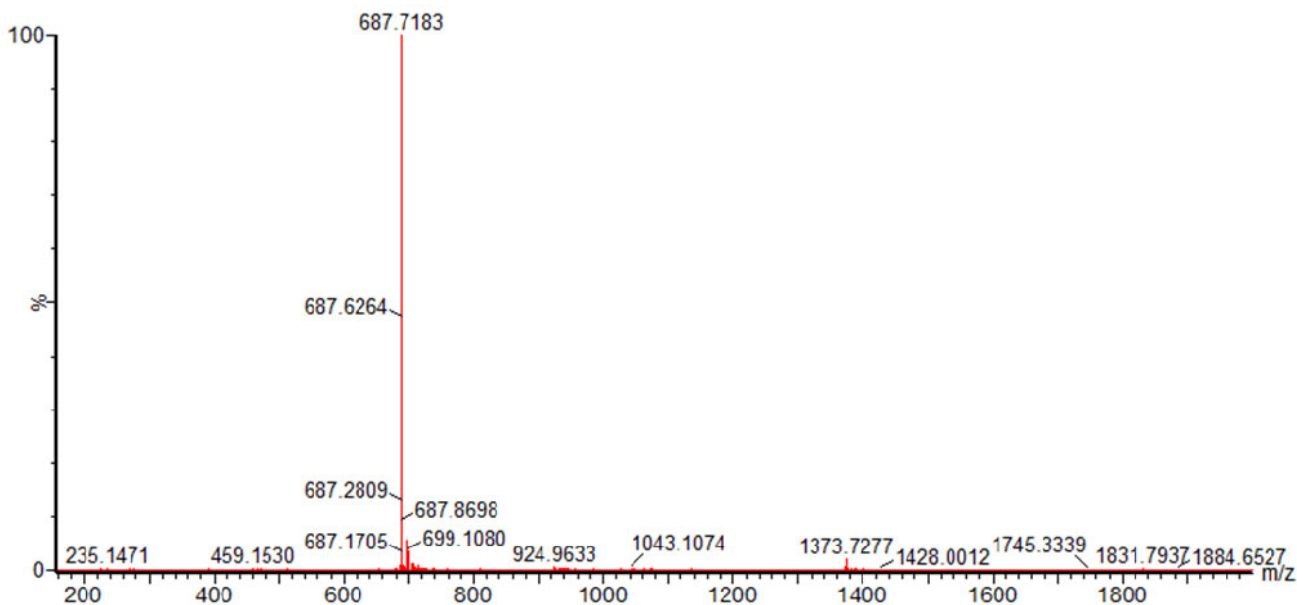

R$_t$ = 7.03 min (3-60% B over 10 min, 0.5 mL/min). m/z (ESI-MS): [M+2H$^+$]$^{2+}$ calculated mass = 687.2, observed = 697.7

Final sequence: 5-FITC-Ahx- AEGFPApTV-OH (FITC-**3**)

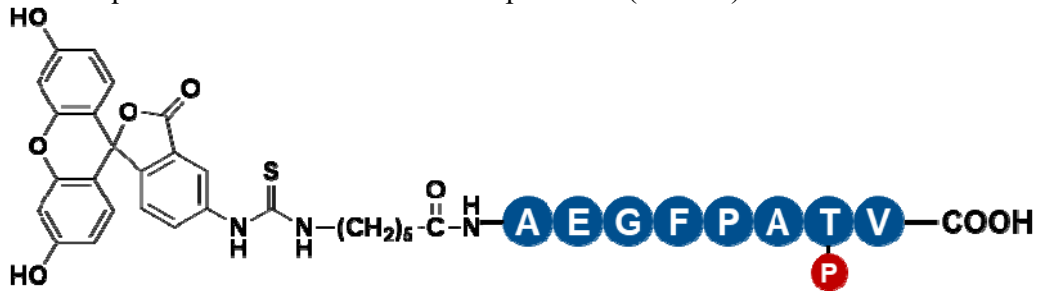

*FAM-**4** sequence*

Sequence: 5-FAM-GSLSARKLpSLQER-OH

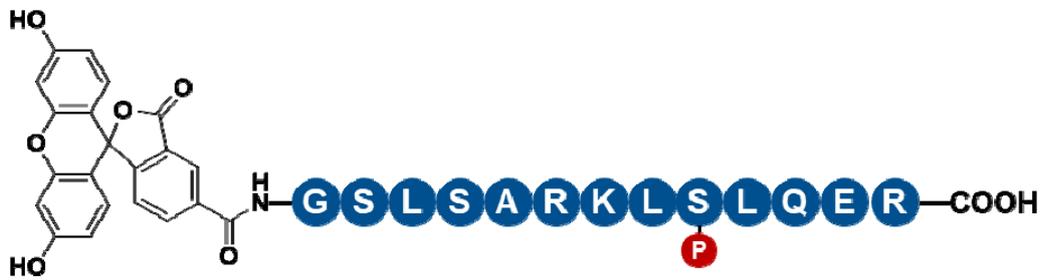

## IVe) pK$_a$ determination

*Protocol Summary*

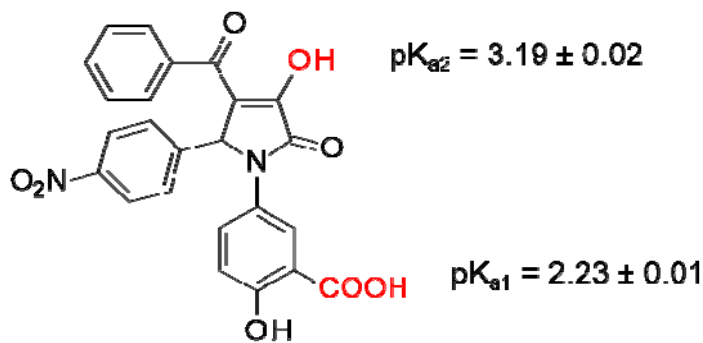

pK$_a$s are determined using the SiriusT3 instrument from Sirius Analytical by performing an acid/base titration. The technique uses an *in situ* UV probe to measure the UV absorbance profile of the compound at each pH point during the titration. Measured pKa values are reported as mean ± SEM.

*Experimental procedure*

The sample pK$_a$s are investigated using the fast UV-metric method. This involves measuring the UV absorbance profile at each pH point during an acid/base titration using an *in situ* UV probe in the titration cell of a SiriusT3 instrument. Each sample is titrated in a triple titration over a nominal pH range of 2 to 12 in approximately 50, 40, 30 % methanol. Sample concentrations are typically in the range 35 - 15 µM. All titrations are carried out at 25 °C.

The pK$_a$(s) are determined by monitoring the change in UV absorbance with pH as the compound undergoes ionisation. This information is used to produce a 3D matrix of pH *vs.* Wavelength *vs.* Absorbance data. A mathematical technique called Target Factor Analysis is applied to the matrix to produce molar absorbance profiles for the different light absorbing species present in solution and also a "Distribution of Species" plot showing how the proportion of each species varies with pH. Sample pK$_a$s are extrapolated to aqueous conditions using the Yasuda-Shedlovsky method.

The method requires that the sample compound possesses a chromophore and that changes in ionisation influence the absorbance spectrum of the compound. This procedure will measure pK$_a$ values in the range 2-12.

**IVf) Spectroscopic data of products**

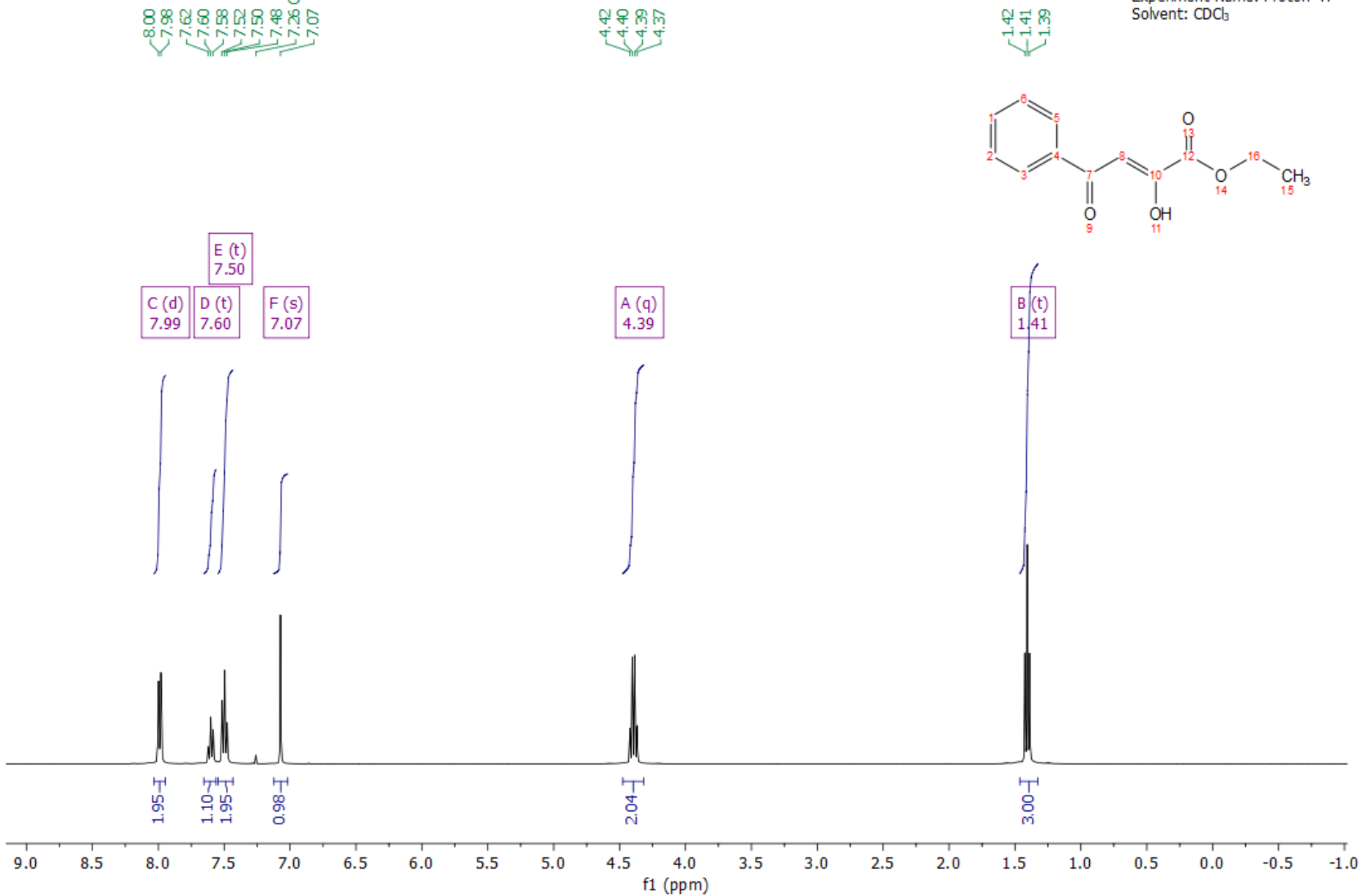

Ethyl 2,4-dioxo-4-phenyl-butanoate

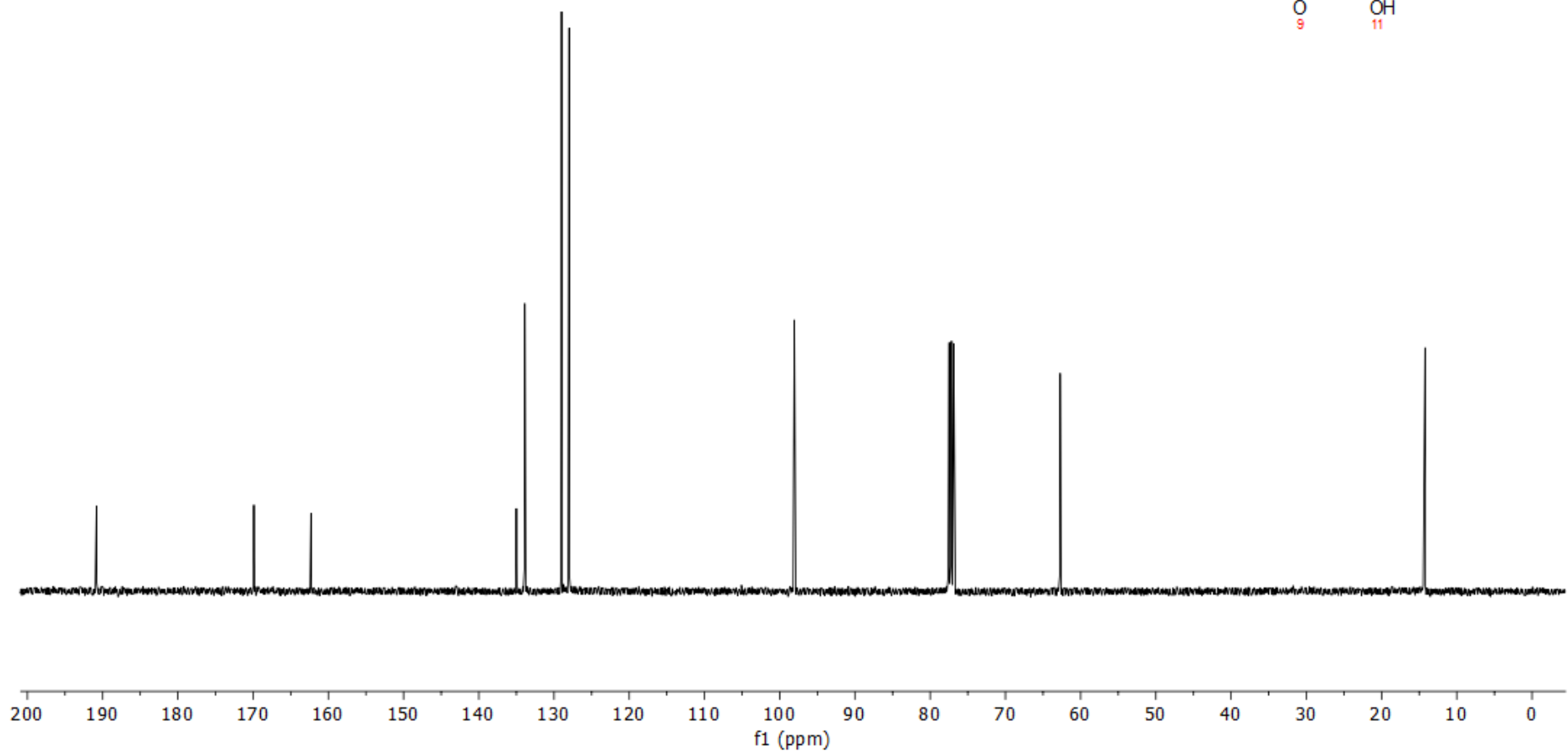

Pyrrolidone1, 5-[3-benzoyl-4-hydroxy-2-(4-nitrophenyl)-5-oxo-2H-pyrrol-1-yl]-2-hydroxy-benzoic acid (**2**)

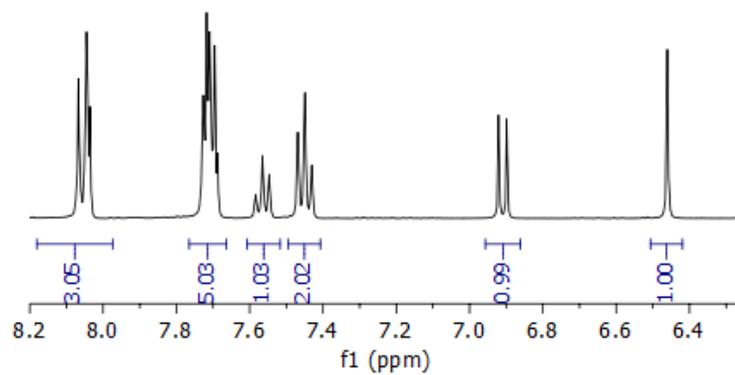
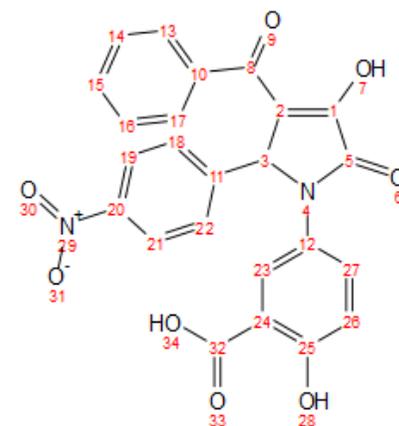
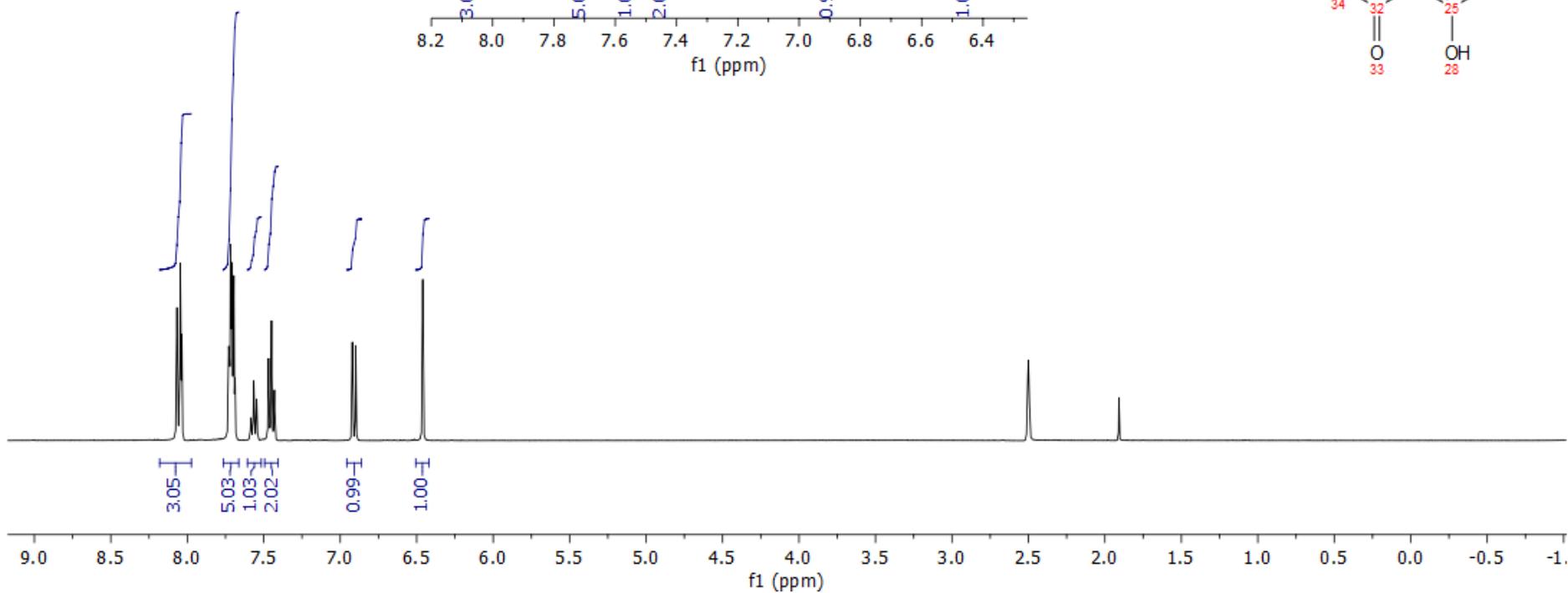

Pyrrolidone1, 5-[3-benzoyl-4-hydroxy-2-(4-nitrophenyl)-5-oxo-2H-pyrrol-1-yl]-2-hydroxy-benzoic acid (**2**)

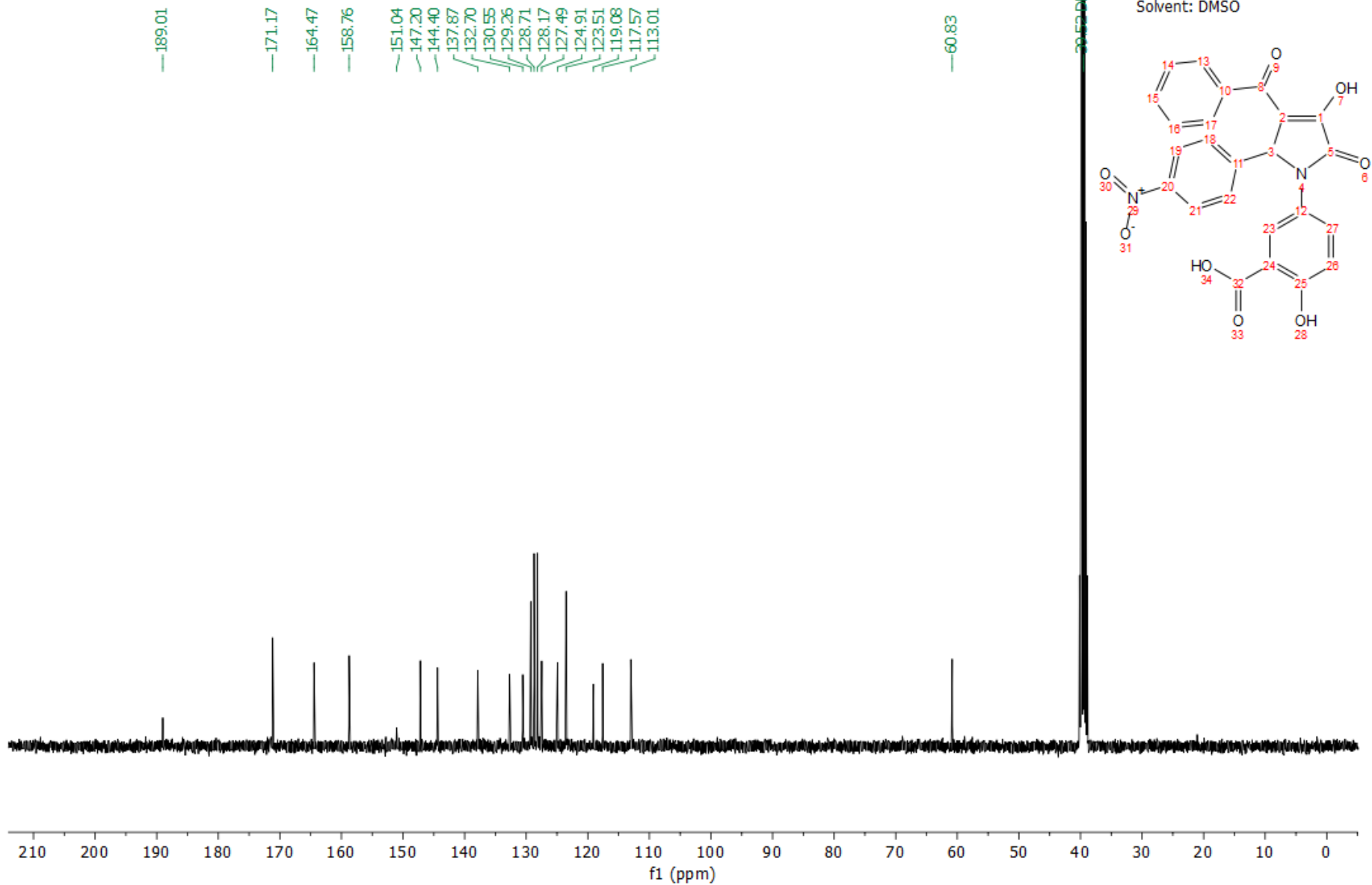

Frequency: 100.61 MHz
Experiment Name: ¹³C Carbon
Solvent: DMSO

Peaks: 189.01, 171.17, 164.47, 158.76, 151.04, 147.20, 144.40, 137.87, 132.70, 130.55, 129.26, 128.71, 128.17, 127.49, 124.91, 123.51, 119.08, 117.57, 113.01, 60.83

S76

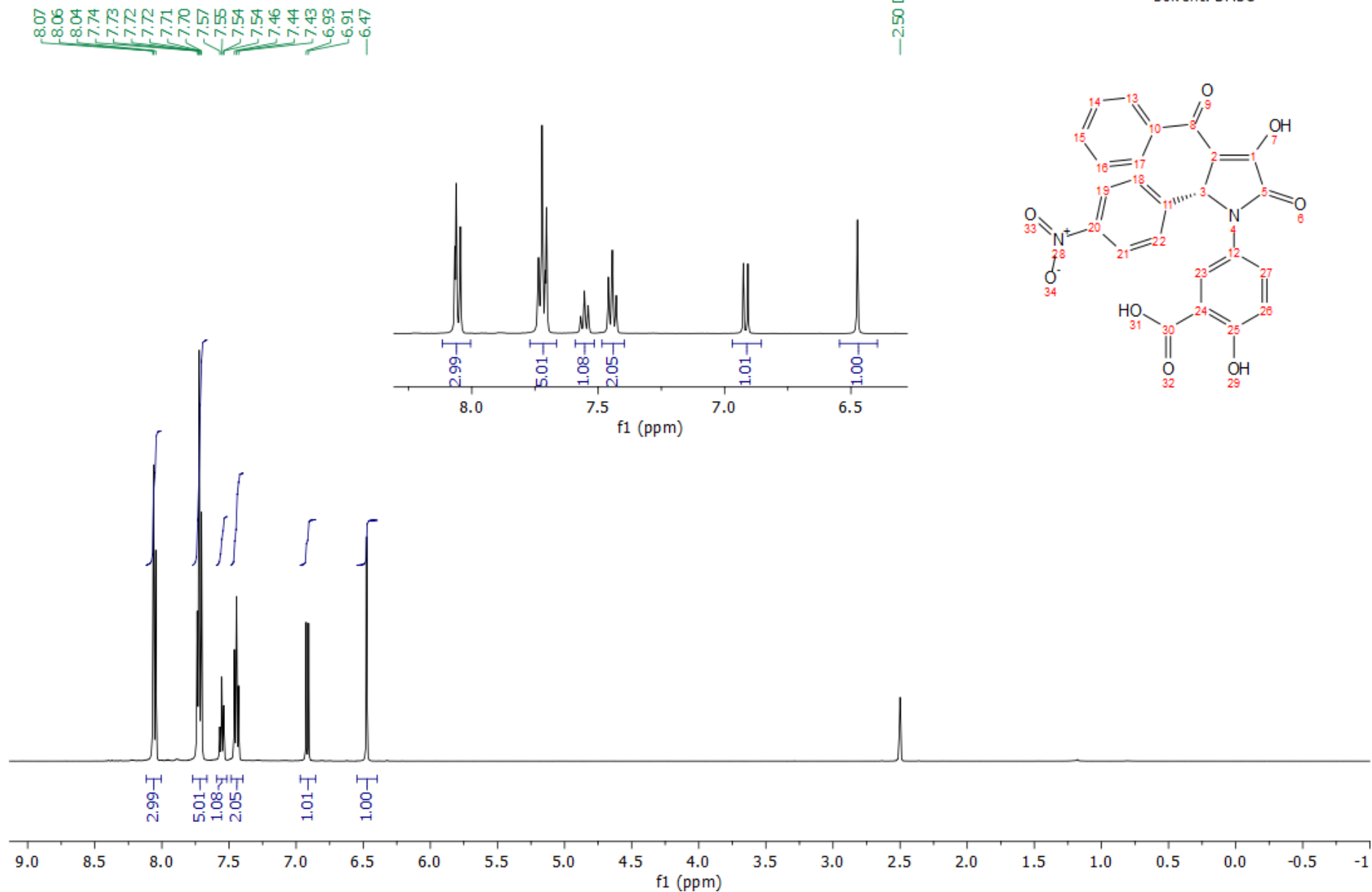

(R)-5-[3-benzoyl-4-hydroxy-2-(4-nitrophenyl)-5-oxo-2H-pyrrol-1-yl]-2-hydroxy-benzoic acid ((R)-2)

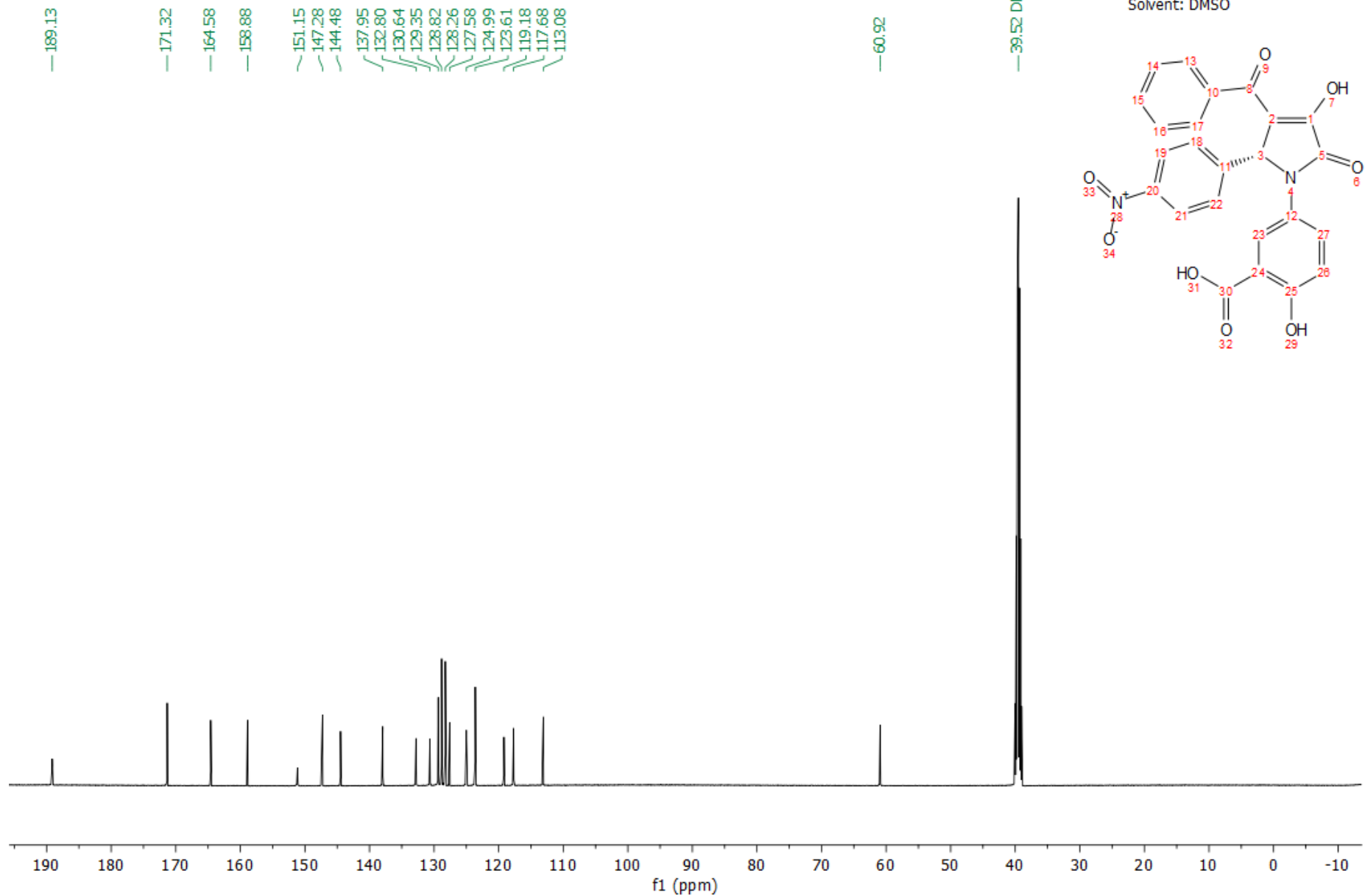

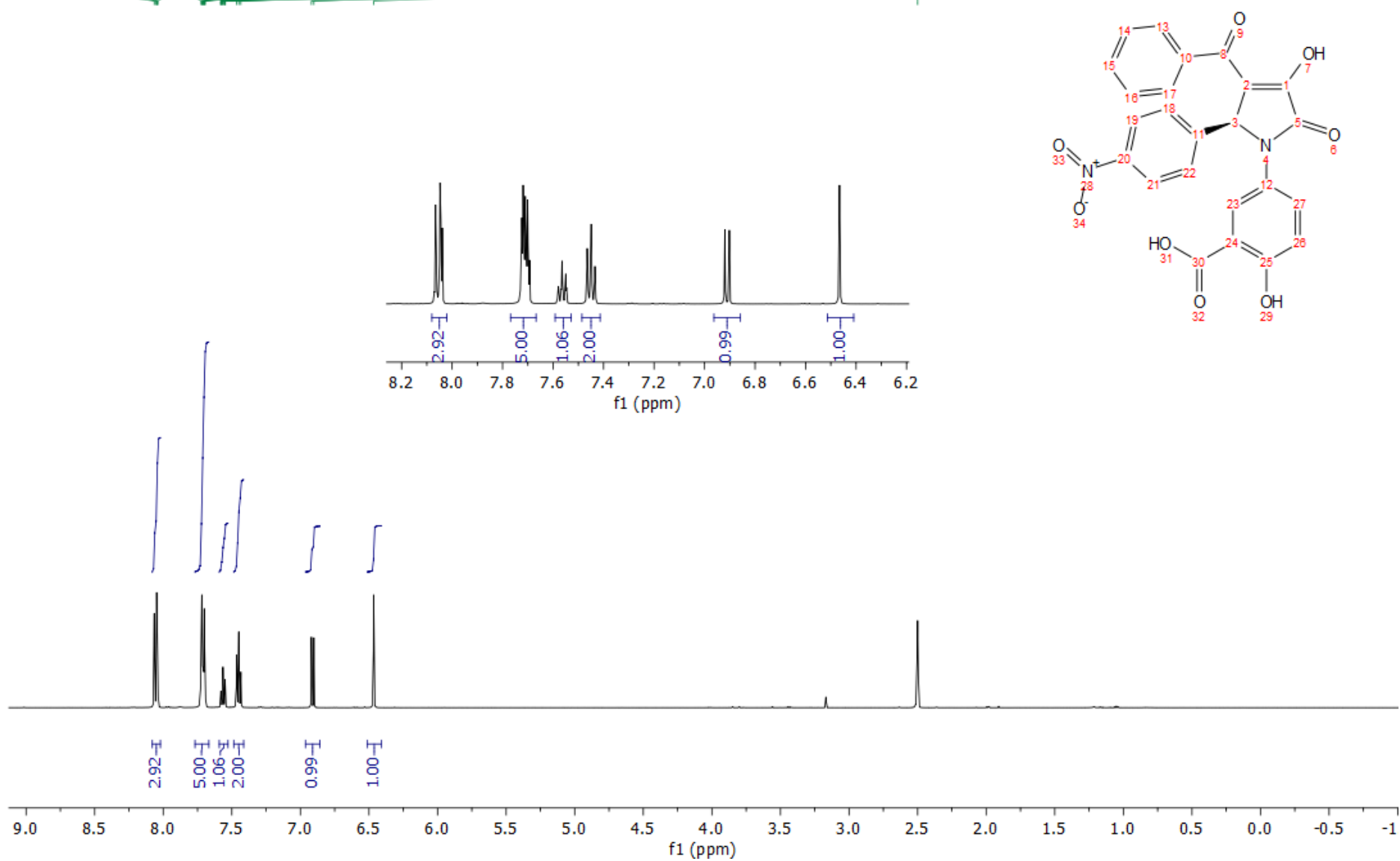

(S)-5-[3-benzoyl-4-hydroxy-2-(4-nitrophenyl)-5-oxo-2H-pyrrol-1-yl]-2-hydroxy-benzoic acid ((S)-2)

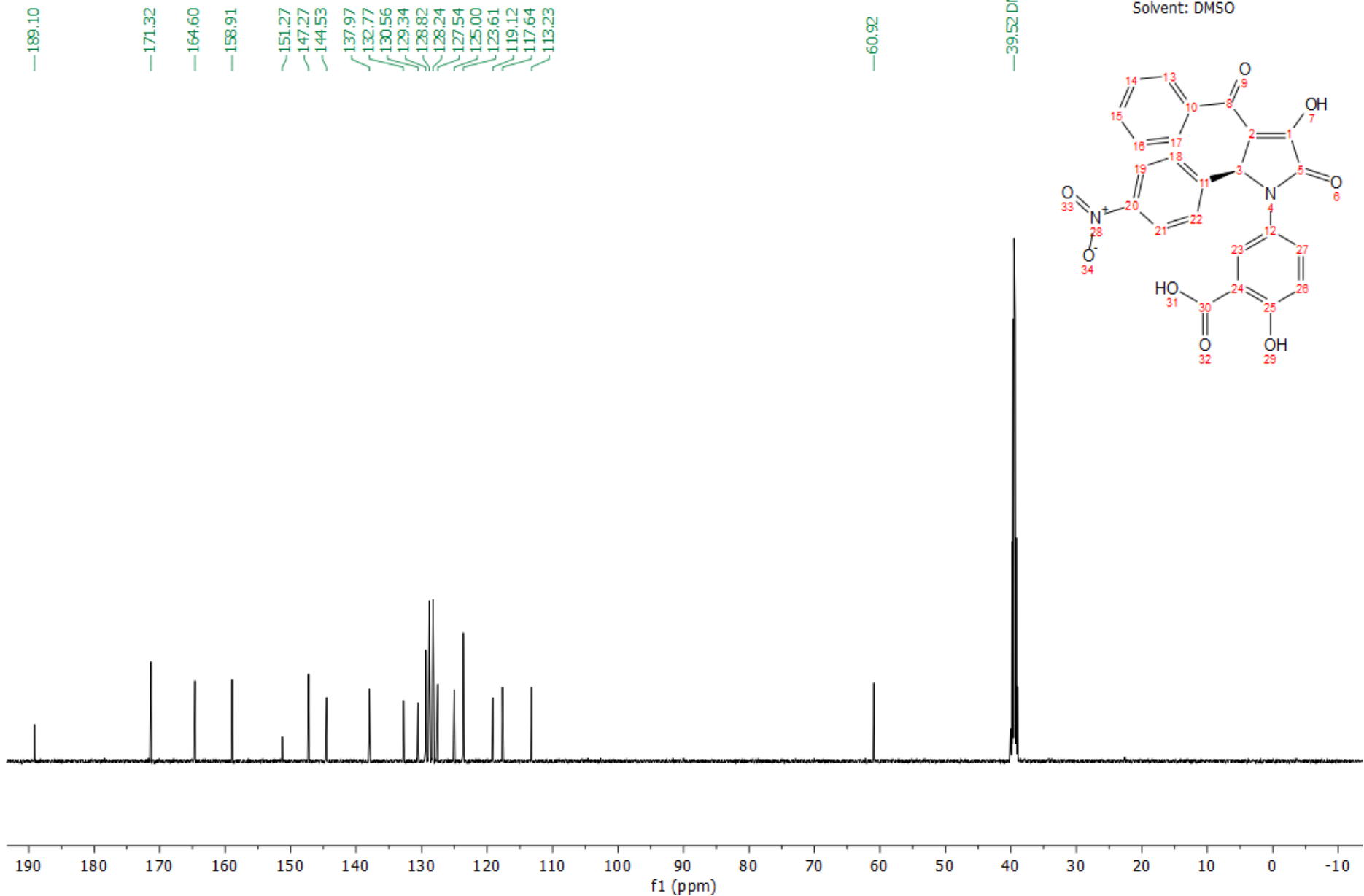

2-hydroxy-5-[4-(4-nitrophenyl)-6-oxo-3-phenyl-1,4-dihydropyrrolo[3,4-c]pyrazol-5-yl]benzoic acid (**5**)

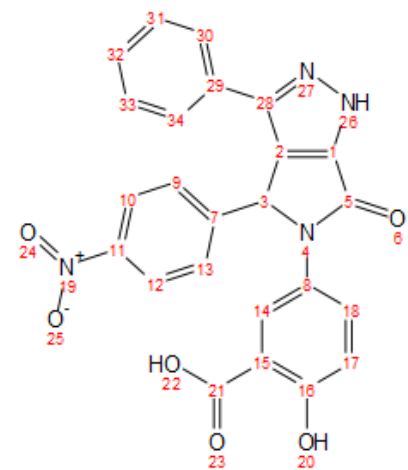
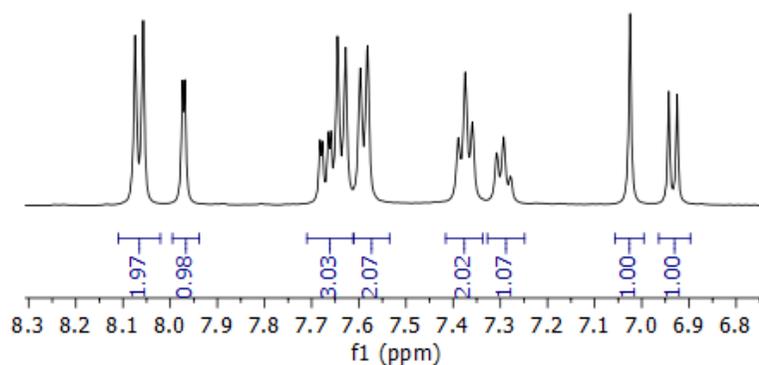
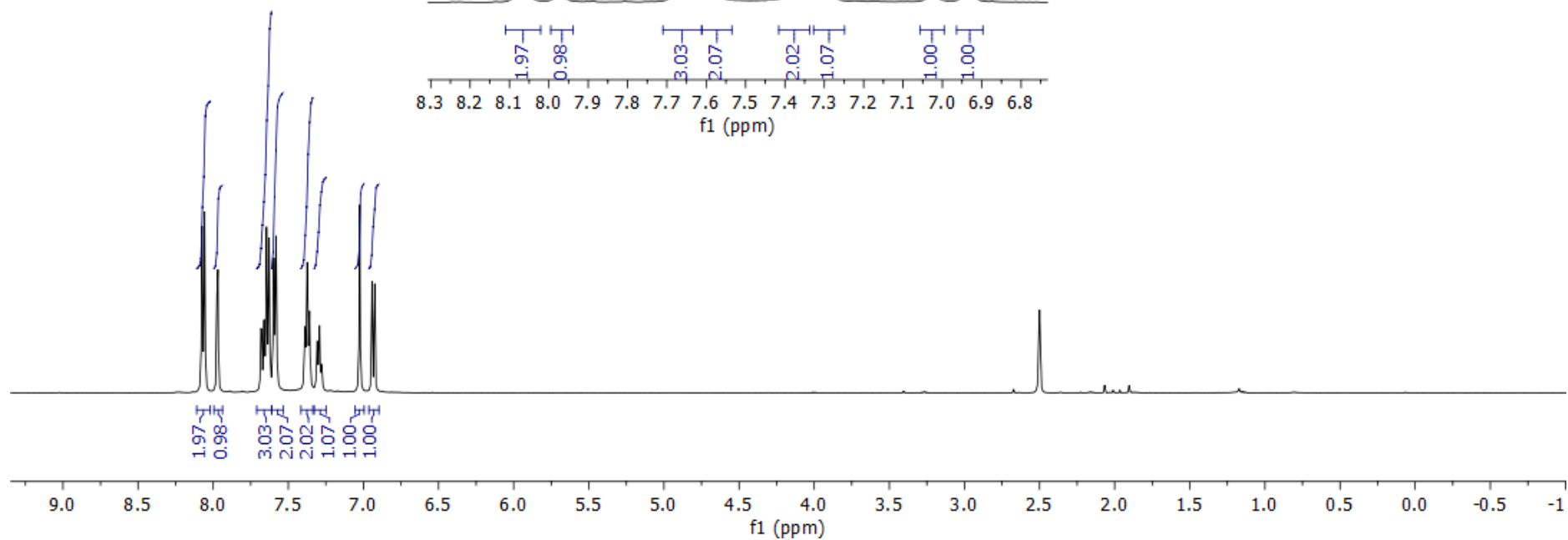

2-hydroxy-5-[4-(4-nitrophenyl)-6-oxo-3-phenyl-1,4-dihydropyrrolo[3,4-c]pyrazol-5-yl]benzoic acid (**5**)

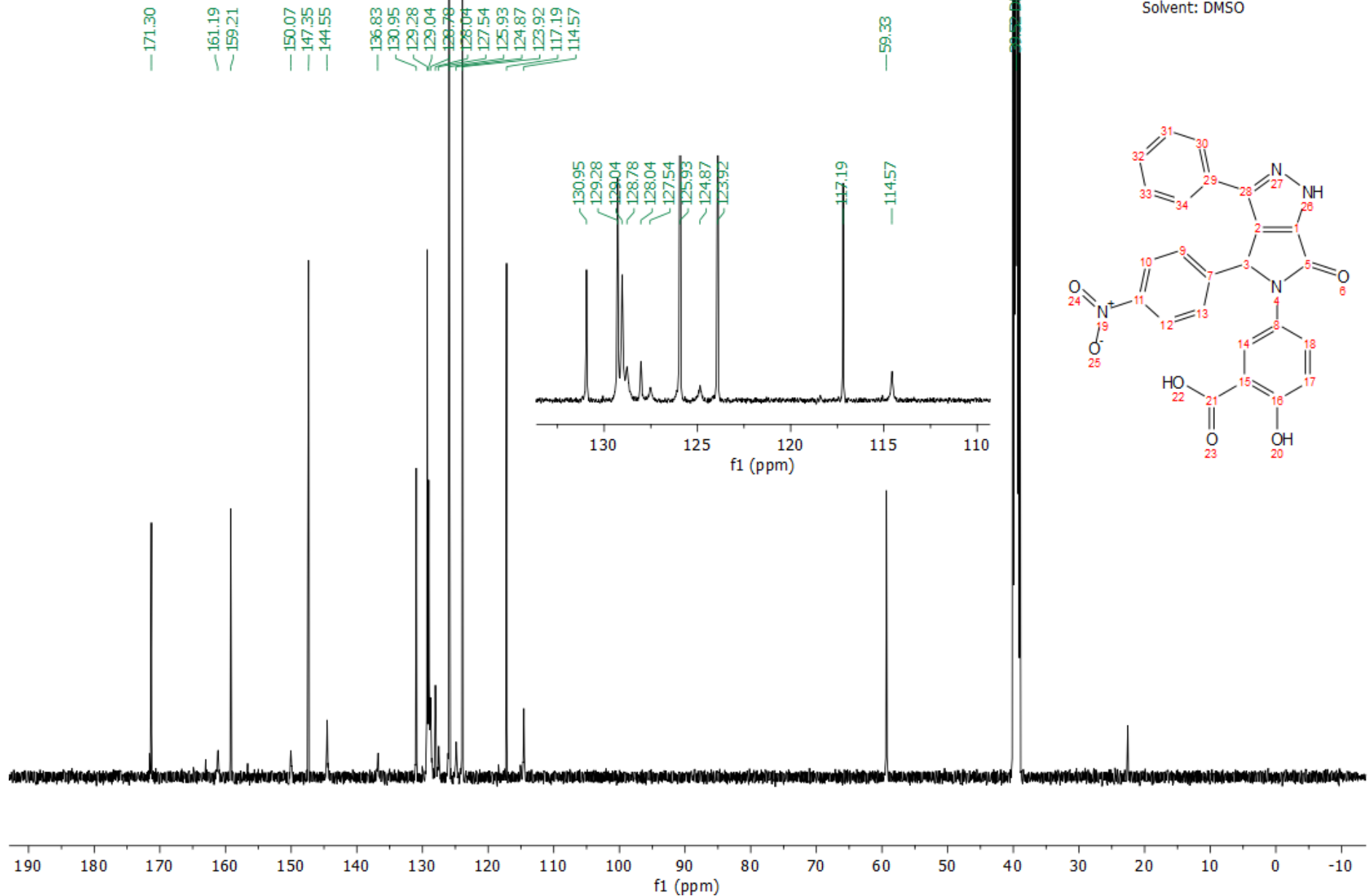

(R)-2-hydroxy-5-[4-(4-nitrophenyl)-6-oxo-3-phenyl-1,4-dihydropyrrolo[3,4-c]pyrazol-5-yl]benzoic acid (**(R)-5**)

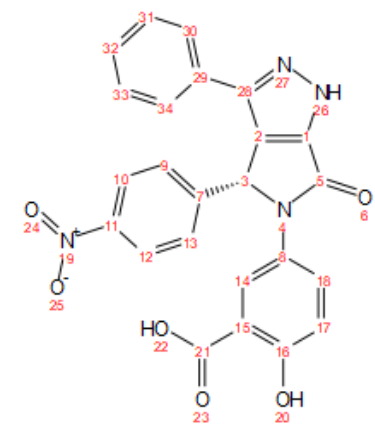
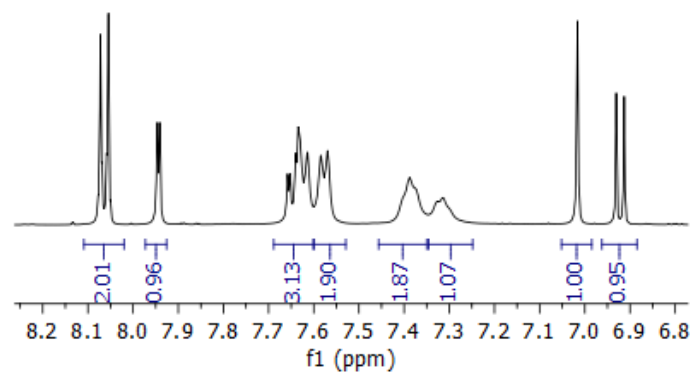
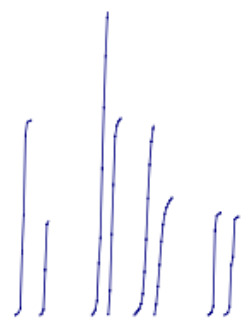

(R)-2-hydroxy-5-[4-(4-nitrophenyl)-6-oxo-3-phenyl-1,4-dihydropyrrolo[3,4-c]pyrazol-5-yl]benzoic acid ((R)-5)

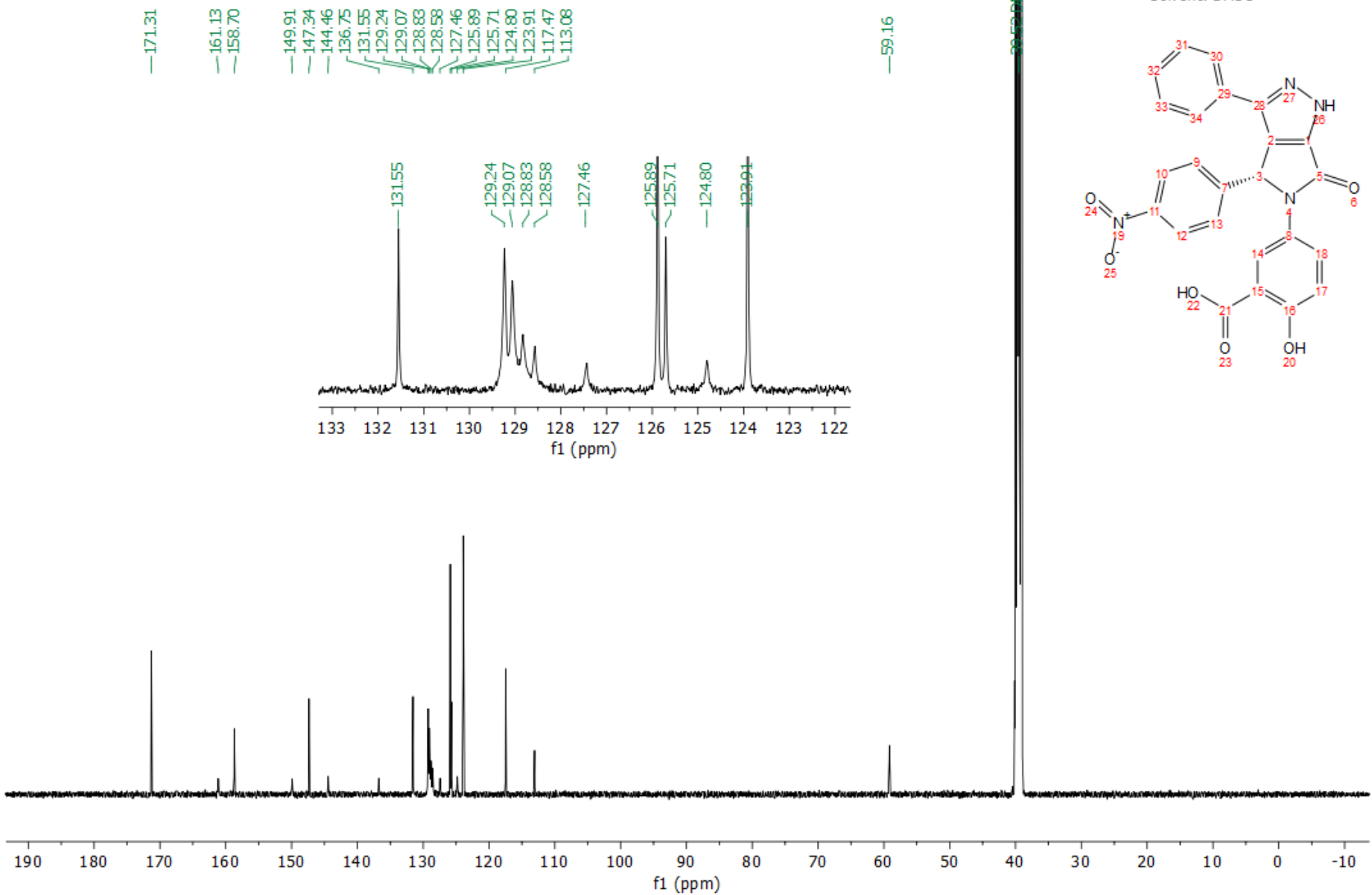

5-[3-[amino(phenyl)methylene]-2-(4-nitrophenyl)-4,5-dioxo-pyrrolidin-1-yl]-2-hydroxy-benzoic acid (6)

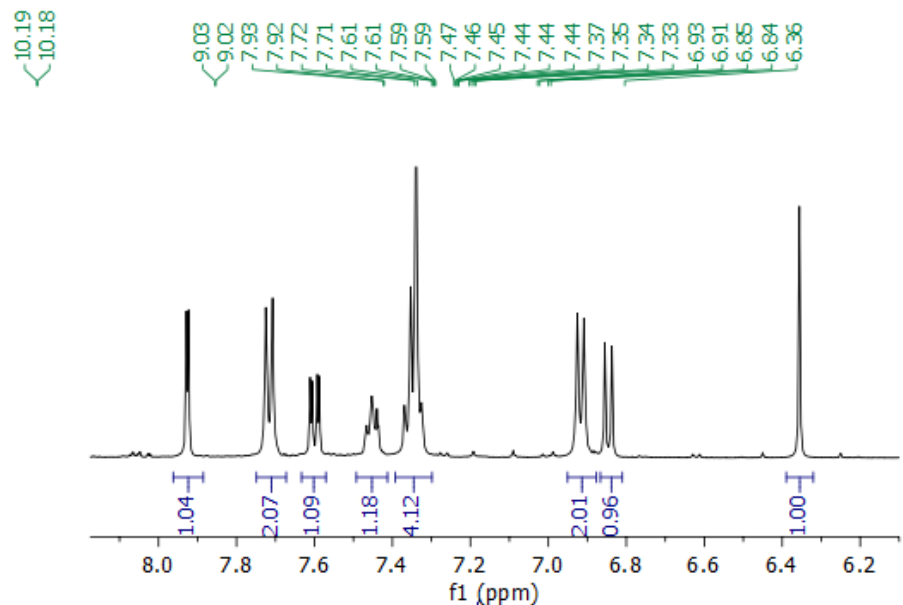
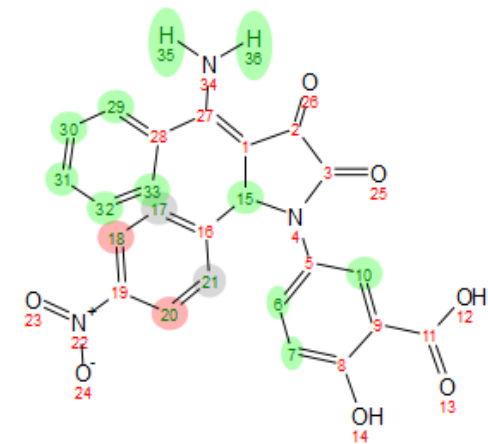
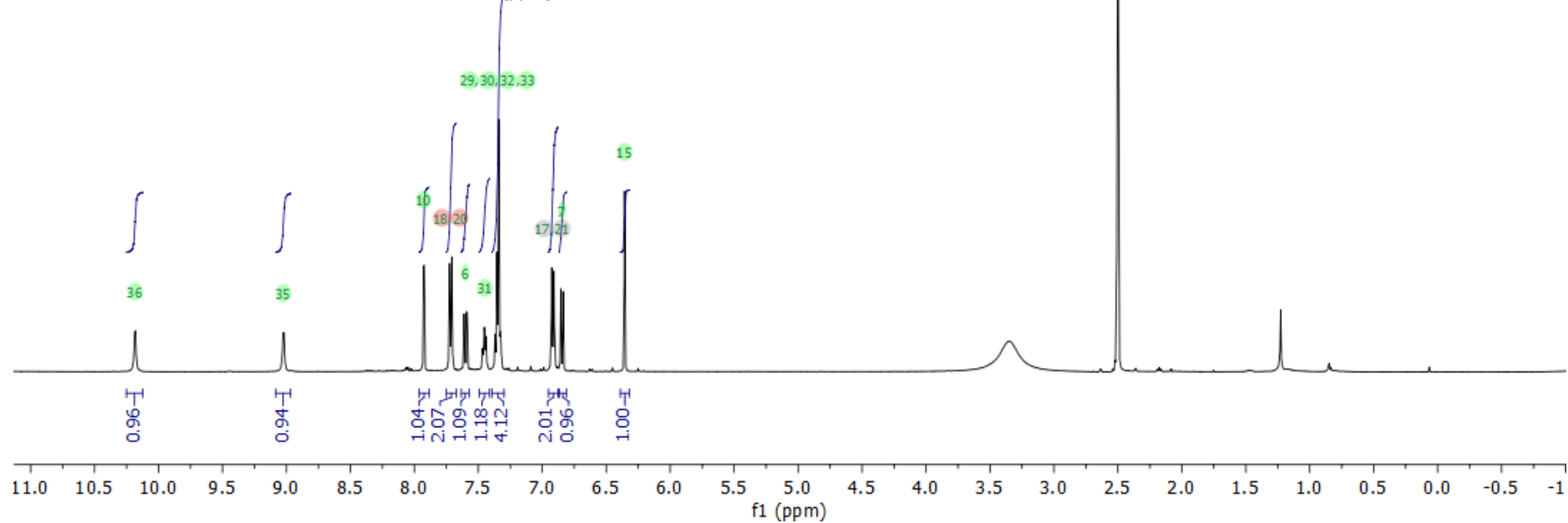

5-[3-[amino(phenyl)methylene]-2-(4-nitrophenyl)-4,5-dioxo-pyrrolidin-1-yl]-2-hydroxy-benzoic acid (6)

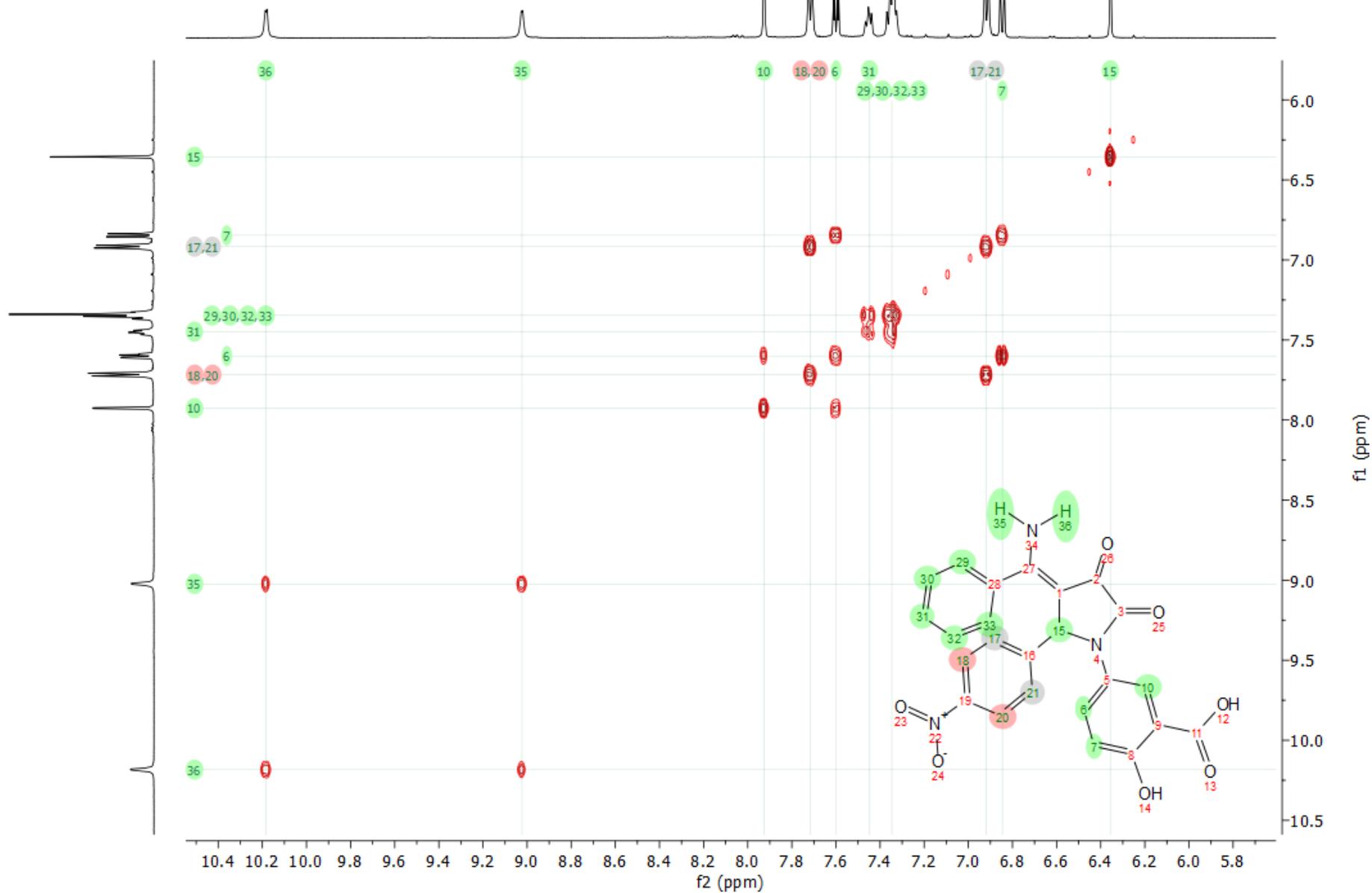

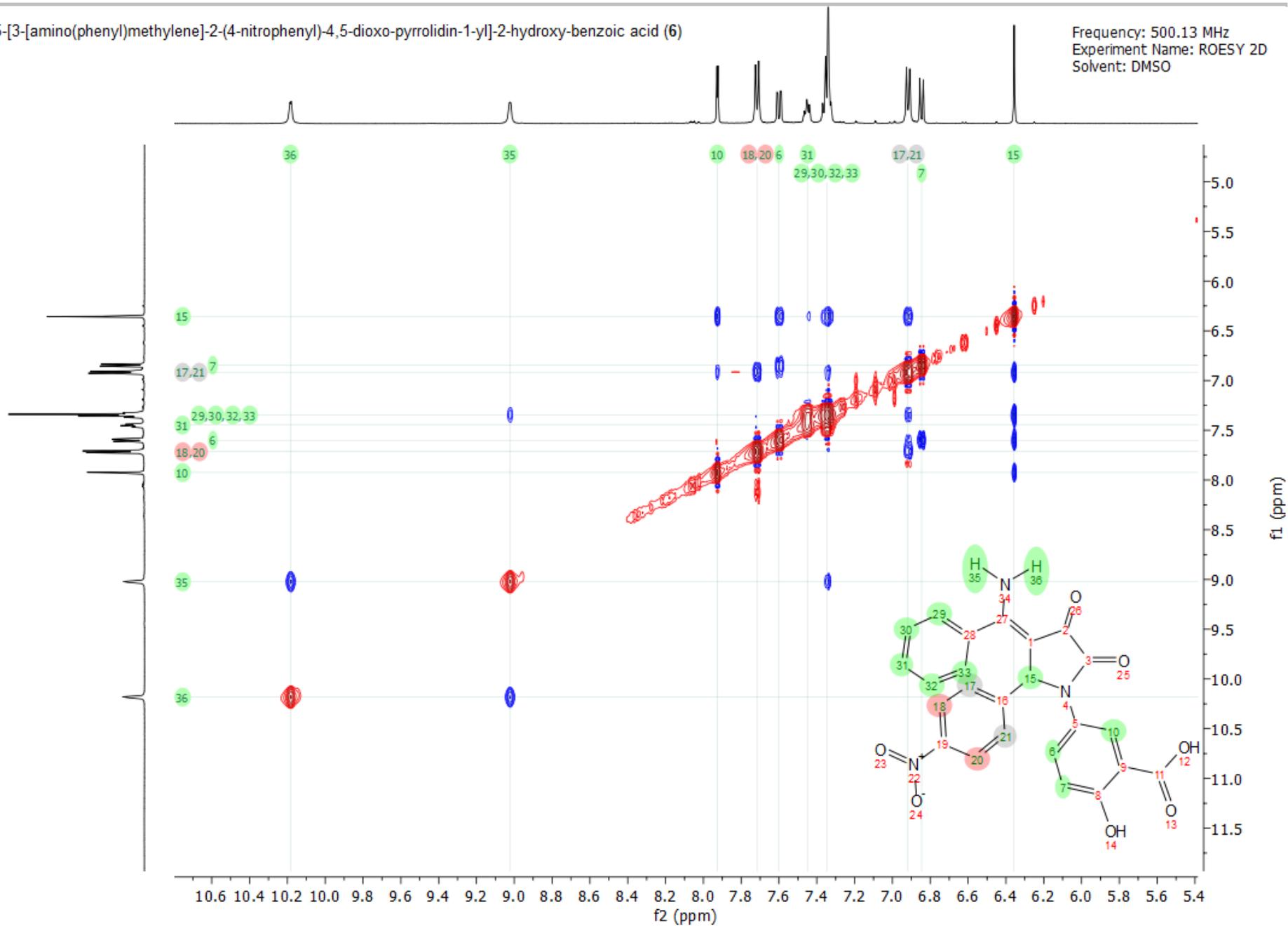

5-[3-[amino(phenyl)methylene]-2-(4-nitrophenyl)-4,5-dioxo-pyrrolidin-1-yl]-2-hydroxy-benzoic acid (**6**)

5-[3-[amino(phenyl)methylene]-2-(4-nitrophenyl)-4,5-dioxo-pyrrolidin-1-yl]-2-hydroxy-benzoic acid (6)

Frequency: 150.90 MHz
Experiment Name: ¹³C Carbon
Solvent: DMSO

—178.00
—171.13
—164.05
—163.17
—158.93
—146.64
—146.27
—133.49
—130.86
—130.52
—128.56
—128.44
—127.85
—127.63
—125.25
—122.77
—117.31
—113.11
—105.53
—58.89

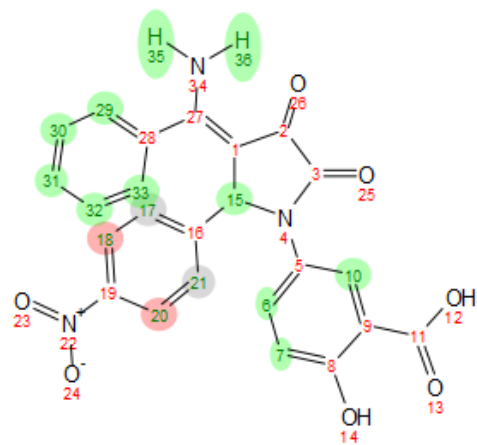

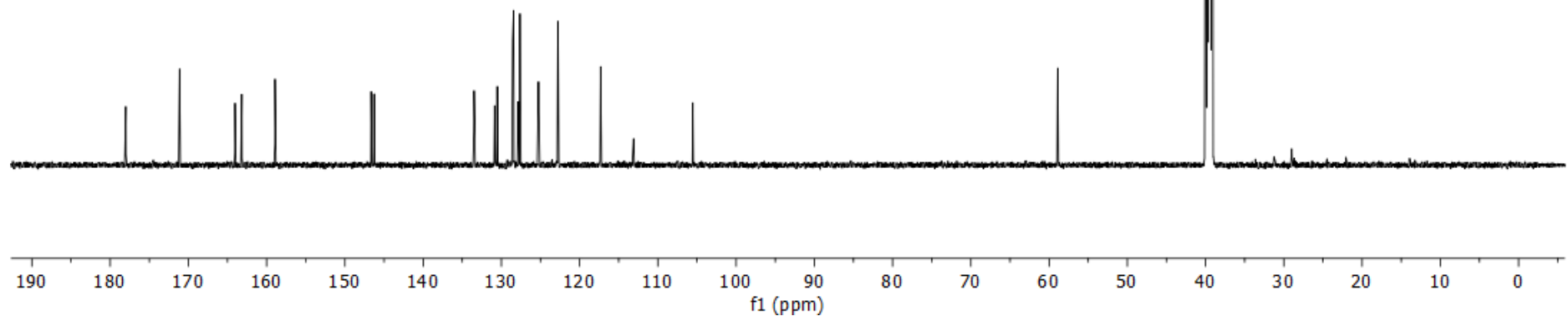

S88

(*R*)-5-[3-[amino(phenyl)methylene]-2-(4-nitrophenyl)-4,5-dioxo-pyrrolidin-1-yl]-2-hydroxy-benzoic acid (**(*R*)-6**)

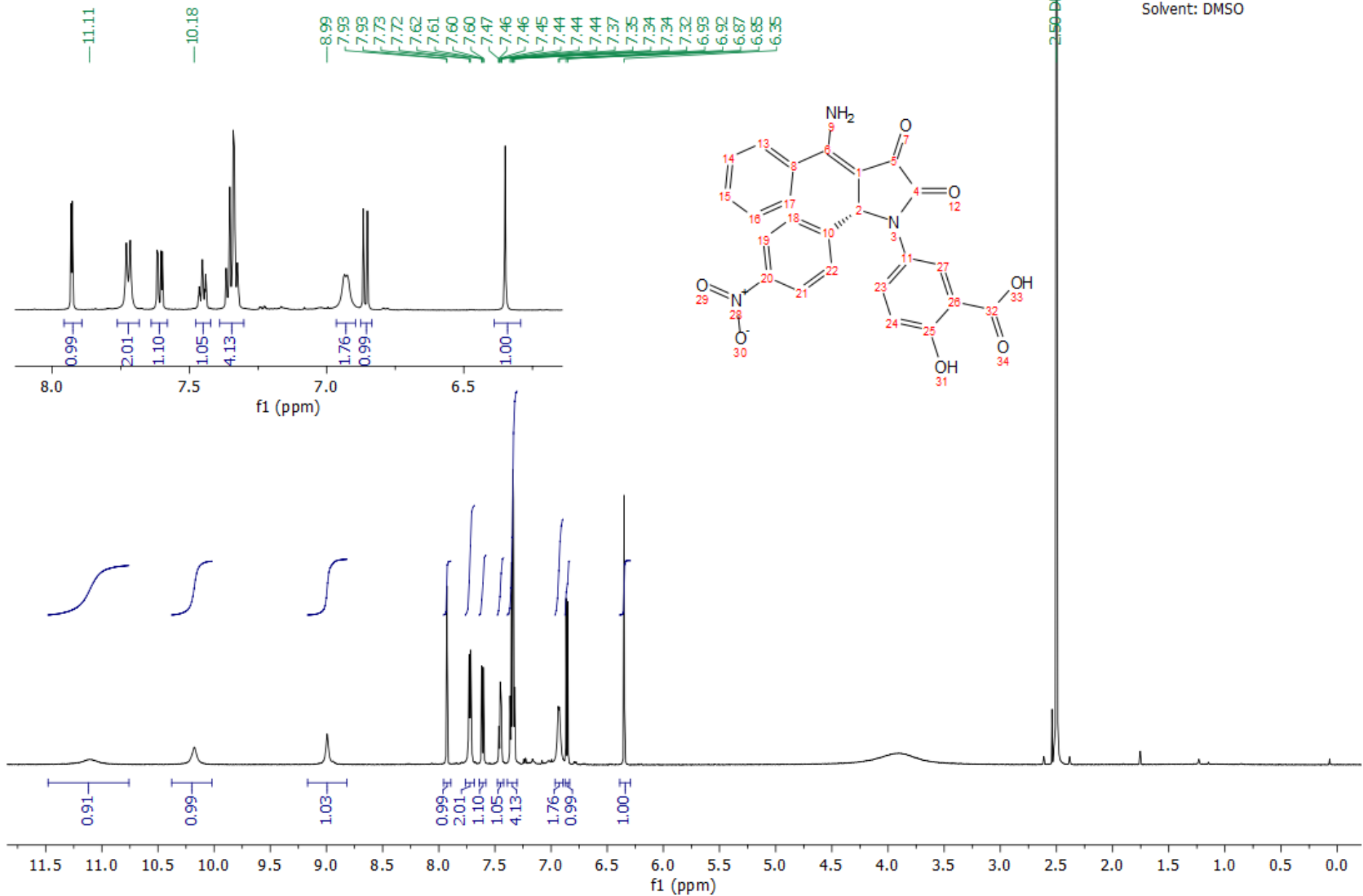

(R)-5-[3-[amino(phenyl)methylene]-2-(4-nitrophenyl)-4,5-dioxo-pyrrolidin-1-yl]-2-hydroxy-benzoic acid ((R)-6)

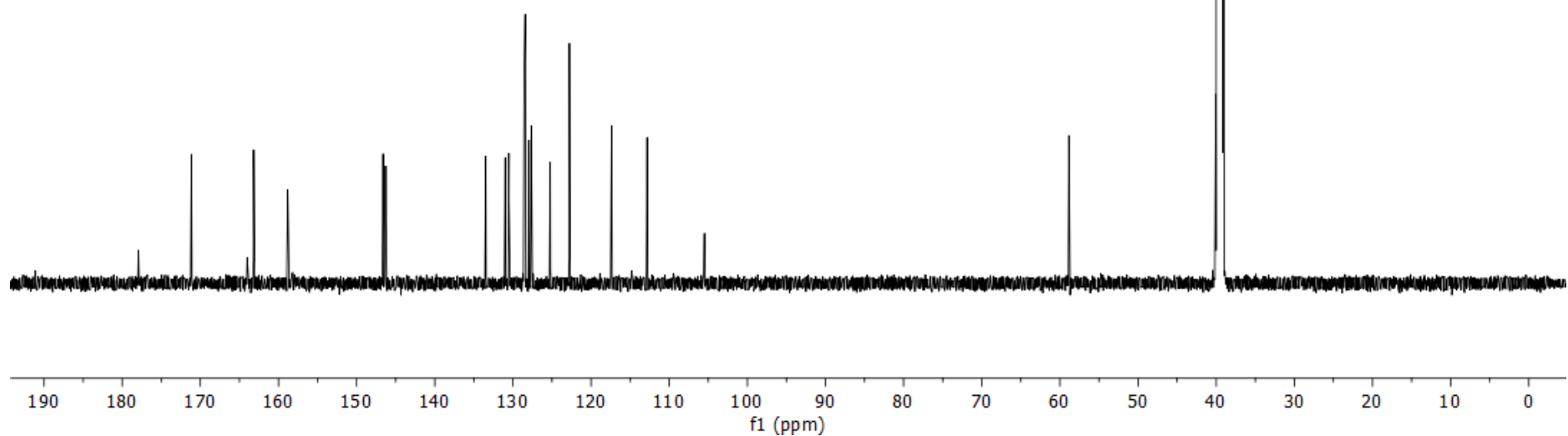

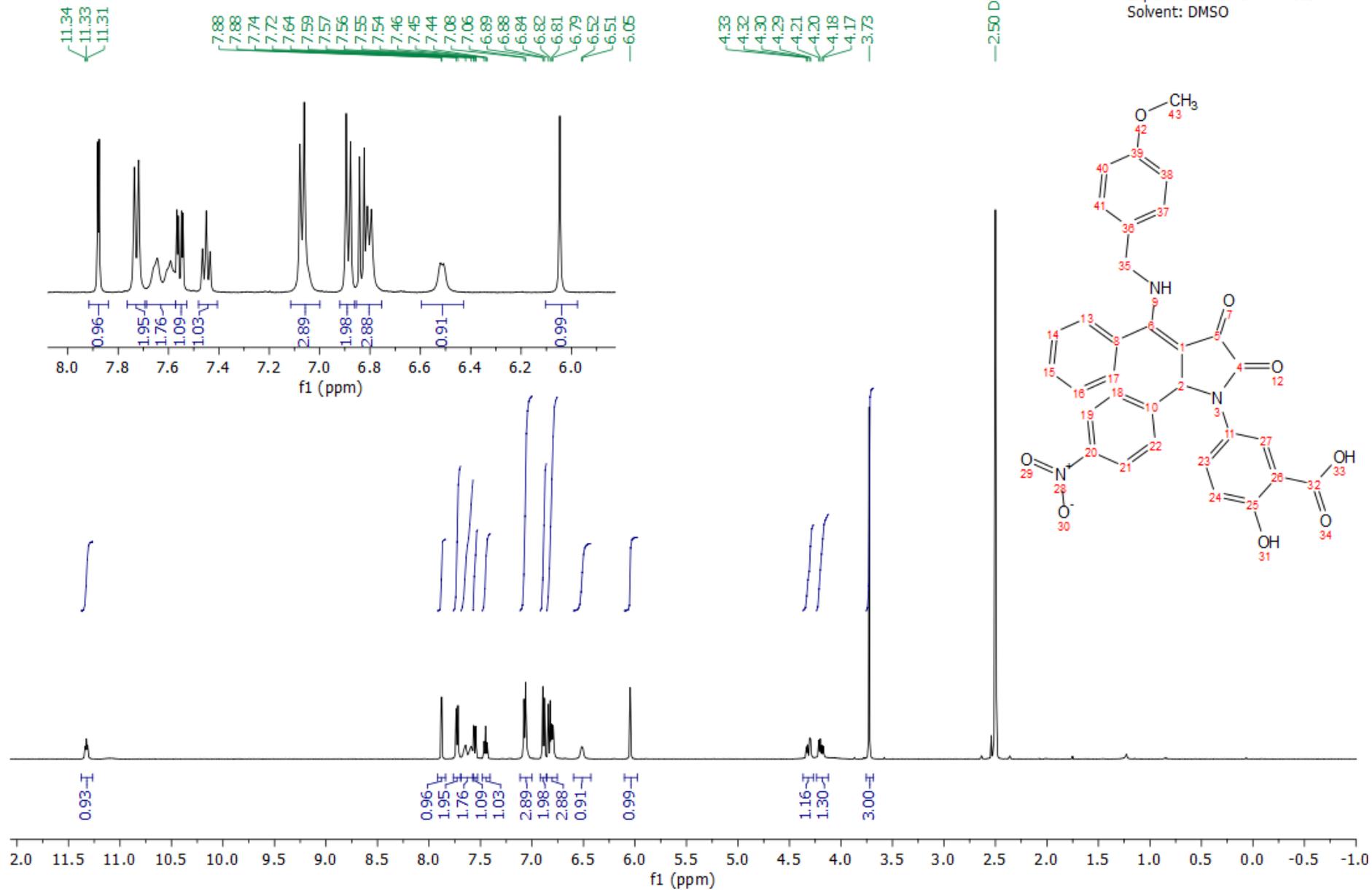

2-hydroxy-5-[3-[[(4-methoxyphenyl)methylamino]-phenyl-methylene]-2-(4-nitrophenyl)-4,5-dioxo-pyrrolidin-1-yl]benzoic acid (**7**)

Frequency: 125.76 MHz
Experiment Name: ¹³C Carbon
Solvent: DMSO

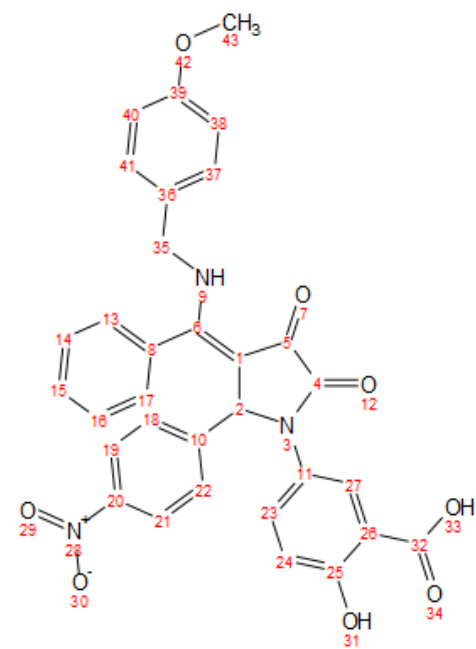

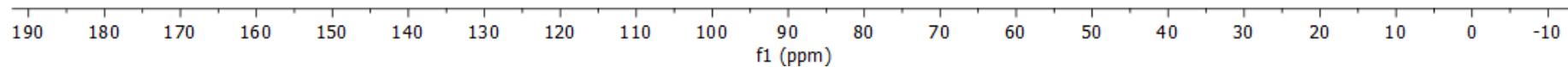

# Sample ID (*R*)-7

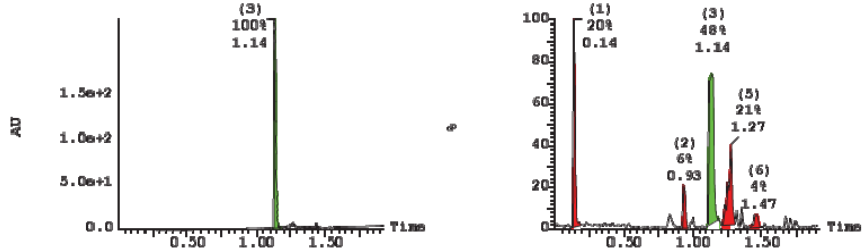
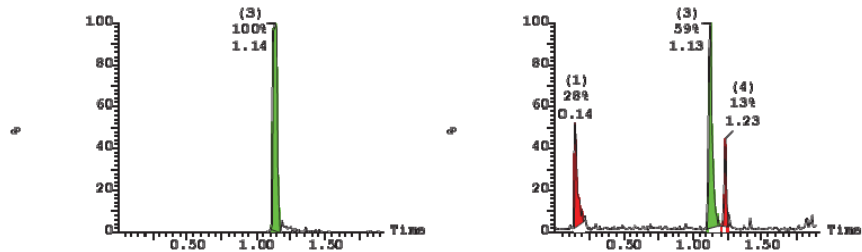
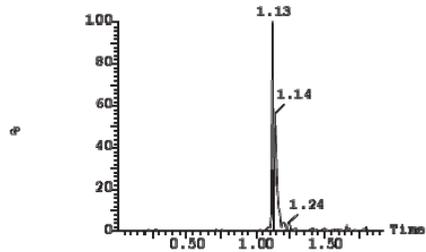
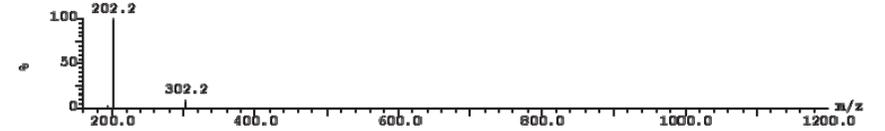
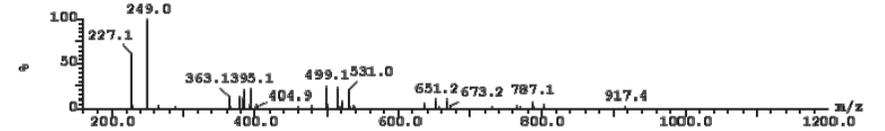
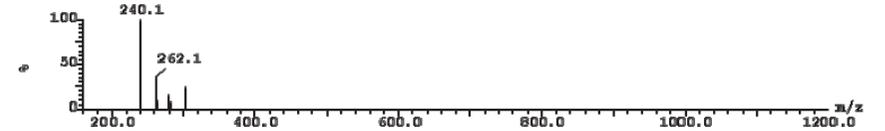
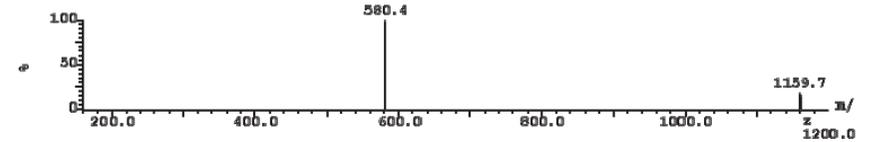

S93

Method: C:\MassLynx\L2L_LC_RM_pH3_2min.olp pH: See Method  Column: HSS C18 1.8μ 2.1x50 mm
Printed: Tue Oct 01 15:39:47 2019

Sample Report (continued):

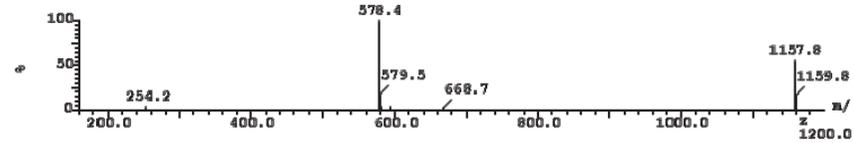

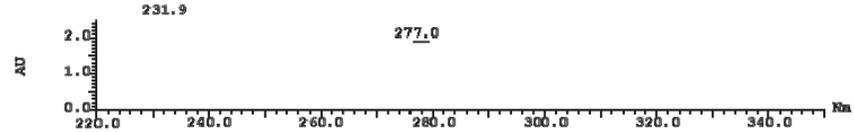

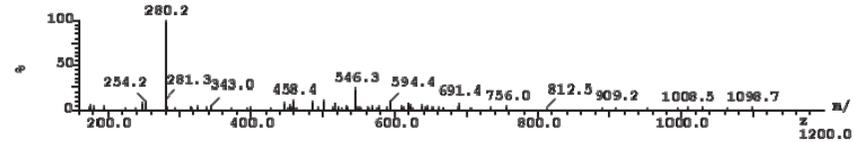

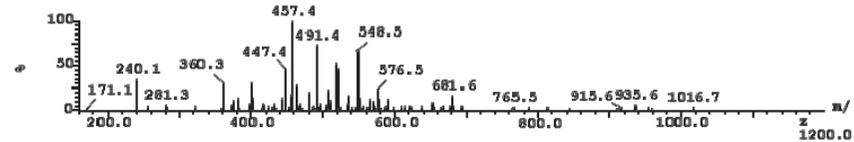

Method: C:\MassLynx\L2L_LC_RM_pH3_2min.olp pH: See Method  Column: HSS C18 1.8μ 2.1x50 mm
Printed: Tue Oct 01 15:39:47 2019

Sample Report (continued):

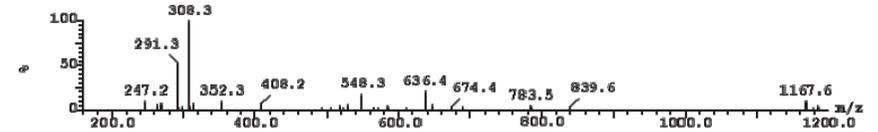

5-[3-benzoyl-4-[(4-methoxyphenyl)methylamino]-2-(4-nitrophenyl)-5-oxo-2H-pyrrol-1-yl]-2-hydroxy-benzoic acid (**8**).

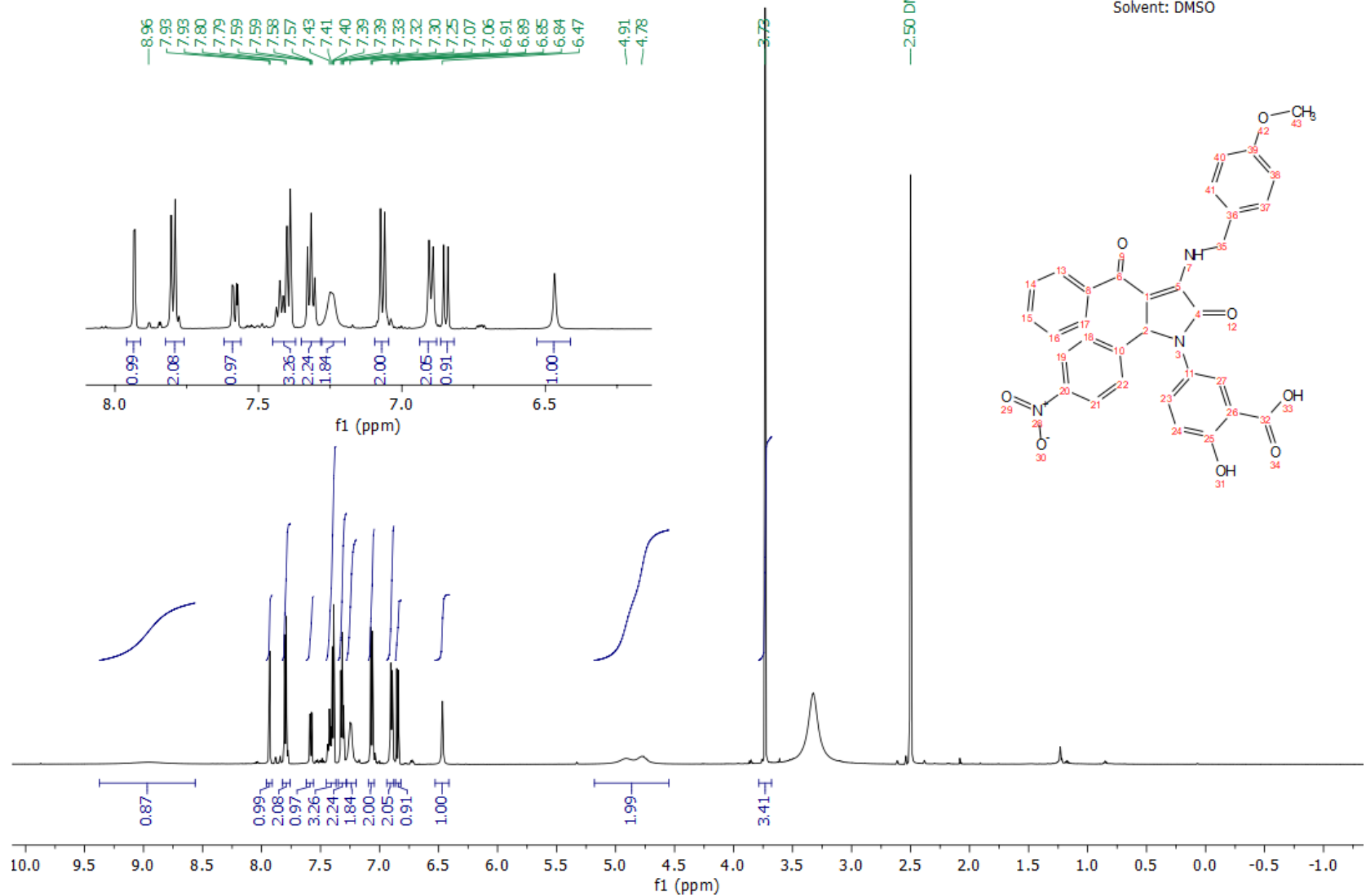

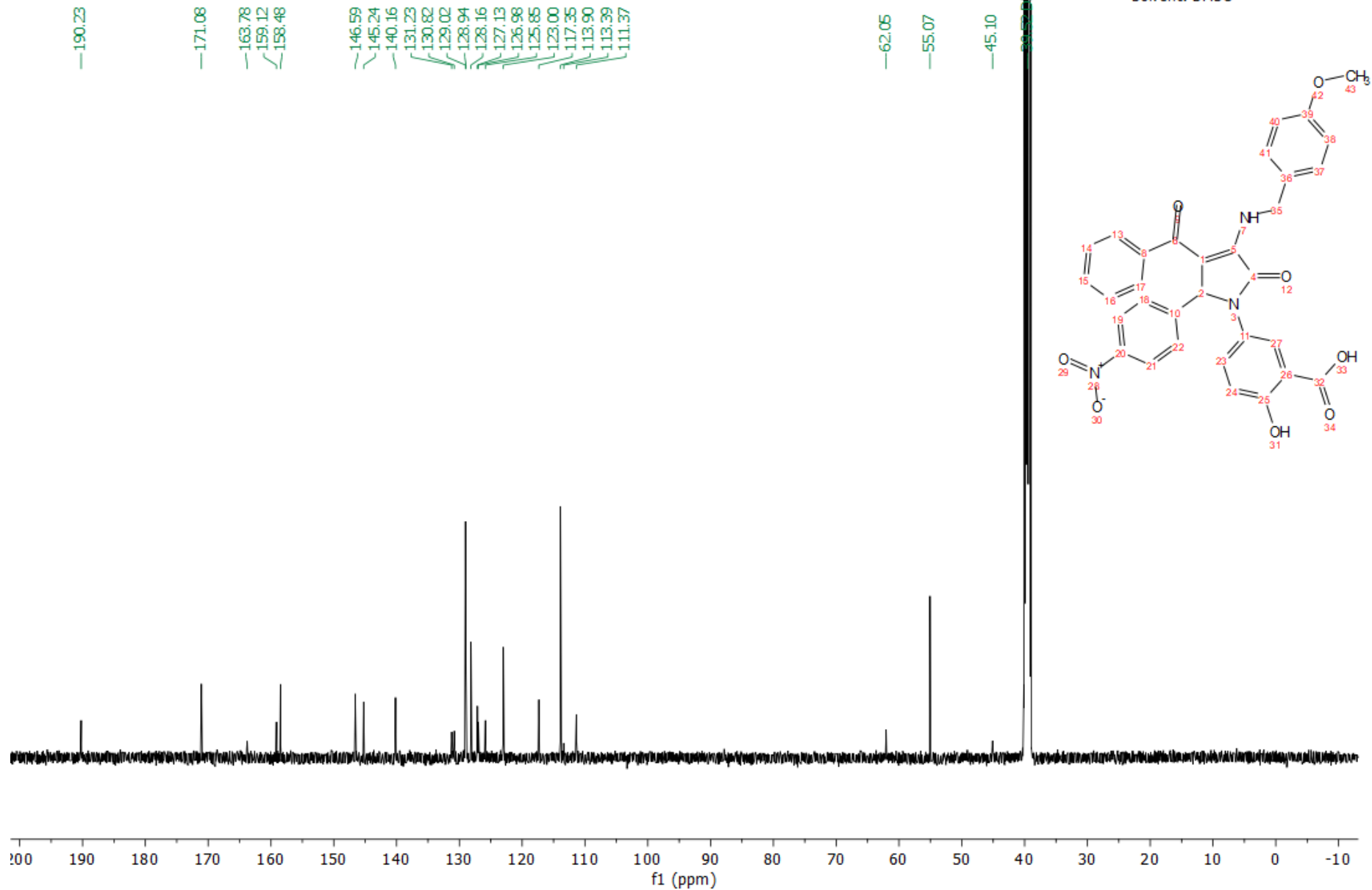

5-[3-benzoyl-4-[(4-methoxyphenyl)methylamino]-2-(4-nitrophenyl)-5-oxo-2H-pyrrol-1-yl]-2-hydroxy-benzoic acid (**8**).

# Sample ID (*R*)-8

Method: C:\MassLynx\L2L_LC_RM_pH3_2min.olp  Column: HSS C18 1.8µ 2.1x50 mm

Printed: Tue Oct 01 15:41:55 2019

Sample Report:

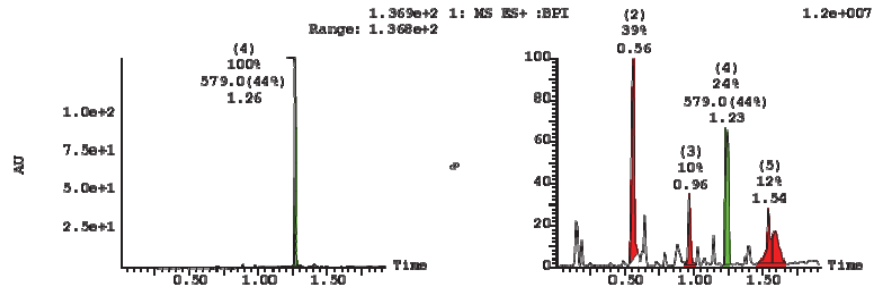

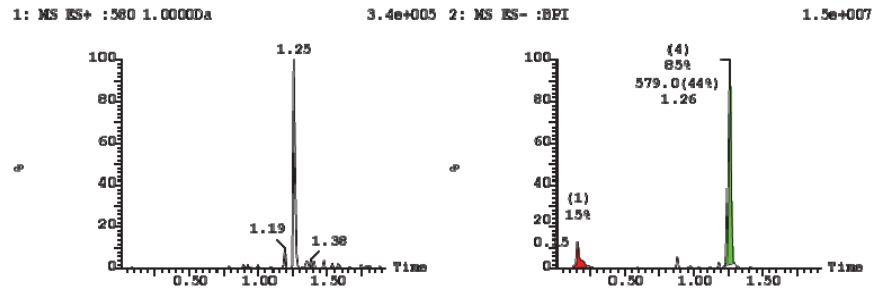

Method: C:\MassLynx\L2L_LC_RM_pH3_2min.olp  Column: HSS C18 1.8µ 2.1x50 mm

Printed: Tue Oct 01 15:41:55 2019

Sample Report (continued):

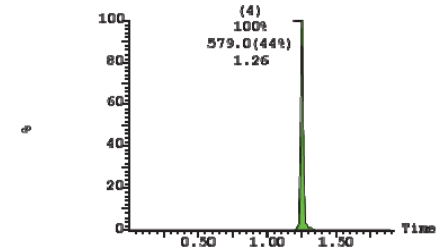

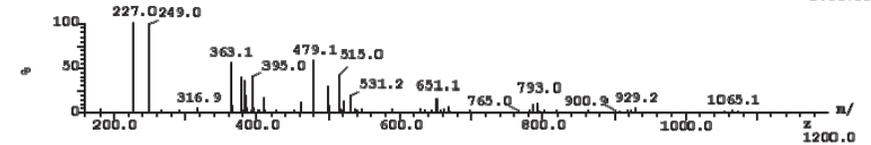

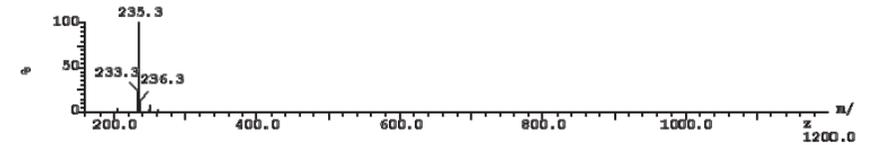

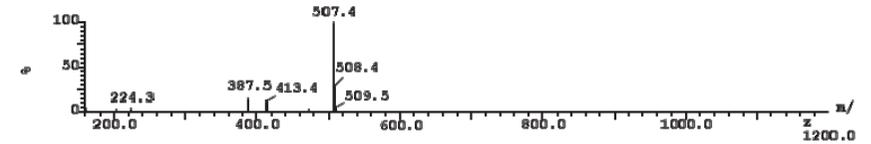

Column: HSS C18 1.8µ 2.1x50 mm

Printed: Tue Oct 01 15:41:55 2019

Sample Report (continued):

```
Peak ID   Time    Mass Found
   4      1.23    Not Found
4:(Time: 1.26) Combine (114:124-(107:111+131:136))         1:MS ES+
                                                          1.4e+006
```

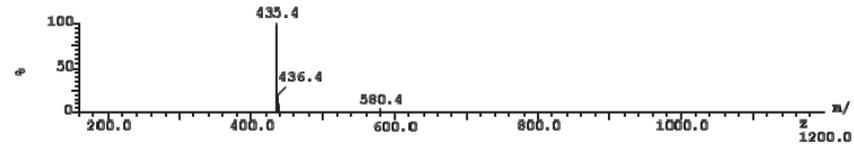

```
Peak ID Time Mass Found
   4    1.23   578
4:(Time: 1.26) Combine (113:123-(104:108+135:140))         2:MS ES-
                                                          2.4e+006
```

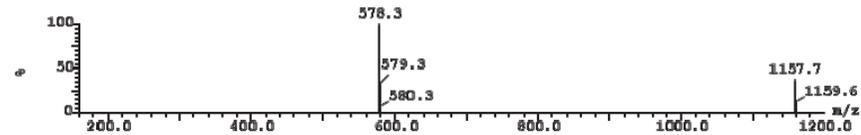

```
Peak ID   Time    Mass Found
   4      1.23    Not Found
4:(Time: 1.26) Combine (2985)                            3:UV Detector
                                                           1.891 AU
```

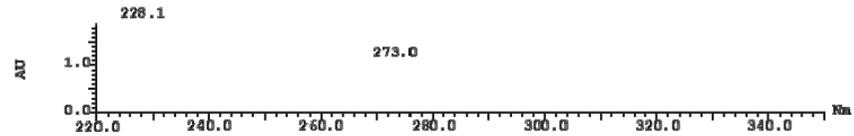

```
Peak ID   Time    Mass Found
   5      1.54    Not Found
5:(Time: 1.54) Combine (140:150-(128:132+156:161))         1:MS ES+
                                                          1.2e+006
```

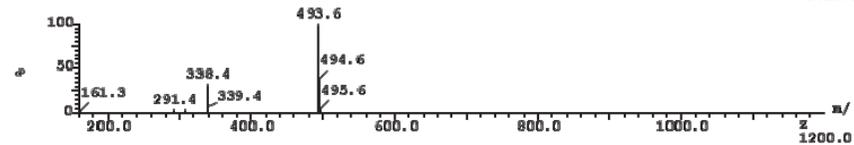

Column: HSS C18 1.8µ 2.1x50 mm

Printed: Tue Oct 01 15:41:55 2019

Sample Report (continued):

```
Peak ID   Time    Mass Found
   6      1.58    Not Found
6:(Time: 1.58) Combine (144:154-(138:142+165:170))         1:MS ES+
                                                          1.2e+006
```

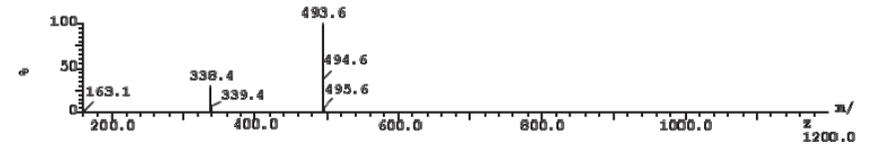

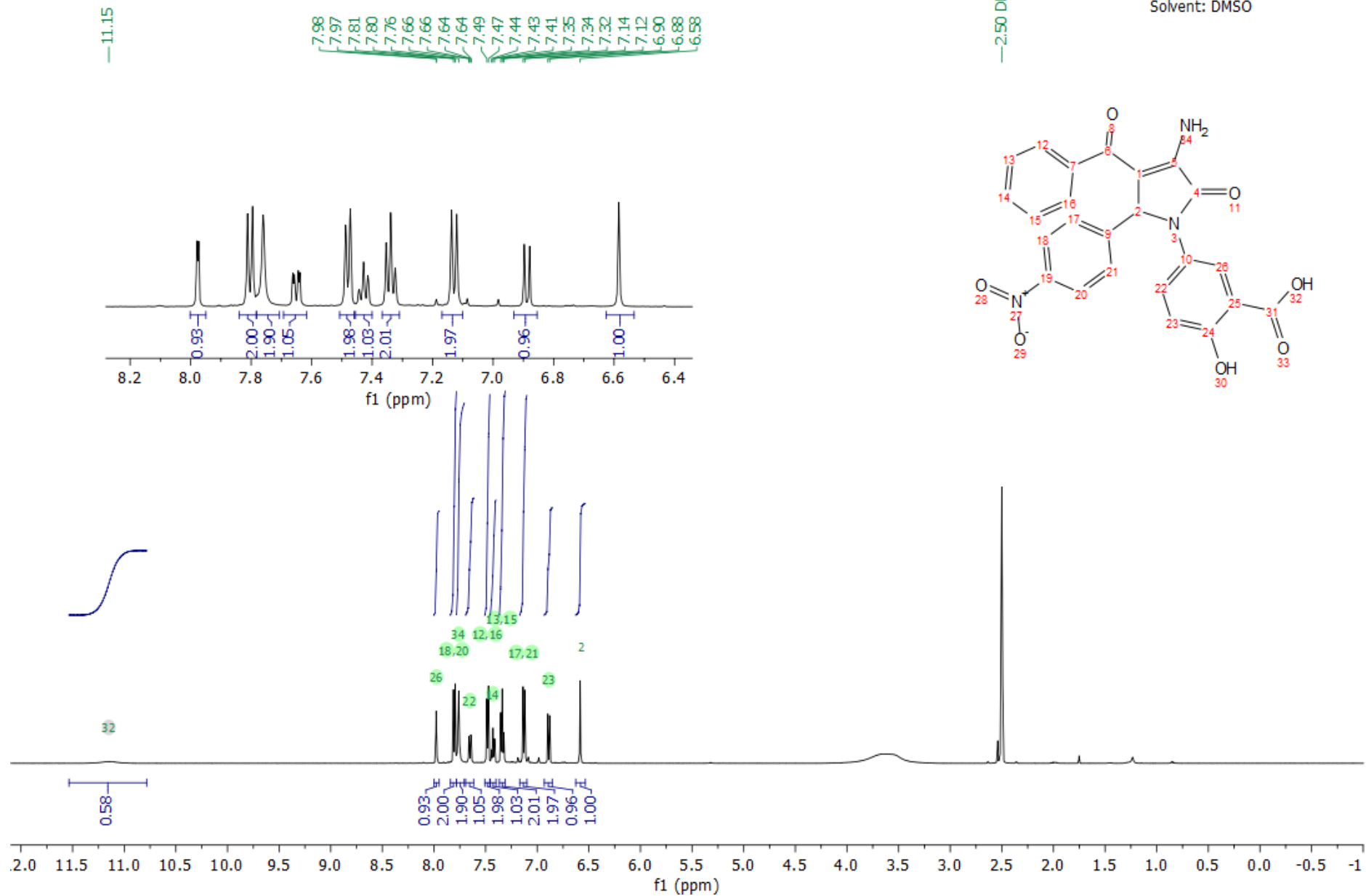

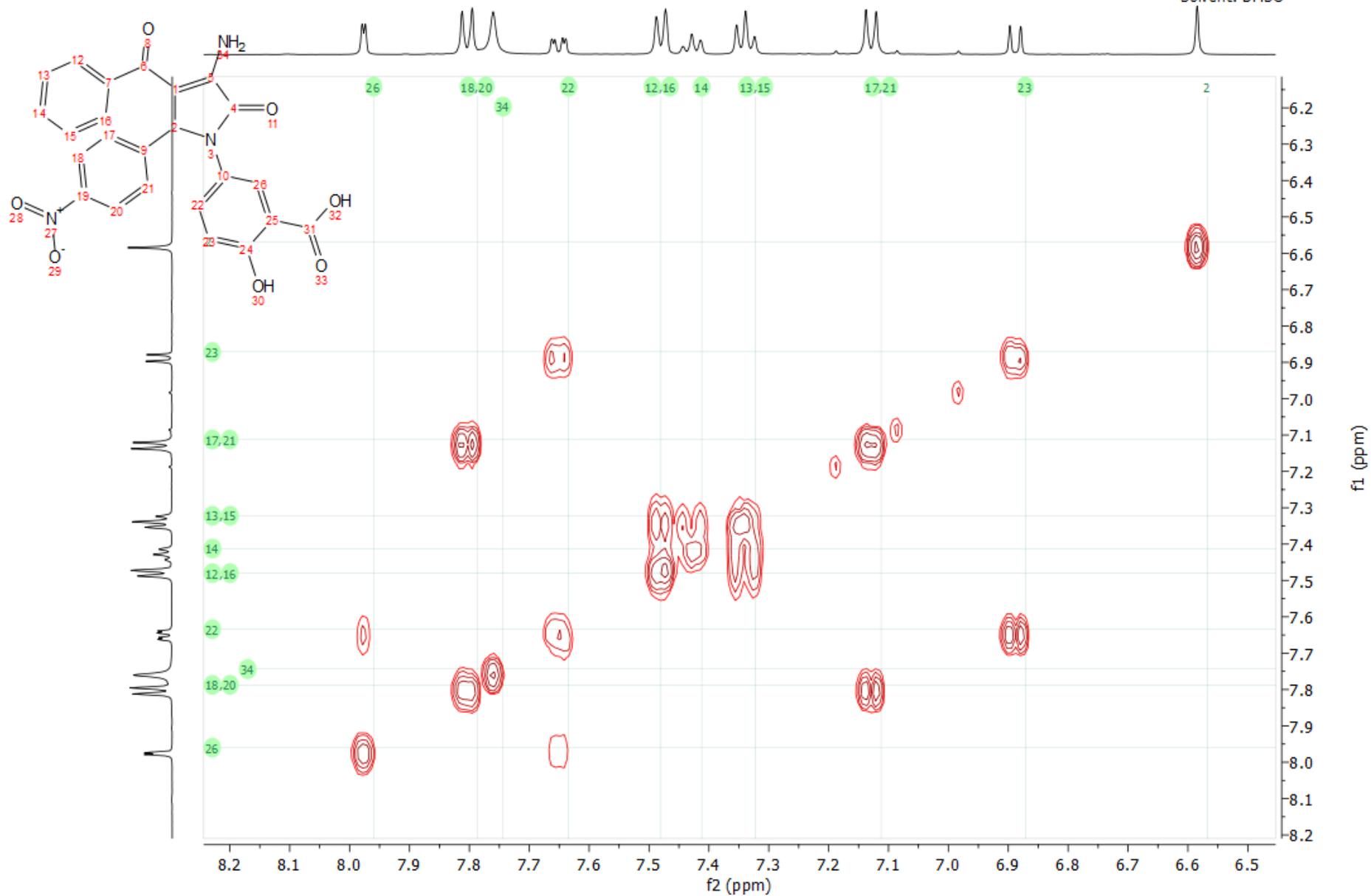

5-[4-amino-3-benzoyl-2-(4-nitrophenyl)-5-oxo-2H-pyrrol-1-yl]-2-hydroxy-benzoic acid (9)

190.40, 171.19, 163.74, 158.92, 148.27, 146.61, 145.29, 140.12, 131.26, 130.73, 128.99, 128.23, 127.31, 126.88, 125.48, 123.05, 117.54, 112.96, 110.22, 62.13

Frequency: 125.76 MHz
Experiment Name: $^{13}$C Carbon
Solvent: DMSO

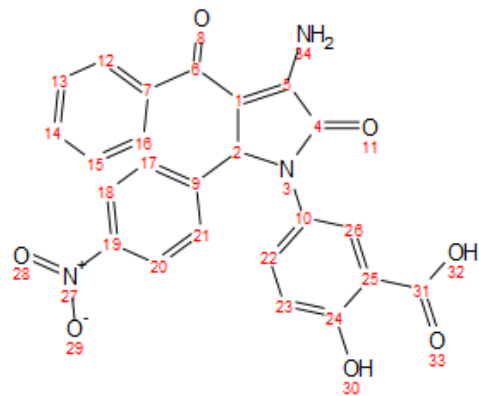

(R)-5-[4-amino-3-benzoyl-2-(4-nitrophenyl)-5-oxo-2H-pyrrol-1-yl]-2-hydroxy-benzoic acid ((R)-9)

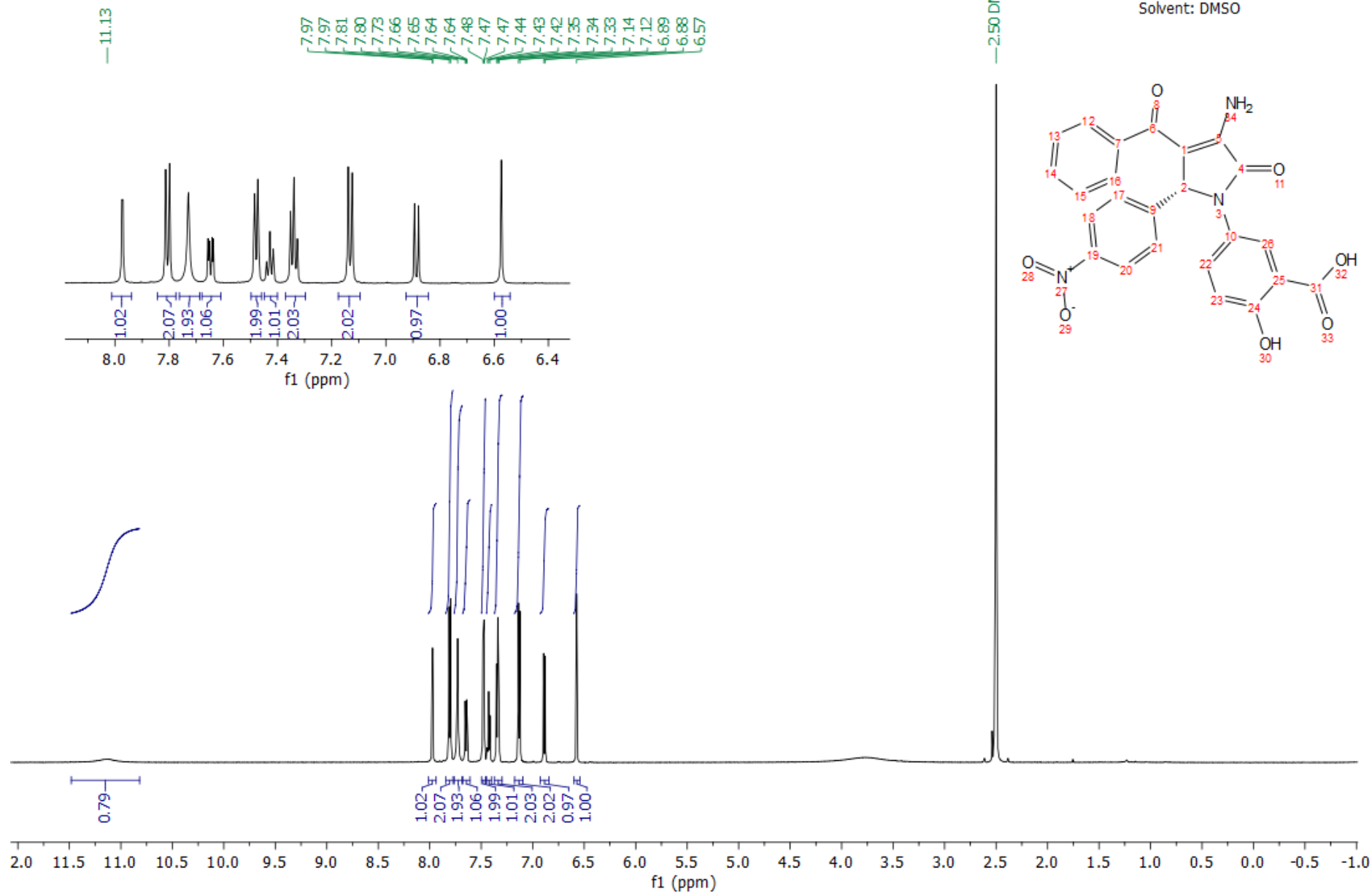

(R)-5-[4-amino-3-benzoyl-2-(4-nitrophenyl)-5-oxo-2H-pyrrol-1-yl]-2-hydroxy-benzoic acid ((R)-9)

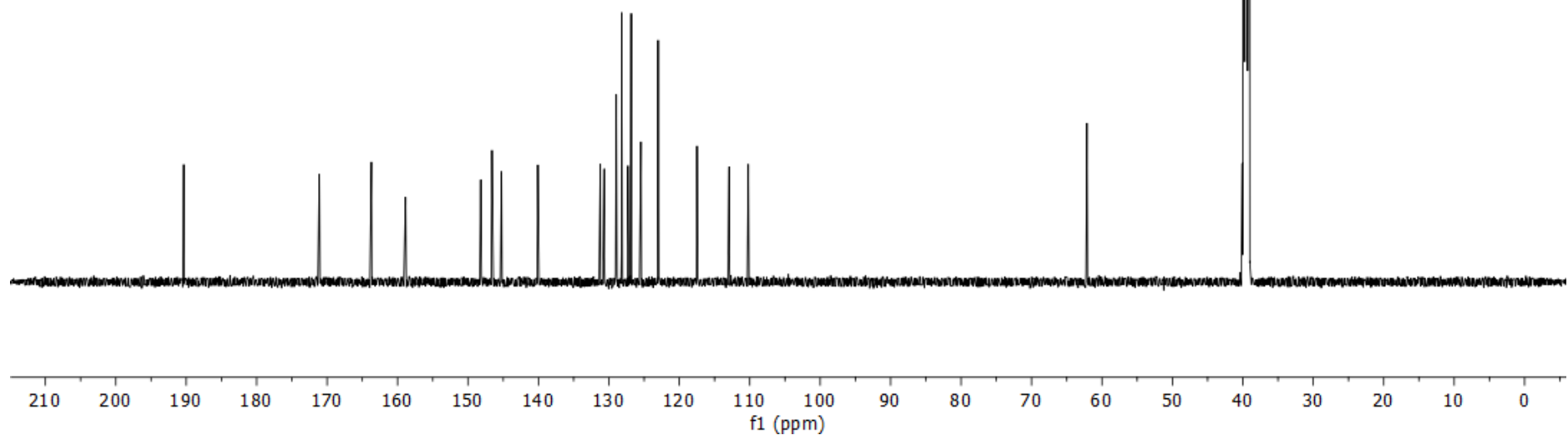

Enantiomeric purity for (*R*)-5-[3-[amino(phenyl)methylene]-2-(4-nitrophenyl)-4,5-dioxo-pyrrolidin-1-yl]-2-hydroxy-benzoic acid ((*R*)-6).

| SSL Report | Request type | | | |
|---|---|---|---|---|

**Sample Information**

| Sample ID | EN Number | Incoming amount | Compound name | Project name |
|---|---|---|---|---|
| (*R*)-6 | | 1 mg | | |

**Analytical Conditions**

| Column | Dimension (mm) | Particle Size (µm) |
|---|---|---|
| Chiralpak IH | 150 x 4.6 | 3 |

| Mobile Phase | | |
|---|---|---|
| 25% EtOH/FA 100/20mM in $CO_2$, 120 bar | | |
| Flow (ml/min) | Detection | Temperature (C) |
| 3.5 | 247 | 40 |
| Gradient | Instrument | Sample concentration (mg/ml) |
| | | 2 mg/mL |

| Sample ID | Purity | Chiral Purity | Weight |
|---|---|---|---|
| Comments | | 91.8 ee | |
| Sample dissolved in EtOH | | | |

Sample Name: (*R*)-6

| | Column: | | Column ID: | ChiralpakIH |
|---|---|---|---|---|
| | Mobile Phase A: | $CO_2$ | Column Dimension: | 150 * 4.6 |
| | Mobile Phase B: | EtOH/FA 100/20mM | Particle Size: | 3 |
| | Gradient: | 25% B, 120 bar | Injection volume: | 8.00 ul |
| | | | Flow : Wavelength: | 3.5 ml/min |
| | Temperature: | 40°C | | PDA Spectrum PDA |
| | Sample Concentration: | Sample in EtOH | Vial: | 2:f,1 |

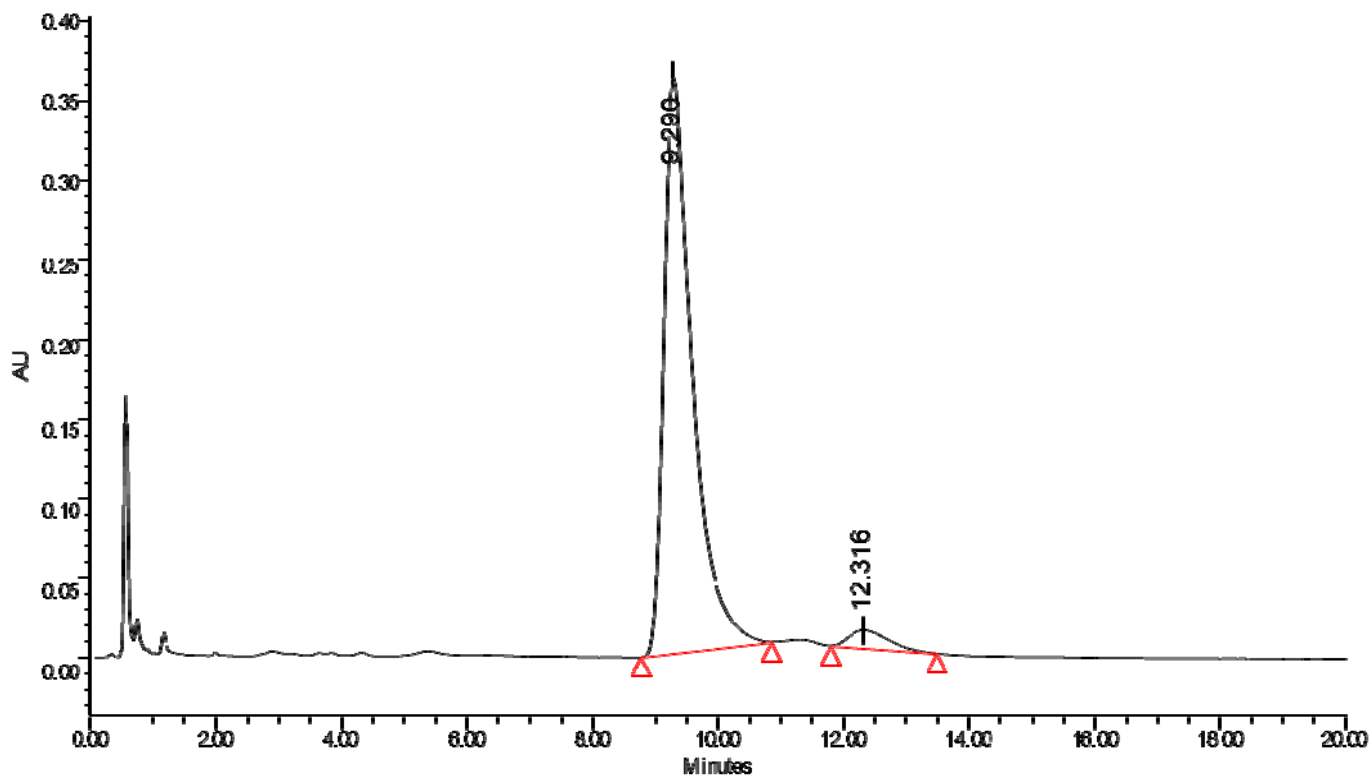

| | Retention Time | Area | % Area | K | N | USP Resolution |
|---|---|---|---|---|---|---|
| 1 | 9.290 | 11626356 | 95.91 | 0.000 | 1929.3 | |
| 2 | 12.316 | 495178 | 4.09 | 0.328 | 1793.5 | 2.983 |

ee = 91.8

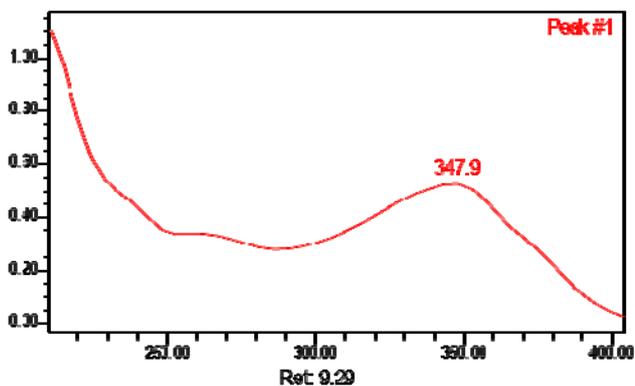

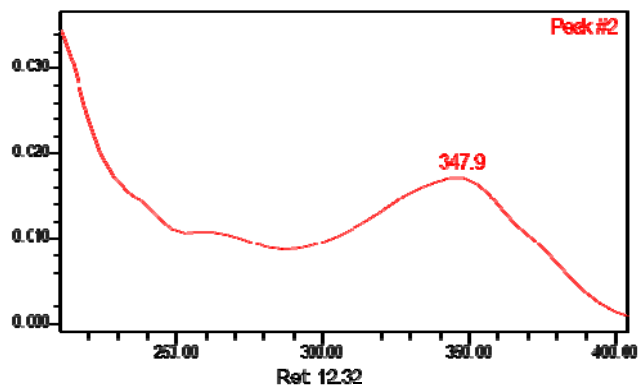

S105

## Sample Name: rac-6

| | | | | |
|---|---|---|---|---|
| Column: | | Column ID: | ChiralpakIH | |
| Mobile Phase A: | CO₂ | Column Dimension: | 150 * 4.6 | |
| Mobile Phase B: | EtOH/FA 100/20mM | Particle Size: | 3 | |
| Gradient: | 25% B, 120 bar | Injection volume: | 8.00 ul | |
| | | Flow: | 3.5 ml/min | |
| Temperature: | 40°C | Wavelength: | PDA Spectrum PDA | |
| Sample Concentration: | Sample in EtOH | Vial: | 1:c,2 | |

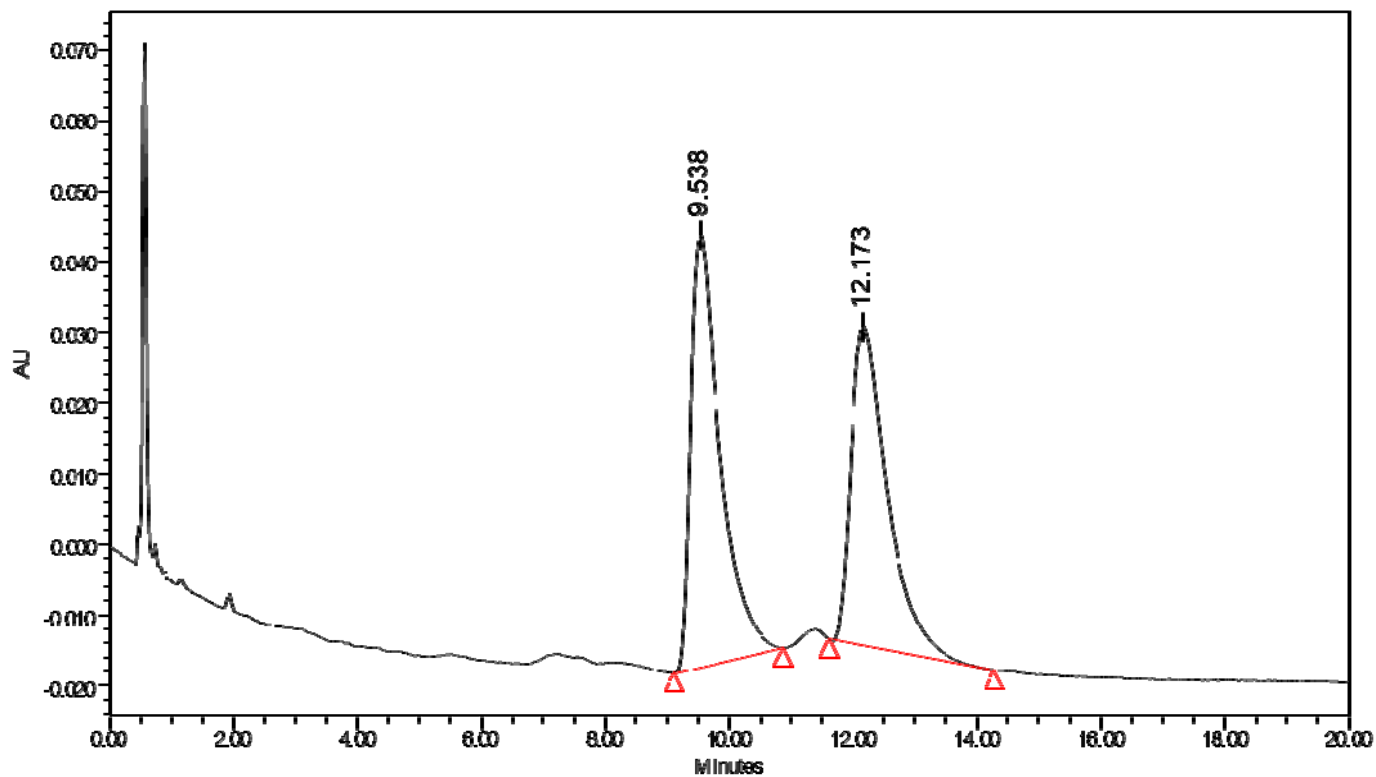

| | Retention Time | Area | % Area | K | N | USP Resolution | Width @50% |
|---|---|---|---|---|---|---|---|
| 1 | 9.538 | 2016809 | 52.08 | 0.000 | 2106.7 | | 4.699859e-001 |
| 2 | 12.173 | 1855363 | 47.92 | 0.276 | 2151.7 | 2.663 | 6.183230e-001 |

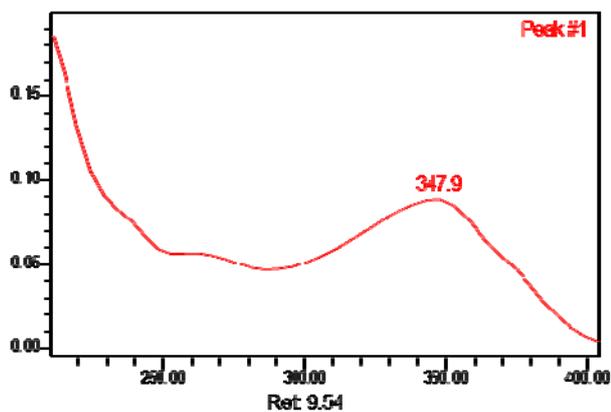

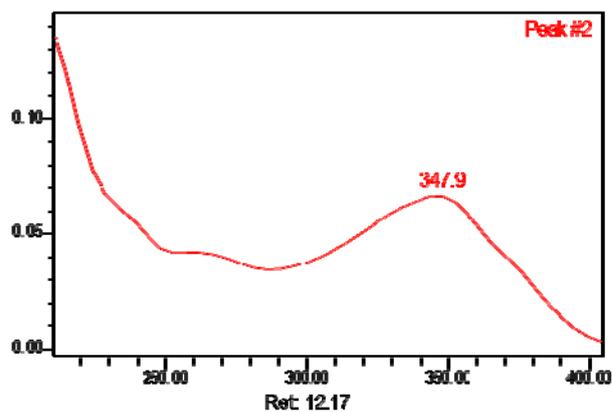

Enantiomeric purity for **(*R*)-5-[4-amino-3-benzoyl-2-(4-nitrophenyl)-5-oxo-2H-pyrrol-1-yl]-2-hydroxy-benzoic acid ((*R*)-9).**

| SSL Report | Request type | | |
|---|---|---|---|

**Sample Information**

| Sample ID | EN Number | Incoming amount | Compound name | Project name |
|---|---|---|---|---|
| (*R*)-9 | | 1 mg | | |

**Analytical Conditions**

| Column | Dimension (mm) | Particle Size (μm) |
|---|---|---|
| Chiralpak IH | 150 x 4.6 | 3 |
| Mobile Phase | | |
| 35% EtOH/FA 100/20mM in $CO_2$, 120 bar | | |
| Flow (ml/min) | Detection | Temperature (C) |
| 3.5 | 247 | 40 |
| Gradient | Instrument | Sample concentration (mg/ml) |
| | | 2 mg/mL |

| Sample ID | Purity | Chiral Purity | Weight |
|---|---|---|---|
| Comments | | 89.9 ee | |
| Sample dissolved in EtOH | | | |

**Sample Name:** (*R*)-9

| | |
|---|---|
| Column: | x |
| Mobile Phase A: | CO₂ |
| Mobile Phase B: | EtOH/FA 100/20 mM |
| Gradient: | 35% B, 120 bar |
| Temperature: | 40°C |
| Sample Concentration: | Sample in EtOH |
| Column ID: | Chiralpak IH |
| Column Dimension: | 150 * 4.6 |
| Particle Size: | 3 |
| Injection volume: | 4.00 ul |
| Flow: | 3.5 ml/min |
| Wavelength: | PDA Spectrum PDA |
| Vial: | 2:f,2 |

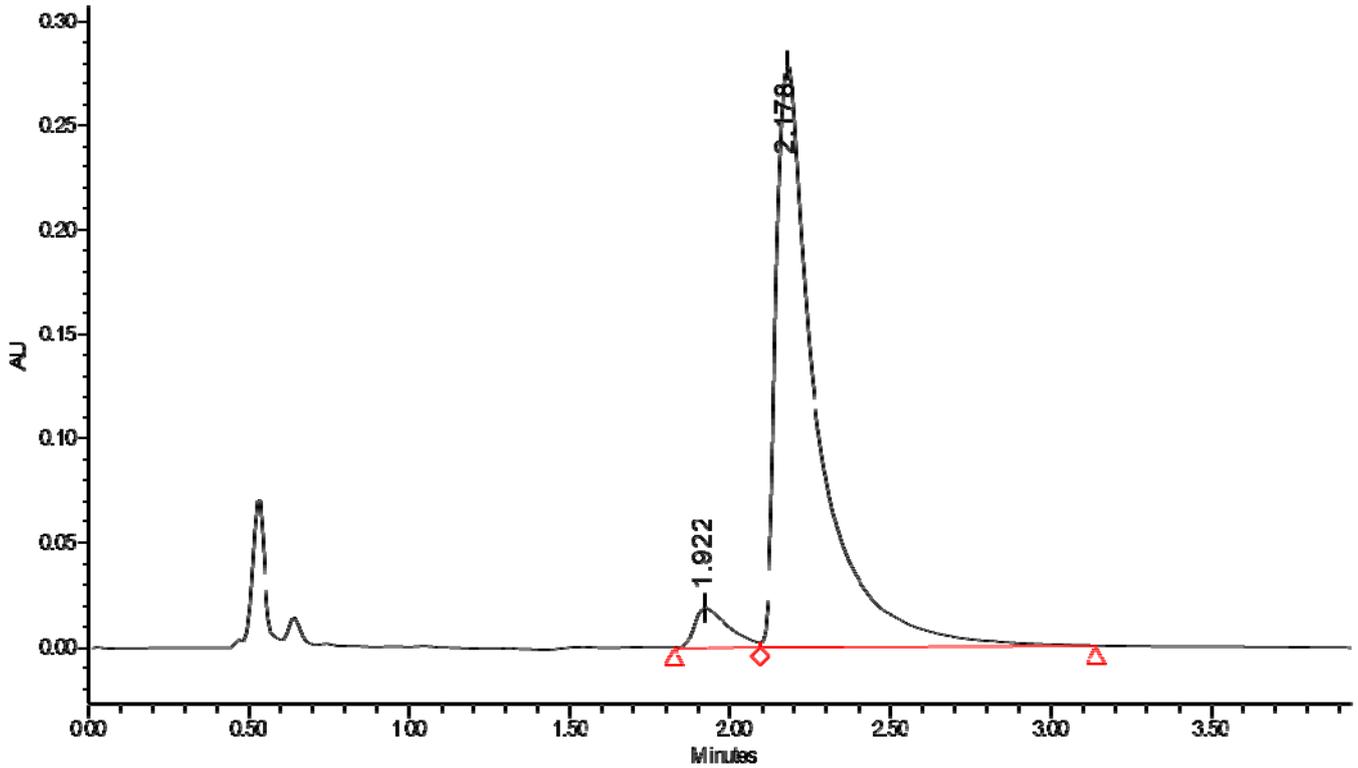

| | Retention Time | Area | % Area | k' | N | USP Resolution |
|---|---|---|---|---|---|---|
| 1 | 1.922 | 133211 | 5.12 | 0.000 | 1528.0 | |
| 2 | 2.178 | 2468454 | 94.88 | 0.134 | 1943.5 | 1.227 |

ee = 89.8

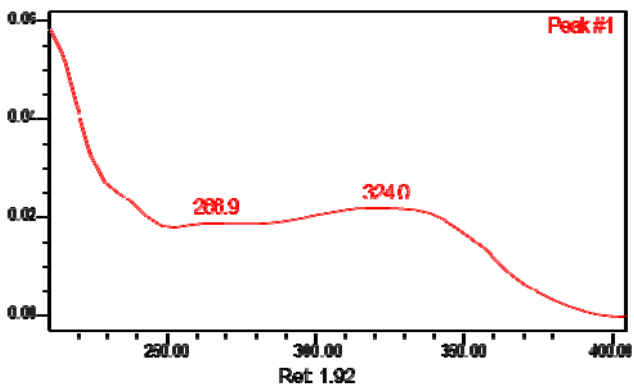

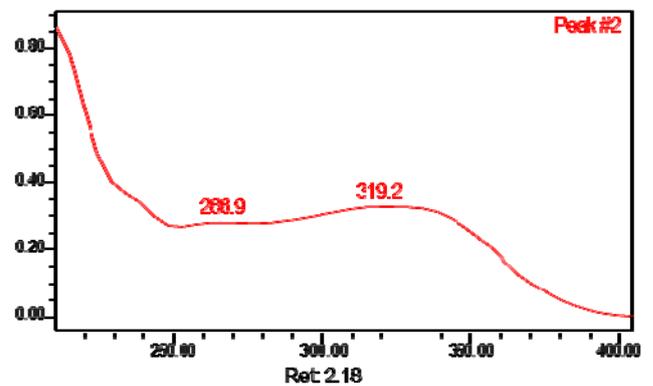

## Sample Name: rac-9

| | |
|---|---|
| Column: | x |
| Mobile Phase A: | CO2 |
| Mobile Phase B: | EtOH/FA 100/20 mM |
| Gradient: | 30% B, 120 bar |
| Temperature: | 40°C |
| Sample Concentration: | Sample in EtOH |
| Column ID: | Chiralpak IH |
| Column Dimension: | 150 * 4.6 |
| Particle Size: | 3 |
| Injection volume: | 8.00 ul |
| Flow: | 3.5 ml/min |
| Wavelength: | PDA Spectrum PDA |
| Vial: | 1:c,3 |

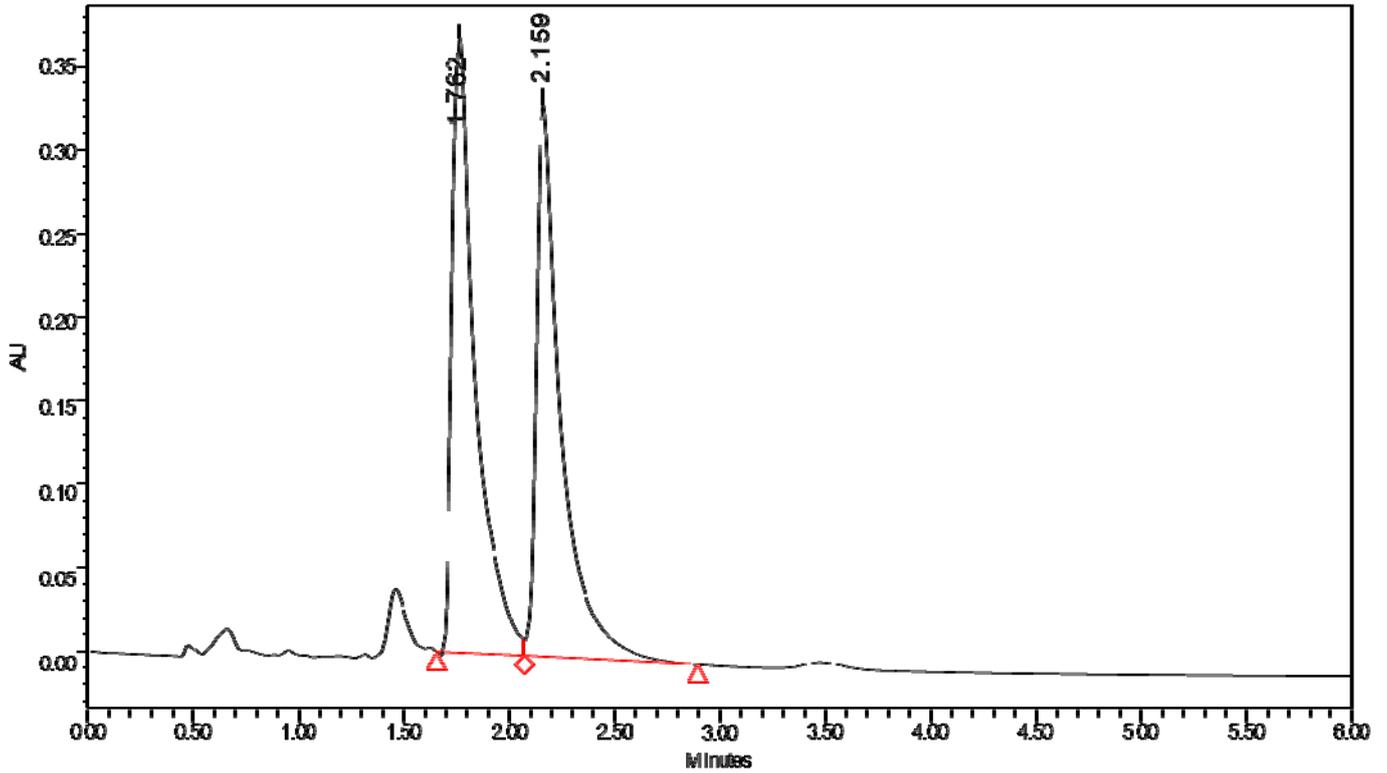

| | Retention Time | Area | %Area | K' | N | USP Resolution | Width @ 50% |
|---|---|---|---|---|---|---|---|
| 1 | 1.762 | 2851925 | 50.90 | 0.000 | 1479.3 | | 1.033490e-001 |
| 2 | 2.159 | 2751068 | 49.10 | 0.225 | 1996.0 | 1.961 | 1.073459e-001 |

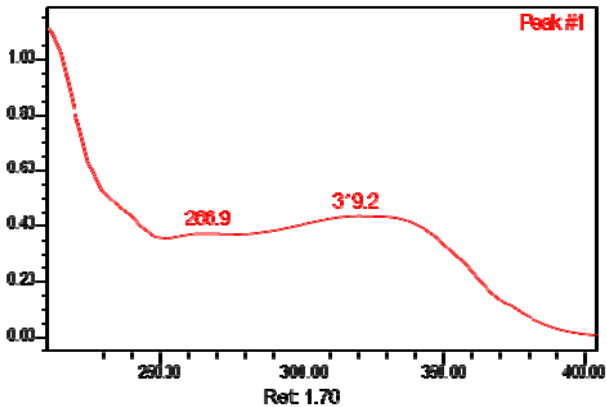

Peak #1, 266.9, 319.2, Ret 1.70

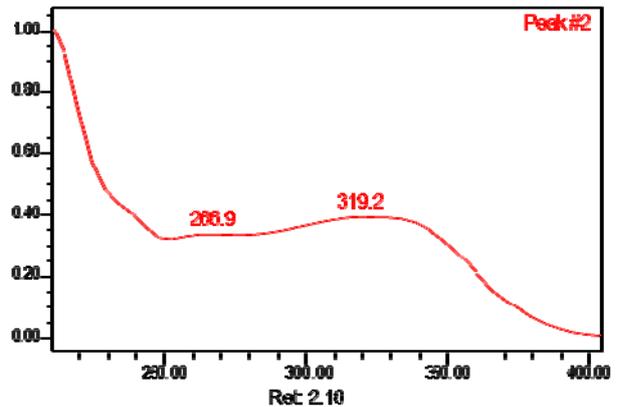

Peak #2, 266.9, 319.2, Ret 2.10

## V. Computational details

All calculations were performed within the Schrödinger Small-Molecule Drug Discovery Suite 2019-2.[12] Initial geometries were derived from MCMM[13] conformational searches in Macromodel version 12.4 using the OPLS3e[14] force field in combination with the GB/SA continuum solvation model for water.[15] Density functional theory calculations were performed using Jaguar version 10.4.[16] Structures were optimized using the B3LYP-D3[17] *a posteriori*-corrected hybrid functional[18] with the 6-31G**+ basis set and the PBF solvation model[19] for water. Normal-mode analysis were used to estimate the Gibbs free energies. Final energies were calculated using B3LYP-D3/6-311G**+ with the PBF solvation model for water and with M06-2X-D3/6-31**+ together with SM6.[20,21] M06-2X-D3/6-311**+ together with PBF (water) energies were calculated to compare the two functionals. For comparative purposes, some combinations were also included.

For each compound, six different conformations were examined in detail starting from the conformation as found for (*R*)-**2** in the crystal structure. The torsion changed next was then the bond to the exocyclic carbonyl in combination with the attached phenyl group resulting in two different anti-conformations. For these three variants the salicylate was also rotated 180°. Structures of the optimized conformations are illustrated in Supplementary Figure 23.

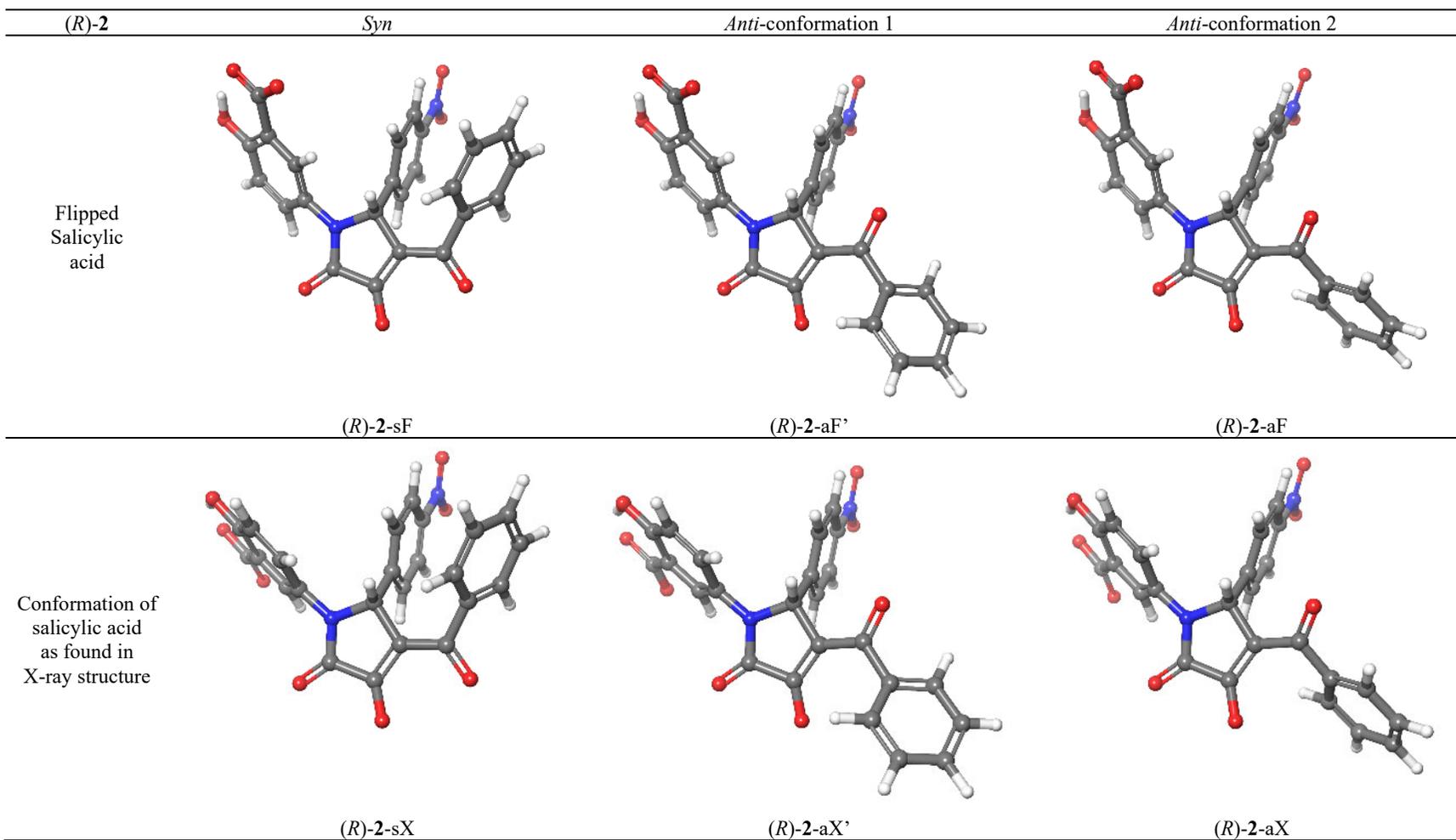

**Supplementary Figure 23**. Optimized geometries of the 6 conformations of (*R*)-**2** studied in detail illustrating the naming scheme also employed for compounds (*R*)-**6** and (*R*)-**9**.

**Supplementary Table 2.** Conformational energies in Hartrees for ligands (*R*)-**2**, (*R*)-**6**, (*R*)-**9** calculated using different functionals, basis sets and solvation models.

| | | | | B3LYP-D3 | | | | | M06-2X_D3 | | | |
|---|---|---|---|---|---|---|---|---|---|---|---|---|
| | | | | 6-31+G** | | | 6-311+G** | | 6-31+G** | | 6-311+G** | |
| Name of conformation | Compound | Exocyclic carbonyl | Salicylic acid | Gas Phase Energy | PBF Solution Phase Energy | Total Free Energy (au) 298.15K 1.0atm | Gas Phase Energy | PBF Solution Phase Energy | Gas Phase Energy | SM6 Solution Phase Energy | Gas Phase Energy | PBF Solution Phase Energy |
| (*R*)-**2**-sF | (*R*)-**2** | *syn* | Flipped | -1634.453750 | -1634.738383 | -1634.452014 | -1634.816852 | -1635.102297 | -1633.751313 | -1634.022958 | -1634.139214 | -1634.427444 |
| **(*R*)-2-sX** | (*R*)-**2** | *syn* | X-ray | -1634.447095 | -1634.736970 | -1634.451019 | -1634.809929 | -1635.100626 | -1633.744478 | -1634.023912 | -1634.131973 | -1634.425593 |
| (*R*)-**2**-aF' | (*R*)-**2** | *anti* | Flipped | -1634.462425 | -1634.735341 | -1634.449469 | -1634.825293 | -1635.099054 | -1633.759245 | -1634.025238 | -1634.146631 | -1634.423957 |
| (*R*)-**2**-aX' | (*R*)-**2** | *anti* | X-ray | -1634.457409 | -1634.734818 | -1634.448884 | -1634.820089 | -1635.098392 | -1633.754053 | -1634.026338 | -1634.141105 | -1634.422973 |
| (*R*)-**2**-aF | (*R*)-**2** | *anti* | Flipped | -1634.461160 | -1634.735939 | -1634.449564 | -1634.823698 | -1635.099807 | -1633.758193 | -1634.026363 | -1634.145333 | -1634.424888 |
| **(*R*)-2-aX** | (*R*)-**2** | *anti* | X-ray | -1634.457272 | -1634.736693 | -1634.450466 | -1634.819795 | -1635.100285 | -1633.754083 | -1634.027758 | -1634.141087 | -1634.425030 |
| (*R*)-**6**-sF | (*R*)-**6** | *syn* | Flipped | -1615.185874 | -1615.323308 | -1615.012390 | -1615.540117 | -1615.678202 | -1614.483202 | -1614.618400 | -1614.861152 | -1615.001336 |
| **(*R*)-6-sX** | (*R*)-**6** | *syn* | X-ray | -1615.177281 | -1615.322817 | -1615.012520 | -1615.531434 | -1615.677671 | -1614.474501 | -1614.619681 | -1614.852279 | -1615.000525 |
| (*R*)-**6**-aF' | (*R*)-**6** | *anti* | Flipped | -1615.167250 | -1615.312714 | -1615.002754 | -1615.521362 | -1615.667778 | -1614.463804 | -1614.604592 | -1614.841471 | -1614.990533 |
| (*R*)-**6**-aX' | (*R*)-**6** | *anti* | X-ray | -1615.164794 | -1615.312751 | -1615.003840 | -1615.518988 | -1615.667632 | -1614.461861 | -1614.608765 | -1614.839500 | -1614.990146 |
| (*R*)-**6**-aF | (*R*)-**6** | *anti* | Flipped | -1615.168031 | -1615.312706 | -1615.002505 | -1615.522149 | -1615.667602 | -1614.464640 | -1614.604511 | -1614.842350 | -1614.990532 |
| **(*R*)-6-aX** | (*R*)-**6** | *anti* | X-ray | -1615.161596 | -1615.313195 | -1615.003560 | -1615.515683 | -1615.667678 | -1614.458256 | -1614.607089 | -1614.835745 | -1614.990340 |
| (*R*)-**9**-sF | (*R*)-**9** | *syn* | Flipped | -1615.199092 | -1615.323821 | -1615.012974 | -1615.553356 | -1615.678701 | -1614.496256 | -1614.617066 | -1614.874292 | -1615.002241 |
| **(*R*)-9-sX** | (*R*)-**9** | *syn* | X-ray | -1615.193073 | -1615.323567 | -1615.013153 | -1615.547163 | -1615.678433 | -1614.490318 | -1614.617754 | -1614.868032 | -1615.001661 |
| (*R*)-**9**-aF' | (*R*)-**9** | *anti* | Flipped | -1615.190414 | -1615.318138 | -1615.007994 | -1615.544177 | -1615.672666 | -1614.487504 | -1614.611874 | -1614.864807 | -1614.995877 |
| (*R*)-**9**-aX' | (*R*)-**9** | *anti* | X-ray | -1615.183424 | -1615.315676 | -1615.005847 | -1615.537414 | -1615.670454 | -1614.480460 | -1614.610151 | -1614.857859 | -1614.993418 |
| (*R*)-**9**-aF | (*R*)-**9** | *anti* | Flipped | -1615.189880 | -1615.317617 | -1615.007372 | -1615.543811 | -1615.672397 | -1614.487248 | -1614.612164 | -1614.864834 | -1614.995873 |
| **(*R*)-9-aX** | (*R*)-**9** | *anti* | X-ray | -1615.187412 | -1615.318914 | -1615.008795 | -1615.541159 | -1615.673487 | -1614.484769 | -1614.613301 | -1614.862003 | -1614.996798 |

a) Entries in bold are discussed in the main manuscript and corresponds to the conformation of the salicylate and benzoyl as found in the crystal structure.

**Supplementary Table 3**. Relative conformational energies in kcal mol$^{-1}$ for ligands (*R*)-**2**, (*R*)-**6**, (*R*)-**9** calculated using different functionals, basis sets and solvation models.

| | | | B3LYP-D3 | | | | | | M06-2X_D3 | | | | | |
|---|---|---|---|---|---|---|---|---|---|---|---|---|---|---|
| | | | 6-31+G** | | | 6-311+G** | | | 6-31+G** | | | 6-311+G** | | |
| | | | Gas Phase Energy | Solution Phase Energy | Total Free Energy (au) 298.15K 1.0atm | Gas Phase Energy | Solution Phase Energy | Total Free Energy (au) 298.15K 1.0atm | Gas Phase Energy | SM6 Solution Phase Energy | Total Free Energy (au) 298.15K 1.0atm | Gas Phase Energy | PBF Solution Phase Energy | Total Free Energy (au) 298.15K 1.0atm |
| (*R*)-**2** | syn | Flipped | 0.0 | 0.0 | 0.0 | 0.0 | 0.0 | 0.0 | 0.0 | 0.0 | 0.0 | 0.0 | 0.0 | 0.0 |
| (*R*)-**2** | syn | X-ray | 4.2 | 0.9 | 0.6 | 4.3 | 1.0 | **0.8** | 4.3 | -0.6 | **-0.9** | 4.5 | 1.2 | 1.2 |
| (*R*)-**2** | anti | Flipped | -5.4 | 1.9 | 1.6 | -5.3 | 2.0 | 1.7 | -5.0 | -1.4 | -1.7 | -4.7 | 2.2 | 2.2 |
| (*R*)-**2** | anti | X-ray | -2.3 | 2.2 | 2.0 | -2.0 | 2.5 | 2.2 | -1.7 | -2.1 | -2.4 | -1.2 | 2.8 | 2.8 |
| (*R*)-**2** | anti | Flipped | -4.6 | 1.5 | 1.5 | -4.3 | 1.6 | 1.6 | -4.3 | -2.1 | -2.1 | -3.8 | 1.6 | 1.6 |
| (*R*)-**2** | anti | X-ray | -2.2 | 1.1 | 1.0 | -1.8 | 1.3 | **1.2** | -1.7 | -3.0 | **-3.1** | -1.2 | 1.5 | 1.5 |
| (*R*)-**6** | syn | Flipped | -5.4 | -0.3 | 0.1 | -5.4 | -0.3 | 0.1 | -5.5 | 0.8 | 1.2 | -5.6 | -0.5 | -0.5 |
| (*R*)-**6** | syn | X-ray | 0.0 | 0.0 | 0.0 | 0.0 | 0.0 | **0.0** | 0.0 | 0.0 | **0.0** | 0.0 | 0.0 | 0.0 |
| (*R*)-**6** | anti | Flipped | 6.3 | 6.3 | 6.1 | 6.3 | 6.2 | 6.0 | 6.7 | 9.5 | 9.3 | 6.8 | 6.3 | 6.3 |
| (*R*)-**6** | anti | X-ray | 7.8 | 6.3 | 5.4 | 7.8 | 6.3 | 5.4 | 7.9 | 6.8 | 6.0 | 8.0 | 6.5 | 6.5 |
| (*R*)-**6** | anti | Flipped | 5.8 | 6.3 | 6.3 | 5.8 | 6.3 | 6.3 | 6.2 | 9.5 | 9.5 | 6.2 | 6.3 | 6.3 |
| (*R*)-**6** | anti | X-ray | 9.8 | 6.0 | 5.6 | 9.9 | 6.3 | **5.9** | 10.2 | 7.9 | **7.5** | 10.4 | 6.4 | 6.4 |
| (*R*)-**9** | syn | Flipped | -3.8 | -0.2 | 0.1 | -3.9 | -0.2 | 0.1 | -3.7 | 0.4 | 0.7 | -3.9 | -0.4 | -0.4 |
| (*R*)-**9** | syn | X-ray | 0.0 | 0.0 | 0.0 | 0.0 | 0.0 | **0.0** | 0.0 | 0.0 | **0.0** | 0.0 | 0.0 | 0.0 |
| (*R*)-**9** | anti | Flipped | 1.7 | 3.4 | 3.2 | 1.9 | 3.6 | 3.4 | 1.8 | 3.7 | 3.5 | 2.0 | 3.6 | 3.6 |
| (*R*)-**9** | anti | X-ray | 6.1 | 5.0 | 4.6 | 6.1 | 5.0 | 4.6 | 6.2 | 4.8 | 4.4 | 6.4 | 5.2 | 5.2 |
| (*R*)-**9** | anti | Flipped | 2.0 | 3.7 | 3.6 | 2.1 | 3.8 | 3.7 | 1.9 | 3.5 | 3.4 | 2.0 | 3.6 | 3.6 |
| (*R*)-**9** | anti | X-ray | 3.6 | 2.9 | 2.7 | 3.8 | 3.1 | **2.9** | 3.5 | 2.8 | **2.6** | 3.8 | 3.1 | 3.1 |

a) Values in bold are included in the main manuscript and corresponds to the conformation of the salicylate and benzoyl moities as found in the crystal structure.

**Coordinates in Angstrom of the optimized structures optimized by B3LYP-D3/6-31G**+ in combination with the PBF model for water.**

```
48
(R)-2-sF
C     0.17700   -63.60880   -26.05090
C     0.73540   -62.43230   -25.30240
C     1.10920   -62.57440   -23.95780
C     1.59890   -61.49520   -23.22580
C     2.93590   -61.10800   -21.16950
C     2.76240   -61.44840   -19.68970
C     1.58870   -62.20850   -19.57250
C     0.99370   -62.66890   -18.35410
C    -0.23380   -63.54020   -18.44890
C    -0.20500   -64.76910   -19.12460
C    -1.36040   -65.55120   -19.21610
C    -2.55800   -65.09950   -18.65160
C    -2.59020   -63.87590   -17.97160
C    -1.43010   -63.10680   -17.85730
C     0.96880   -62.40660   -20.94840
C    -0.46160   -61.89730   -21.09610
C    -1.46610   -62.74140   -21.58380
C    -2.77570   -62.28590   -21.72140
C    -3.06110   -60.96370   -21.37540
C    -2.07670   -60.09240   -20.89960
C    -0.77610   -60.57200   -20.76110
C     1.73210   -60.24090   -23.84840
C     1.38880   -60.08570   -25.18610
C     0.88440   -61.16980   -25.92190
N     1.88110   -61.64830   -21.83710
N    -4.43410   -60.47430   -21.51430
O    -0.16740   -63.41400   -27.27410
O     0.06650   -64.71940   -25.46480
O     3.87220   -60.45830   -21.65830
O     3.60850   -61.05570   -18.83730
O     1.41470   -62.35030   -17.21110
O    -5.30310   -61.25950   -21.91240
O    -4.67080   -59.29470   -21.22700
O     0.54640   -60.96060   -27.22240
H     0.99340   -63.55070   -23.50020
H     0.72050   -65.11810   -19.57540
H    -1.32810   -66.50530   -19.73590
H    -3.46170   -65.69720   -18.73980
H    -3.51930   -63.52070   -17.53350
H    -1.45400   -62.15510   -17.33310
H     0.99810   -63.45990   -21.25240
H    -1.23340   -63.76830   -21.84830
H    -3.55740   -62.93960   -22.09180
H    -2.32220   -59.06860   -20.64390
H     0.00000   -59.91130   -20.38440
H     2.09280   -59.39020   -23.28050
H     1.48550   -59.12120   -25.67470
H     0.20640   -61.84080   -27.56450
48
(R)-2-sX
C     0.71200   -58.89140   -25.77060
C     1.00660   -60.23620   -25.16740
C     1.35780   -60.33460   -23.81360
C     1.62500   -61.57100   -23.23400
C     2.93690   -61.10710   -21.16880
C     2.76100   -61.45020   -19.68160
C     1.58980   -62.21890   -19.57340
C     0.98920   -62.70840   -18.36950
C    -0.25320   -63.55520   -18.51280
C    -0.23400   -64.76640   -19.22160
C    -1.40570   -65.51260   -19.37710
C    -2.61050   -65.04370   -18.84310
C    -2.63370   -63.84060   -18.12730
C    -1.45800   -63.10670   -17.95030
C     0.95910   -62.38710   -20.94700
C    -0.44980   -61.81590   -21.08830
C    -1.46640   -62.58380   -21.66810
C    -2.75860   -62.07810   -21.79710
C    -3.01230   -60.78160   -21.34400
C    -2.01070   -59.98390   -20.77660
C    -0.73080   -60.51200   -20.65290
C     1.53920   -62.74080   -24.00600
C     1.18030   -62.66690   -25.34720
C     0.90980   -61.42210   -25.93600
N     1.89450   -61.66570   -21.83500
N    -4.36810   -60.24120   -21.45970
O     0.81220   -57.85760   -25.05660
O     0.36630   -58.86990   -27.00830
O     3.86600   -60.45100   -21.66150
O     3.60150   -61.05780   -18.82430
O     1.40920   -62.44150   -17.21310
O    -5.24770   -60.95690   -21.95410
O    -4.58050   -59.09170   -21.05610
O     0.55520   -61.40120   -27.24770
H     1.40250   -59.43170   -23.21390
H     0.69810   -65.12730   -19.64950
H    -1.38060   -66.45190   -19.92380
H    -3.52650   -65.61230   -18.98280
H    -3.56850   -63.47260   -17.71240
H    -1.47640   -62.16920   -17.40100
H     0.93880   -63.43540   -21.26470
H    -1.25580   -63.59520   -22.00890
H    -3.55160   -62.67350   -22.23560
H    -2.23630   -58.97920   -20.43780
H     0.05390   -59.90980   -20.20380
H     1.74760   -63.70690   -23.55370
H     1.10290   -63.56370   -25.95570
H     0.39590   -60.43610   -27.47630
48
(R)-2-aF'
C     0.25050   -63.61490   -26.08730
C     0.77960   -62.43780   -25.32030
C     1.14180   -62.58950   -23.97380
C     1.61710   -61.51200   -23.23150
C     2.92900   -61.10300   -21.16880
C     2.74380   -61.41170   -19.68060
C     1.59760   -62.22000   -19.57750
C     0.98130   -62.83090   -18.44280
C     1.60930   -62.74060   -17.08150
C     2.94280   -63.11630   -16.86700
C     3.47980   -63.10130   -15.57730
C     2.69610   -62.68260   -14.49510
C     1.36590   -62.29970   -14.70480
C     0.81930   -62.34490   -15.99080
C     0.95670   -62.37940   -20.94130
C    -0.43100   -61.75660   -21.06160
C    -1.53240   -62.53050   -21.44670
C    -2.79030   -61.95000   -21.60080
C    -2.92590   -60.57850   -21.36730
C    -1.84240   -59.78100   -20.98030
C    -0.59880   -60.38290   -20.82820
C     1.74460   -60.25150   -23.84210
C     1.40940   -60.08620   -25.17990
C     0.91990   -61.16830   -25.92710
N     1.89660   -61.67420   -21.84390
N    -4.23790   -59.95430   -21.53870
O    -0.07600   -63.41260   -27.31360
O     0.14710   -64.73330   -25.51480
O     3.86690   -60.45310   -21.65440
O     3.55990   -60.94580   -18.83910
O    -0.10280   -63.46730   -18.54930
O    -5.19240   -60.66750   -21.87040
O    -4.34060   -58.73630   -21.34670
O     0.58730   -60.94900   -27.22700
H     1.03340   -63.56980   -23.52210
H     3.55340   -63.42790   -17.70840
H     4.50930   -63.41100   -15.41790
H     3.11860   -62.65740   -13.49390
H     0.75460   -61.97230   -13.86800
H    -0.21810   -62.06570   -16.15540
H     0.89900   -63.43740   -21.22650
H    -1.40880   -63.59470   -21.62770
H    -3.64640   -62.54280   -21.90230
H    -1.97520   -58.71880   -20.80910
H     0.25270   -59.77680   -20.53180
H     2.09570   -59.40300   -23.26510
H     1.50220   -59.11550   -25.65940
H     0.25610   -61.82890   -27.57920
48
(R)-2-aX'
C     1.24190   -59.03530   -25.97890
C     1.34130   -60.36570   -25.28750
C     1.59820   -60.42260   -23.90960
C     1.68980   -61.64530   -23.25060
C     2.92830   -61.10020   -21.16890
C     2.73960   -61.41250   -19.67700
C     1.60130   -62.23300   -19.57990
C     0.99920   -62.90030   -18.46750
C     1.68660   -62.97540   -17.13510
C     3.04790   -63.29900   -17.03070
C     3.64020   -63.45970   -15.77540
C     2.88250   -63.27500   -14.61230
C     1.52440   -62.94650   -14.70950
C     0.92620   -62.81280   -15.96550
C     0.94370   -62.35360   -20.93810
C    -0.40690   -61.64790   -21.05020
C    -1.53280   -62.33100   -21.52590
C    -2.75650   -61.67720   -21.66680
C    -2.83400   -60.32430   -21.32500
C    -1.72390   -59.61480   -20.85090
C    -0.51480   -60.28800   -20.71750
C     1.51320   -62.84620   -23.96740
C     1.24450   -62.80560   -25.33060
C     1.15930   -61.57560   -26.00030
N     1.91270   -61.69850   -21.84320
N    -4.11250   -59.62620   -21.46890
O     0.99270   -59.05020   -27.23980
O     1.40220   -57.97730   -25.31350
O     3.86080   -60.44540   -21.65620
O     3.54790   -60.94330   -18.82970
O    -0.11620   -63.47890   -18.58800
O    -5.08620   -60.25920   -21.89520
O    -4.16950   -58.43030   -21.15800
O     0.89950   -61.59170   -27.33420
H     1.71610   -59.49540   -23.35940
H     3.63690   -63.43860   -17.93120
H     4.69120   -63.72790   -15.70520
H     3.34630   -63.39100   -13.63590
H     0.93250   -62.80300   -13.80930
H    -0.13200   -62.57920   -16.04770
H     0.82110   -63.40380   -21.22900
H    -1.45560   -63.38310   -21.78660
H    -3.63140   -62.20070   -22.03590
H    -1.81070   -58.56450   -20.59690
H     0.35390   -59.74810   -20.35120
H     1.58860   -63.79640   -23.45670
H     1.10400   -63.72290   -25.89580
H     0.87910   -60.62910   -27.62240
48
(R)-2-aF
C    -0.13090   -63.34290   -26.02850
C     0.52670   -62.22690   -25.26630
C     0.96310   -62.44380   -23.95090
C     1.54950   -61.42280   -23.20640
C     2.93560   -61.11380   -21.16820
C     2.77770   -61.46900   -19.68940
C     1.58870   -62.20230   -19.57660
C     0.97720   -62.82610   -18.43470
C     1.20420   -62.27080   -17.06110
C     1.35310   -60.88950   -16.85410
C     1.46950   -60.37710   -15.56060
C     1.45470   -61.24350   -14.46090
C     1.30810   -62.62250   -14.65910
C     1.17110   -63.13210   -15.95250
C     0.97710   -62.41130   -20.94780
C    -0.45480   -61.90240   -21.05390
C    -1.50760   -62.78870   -21.31270
```

| | | | | | | | | | | |
|---|---|---|---|---|---|---|---|---|---|---|
| C | -2.81880 | -62.32440 | -21.41850 | H | -2.03940 | -58.78300 | -20.76040 | C | -2.72640 | -62.01520 | -21.92350 |
| C | -3.05670 | -60.95590 | -21.26590 | H | 0.22460 | -59.78260 | -20.58060 | C | -2.97250 | -60.71420 | -21.48110 |
| C | -2.02320 | -60.04700 | -21.00620 | H | 1.31650 | -63.59740 | -23.59000 | C | -1.98580 | -59.94030 | -20.85890 |
| C | -0.72540 | -60.53320 | -20.89810 | H | 0.59040 | -63.27980 | -25.94820 | C | -0.72440 | -60.49520 | -20.67500 |
| C | 1.72280 | -60.15420 | -23.78990 | H | 0.34630 | -60.03640 | -27.37100 | C | 1.54200 | -62.71720 | -24.01920 |
| C | 1.31660 | -59.92630 | -25.09900 | 50 | | | | C | 1.17330 | -62.62630 | -25.35690 |
| C | 0.70800 | -60.94850 | -25.84350 | (R)-6-sF | | | | C | 0.87580 | -61.37630 | -25.92320 |
| N | 1.87430 | -61.64190 | -21.83730 | C | -0.03490 | -63.47270 | -26.01580 | N | 1.89770 | -61.66900 | -21.83140 |
| N | -4.42710 | -60.45420 | -21.38680 | C | 0.58580 | -62.32560 | -25.26740 | N | -4.30590 | -60.13790 | -21.67520 |
| O | -0.55070 | -63.07070 | -27.21330 | C | 1.02940 | -62.51260 | -23.95060 | O | 0.27170 | -58.81650 | -26.94580 |
| O | -0.24440 | -64.47700 | -25.49170 | C | 1.56790 | -61.46010 | -23.21370 | O | 0.68660 | -57.83380 | -24.97190 |
| O | 3.87110 | -60.46050 | -21.65580 | C | 2.94360 | -61.10810 | -21.17410 | O | 3.86870 | -60.44650 | -21.66290 |
| O | 3.66710 | -61.12530 | -18.85640 | C | 2.77590 | -61.45540 | -19.69310 | O | 3.57540 | -61.07400 | -18.80770 |
| O | 0.19380 | -63.80260 | -18.57190 | C | 1.58340 | -62.20990 | -19.57060 | O | -5.16890 | -60.82620 | -22.23110 |
| O | -5.33470 | -61.26350 | -21.61170 | C | 1.02860 | -62.66080 | -18.36630 | O | -4.51270 | -58.98690 | -21.27380 |
| O | -4.62280 | -59.23910 | -21.25940 | C | -0.26300 | -63.40680 | -18.38890 | O | 0.51490 | -61.34050 | -27.23200 |
| O | 0.30180 | -60.66320 | -27.11040 | C | -0.36110 | -64.64710 | -19.03900 | H | 1.35040 | -59.42270 | -23.17180 |
| H | 0.81420 | -63.42880 | -23.52200 | C | -1.60080 | -65.28250 | -19.14380 | H | 0.60050 | -65.11010 | -19.60020 |
| H | 1.35860 | -60.21720 | -17.70560 | C | -2.74740 | -64.67590 | -18.61920 | H | -1.56740 | -66.28640 | -19.86480 |
| H | 1.56900 | -59.30520 | -15.41070 | C | -2.64980 | -63.44340 | -17.96410 | H | -3.63980 | -65.31480 | -18.88760 |
| H | 1.55170 | -60.84540 | -13.45400 | C | -1.40970 | -62.81220 | -17.83840 | H | -3.52670 | -63.18980 | -17.59920 |
| H | 1.29500 | -63.29600 | -13.80630 | C | 0.96370 | -62.41190 | -20.94730 | H | -1.35630 | -62.01630 | -17.32120 |
| H | 1.03850 | -64.19810 | -16.11310 | C | -0.46680 | -61.91470 | -21.12660 | H | 0.92800 | -63.43540 | -21.26160 |
| H | 1.00390 | -63.47100 | -21.23320 | C | -1.44740 | -62.76920 | -21.64530 | H | -1.24590 | -63.55980 | -22.07280 |
| H | -1.30460 | -63.84890 | -21.43700 | C | -2.74650 | -62.31390 | -21.86050 | H | -3.50740 | -62.59050 | -22.40750 |
| H | -3.63770 | -63.00520 | -21.62350 | C | -3.04210 | -60.98320 | -21.55880 | H | -2.20630 | -58.93080 | -20.53130 |
| H | -2.23670 | -58.98970 | -20.89640 | C | -2.08020 | -60.10430 | -21.04660 | H | 0.05090 | -59.90560 | -20.19340 |
| H | 0.08830 | -59.84090 | -20.69960 | C | -0.79240 | -60.58260 | -20.83100 | H | 1.77100 | -63.68700 | -23.58520 |
| H | 2.16020 | -59.34670 | -23.21360 | C | 1.69340 | -60.19110 | -23.80740 | H | 1.10960 | -63.51370 | -25.98080 |
| H | 1.44230 | -58.94800 | -25.55530 | C | 1.28140 | -59.99280 | -25.11950 | H | 0.33780 | -60.37610 | -27.44860 |
| H | -0.11920 | -61.50330 | -27.46100 | C | 0.71350 | -61.04500 | -25.85490 | H | 1.15460 | -62.87790 | -16.34070 |
| 48 | | | | N | 1.88150 | -61.64590 | -21.83310 | H | 2.40730 | -61.90950 | -17.06820 |
| (R)-2-aX | | | | N | -4.40200 | -60.48930 | -21.78830 | N | 1.56660 | -62.47460 | -17.17760 |
| C | 1.00690 | -58.61620 | -25.64900 | O | -0.46550 | -63.22770 | -27.20200 | 50 | | | |
| C | 1.09950 | -60.00830 | -25.09120 | O | -0.11140 | -64.60310 | -25.46320 | (R)-6-aF' | | | |
| C | 1.49330 | -60.20190 | -23.75970 | O | 3.87460 | -60.45400 | -21.66070 | C | -0.12350 | -63.18310 | -26.08380 |
| C | 1.57540 | -61.48210 | -23.21790 | O | 3.58860 | -61.08370 | -18.81690 | C | 0.50380 | -62.08610 | -25.27100 |
| C | 2.93710 | -61.11350 | -21.16680 | O | -5.25180 | -61.28010 | -22.21310 | C | 0.96610 | -62.35860 | -23.97550 |
| C | 2.78100 | -61.47610 | -19.68480 | O | -4.64490 | -59.30120 | -21.54630 | C | 1.52160 | -61.35720 | -23.18560 |
| C | 1.59160 | -62.21530 | -19.57680 | O | 0.29330 | -60.78510 | -27.12140 | C | 2.94530 | -61.11570 | -21.17380 |
| C | 0.99530 | -62.89070 | -18.45800 | H | 0.92180 | -63.50020 | -23.51590 | C | 2.79750 | -61.47320 | -19.69230 |
| C | 1.29680 | -62.45400 | -17.05730 | H | 0.52590 | -65.10890 | -19.46480 | C | 1.58010 | -62.20320 | -19.57000 |
| C | 1.47330 | -61.09750 | -16.74060 | H | -1.67460 | -66.24100 | -19.65060 | C | 0.97990 | -62.63300 | -18.37720 |
| C | 1.66240 | -60.70120 | -15.41500 | H | -3.71500 | -65.15970 | -18.72520 | C | 1.21460 | -61.88720 | -17.11460 |
| C | 1.69770 | -61.65930 | -14.39490 | H | -3.53980 | -62.96700 | -17.56170 | C | 1.42870 | -62.55380 | -15.89650 |
| C | 1.52650 | -63.01490 | -14.70350 | H | -1.33600 | -61.84270 | -17.35300 | C | 1.63490 | -61.81800 | -14.72510 |
| C | 1.31380 | -63.40830 | -16.02670 | H | 1.00850 | -63.46670 | -21.24420 | C | 1.60650 | -60.41890 | -14.75700 |
| C | 0.96260 | -62.38620 | -20.94400 | H | -1.20170 | -63.79890 | -21.88620 | C | 1.37020 | -59.75120 | -15.96510 |
| C | -0.43690 | -61.78870 | -21.04050 | H | -3.50930 | -62.97160 | -22.26130 | C | 1.18200 | -60.48040 | -17.13740 |
| C | -1.53710 | -62.59310 | -21.35930 | H | -2.33930 | -59.07470 | -20.82750 | C | 0.97500 | -62.42000 | -20.95040 |
| C | -2.81060 | -62.04420 | -21.46080 | H | -0.03300 | -59.91380 | -20.43480 | C | -0.45680 | -61.90710 | -21.05150 |
| C | -2.97200 | -60.67360 | -21.23980 | H | 2.09590 | -59.36120 | -23.23690 | C | -1.51110 | -62.78890 | -21.31520 |
| C | -1.88910 | -59.84420 | -20.92160 | H | 1.36850 | -59.01490 | -25.58520 | C | -2.82290 | -62.32050 | -21.40730 |
| C | -0.62530 | -60.41380 | -20.82340 | H | -0.09460 | -61.64220 | -27.46890 | C | -3.05420 | -60.95420 | -21.24060 |
| C | 1.25480 | -62.59690 | -24.00970 | H | 1.15210 | -62.76600 | -16.32110 | C | -2.01790 | -60.04860 | -20.97980 |
| C | 0.84860 | -62.42540 | -25.32850 | H | 2.43050 | -61.85890 | -17.08400 | C | -0.72040 | -60.53780 | -20.88240 |
| C | 0.76540 | -61.13750 | -25.87830 | N | 1.57300 | -62.40210 | -17.17110 | C | 1.63640 | -60.05080 | -23.69620 |
| N | 1.89090 | -61.66500 | -21.83940 | 50 | | | | C | 1.20330 | -59.76630 | -24.98500 |
| N | -4.30690 | -60.08290 | -21.34820 | (R)-6-sX | | | | C | 0.62690 | -60.77140 | -25.77860 |
| O | 1.32350 | -57.63620 | -24.92240 | C | 0.61790 | -58.85370 | -25.70920 | N | 1.87400 | -61.63400 | -21.83100 |
| O | 0.60100 | -58.49850 | -26.86240 | C | 0.95220 | -60.20230 | -25.13390 | N | -4.42660 | -60.44700 | -21.35010 |
| O | 3.86460 | -60.44980 | -21.65710 | C | 1.31840 | -60.31700 | -23.78520 | O | -0.55570 | -62.86870 | -27.25200 |
| O | 3.67130 | -61.13580 | -18.85320 | C | 1.61480 | -61.55780 | -23.22970 | O | -0.20260 | -64.34390 | -25.59920 |
| O | 0.17280 | -63.83100 | -18.63370 | C | 2.94250 | -61.10440 | -21.17360 | O | 3.87240 | -60.45690 | -21.66050 |
| O | -5.25420 | -60.81870 | -21.65230 | C | 2.76930 | -61.45040 | -19.68700 | O | 3.67360 | -61.15440 | -18.86530 |
| O | -4.43500 | -58.87230 | -21.13160 | C | 1.58680 | -62.22260 | -19.57290 | O | -5.33610 | -61.25460 | -21.56810 |
| O | 0.35570 | -61.01990 | -27.16990 | C | 1.03530 | -62.71130 | -18.38160 | O | -4.61440 | -59.23220 | -21.22010 |
| H | 1.71700 | -59.33420 | -23.14910 | C | -0.23750 | -63.48760 | -18.44150 | O | 0.19490 | -60.43480 | -27.02160 |
| H | 1.44430 | -60.35400 | -17.53000 | C | -0.30130 | -64.69320 | -19.15940 | H | 0.86470 | -63.37080 | -23.59960 |
| H | 1.78110 | -59.64710 | -15.17770 | C | -1.52230 | -65.35490 | -19.30670 | H | 1.46620 | -63.64110 | -15.86870 |
| H | 1.85270 | -61.35070 | -13.36410 | C | -2.68630 | -64.81080 | -18.75200 | H | 1.82310 | -62.33660 | -13.78890 |
| H | 1.55280 | -63.76070 | -13.91330 | C | -2.62320 | -63.61630 | -18.02640 | H | 1.76410 | -59.84900 | -13.84450 |
| H | 1.16070 | -64.45550 | -16.27210 | C | -1.40110 | -62.95810 | -17.86160 | H | 1.33420 | -58.66490 | -15.99320 |
| H | 0.92230 | -63.44340 | -21.23400 | C | 0.95110 | -62.38730 | -20.94500 | H | 0.99900 | -59.96280 | -18.07430 |
| H | -1.39560 | -63.65610 | -21.53280 | C | -0.44960 | -61.80230 | -21.10480 | H | 1.01180 | -63.47450 | -21.25930 |
| H | -3.67240 | -62.66150 | -21.71020 | C | -1.45330 | -62.55000 | -21.73150 | H | -1.31270 | -63.84770 | -21.46320 |

| | | | | | | | | | | |
|---|---|---|---|---|---|---|---|---|---|---|
| H | -3.64490 | -62.99680 | -21.61420 | C | 1.42450 | -62.54970 | -15.89730 | H | 2.05150 | -62.75360 | -13.84180 |
| H | -2.22770 | -58.99140 | -20.86140 | C | 0.97770 | -62.42220 | -20.95070 | H | 1.57350 | -63.89000 | -15.99640 |
| H | 0.09500 | -59.84510 | -20.68880 | C | -0.45810 | -61.92160 | -21.05230 | H | 0.90980 | -63.43810 | -21.26740 |
| H | 2.04940 | -59.25900 | -23.08020 | C | -1.50500 | -62.81480 | -21.30300 | H | -1.43130 | -63.60570 | -21.55230 |
| H | 1.28250 | -58.75950 | -25.38670 | C | -2.82140 | -62.35990 | -21.38960 | H | -3.69250 | -62.56980 | -21.65660 |
| H | -0.19150 | -61.27110 | -27.41830 | C | -3.06450 | -60.99490 | -21.23070 | H | -1.97460 | -58.73550 | -20.67500 |
| H | -0.44140 | -63.84510 | -17.50280 | C | -2.03520 | -60.07740 | -20.98520 | H | 0.26980 | -59.77260 | -20.54450 |
| H | 0.00390 | -64.31750 | -19.09760 | C | -0.73290 | -60.55330 | -20.89230 | H | 1.64320 | -63.54570 | -23.71480 |
| N | 0.11240 | -63.65540 | -18.33420 | C | 1.68810 | -60.06330 | -23.71950 | H | 0.95230 | -63.17600 | -26.08240 |
| 50 | | | | C | 1.23390 | -59.77350 | -25.00010 | H | 0.28360 | -59.91100 | -27.30470 |
| (R)-6-aX' | | | | C | 0.56950 | -60.75020 | -25.75830 | H | -0.44730 | -63.90460 | -17.54980 |
| C | 1.26680 | -58.87280 | -25.86470 | N | 1.87190 | -61.63120 | -21.83160 | H | -0.07940 | -64.28690 | -19.18680 |
| C | 1.34490 | -60.23220 | -25.22670 | N | -4.44190 | -60.50270 | -21.32930 | N | 0.09140 | -63.68760 | -18.38390 |
| C | 1.59950 | -60.34700 | -23.85280 | O | -0.79140 | -62.78960 | -27.15700 | 50 | | | |
| C | 1.67770 | -61.59710 | -23.24410 | O | -0.50190 | -64.25170 | -25.47790 | (R)-9-sF | | | |
| C | 2.93420 | -61.09780 | -21.17440 | O | 3.87330 | -60.45730 | -21.66070 | C | -0.14330 | -63.42680 | -25.98720 |
| C | 2.74940 | -61.39100 | -19.67890 | O | 3.67650 | -61.15740 | -18.86650 | C | 0.50430 | -62.28930 | -25.24650 |
| C | 1.59450 | -62.23920 | -19.57430 | O | -5.34620 | -61.32010 | -21.53030 | C | 0.98520 | -62.49160 | -23.94550 |
| C | 1.10290 | -62.84860 | -18.42200 | O | -4.64000 | -59.28880 | -21.20610 | C | 1.54420 | -61.44670 | -23.21260 |
| C | 1.83100 | -62.73640 | -17.12640 | O | 0.11870 | -60.40380 | -26.99320 | C | 2.93580 | -61.11410 | -21.17470 |
| C | 3.13960 | -63.22660 | -17.00770 | H | 0.70910 | -63.32550 | -23.54640 | C | 2.75940 | -61.46330 | -19.72210 |
| C | 3.82550 | -63.10180 | -15.79930 | H | 1.01940 | -59.95460 | -18.07360 | C | 1.59790 | -62.19930 | -19.57760 |
| C | 3.21430 | -62.47250 | -14.70640 | H | 1.39330 | -58.66020 | -15.99760 | C | 1.02010 | -62.61430 | -18.32570 |
| C | 1.90760 | -61.98640 | -14.82150 | H | 1.81900 | -59.84960 | -13.85040 | C | -0.26590 | -63.39190 | -18.36290 |
| C | 1.21040 | -62.12590 | -16.02660 | H | 1.83510 | -62.33690 | -13.79250 | C | -0.35670 | -64.62530 | -19.02730 |
| C | 0.93900 | -62.35260 | -20.93780 | H | 1.44760 | -63.63620 | -15.87100 | C | -1.58010 | -65.29820 | -19.09260 |
| C | -0.40040 | -61.62050 | -21.03520 | H | 1.02190 | -63.47700 | -21.25760 | C | -2.72380 | -64.73150 | -18.52030 |
| C | -1.54070 | -62.27580 | -21.51390 | H | -1.29750 | -63.87270 | -21.44550 | C | -2.63600 | -63.50400 | -17.85310 |
| C | -2.75390 | -61.59320 | -21.63150 | H | -3.63820 | -63.04530 | -21.58550 | C | -1.40890 | -62.84320 | -17.75980 |
| C | -2.79830 | -60.24660 | -21.26800 | H | -2.25420 | -59.02130 | -20.87390 | C | 0.96740 | -62.41010 | -20.94780 |
| C | -1.67230 | -59.56490 | -20.79050 | H | 0.07690 | -59.85230 | -20.70820 | C | -0.47600 | -61.92340 | -21.09000 |
| C | -0.47710 | -60.26420 | -20.67530 | H | 2.16780 | -59.28860 | -23.13210 | C | -1.45280 | -62.78320 | -21.60650 |
| C | 1.48770 | -62.76170 | -24.00460 | H | 1.36430 | -58.78030 | -25.42140 | C | -2.76500 | -62.34560 | -21.77220 |
| C | 1.22050 | -62.66830 | -25.36560 | H | -0.33730 | -61.21710 | -27.36270 | C | -3.07300 | -61.02720 | -21.43180 |
| C | 1.15090 | -61.41120 | -25.98680 | H | -0.47860 | -63.80650 | -17.50410 | C | -2.11000 | -60.14270 | -20.93210 |
| N | 1.91710 | -61.69990 | -21.83980 | H | -0.03490 | -64.29710 | -19.09410 | C | -0.80780 | -60.60300 | -20.76700 |
| N | -4.06600 | -59.51680 | -21.39260 | N | 0.08920 | -63.63820 | -18.33020 | C | 1.65660 | -60.17170 | -23.79500 |
| O | 1.01960 | -58.83340 | -27.12470 | 50 | | | | C | 1.20980 | -59.95910 | -25.09330 |
| O | 1.44290 | -57.84600 | -25.15580 | (R)-6-aX | | | | C | 0.61790 | -61.00240 | -25.82320 |
| O | 3.86200 | -60.44030 | -21.65950 | C | 0.63180 | -58.53360 | -25.45930 | N | 1.87370 | -61.64890 | -21.83930 |
| O | 3.51690 | -60.90560 | -18.83840 | C | 0.93200 | -59.93180 | -24.99530 | N | -4.44900 | -60.55380 | -21.60300 |
| O | -5.05320 | -60.12190 | -21.82250 | C | 1.31770 | -60.15720 | -23.66770 | O | -0.60650 | -63.16990 | -27.15800 |
| O | -4.09120 | -58.32670 | -21.06100 | C | 1.57430 | -61.44590 | -23.20880 | O | -0.20940 | -64.56120 | -25.44160 |
| O | 0.89540 | -61.37170 | -27.31970 | C | 2.94720 | -61.11570 | -21.17430 | O | 3.87210 | -60.45740 | -21.64630 |
| H | 1.73160 | -59.44290 | -23.26830 | C | 2.80170 | -61.48280 | -19.68880 | O | 1.52920 | -62.28590 | -17.22620 |
| H | 3.61820 | -63.69370 | -17.86410 | C | 1.58420 | -62.21910 | -19.57080 | O | -5.29490 | -61.34650 | -22.03250 |
| H | 4.83900 | -63.48510 | -15.71270 | C | 0.99380 | -62.69370 | -18.39120 | O | -4.70860 | -59.38110 | -21.30990 |
| H | 3.75580 | -62.36160 | -13.77030 | C | 1.29110 | -62.04440 | -17.08930 | O | 0.16080 | -60.72820 | -27.07340 |
| H | 1.43120 | -61.49470 | -13.97720 | C | 1.28380 | -60.64020 | -17.00780 | H | 0.88770 | -63.48340 | -23.51750 |
| H | 0.19940 | -61.73660 | -16.12030 | C | 1.54000 | -60.00490 | -15.79450 | H | 0.52490 | -65.06020 | -19.49110 |
| H | 0.80970 | -63.40010 | -21.24810 | C | 1.82120 | -60.76570 | -14.65220 | H | -1.64470 | -66.25410 | -19.60580 |
| H | -1.48590 | -63.32160 | -21.80740 | C | 1.82740 | -62.16330 | -14.72610 | H | -3.68050 | -65.24190 | -18.59560 |
| H | -3.64160 | -62.09200 | -22.00340 | C | 1.55270 | -62.80470 | -15.93800 | H | -3.52410 | -63.05960 | -17.41160 |
| H | -1.73540 | -58.51680 | -20.52020 | C | 0.95680 | -62.38800 | -20.94610 | H | -1.34030 | -61.88560 | -17.25070 |
| H | 0.40300 | -59.74390 | -20.30720 | C | -0.43320 | -61.76440 | -21.02740 | H | 1.01370 | -63.46770 | -21.23650 |
| H | 1.55530 | -63.73770 | -23.53130 | C | -1.54950 | -62.54470 | -21.34620 | H | -1.19950 | -63.80520 | -21.87050 |
| H | 1.07090 | -63.56020 | -25.96750 | C | -2.82150 | -61.97190 | -21.41280 | H | -3.52990 | -63.00870 | -22.15980 |
| H | 0.88600 | -60.39930 | -27.57350 | C | -2.94810 | -60.60390 | -21.16540 | H | -2.37980 | -59.12370 | -20.67920 |
| H | -0.30780 | -64.05600 | -17.53980 | C | -1.84660 | -59.79730 | -20.85360 | H | -0.04920 | -59.93090 | -20.37520 |
| H | -0.63790 | -63.70860 | -19.18920 | C | -0.59230 | -60.38980 | -20.78270 | H | 2.07370 | -59.34870 | -23.22460 |
| N | -0.03090 | -63.56970 | -18.38780 | C | 1.44400 | -62.54030 | -24.07650 | H | 1.28510 | -58.97650 | -25.55050 |
| 50 | | | | C | 1.05890 | -62.33920 | -25.39770 | H | -0.24210 | -61.57920 | -27.41840 |
| (R)-6-aF | | | | C | 0.80170 | -61.04120 | -25.86680 | H | 3.56240 | -61.25030 | -17.84340 |
| C | -0.35040 | -63.11200 | -25.99290 | N | 1.89480 | -61.66350 | -21.83260 | N | 3.66240 | -61.04320 | -18.83300 |
| C | 0.37900 | -62.04060 | -25.22420 | N | -4.27660 | -59.98680 | -21.23820 | H | 4.46880 | -60.50750 | -19.14520 |
| C | 0.86550 | -62.32760 | -23.94050 | O | 0.27900 | -58.39010 | -26.68650 | 50 | | | |
| C | 1.50770 | -61.35180 | -23.18280 | O | 0.72990 | -57.57900 | -24.64370 | (R)-9-sX | | | |
| C | 2.94570 | -61.11650 | -21.17460 | O | 3.86370 | -60.44350 | -21.66200 | C | 0.42320 | -58.64990 | -25.47560 |
| C | 2.79890 | -61.47550 | -19.69170 | O | 3.68080 | -61.16760 | -18.86480 | C | 0.78140 | -60.02970 | -24.99830 |
| C | 1.57820 | -62.20270 | -19.56820 | O | -5.24550 | -60.70670 | -21.50350 | C | 1.24120 | -60.21420 | -23.68830 |
| C | 0.96860 | -62.62420 | -18.37670 | O | -4.37030 | -58.77230 | -21.03030 | C | 1.55490 | -61.48420 | -23.21530 |
| C | 1.21160 | -61.88070 | -17.11390 | O | 0.42630 | -60.89490 | -27.16380 | C | 2.93630 | -61.11330 | -21.17470 |
| C | 1.20040 | -60.47370 | -17.13780 | H | 1.39940 | -59.30930 | -22.99610 | C | 2.75670 | -61.46250 | -19.71850 |
| C | 1.41000 | -59.74660 | -15.96810 | H | 1.07080 | -60.05140 | -17.89470 | C | 1.60000 | -62.20840 | -19.57970 |
| C | 1.64350 | -60.41720 | -14.76070 | H | 1.52390 | -58.91950 | -15.73920 | C | 1.01790 | -62.63740 | -18.33500 |
| C | 1.64920 | -61.81630 | -14.72780 | H | 2.03380 | -60.26900 | -13.70870 | C | -0.27120 | -63.40920 | -18.38970 |

S116

| | | | | | | | | | | | |
|---|---|---|---|---|---|---|---|---|---|---|---|
| C | -0.37170 | -64.61770 | -19.09740 | H | 0.88320 | -63.53800 | -23.48560 | C | 2.77360 | -61.48920 | -19.72320 |
| C | -1.59790 | -65.28330 | -19.17960 | H | 3.56120 | -63.74850 | -18.13240 | C | 1.59920 | -62.19100 | -19.57940 |
| C | -2.73550 | -64.73390 | -18.57880 | H | 4.73850 | -64.24990 | -16.00380 | C | 0.95840 | -62.77550 | -18.40870 |
| C | -2.63820 | -63.53210 | -17.86740 | H | 3.62020 | -63.79450 | -13.82470 | C | 1.03930 | -62.07300 | -17.09180 |
| C | -1.40810 | -62.87890 | -17.75950 | H | 1.31300 | -62.85940 | -13.78150 | C | 1.15100 | -60.67250 | -17.02350 |
| C | 0.95720 | -62.39190 | -20.94610 | H | 0.11530 | -62.40900 | -15.91960 | C | 1.11570 | -60.02210 | -15.78850 |
| C | -0.45750 | -61.83490 | -21.08240 | H | 0.90930 | -63.44100 | -21.20580 | C | 0.97770 | -60.76510 | -14.61070 |
| C | -1.46660 | -62.61740 | -21.65640 | H | -1.37400 | -63.61830 | -21.68290 | C | 0.86540 | -62.16100 | -14.67090 |
| C | -2.76040 | -62.11840 | -21.79800 | H | -3.61470 | -62.59500 | -22.02050 | C | 0.88690 | -62.81150 | -15.90500 |
| C | -3.02170 | -60.81780 | -21.36250 | H | -2.03390 | -58.75840 | -20.84070 | C | 0.97950 | -62.41650 | -20.95000 |
| C | -2.02860 | -60.00750 | -20.79990 | H | 0.20140 | -59.78470 | -20.50790 | C | -0.46790 | -61.95380 | -21.06520 |
| C | -0.74610 | -60.52730 | -20.66380 | H | 2.20320 | -59.43510 | -23.30850 | C | -1.48300 | -62.87980 | -21.33430 |
| C | 1.40740 | -62.60280 | -24.05010 | H | 1.53910 | -59.13270 | -25.68110 | C | -2.80700 | -62.46310 | -21.47430 |
| C | 0.94450 | -62.44240 | -25.35180 | H | 0.03680 | -61.77470 | -27.51820 | C | -3.09340 | -61.10260 | -21.34410 |
| C | 0.62440 | -61.16310 | -25.83360 | H | 3.56840 | -61.00850 | -17.85580 | C | -2.09970 | -60.15570 | -21.06960 |
| N | 1.88640 | -61.66560 | -21.83690 | N | 3.63190 | -60.90660 | -18.86150 | C | -0.78720 | -60.59350 | -20.92810 |
| N | -4.38040 | -60.28500 | -21.49380 | H | 4.39020 | -60.33780 | -19.22990 | C | 1.67660 | -60.10820 | -23.75830 |
| O | -0.03120 | -58.54700 | -26.67250 | 50 | | | | C | 1.21330 | -59.84660 | -25.04210 |
| O | 0.57680 | -57.67050 | -24.69730 | (R)-9-aX' | | | | C | 0.55250 | -60.84240 | -25.77820 |
| O | 3.86690 | -60.45060 | -21.64830 | C | 1.16750 | -59.04800 | -25.97420 | N | 1.86450 | -61.63960 | -21.84040 |
| O | 1.52490 | -62.32900 | -17.22970 | C | 1.27650 | -60.37830 | -25.28290 | N | -4.47800 | -60.64880 | -21.50860 |
| O | -5.25140 | -61.00890 | -21.98900 | C | 1.57620 | -60.43270 | -23.91340 | O | -0.81600 | -62.91240 | -27.12470 |
| O | -4.60070 | -59.13380 | -21.10060 | C | 1.67470 | -61.65430 | -23.25370 | O | -0.43860 | -64.36360 | -25.45470 |
| O | 0.15980 | -61.05850 | -27.10530 | C | 2.92460 | -61.10540 | -21.17350 | O | 3.87100 | -60.46260 | -21.64240 |
| H | 1.32540 | -59.35250 | -23.03470 | C | 2.73370 | -61.43000 | -19.71120 | O | 0.27390 | -63.81310 | -18.52520 |
| H | 0.50570 | -65.04050 | -19.57970 | C | 1.61540 | -62.22570 | -19.58650 | O | -5.34910 | -61.49190 | -21.74930 |
| H | -1.66870 | -66.22070 | -19.72520 | C | 1.04380 | -62.93660 | -18.45770 | O | -4.71640 | -59.44060 | -21.40080 |
| H | -3.69450 | -65.23850 | -18.66510 | C | 1.90940 | -63.38540 | -17.32290 | O | 0.08740 | -60.52130 | -27.01460 |
| H | -3.52150 | -63.10160 | -17.40300 | C | 3.22090 | -63.83510 | -17.55390 | H | 0.73360 | -63.37730 | -23.52360 |
| H | -1.33230 | -61.94070 | -17.21630 | C | 3.96630 | -64.39070 | -16.51160 | H | 1.23290 | -60.08960 | -17.93530 |
| H | 0.95170 | -63.44520 | -21.25210 | C | 3.41360 | -64.48540 | -15.22970 | H | 1.18560 | -58.93830 | -15.74680 |
| H | -1.24820 | -63.62780 | -21.98800 | C | 2.10880 | -64.03120 | -14.99130 | H | 0.94930 | -60.25820 | -13.64950 |
| H | -3.54810 | -62.72240 | -22.23410 | C | 1.35520 | -63.49500 | -16.03610 | H | 0.75310 | -62.73710 | -13.75650 |
| H | -2.26100 | -58.99950 | -20.47570 | C | 0.93890 | -62.34880 | -20.93760 | H | 0.78100 | -63.89080 | -15.96250 |
| H | 0.03310 | -59.91240 | -20.22200 | C | -0.41120 | -61.64610 | -21.06600 | H | 1.03830 | -63.47980 | -21.21740 |
| H | 1.64110 | -63.59620 | -23.67570 | C | -1.50760 | -62.32080 | -21.61560 | H | -1.24130 | -63.93320 | -21.44480 |
| H | 0.81510 | -63.29860 | -26.00820 | C | -2.72170 | -61.66290 | -21.81170 | H | -3.59700 | -63.17350 | -21.69070 |
| H | -0.02800 | -60.08400 | -27.25630 | C | -2.81810 | -60.31810 | -21.44770 | H | -2.35230 | -59.10560 | -20.97670 |
| H | 3.54530 | -61.23730 | -17.83490 | C | -1.73870 | -59.61990 | -20.89410 | H | -0.00420 | -59.86800 | -20.72540 |
| N | 3.65090 | -61.03330 | -18.82480 | C | -0.53700 | -60.29540 | -20.70890 | H | 2.14990 | -59.31920 | -23.18440 |
| H | 4.45260 | -60.48700 | -19.13160 | C | 1.46580 | -62.85110 | -23.95960 | H | 1.33120 | -58.86010 | -25.48220 |
| 50 | | | | C | 1.15410 | -62.81760 | -25.31350 | H | -0.36790 | -61.34390 | -27.36410 |
| (R)-9-aF' | | | | C | 1.05640 | -61.58840 | -25.98440 | H | 3.76600 | -61.40240 | -17.89950 |
| C | -0.01230 | -63.55010 | -26.00840 | N | 1.90930 | -61.70700 | -21.84670 | N | 3.75860 | -61.13750 | -18.87790 |
| C | 0.62810 | -62.40320 | -25.27930 | N | -4.08660 | -59.61390 | -21.65610 | H | 4.55750 | -60.61980 | -19.23410 |
| C | 1.03490 | -62.56760 | -23.94800 | O | 0.85690 | -59.06180 | -27.22050 | 50 | | | |
| C | 1.59930 | -61.51260 | -23.23430 | O | 1.38040 | -57.99280 | -25.31930 | (R)-9-aX | | | |
| C | 2.92510 | -61.10800 | -21.17470 | O | 3.85860 | -60.44300 | -21.64150 | C | 0.73540 | -58.54650 | -25.51690 |
| C | 2.73820 | -61.42720 | -19.71380 | O | -0.15860 | -63.27330 | -18.47300 | C | 0.94570 | -59.94920 | -25.01950 |
| C | 1.61040 | -62.21240 | -19.58170 | O | -5.02140 | -60.22740 | -22.18480 | C | 1.36950 | -60.16500 | -23.70190 |
| C | 1.00260 | -62.84610 | -18.43200 | O | -4.17020 | -58.43420 | -21.29570 | C | 1.54920 | -61.45560 | -23.21370 |
| C | 1.77120 | -63.02450 | -17.15920 | O | 0.74680 | -61.60760 | -27.30660 | C | 2.93310 | -61.11900 | -21.17420 |
| C | 3.07720 | -63.54160 | -17.18060 | H | 1.71640 | -59.50410 | -23.37060 | C | 2.77440 | -61.49270 | -19.72060 |
| C | 3.73770 | -63.82630 | -15.98130 | H | 3.64220 | -63.78690 | -18.55350 | C | 1.60300 | -62.20350 | -19.58120 |
| C | 3.10500 | -63.57570 | -14.75700 | H | 4.97070 | -64.75950 | -16.70330 | C | 0.97850 | -62.84870 | -18.43570 |
| C | 1.80460 | -63.05240 | -14.73200 | H | 3.99380 | -64.91970 | -14.42000 | C | 1.12220 | -62.26700 | -17.06760 |
| C | 1.13290 | -62.79150 | -15.92890 | H | 1.67840 | -64.10800 | -13.99640 | C | 1.25020 | -60.87970 | -16.87800 |
| C | 0.95400 | -62.37690 | -20.94400 | H | 0.33330 | -63.16500 | -15.86830 | C | 1.26630 | -60.34510 | -15.58770 |
| C | -0.44000 | -61.77050 | -21.07150 | H | 0.82100 | -63.40450 | -21.21050 | C | 1.16740 | -61.19220 | -14.47820 |
| C | -1.51710 | -62.55750 | -21.49620 | H | -1.41570 | -63.36640 | -21.89570 | C | 1.04270 | -62.57630 | -14.66030 |
| C | -2.77760 | -61.99320 | -21.68540 | H | -3.57370 | -62.17720 | -22.24130 | C | 1.01030 | -63.11060 | -15.94830 |
| C | -2.93880 | -60.62600 | -21.44440 | H | -1.84100 | -58.57510 | -20.62360 | C | 0.96470 | -62.39160 | -20.94630 |
| C | -1.88030 | -59.81710 | -21.01640 | H | 0.31070 | -59.76260 | -20.28740 | C | -0.45010 | -61.83200 | -21.05260 |
| C | -0.63180 | -60.40200 | -20.83180 | H | 1.54410 | -63.80690 | -23.44850 | C | -1.51590 | -62.66690 | -21.40730 |
| C | 1.78140 | -60.26670 | -23.86220 | H | 0.98430 | -63.73510 | -25.87000 | C | -2.80770 | -62.15530 | -21.53890 |
| C | 1.40600 | -60.09270 | -25.18850 | H | 0.71970 | -60.64810 | -27.60120 | C | -3.00990 | -60.79240 | -21.31160 |
| C | 0.81910 | -61.14790 | -25.90450 | H | 3.52020 | -61.01830 | -17.83630 | C | -1.96280 | -59.93420 | -20.95510 |
| N | 1.89170 | -61.68000 | -21.84840 | N | 3.61670 | -60.91810 | -18.84020 | C | -0.68440 | -60.46610 | -20.82590 |
| N | -4.25430 | -60.01640 | -21.65350 | H | 4.36590 | -60.32310 | -19.18480 | C | 1.29210 | -62.56200 | -24.03880 |
| O | -0.39660 | -63.32980 | -27.21420 | 50 | | | | C | 0.86030 | -62.36900 | -25.34640 |
| O | -0.14720 | -64.65770 | -25.42350 | (R)-9-aF | | | | C | 0.68500 | -61.06940 | -25.84680 |
| O | 3.86580 | -60.45250 | -21.63990 | C | -0.33560 | -63.22030 | -25.97250 | N | 1.88130 | -61.66690 | -21.84060 |
| O | -0.15850 | -63.30260 | -18.50100 | C | 0.38140 | -62.13170 | -25.22230 | N | -4.36060 | -60.24000 | -21.45720 |
| O | -5.18330 | -60.73720 | -22.03560 | C | 0.87730 | -62.38490 | -23.93640 | O | 0.33920 | -58.41130 | -26.73040 |
| O | -4.38230 | -58.80520 | -21.44050 | C | 1.50990 | -61.38730 | -23.19780 | O | 0.94850 | -57.57850 | -24.73860 |
| O | 0.44430 | -60.91720 | -27.18980 | C | 2.93260 | -61.12010 | -21.17350 | O | 3.86470 | -60.45260 | -21.64350 |

| | | | |
|---|---:|---:|---:|
| O | 0.25770 | -63.85430 | -18.60790 |
| O | -5.27650 | -61.00270 | -21.78480 |
| O | -4.52690 | -59.03350 | -21.24730 |
| O | 0.25690 | -60.93150 | -27.12830 |
| H | 1.53620 | -59.30790 | -23.05840 |
| H | 1.30240 | -60.21680 | -17.73650 |
| H | 1.34510 | -59.27020 | -15.44950 |
| H | 1.17950 | -60.77540 | -13.47440 |
| H | 0.96200 | -63.23360 | -13.79880 |
| H | 0.89230 | -64.17950 | -16.10090 |
| H | 0.95340 | -63.45360 | -21.22140 |
| H | -1.33930 | -63.72340 | -21.58860 |
| H | -3.63770 | -62.79500 | -21.81780 |
| H | -2.14930 | -58.87900 | -20.78990 |
| H | 0.13820 | -59.80870 | -20.55710 |
| H | 1.42200 | -63.57010 | -23.65410 |
| H | 0.64930 | -63.21460 | -25.99540 |
| H | 0.18290 | -59.94410 | -27.29500 |
| H | 3.77840 | -61.40590 | -17.90310 |
| N | 3.76170 | -61.13940 | -18.88080 |
| H | 4.55890 | -60.62180 | -19.24220 |

# VI. Supplementary References

1. Neves, J. F. *et al.* Backbone chemical shift assignments of human 14-3-3σ. **13**, 103-107, doi:10.1007/s12104-018-9860-1 (2019).
2. Burkhardt, A. *et al.* Status of the crystallography beamlines at PETRA III. *The European Physical Journal Plus* **131**, 56, doi:10.1140/epjp/i2016-16056-0 (2016).
3. Meents, A. *et al. Development of an in-vacuum x-ray microscope with cryogenic sample cooling for beamline P11 at PETRA III*. Vol. 8851 OPO (SPIE, 2013).
4. Winter, G. xia2: an expert system for macromolecular crystallography data reduction. *Journal of Applied Crystallography* **43**, 186-190, doi:doi:10.1107/S0021889809045701 (2010).
5. McCoy, A. J. *et al.* Phaser crystallographic software. *Journal of Applied Crystallography* **40**, 658-674, doi:doi:10.1107/S0021889807021206 (2007).
6. Adams, P. D. *et al.* PHENIX: a comprehensive Python-based system for macromolecular structure solution. *Acta Crystallographica Section D* **66**, 213-221, doi:doi:10.1107/S0907444909052925 (2010).
7. Emsley, P. & Cowtan, K. Coot: model-building tools for molecular graphics. *Acta Crystallographica Section D* **60**, 2126-2132, doi:doi:10.1107/S0907444904019158 (2004).
8. Chen, V. B. *et al.* MolProbity: all-atom structure validation for macromolecular crystallography. *Acta Crystallographica Section D* **66**, 12-21, doi:doi:10.1107/S0907444909042073 (2010).
9. Karplus, P. A. & Diederichs, K. Linking Crystallographic Model and Data Quality. *Science* **336**, 1030-1033, doi:10.1126/science.1218231 (2012).
10. Rose, R. *et al.* Identification and Structure of Small-Molecule Stabilizers of 14–3–3 Protein–Protein Interactions. **49**, 4129-4132, doi:10.1002/anie.200907203 (2010).
11. Richter, A., Rose, R., Hedberg, C., Waldmann, H. & Ottmann, C. An Optimised Small-Molecule Stabiliser of the 14-3-3–PMA2 Protein–Protein Interaction. *Chemistry – A European Journal* **18**, 6520-6527, doi:10.1002/chem.201103761 (2012).
12. Small-Molecule Drug Discovery Suite v. 2019-2 (Schrödinger, LLC, New York, NY, 2019).
13. Chang, G., Guida, W. C. & Still, W. C. An internal-coordinate Monte Carlo method for searching conformational space. *Journal of the American Chemical Society* **111**, 4379-4386, doi:10.1021/ja00194a035 (1989).
14. Roos, K. *et al.* OPLS3e: Extending Force Field Coverage for Drug-Like Small Molecules. *Journal of Chemical Theory and Computation* **15**, 1863-1874, doi:10.1021/acs.jctc.8b01026 (2019).
15. Still, W. C., Tempczyk, A., Hawley, R. C. & Hendrickson, T. Semianalytical treatment of solvation for molecular mechanics and dynamics. *Journal of the American Chemical Society* **112**, 6127-6129, doi:10.1021/ja00172a038 (1990).
16. Bochevarov, A. D. *et al.* Jaguar: A high-performance quantum chemistry software program with strengths in life and materials sciences. *International Journal of Quantum Chemistry* **113**, 2110-2142, doi:10.1002/qua.24481 (2013).
17. Grimme, S., Antony, J., Ehrlich, S. & Krieg, H. A consistent and accurate ab initio parametrization of density functional dispersion correction (DFT-D) for the 94 elements H-Pu. *The Journal of Chemical Physics* **132**, 154104, doi:10.1063/1.3382344 (2010).
18. Becke, A. D. Density-functional thermochemistry. III. The role of exact exchange. *The Journal of Chemical Physics* **98**, 5648-5652, doi:10.1063/1.464913 (1993).
19. Marten, B. *et al.* New Model for Calculation of Solvation Free Energies:  Correction of Self-Consistent Reaction Field Continuum Dielectric Theory for Short-Range Hydrogen-Bonding Effects. *The Journal of Physical Chemistry* **100**, 11775-11788, doi:10.1021/jp953087x (1996).
20. Kelly, C. P., Cramer, C. J. & Truhlar, D. G. SM6:  A Density Functional Theory Continuum Solvation Model for Calculating Aqueous Solvation Free Energies of Neutrals, Ions, and Solute−Water Clusters. *Journal of Chemical Theory and Computation* **1**, 1133-1152, doi:10.1021/ct050164b (2005).
21. Zhao, Y. & Truhlar, D. G. The M06 suite of density functionals for main group thermochemistry, thermochemical kinetics, noncovalent interactions, excited states, and transition elements: two new functionals and systematic testing of four M06-class functionals and 12 other functionals. *Theoretical Chemistry Accounts* **120**, 215-241, doi:10.1007/s00214-007-0310-x (2008).